\documentclass[12pt,a4paper]{article}
\pdfoutput=1
\usepackage[pdftex]{graphics}
\usepackage{jheppub}
\usepackage{amsfonts,amsmath,amssymb}
\usepackage{mathrsfs}
\usepackage{graphicx}
\graphicspath{{./Figures/}}
\usepackage{slashed}
\usepackage{dsfont}
\usepackage{mathtools}
\usepackage{bm}
\usepackage{tikz}
\usepackage{wasysym} % smiley :)
\usepackage{tikz}
\usetikzlibrary{matrix}
\usetikzlibrary{shapes.arrows,chains}
\usepackage{multirow}
\usepackage{enumitem}
\usepackage{epsdice}
 
\makeatletter
\def\@fpheader{\relax}
\makeatother

\def\CA{{\cal A}}

\def\CF{{\cal F}}

\def\CH{{\cal H}}
\def\CI{{\cal I}}
\def\CJ{{\cal J}}
\def\CK{{\cal K}}
\def\CL{{\cal L}}
\def\CM{{\cal M}}
\def\CN{{\cal N}}

\def\CT{{\cal T}}
\def\CX{{\cal X}}
\def\a{\alpha}
\def\b{\beta}
\def\g{\gamma}
\def\d{\delta}
\def\e{\epsilon}
\def\l{\lambda}

\def\vth{\vartheta}
\def\m{\mu}

\def\vph{\varphi}

\def\G{\Gamma}

\def\vare{\varepsilon}
\def\dd{d}

\newcommand{\scA}{\ensuremath{\mathcal{A}}}
\newcommand{\scH}{\ensuremath{\mathcal{H}}}

\newcommand{\sfY}{\mathsf{Y}}
\newcommand{\sfy}{\ensuremath{\mathsf{y}}}
\newcommand{\sfu}{\ensuremath{\mathsf{u}}}
\newcommand{\sfp}{\ensuremath{\mathsf{p}}}

\newcommand{\frakL}{\mathfrak{L}}
\newcommand{\frakX}{\mathfrak{X}}
\newcommand{\frakM}{\mathfrak{M}}
\newcommand{\frakP}{\mathfrak{P}}
\newcommand{\bfrakM}{\mathfrak{M}'}

\newcommand{\be}{\begin{equation}}
\newcommand{\ee}{\end{equation}}

\newcommand{\hatM}{\hat{M}}
\newcommand{\bhbar}{\hbar}

\newcommand{\bareta}{\bar{\eta}}

\def\Tr{{\rm Tr}}

\def\half{\frac{1}{2}}

\newcommand{\nn}{\nonumber}

\title{Aspects of Defects in 3d--3d Correspondence}

\preprint{IPMU15-0176}

\author[\epsdice{1}]{Dongmin Gang,}
\author[\epsdice{2}, \epsdice{3}]{Nakwoo Kim,}
\author[\epsdice{1}, \epsdice{4}]{Mauricio Romo,}
\author[\epsdice{1}, \epsdice{4}]{\\ and Masahito Yamazaki}

\affiliation[\epsdice{1}]{Kavli Institute for the Physics and Mathematics of the Universe (WPI),\\
University of Tokyo, Chiba 277-8583, Japan}
\affiliation[\epsdice{2}]{Department of Physics and Research Institute of Basic Science,\\
Kyung Hee University, Seoul 02447, Korea}
\affiliation[\epsdice{3}]{School of Physics, Korea Institute for Advanced Study, Seoul 02455, Korea}
\affiliation[\epsdice{4}]{School of Natural Sciences, Institute for Advanced Study, Princeton, NJ 08540, USA}

\abstract{In this paper we study supersymmetric co-dimension 2 and 4 defects in the compactification of the 6d $(2,0)$ theory of type $A_{N-1}$ on a 3-manifold $M$. The so-called 3d--3d correspondence is a relation between complexified Chern-Simons theory (with gauge group $SL(N, \mathbb{C})$) on $M$ and a 3d $\mathcal{N}=2$ theory $T_{N}[M]$. We establish a dictionary for this correspondence in the presence of supersymmetric defects, which are knots/links inside the 3-manifold. Our study employs a number of different methods: state-integral models for complex Chern-Simons theory, cluster algebra techniques, domain wall theory $T[SU(N)]$, 5d $\mathcal{N}=2$ SYM, and also supergravity analysis through holography. These methods are complementary and we find agreement between them. In some cases the results lead to highly non-trivial predictions on the partition function. Our discussion includes a general expression for the cluster partition function, in particular for non-maximal punctures and $N>2$. We also highlight the non-Abelian description of the 3d $\mathcal{N}=2$ $T_N[M]$ theory with defect included, as well as its Higgsing prescription and the resulting `refinement' in complex CS theory. This paper is a companion to our shorter paper \cite{Gang:2015bwa}, which summarizes our main results.
}

\begin{document}
\maketitle

%%%%%%%%%%%%%%%%%%%%%%%%%%%%%%%%%%%%%%%%%%%%%
%%%%%%%%%%%%%              body begins here               %%%%%%%%%%%%%%
%%%%%%%%%%%%%%%%%%%%%%%%%%%%%%%%%%%%%%%%%%%%%

%%%%%%%%%%%%%%%%%%%%%%%%%%%%%%%%%%%%%%%%%%%%%
\section{Introduction and Outline} \label{sec : introduction}
%%%%%%%%%%%%%%%%%%%%%%%%%%%%%%%

We have learned over the past few years that compactification of M5-branes on various manifolds
generates a class of lower-dimensional supersymmetric field theories labeled by the geometrical data.
This has led to fruitful interplay between the physics of supersymmetric gauge theories (and in 
particular their non-perturbative dualities) and the geometry of the compactification manifolds (see {\it e.g.}\,\cite{Teschner:2014oja} and references therein).

When we choose to compactify on a 3-manifold $M$, we have the correspondence 
between complex Chern-Simons (CS) theory on $M$ and 
3d $\mathcal{N}=2$ theory $T[M]$. This has been worked out in a number of papers \cite{Terashima:2011qi,Terashima:2011xe,Dimofte:2010tz,Dimofte:2011jd,Dimofte:2011ju,Cecotti:2011iy}, and the
appearance of complex Chern-Simons theory has recently been derived in \cite{Cordova:2013cea,Lee:2013ida} (see also \cite{Yagi:2013fda}).

In this paper we include supersymmetric defects to this story, inherited from co-dimension $2$ and co-dimension $4$
defects in the 6d $(2,0)$ theory. This is particularly interesting for us, since on the one hand
such generalizations allow us to look more closely into the dictionary of the 3d--3d correspondence, 
for example on the relation between Abelian versus non-Abelian description of the 
3d $\mathcal{N}=2$ theory $T[M]$. On the other hand, the partition functions of our theories with defects can be computed by a number of different methods,
hence our setup is ideal for developing computational tools and doing consistency checks between them.

In the rest of this introduction we provide more detailed outline of this paper.

%%%%%%%%%%%%%%%%%%%%%%%%%%%%%%%%%%%%%%%%%%%%%%%%%%%
\subsection{M5-branes on 3-manifolds}

Let us consider $N>1$ M5-branes, whose low energy world-volume theory is the 6d $A_{N-1}$ (2,0) theory.
We wrap the M5-branes on a closed 
3-manifold $\hat{M}$:
\begin{align}
&\textrm{ $N$ M5s on }\overbrace{\mathbb{R}^{1,2}}^{1,2,3}\; \times \overbrace{\hatM}^{3,4,5}  \label{(2,0) on hatM}\;.
\end{align} 
Since $\hat{M}$ is a curved manifold, 
we perform a partial topological twisting along 
$\hatM$, and  turn on an R-symmetry flux mixing the $SO(3)$ connection on $\hatM$ with
an $SO(3)$ current inside $SO(5)$ R-symmetry of 6d $(2,0)$ theory.
The resulting theory has four supercharges with the remaining $SO(2)$ R-symmetry.
Thus 
such a compactification generates a 3d $\mathcal{N}=2$ theory, which we denote by $T_N[\hatM]$.
The 3d--3d correspondence relates\footnote{This has generalizations to 
other gauge groups $G$, as is clear from the derivation of \cite{Cordova:2013cea,Lee:2013ida}.
The same comment applies to our discussion in Sec.~\ref{codim_4_as_Wilson}.
}
\begin{align}
\textrm{3d $\mathcal{N}=2$ theory $T_{N}[\hat{M}]$} \quad\Longleftrightarrow \quad
\textrm{$SL(N)$ CS theory on $\hat{M}$} \ .
\end{align}
We will comment on more precise versions of this relation momentarily.

%%%%%%%%%%%%%%%%%%%%%%%%%%%%%%%%%%%%%%%%%%%%%%%%%%%
\subsection{Supersymmetric Defects}

We would like to add defects to the system \eqref{(2,0) on hatM} now.
The defects will be described by M2 and M5-branes in M-theory.
In order to preserve supersymmetry, these defect M-branes should be either
co-dimension $2$ or co-dimension $4$ inside the original $N$ M5-branes which give rise to $(2,0)$ theory.
We can still have several choices as to how to split the 6 dimensions
into the $\mathbb{R}^{1,2}$ directions on which 3d $\mathcal{N}=2$ theory lives
and the 3-manifold directions.

Their configurations in the context of the 3d--3d correspondence will be given more concretely
 below in this subsection, and let us
emphasize that we can also consider the composite of different types of defects.
Note that  there exist other types of supersymmetric defects as well, e.g.
domain walls which will not be explored in this paper and we leave them for future work.

Our co-dimension 2 and co-dimension 4 defects will be discussed in more detail later in Sec.~\ref{sec.3d3d_defect}.
We here provide a summary of their properties.\footnote{In this paper, co-dimensions always refer to
co-dimensions inside the 6d theory. In 3d--3d correspondence, we have two `3d' directions, and we also consider compactification of 6d theory to 5d $\mathcal{N}=2$ SYM. In each of these cases the co-dimensions
in these (3d or 5d) spaces will be different from those in 6d.}\footnote{There are many discussions of supersymmetric defects in the compactifications of 6d $(2,0)$ theory of 5d $\mathcal{N}=2$ SYM. Our viewpoint of supersymmetric defects is somewhat close to that in \cite{Bullimore:2014upa} for the case of the $S^5$ partition functions of 5d $\mathcal{N}=2$ SYM.
}

\paragraph{Co-dimension $2$ Defects}

The brane configuration is 
\begin{align}
\begin{array}{cccc|ccc|ccc}
&  \multicolumn{3}{c}{\overbrace{ }^{\mathbb{R}^{1,2}}}    &  \multicolumn{3}{c}{\overbrace{ }^{\hatM}}   &  &  
\\
\textrm{$N$ M5: } & 0 & 1 & 2 & 3 & 4 & 5 &   &  &   
\\
\textrm{Defect M5: } & 0 & 1 & 2 & 3 & &  &   & 7 & 8 
\end{array}
%\;.
\label{codim_2_brane}
\end{align}  
For the 6d $A_{N-1}$ $(2,0)$ theory, the  co-dimension 2 defect is labelled  by  an embedding $\rho: SU(2)\to SU(N)$ or equivalently a partition $[n_1, \ldots, n_s]$ of $N$.
Let denote by $K$ the trajectory of the defect inside $\hatM$. 

Since the defect fills the whole  $\mathbb{R}^{1,2}$,  the effect of this defect is to replace the 3d $\mathcal{N}=2$ theory
$T_N[\hatM]$ by a new theory\footnote{The supercharges preserved by the defect M5 coincides with those preserved under topological twisting along $\hat{M}$.}, 
which we denote by $T_N[\hat{M}\backslash K, \rho]$. 
Geometrically, this is to replace a closed 3-manifold $\hatM$ by a knot/link complement,
which we denote by 
\begin{align}
M:=\hat{M}\backslash K \;.
\label{M_def}
\end{align}
In the $SL(N)$ CS theory, the defect will be a loop defect around the knot $K$.  
We propose that  the loop defect of type $\rho$ can be identified with monodromy  defect associated to Levi-subgroup $\mathbb{L}^{(\rho)}$ of $SL(N)$:
\begin{align}
\mathbb{L}^{(\rho)}:= S\left[\bigotimes_i  GL(n_i) \right]  \subset SL(N) \;, \label{Levi-subgroup}
\end{align}
and the generalization of the 3d--3d correspondence with this defect is
\begin{align}
\begin{split}
&\textrm{3d $\mathcal{N}=2$ theory $T_{N}[\hat{M} \backslash K,\rho]$} \quad\Longleftrightarrow 
\\
& \quad \textrm{$SL(N)$ CS theory on $\hat{M}$ with a monodromy defect of type $\rho$ around $K$\;.  }
\end{split}
\label{3d3d_codim_2}
\end{align}
Detailed description for the monodromy defect will be given in  sec. \ref{sec : co-dimension 2},
and we will give an explicit example of the $T_N[M, \rho]$ in Sec.~\ref{sec.Xe_rule}.

\paragraph{Co-dimension $4$ Defects}
Let us next consider co-dimension $4$ defects.
The relevant brane configuration is
\begin{align}
\begin{array}{cccc|ccc|ccccc}
&  \multicolumn{3}{c}{\overbrace{ }^{\mathbb{R}^{1,2}}}    &  \multicolumn{3}{c}{\overbrace{ }^{\hatM}}   &  &  &
\\
\textrm{$N$ M5: } & 0 & 1 & 2 & 3 & 4 & 5 &   &  &  &  &
\\
\textrm{Defect M2: } & 0 &  &  & 3 & &  & 6  &  &  & &
\\
\textrm{Defect M5: } & 0 &  &  & 3 & &  &   & 7 & 8 & 9 &\sharp
\end{array}
\label{codim_4_brane} 
\end{align}

\begin{figure}[htbp]
\begin{center}
   \includegraphics[width=.3\textwidth]{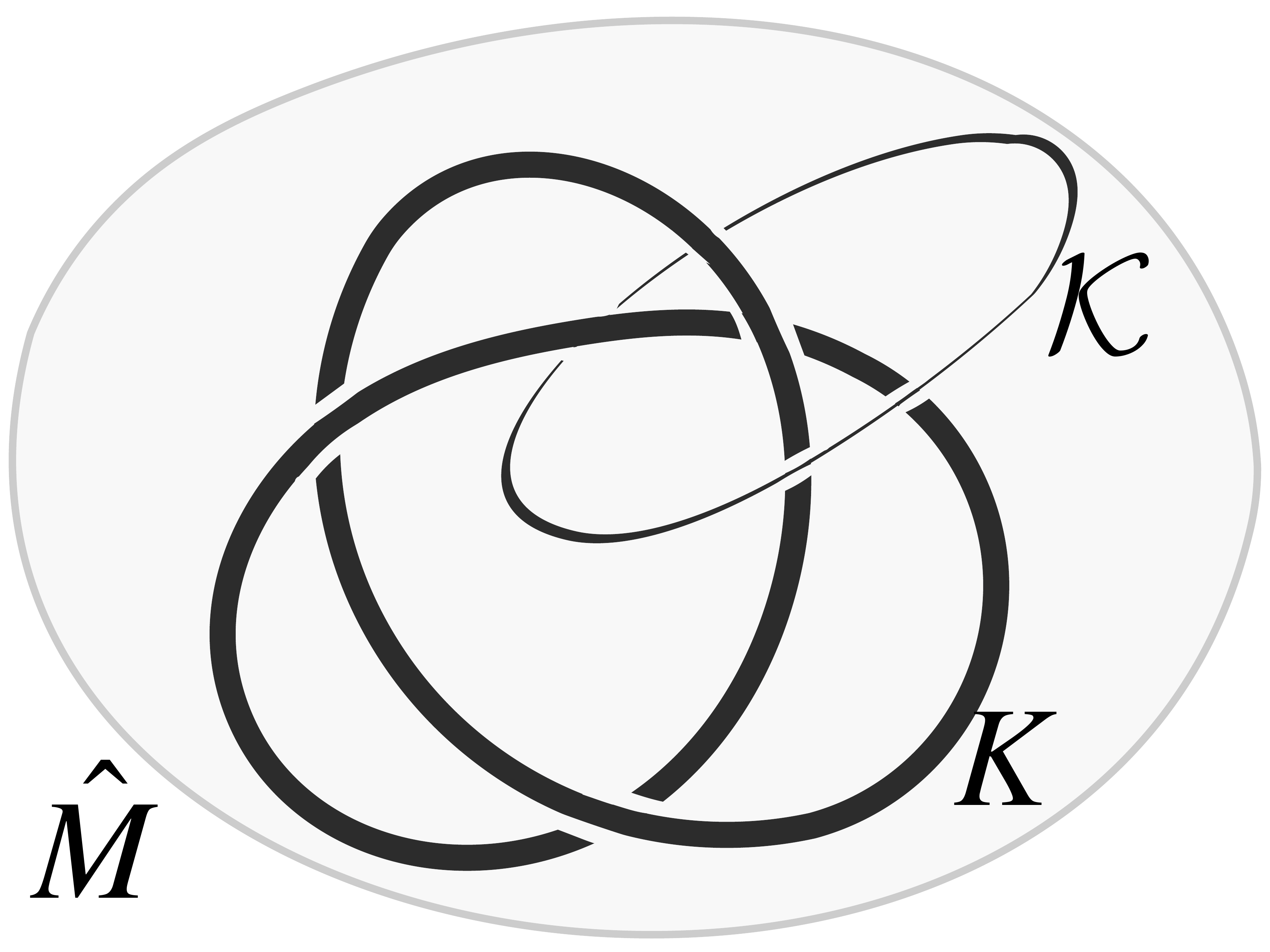}
   \end{center}
   \caption{Inside a closed 3-manifold $\hatM$, we in general simultaneously include a co-dimension $2$ defect along $K$, and then
   a co-dimension 4 defect along $\CK$. The two knots, $K$ and $\CK$, can be mutually knotted inside $\hatM$.}
    \label{fig:K_CK}
\end{figure}

The difference from our previous case in eq.~\eqref{codim_2_brane} is that 
we could have either an M2-brane or an M5-brane.
In both cases, the defect is a 1d line-like defect both in the 3d $\mathcal{N}=2$ theory 
as well as the $SL(N)$ Chern-Simons theory.  The defect is specified by a
finite-dimensional unitary representation $R$ of $SU(N)$,
and as we will see in Sec.~\ref{sec:sugra} in the large $N$ limit,
the difference between the M2-branes and M5-branes is accounted for the choice of the 
representation $R$. 
The generalization of the 3d--3d correspondence with this defect inserted is proposed to be
\begin{align}
&\textrm{Supersymmetric loop operator labeled by a representation $R$ and $\mathcal{K}$ in $T_{N}[\hat{M}]$ } \nonumber
\\
&\quad \Longleftrightarrow  \textrm{Wilson loop in representation $R$ along $\mathcal{K}$ in $SL(N)$ CS theory on $\hatM$ } \,,
\label{3d3d_codim_4}
\end{align}
where on the right hand side the $SU(N)$ representation $R$ is naturally complexified to a representation of $SL(N)$.
The correspondence can be made more concrete by putting the 3d $\CN=2$ theory on a curved background, for example $S^1\times S^2$ or $S^3/\mathbb{Z}_k$, while preserving certain rigid supersymmetries.
On those curved backgrounds, there are two supersymmetric cycles: considering these 3-manifolds as $S^1$ bundle over $S^2$, these cycles wrap the fiber $S^1$ located at the north/south poles of the base $S^2$. These two choices correspond to the choice of either holomorphic or anti-holomorphic Wilson loop in the CS theory. 
We also consider  the co-dimension 4 defect  in the presence of co-dimension 2 along $K$ in $\hat{M}$. In the case, the co-dimension 4 can be considered as  a knot $\mathcal{K}$ in the knot complement $M:=\hat{M}\backslash K$ (notice the difference between $K$ and $\mathcal{K}$, see Fig.~\ref{fig:K_CK}).

If we follow the proposals in \cite{Dimofte:2011jd,Dimofte:2011ju}, the theory $T_{N}[M]$ in general 
does not have a gauge group $SU(N)$. So it is not immediately obvious
why (a subset of) Wilson loops in $T_{N}[M]$ should be labeled by the representation $R$ of $SU(N)$.
We will see later that this fact indeed gives a non-trivial hint as to the theory $T_{N}[M]$, especially 
on the non-Abelian nature of the gauge group.

%%%%%%%%%%%%%%%%%%%%%%%%%%%%%%%%%%%%%%%%%
\subsection{Computational Methods}\label{sec.computational}

In order to better understand eq.~\eqref{3d3d_codim_2} and eq.~\eqref{3d3d_codim_4}, 
a useful quantity to compute is the partition function of  the theories in the presence of  defects\footnote{
It should be noted, however, most of the ingredients, for example the construction of the theory $T_N[M, \rho]$ for a given $\rho$, works for directly at the Lagrangian level, and in the end does not necessarily need the analysis of the specific partition functions.}.

We will use a number of complementary computational methods, each of which has its own virtues and limitations.
Whenever more than one result is available, we will check consistencies between them, and in some cases 
such checks leads to new mathematical conjectures.

\paragraph{State Integral Model}

One method is to use an ideal triangulation of the 3-manifold,
and compute the partition function from the state integral construction of 3-manifolds.
In this formalism, only the case of special $\rho$ (the so-called `maximal' case, as we will explain)
has been considered in \cite{Hikami:2001en,Hikami:2006cv,Dimofte:2009yn,Dimofte:2011gm} for $N=2$, and \cite{Dimofte:2013iv} for $N>2$
(see also \cite{BFG,GGZ,2011arXiv1111.2828G,Garoufalidis:2013upa}),
where the latter is based on the `octahedral' decomposition of the ideal tetrahedron.
We also extend the existing construction of state-integral models to include co-dimension 4 defects.

\paragraph{Cluster Partition Function}

Another method is formulating our 3-manifold problem in terms of 
quivers and their mutations, and compute its partition function
(cluster partition function) in the formalism of \cite{Terashima:2013fg}.
In this formalism, a co-dimension 2 defect corresponds to a change of the quiver and its mutation sequence.
We will work out an example of the co-dimension 2 defect with  non-maximal $\rho$.
A co-dimension 4 defect, in contrast, corresponds to a generalization of the cluster partition function 
with Wilson line insertion. We also work out this generalization in this paper.
We will point out that that the ``mutation network'' of \cite{Terashima:2013fg} 
is a generalization of the octahedron decomposition.

\paragraph{Non-Abelian Description of $T_N[M, \rho]$}

In general, the only known descriptions of $T_N[M, \rho]$ is in terms of Abelian gauge groups,
which (as we will comment below) is insufficient for the full description of defects.
Fortunately, however, there are some known non-Abelian descriptions.
One description of the 3d $\mathcal{N}=2$ theory $T_N[M, \rho]$ is to describe it as a 
1/2 BPS boundary condition (or domain wall) 
for the 4d $\mathcal{N}=2$ theory $T_N[\Sigma]$ associated with a punctured Riemann surface $\Sigma$.
In particular, for the cases discussed in \cite{Terashima:2011qi,Terashima:2011xe},
the 3d $\mathcal{N}=2$ theory is already known to be constructed from 
$\mathcal{N}=2$ mass-deformation of the 3d $\mathcal{N}=4$ theory known as $T[SU(N)]$ (see \cite{Gaiotto:2008ak} for the properties of this theory).
We can then directly compute their partition functions by supersymmetric localization.

\paragraph{Defects in 5d $\mathcal{N}=2$ SYM}

When we wish to compute $S^3$ or $S^1\times S^2$ partition function 
of $T_N[M]$ theory, we can take advantage of the $S^1$ direction of the 3d geometry and reduce the 6d $(2,0)$ theory along this $S^1$.
The resulting theory is then 5d $\mathcal{N}=2$ SYM, and  the defects of the 6d theory is represented by defects of the 5d $\CN=2$ SYM, whose partition function (for co-dimension 4 defect) we can compute directly using supersymmetric localization, generalizing the results of \cite{Cordova:2013cea,Lee:2013ida}.
We also propose a Higgsing prescription for the co-dimension 2 defects in terms of $T[SU(N)]$ theory and its generalizations,
and comment on its implications to complex Chern-Simons theory.

\paragraph{Holographic Dual}

Finally, we can study the large $N$ limit of our systems using the dual supergravity solution.
We include the probe M2 and M5-branes, corresponding to the brane setup of eqs.~\eqref{codim_2_brane}
and \eqref{codim_4_brane}, to the M-theory background of \cite{Gang:2014qla,Gang:2014ema}.
This generates a number of conjectures in the large $N$ asymptotic of the partition functions.

%%%%%%%%%%%%%%%%%%%%%%%%%%%%%%%%%%%%%%%%
\bigskip

The organization of the rest of this paper is as follows.
In Sec.~\ref{sec.3d3d_defect} we describe the supersymmetric defects in 3d--3d correspondence in more detail.
We then go on to discuss each of the methods listed in Sec.~\ref{sec.computational}.
In each of the following sections, we start with summary on the background knowledge if necessary, and then
subsequently discuss co-dimension 2 and co-dimension 4 defects.
We shall in turn discuss state integral model (Sec.~\ref{sec.state_integral}), cluster partition function (Sec.~\ref{sec : cluster partition function}), 
$T[SU(N)]$ theory (Sec.~\ref{sec.TSUN}), 5d $\CN=2$ SYM (Sec.~\ref{sec.5dSYM}), holography (Sec.~\ref{sec:sugra}).
The final section (Sec.~\ref{sec.conclusion}) is devoted to summary and outlook. We 
also include several appendices containing technical materials.

%%%%%%%%%%%%%%%%%%%%%%%%%%%%%%%%%%%%%%%%%%
\section{3d--3d Correspondence with Defects}   \label{sec.3d3d_defect} 

Let us comment in more detail the supersymmetric co-dimension 2 and co-dimension 4 defects.

\subsection{Co-dimension 2 Defects }\label{sec : co-dimension 2}

Let us here describe in more detail the co-dimension 2 defects \eqref{codim_2_brane}.
As already stated in the introduction, these defects are labelled by an embedding \cite{Gaiotto:2008ak}
\begin{align}
\rho: SU(2) \rightarrow  SU(N)\;.
\label{rho}
\end{align}
We can also specify $\rho$ by the decomposition of the $N$-dimensional fundamental 
representation $[N]$ of $SU(N)$ into irreducible representations of $SU(2)$:
\begin{align}
[N] \longrightarrow  [n_1]\oplus [n_2]\oplus \ldots \oplus [n_s] \;,
\end{align} where $[n]$ denote the $n$-dimensional irreducible representation of $SU(2)$.
We assume $n_i>0$, and moreover without losing generality choose $n_{i}\ge  n_{i+1}$.
Since $\sum_{i=1}^s n_i=N$, 
\begin{align}
\rho= [n_1, n_2, \ldots, n_s]
\end{align}
is a partition of $N$, or a Young diagram with $N$-boxes.\footnote{In our convention the $i$-th row has $n_i$ boxes.}

The defect is called `maximal' (or `full') and `simple' when 
$\rho=[1]^{N}\!:=[1,1,\ldots, 1]$  and  $\rho=  [N-1, 1]$ respectively.  The number of defect M5-branes
is given by the length $s=:\ell(\rho)$, namely the number of columns of $\rho$.

One way to understand the appearance of $\rho$ is as follows (we will also provide below further explanation from the 
viewpoint of complex Chern-Simons theory). We can compactify M-theory to type IIB string theory, by compactifying along the direction $5$ and then T-dualize along the direction $3$. The brane configuration becomes
\begin{align}
\begin{array}{cccc|c|c|cccc}
\textrm{$N$ D3: } & 0 & 1 & 2 &  & 4 & &   &  &   
\\
\textrm{Defect NS5: } & 0 & 1 & 2 & 3 & &  &   & 7 & 8 
\end{array}
%\;.
\label{codim_2_brane_IIA}
\end{align} 
and we have 4d $\mathcal{N}=4$ $SU(N)$ SYM on the $N$ D3-branes.
The 1/2 BPS boundary conditions of 
4d $\mathcal{N}=4$ SYM was studied in \cite{Gaiotto:2008sa,Gaiotto:2008ak},
which classified such boundary conditions under some mild assumptions.
The conclusion there is that the boundary theory on the defect described by the NS5-branes is then a 3d $\mathcal{N}=4$ theory called $T_{\rho}[SU(N)]$.\footnote{In S-dual, this is mapped to a singular boundary condition described by the Nahm pole, determined by the embedding in eq.~\eqref{rho}.}
This theory has $SU(N)\times H_\rho$ global symmetry, 
where $H_{\rho}$ is
defined as the commutant of $\rho\left(SU(2)\right)$
inside $SU(N)$, {\it i.e.}\ $\left[\rho \left( SU(2)\right) , H_\rho \right]$=0. 
To be more explicit,
\begin{align}
H_{\rho} = S\left[ \bigotimes_\a U(l_\a)\right]\;, 
\label{H_rho}
\end{align}
where $l_{\alpha=1,2,\ldots}$ denote the number of times that the number $j$ appears in the partition $\rho$, $\sum_{\a} \a l_\a = N$ (namely $\rho=[N^{l_N}, \ldots, 2^{l_2}, 1^{l_1}]$), and `$S$' on the right hand side means to mod out by the overall $U(1)$ factor.
The theory $T_{\rho}[SU(N)]$ 
couples naturally to the bulk 5d $\CN=2$ SYM
by gauging the $SU(N)$ global symmetry. In brane realizations, $T_{\rho}[SU(N)]$
describes a boundary condition where we have $s$ separate NS5-branes and $n_i$ of the $N$ D4-branes end on the $i$-th NS5-brane.

To study 3d--3d correspondence of the system, we put the 6d $(2,0)$ theory on a closed 3-manifold $\hat{M}$ (along $3, 4, 5$-directions) with a partial topological twisting. The defect M5-branes are located along an knot $K$ in the 3-manifold.  In eleven dimensional M-theory, the topological twisting  is realized as 
\begin{align}
\begin{split}
&\textrm{$N$ M5s   }  \quad\;\;\; :\;\;\;\mathbb{R}^{1,2}\times \hat{ M }\; \quad \quad \subset \quad  \; \mathbb{R}^{1,2}\times T^* \hat{ M }\times \mathbb{R}^2\;,
\\
&\textrm{Defect M5}\;\; :\;\;\; \mathbb{R}^{1,2} \times N^*K  \quad \subset \quad  \; \mathbb{R}^{1,2}\times T^* \hat{ M} \times \mathbb{R}^2\;. 
\end{split}
\label{brane_RTMR}
\end{align}
Here $T^*\hat{M}$ denotes a cotangent bundle over $\hat{M}$ whose fiber  is  along $6,7,8$-directions. The defect M5-brane is located along a knot $K$ along 3rd direction on $\hat{M}$ and $N^* K$ denote the co-normal bundle of the knot in $T^*\hat{M}$.  The effective world-volume theory on $N$ M5-branes is described by 3d $\mathcal{N}=2$ theory, which (as explained in introduction) will be denoted as
\begin{align}
T_{N}[\hat{M}\backslash K,\rho]\; .
\end{align}
When $\rho$ is maximal, we simply denote the theory as  $T_N[\hat{M}\backslash K]$. 
The 3d theory has flavor symmetry $H_{\rho}\subset SU(N)$.

The  3d--3d correspondence with defect $\rho$ was given in eq.~\eqref{3d3d_codim_2}.
Mostly in the literature only the case of $\rho=\textrm{maximal}$ has been studied so far,
with only a few exceptions ({\it e.g.}\,\cite{Terashima:2011qi} discuss the case of $\rho=\textrm{simple}$ for $N>2$,
as will be mentioned in Sec.~\ref{sec.TSUN}).
One of the goals of Sec.~\ref{sec : cluster partition function} is to generalize these works to non-maximal $\rho$.\footnote{One interesting aspect of the 3d--3d correspondence is that the most typical version of the 3d--3d correspondence (this includes almost all the papers on the topic) has $\rho=\textrm{maximal}$, hence it already includes the co-dimension $2$ defects. 
Recall that the co-dimension 2 defect does not break any supersymmetry, and we always have 3d $\mathcal{N}=2$ supersymmetry irrespective of the choice of $\rho$.
} 

There is one interesting aspect in eq.~\eqref{3d3d_codim_2}. Recall that we first start with the geometry $\hatM$, 
and we arrive at the geometry with $K$ removed, namely $M:=\hatM\backslash K$ as defined in 
eq.~\eqref{3d3d_codim_2};
in the end it looks like that partition function is determined solely by the data of $M$. It might then happen that the same geometry $M$
can be obtained by two different ambient manifolds $\hat{M}_1$ and $\hat{M}_2$. In fact,
there are in general infinitely many choices of ambient manifolds $\hat{M}$, related by Dehn surgeries.\footnote{Dehn surgeries can be described as follows.
Let us for example take $M$ to be a knot complement $S^3\backslash K$ in $S^3$.
The boundary of $M$ is a torus $T^2$.
We can close off the boundary of this geometry by gluing a solid torus (whose boundary is also a solid torus),
where the two boundary tori are glued together by an element of $SL(2, \mathbb{Z})$. This is the Dehn filling. 
Dehn surgeries relate different Dehn fillings by first drilling the tubular neighborhood of a knot inside a closed 3-manifold and then 
perform a Dehn filling.
The resulting manifolds have different topologies depending on the choice of the $SL(2, \mathbb{Z})$ element, and for example generically have different hyperbolic volumes.} As we will see later in Sec.~\ref{subsec.examples},
depending on the choice of the ambient manifold we need to change the choice of polarization on the boundary of the 3-manifold, thus changing the associated 3d $\mathcal{N}=2$ theory.\footnote{
In this sense, it might be more precise to denote the theory $T_N[\hatM\backslash K, \rho]$ by $T_N[\hatM, K, \rho]$. 
For the sake of notational simplicity, however, 
we stick with the notation $T_N[\hatM\backslash K, \rho]$.
} More explicitly, the CS path-integral with fixed boundary holonomy along a cycle, say $a(1,0)+b(0,1)$ of boundary torus, corresponds to the partition function of  $N$ M5s on a closed 3-manifold $\hat{M}_{(a,b)}$ with defect M5s along the knot $K_{(a,b)}$:
\begin{align}
\begin{split}
&\hat{M}_{(a,b)} :=\big{(}\textrm{a closed 3-manifold obtained by performing Dehn filling on $M$ }  
\\
&\quad \qquad  \qquad \textrm{which shrinks the cycle $a(1,0)+b(1,0)$} \big{)} \;,
\\
& \textrm{$K_{(a,b)}$ is a knot in $\hat{M}_{(a,b)}$ such that $\hat{M}_{(a,b)}\backslash K_{(a,b)}=M$\;.}  
\end{split}
\label{Dehn_filling}
\end{align}

\paragraph{Complex Chern-Simons Theory}
Let us specify the right hand side of  eq.~\eqref{3d3d_codim_2} more precisely.
The Lagrangian of the complexified Chern-Simons theory is given by\begin{align}
S_\textrm{CS}[\CA,\overline{\CA};\mathbf{\bhbar},\tilde{\bhbar}] = \frac{k+\sigma}{8\pi}\textrm{CS}[\CA] + \frac{k-\sigma}{8\pi}\textrm{CS}[\overline{\CA}]=\frac{i}{2\bhbar} \textrm{CS}[\CA]+\frac{i}{2\tilde{\bhbar}} \textrm{CS}[\overline{\CA}]\;\;, \label{complex CS action}
\end{align}
where the CS functional defined by
\begin{align}
\textrm{CS}[\mathcal{A}]:=\Tr \left( \mathcal{A}\wedge d\CA + \frac{2}3 \CA \wedge \CA \wedge \CA \right)\;.
\end{align}
In eq.~\eqref{complex CS action} we have two CS levels, $k\in \mathbb{Z}$ and $\sigma \in  \mathbb{R}  \textrm{ or } i\mathbb{R}$\footnote{Note that $\sigma$ here is sometimes denoted by $i \sigma$ in the literature.}\cite{Witten:1989ip}.
These parameters are combined into $\bhbar$ and $\tilde{\bhbar}$:
\begin{align}
\bhbar: =\frac{4\pi i}{k+\sigma}\;,\quad \tilde{\bhbar}: =\frac{4\pi i }{k-\sigma} \;.
\label{hbar_def}
\end{align}
These parameters play the role of the ``Planck constant'' in the quantization.
Note, however, that $\bhbar, \tilde{\bhbar}$ are in general not real,
and $\tilde{\bhbar}$ is in general not the complex conjugate of $\bhbar$.

The partition function is defined by the path-integral
\begin{align}
Z^\textrm{CS}= \int_{\mathcal{C}}\, [\mathcal{D}\mathcal{A}] [\mathcal{D} \overline{\CA}] \,\, e^{i  S_\textrm{CS}[\CA,\overline{\CA};\bhbar,\tilde{\bhbar}] } \, .
\label{Z_CS}
\end{align}
Since $\bhbar$ and $\tilde{\bhbar}$ are in general complex, 
the integrand is not bounded and  path-integral is not convergent on the naive integration contour 
where $\overline{\CA}$ is the complex conjugate of $\CA$.  To make sense of the integral, therefore, 
the path-integral should be interpreted as  an infinite dimensional contour integral along a middle-dimensional 
integration cycle $\mathcal{C}$  in the functional space spanned by two independent $SL(N)$ complex 
connections $\CA$ and $\overline{\CA}$ \cite{Witten:2010cx}.
\begin{align}
\begin{split}
&
\mathcal{C} \subset \CM_{SL(N) \textrm{ connection}}
 \\
&
\qquad:=\{ (\CA,\overline{\CA}): \textrm{two $SL(N)$ connections on $M$ with proper b.c.}\}/\sim\;.
\end{split}
\end{align}
with  gauge quotient $\sim$ parametrized by a pair of $U$ and $\bar{U}$ 
\begin{align}
U,\bar{U}\; : \;M \rightarrow SL(N) \;,
\end{align}
which act on $\CA$ and $\bar{\CA}$ respectively. $U$ and $\bar{U}$ should  be related in a proper way such 
that $e^{iS_{CS}}$ is invariant under  large gauge transformation. 
The contour varies depending on the reality of $\sigma$. 
When $\sigma$ is  purely imaginary (or equivalently $\tilde{\bhbar}=-\bhbar^*$ \eqref{hh_reality}) 
the action $S_{\rm CS}$ is real and  the integrand of the path-integral is  bounded along an integration cycle 
where $\overline{\CA}=\CA^{\dagger}$. We propose that this contour, 
possibly with infinitesimal deformation at infinity for convergence,  is the correct integration cycle for the 
3d--3d correspondence:
\begin{align}
\mathcal{C}_{\sigma\in i \mathbb{R}} =  \{ (\CA,\overline{\CA}) : \overline{\CA}= \CA^{\dagger} \}\;. \label{Contour for pure imaginary sigma}
\end{align}
with gauge quotient by
\begin{align}
\bar{U}= U^\dagger\;. \label{two gauge parameter relation}
\end{align}

When $\sigma$ is real, on the other hand, the integrand is not bounded along the cycle $\mathcal{C}_{\sigma \in i \mathbb{R}}$ and we should choose a different contour to make the path-integral convergent. 
We will see in the next section that at the level of the moduli space of flat connections
the choice is $\bar{\mathcal{X}}= \frac{1}{b^2} \mathcal{X}$,
where $\mathcal{X}$ and $\bar{\mathcal{X}}$ are the coordinates of the moduli space of flat connections.
Here the real parameter $b$ is defined by
$b^2:=\frac{\bhbar}{\tilde{\bhbar}}$ (see also eq.~\eqref{real_S3} in the next section).
The correct contour for our path integral should be the extension of 
this to more general (non-flat) connections.\footnote{The most naive choice is $\overline{\CA}= \frac{1}{b^2} \CA^{\dagger}$, 
however need to make sure that the contour choice is consistent with the gauge transformation.}

We can also discuss the choice of the choice of integration contours in terms of the so-called Lefschetz thimbles.
Applying Morse theory to the infinite-dimensional functional space $\CM_{SL(N) \textrm{ connection}}$ with a Morse function $\textrm{Im}(S_{\rm CS})$, 
 \begin{align}
 h:=-\textrm{Im}(S_{\rm CS})\; :\;  \CM_{SL(N) \textrm{ connection}} \rightarrow \mathbb{R}\;,%
 \end{align}
 it can be argued  that any convergent cycle $\mathcal{C}$ as a relative homological cycle can be decomposed into a linear combination of Lefschetz thimbles $\CJ^{(\alpha, \beta)}$  associated to the critical points of $h$ \cite{Witten:2010cx}.  The critical points are given by a pair of flat connections,  $\CA^{(\alpha)}$ and $\CA^{(\beta)} $ :
\begin{align}
\begin{split}
&\frac{\delta h}{\delta (\CA, \overline{\CA})}\big{|} = 0 \; \;\Leftrightarrow \;\; \{ \CA , \overline{\CA}\} = \{\CA^{(\alpha)}, \CA^{(\beta)} \} \;,
\end{split}
\end{align}
and the contour $\mathcal{C}$ is expanded as
\begin{align}
\begin{split}
& \mathcal{C}= \sum_{(\alpha, \beta)} n_{\alpha, \beta} \CJ^{(\a,\b)}\;, \quad n_{\a,\b} \in \mathbb{Z}\;, 
\\
&\CJ^{(\a,\b)} := \big{\{}\textrm{union of all trajectories along upward flow of $h$ }
\\
& \quad \quad \quad \qquad  \textrm{which approach  to the critical point $\{\CA^{(\a)},\CA^{(\b)} \}$ as $t\rightarrow \infty$} \}  \;. 
\end{split}
\end{align}
The upward/downward flows are defined by gradient  of $h$ up to sign along which $h$ is always increasing/decreasing. Here $(\alpha, \beta)$  are labels of $SL(N)$ flat-connections on $M$. The integer coefficients $n_{\alpha, \beta}$ are determined by counting (with sign) trajectories along downward flows of $h$ which start from a point in $\mathcal{C}$ and approach to the saddle point $\{ \CA^{(\a)},\CA^{(\beta)}\}$  as $t\rightarrow \infty$. Decomposition into Lefschetz thimbles is important to study  perturbative expansion of the CS partition function.   
The integer $n_{\a,\b}$ can jump as  $(\bhbar,\tilde{\bhbar})$ varies.
 For  purely imaginary $\sigma$, one can see that
 \begin{align}
 \begin{split}
&n_{\a, \bar{\a}}=1\;, \quad 
\\
&n_{\a, \b}= 0\quad \textrm{if $h(\CA^{(\a)},\CA^{\b}) >0$} \;,
\end{split}
\label{contour_condition}
 \end{align}
where $\CA^{(\bar{\a})}$ is an $SL(N)$ flat connection complex conjugate to $\CA^{\a}$.
The first equation in  eq.~\eqref{contour_condition} follows from the condition that the saddle points $\{\CA^{(\a)} , \CA^{(\bar{\a})}\}$ are located on $\mathcal{C}_{\sigma=i \mathbb{R}}$, and the second one follows from the fact
that $h$ is always zero on the cycle and $h$ always decreases along the downward flow. 

In the most part of this paper, we assume that $\sigma \in i \mathbb{R}$. We will come back to the case $\sigma \in \mathbb{R}$ in  Sec.~\ref{sec:sugra}.

\paragraph{Monodromy Defect}

When we wrote eq.~\eqref{Z_CS} we implicitly assumed that the 3-manifold is closed. 
When the 3-manifold $\hat{M}$ has a defect $K$,
then 
we are instructed to perform a path integral of the complexified gauge connection 
on the 3-manifold (eq.~\eqref{M_def}),
with singular boundary conditions around the defect $K$:
\begin{align}
Z^\textrm{CS}_{\hat{M}\backslash K} (\frakM)= \int_{\textrm{b.c.\ around $K$}} [\mathcal{D}\mathcal{A}] [\mathcal{D} \CA] \,\, e^{i S_\textrm{CS}[\CA,\overline{\CA};\bhbar,\tilde{\bhbar}] } \,.
\label{Z_CS_bc}
\end{align}
Suppose that $K$ is topologically a knot inside $\hat{M}$, the path-integral can be thought as defined on a 3-manifold $M:=\hat{M}\backslash K$ called knot-complement whose  boundary is a torus:
\begin{align}
\partial \left(\hat{M}\backslash K \right) = T^2 \ . 
\end{align}
The knot complement (exterior) can be constructed by removing tubular neighborhood $N_K$ of the knot from $\hat{M}$. 
\begin{align}
\hat{M}\backslash K:=\hat{M} - N_K\;.
\end{align}
In general, we can consider a knot with several disconnected components (in this case, a knot is also called a link).   
When we have $n$ components, we have
\begin{align}
\partial \left(\hat{M}\backslash K \right) =\overbrace{T^2 \cup T^2\cup \cdots \cup T^2 }^{n} . 
\end{align}

Let us begin with the classical CS theory. Since the classical equation of motion gives the 
flat connection condition 
\begin{align}
\mathcal{F}:=d\mathcal{A}+\CA \wedge \CA= 0\;,
\end{align}
the boundary condition should be specified by an $SL(N)$ flat connection 
on $n$ copies of $T^2$:
\begin{align}
\mathrm{Hom} \left( \pi_1\left( \overbrace{T^2 \cup T^2\cup \cdots \cup T^2 }^{n}\right) \to SL(N) \right) / \sim \, ,
\end{align}
where $\sim$ denotes the conjugation by the gauge group.
Note that the fundamental group of $T^2$ is
spanned by two cycles.
In the knot theory literature, the  cycle corresponding to contractable (non-contractible) cycle in the removed solid torus $N_K$  is called the meridian (longitude). In this definition, the longitude is not uniquely determined but only  up to a shift by the meridian.  We denote the meridians (longitudes) for the $a$-th link component
by $\mathfrak{m}^{(a)}$ ($\mathfrak{l}^{(a)}$), for $a=1, \cdots, n$.
Note that since the fundamental group of $T^2$ is commutative, we can use the gauge degrees of freedom to
bring all of them to be the upper triangular form (upper triangular here includes non-trivial entries in the diagonal).

When we go to the quantum theory, due to uncertainly principle it is not possible to 
specify both the holonomies along meridians $\textrm{Hol}(\mathfrak{m}^{(a)})$ and the holonomies along longitudes $\textrm{Hol}(\mathfrak{l}^{(a)})$;
they are canonically conjugate to each other.
It is therefore sufficient to specify only half of them.
This is the choice of the polarization in the quantization.  Fixing holonomy with generic eigenvalues breaks the gauge symmetry $SL(N)$ to its Levi-subgroup $\mathbb{L}^{(\rho)}$ \eqref{Levi-subgroup},  centralizer of $\textrm{Hol}(\mathfrak{m})$. The unbroken Levi-subgroup $\mathbb{L}^{(\rho)}$ determines the type  $\rho$ of the defect \cite{Gukov:2006jk}. This again explains why the defect (for each component of a knot) is labelled by a partition of $N$. 
For example, defect of maximal type ($\rho=[1, \ldots, 1]$) breaks the gauge group maximally, and dimension of residual gauge group is $N-1$.  The defect has continuous parameter $\{\frakM_\a\}_{\a=1}^{\ell(\rho)}$, which corresponds to $\ell(\rho)$ eigenvalues of $\textrm{Hol}(\mathfrak{l})$
(they satisfy one traceless constraint, and hence only $\textrm{rank}(H_\rho)=\ell(\rho)-1$ of them are independent). 

Let us consider the more general case $\rho=[n_1,\ldots, n_s]$.
For generic eigenvalues $\{\frakM_\a\}$ ($\frakM_\a\neq \frakM_\b$ for any $(\a,\b)$), the meridian monodromy $\textrm{Hol}(\mathfrak{m})$ is  given by ($\mathbb{I}_{n\times n}$ denote identity matrix of size $n\times n$)
\begin{align}
\textrm{Hol}(\mathfrak{m}) \in \textrm{orbit of } \left(\begin{array}{cccc}e^{\frakM_1}\mathbb{I}_{n_1} & 0 & \bf{0} & 0 \\0 & e^{\frakM_2}\mathbb{I}_{n_2} & \bf{0} & 0 \\ \bf{0} & \bf{0} & \ldots & \bf{0} \\0 & 0 & \bf{0} & e^{\frakM_s}\mathbb{I}_{n_s} \end{array}\right) \, . 
 \label{co-dimension 2 defects as b.c. in CS theory-1}
\end{align}
The orbit of an element $g\in SL(N)$ is the set of elements in the complex group that are conjugate to $e$. Note that there are still residual Weyl group symmetries for the discrete unbroken gauge group,
and we can assume without generality that $\frakM_\a >  \frakM_\b$ for $\a>\b$.
In the extreme case where all the eigenvalues are trivial, we obtain a closure of a nilpotent orbit:
\begin{align}
\textrm{Hol}(\mathfrak{m}) \in \textrm {closure of orbit of } \rho^t(e^{\sigma_{+}})\;.
 \label{co-dimension 2 defects as b.c. in CS theory-2}
\end{align}
$\rho^t$ denote the transpose partition  of $\rho$ whose corresponding Young diagram is obtained by reflecting the original diagram along its main diagonal. As we will comment more in App.~\ref{app.boundary_holonomy}, it is crucial to have the closure on the right hand side of   
this equation.
In all these cases, the closure of the orbit coincides with the Coulomb branch (or Higgs branch) of the corresponding mass-deformed $T^{\rho}[SU(N)]$ (or $T_{\rho}[SU(N)]$) theory. The $\ell(\rho)-1$ parameters $\frakM_\a$ corresponds to real mass parameters in $T_N[M,\rho]$ coupled to the flavor symmetry $H_\rho$.

\paragraph{3d--3d Correspondence with Defect of Type $\rho$}

The 3d--3d correspondence \eqref{3d3d_codim_2} has several concrete incarnations, 
depending on the partition functions we choose.
Dictionaries found in the literature \cite{Terashima:2011qi,Dimofte:2011ju,Dimofte:2011py,Cordova:2013cea,Lee:2013ida,Dimofte:2014zga}, generalized here with defect $\rho$ included, states
\begin{align}
\begin{split}
&\textrm{($S^3/\mathbb{Z}_k)_b$ partition function of $T_{N}[\hat{M}\backslash K,\rho]$}  
\\
&\quad= \textrm{$SL(N)_{k \in \mathbb{Z}_{>0}, \sigma =k \frac{1-b^2}{1+b^2} \in \mathbb{R} \,\textrm{or}\, i \mathbb{R}}$ CS partition function on $\hat{M}$ }\\
& \qquad \qquad \qquad \textrm{with type $\rho$ defect along $K$ }  \;,
\\
&\textrm{$( S^1\times S^2)_q$ partition function (superconformal index) of $T_{N}[\hat{M}\backslash K,\rho]$}  
\\
&\quad= \textrm{$SL(N)_{k=0, \sigma \in i \mathbb{R} }$ CS partition function on $\hat{M}$ with type $\rho$ defect along $K$ } \;.
\end{split}
  \label{3d-3d relation}
\end{align}
Here $(S^3/\mathbb{Z}_k)_b$ is a one-parameter deformation of the metric of $S^3/\mathbb{Z}_k$ (whose partition function was computed in \cite{Kapustin:2009kz,Gang:2009wy,Jafferis:2010un,Hama:2010av,Hama:2011ea,Imamura:2011wg,Benini:2011nc}),
and $(S^1 \times S^2)_q$ \cite{Kim:2009wb,Imamura:2011su} is a geometry where going around once along $S^1$ is accompanied by a 
rotation along $S^2$. The $(S^1\times S^2)_q$ partition function (superconformal index) has following interpretation as 
the trace over the Hilbert space of $S^2$:
\begin{align}
\textrm{Tr}_{\mathcal{H}_{S^2 (m_a)}}(-1)^{2j_3} q^{j_3 + \frac{R}2} \prod u_a^{F_a}\;,
\label{index_as_trace}
\end{align}
where $j_3$ is a Cartan of $SU(2)$ isometry of $S^2$, $R$ is the R-charge of the 3d $\mathcal{N}=2$ superconformal algebra, $F_a$ denote the Cartan generators of the flavor symmetries of the 3d theory and $u_a$ the associated nugacities. We also turn on background monopole fluxes $m_a$ on $S^2$ coupled to the flavor symmetries and consider generalized superconformal index \cite{Kapustin:2011jm}.  

Note that the levels of the complex CS theory, namely $k$ and $\sigma$,
are translated into the choice of the background geometry $B=(S^1\times S^2)_q, (S^3/\mathbb{Z}_k)_b ,\ldots$
where the 3d  theory $T_{N}[M,\rho]$ is defined: quantized level $k$ is related to the topology of $B$ and $\sigma$ is related to the squashing parameters (such as $q,b,\ldots$) of $B$. Note also that the case of $k=0, \sigma \in \mathbb{R}$ is not covered in eq.~\eqref{3d-3d relation}.

Let us also remark on the reality properties of $\bhbar, \tilde{\bhbar}$.
For $(S^3/\mathbb{Z}_k)_b$, we have (in terms of the parameter $b$ in eq.~\eqref{3d-3d relation})
\begin{align}
(S^3/\mathbb{Z}_k)_b: \quad
\bhbar= \frac{2\pi i}k  (1+b^2) \;, \quad
\tilde{\bhbar}=\frac{2\pi i}k (1+b^{-2}) \;.
\label{real_S3}
\end{align}
As already mentioned, $\sigma=k (1-b^2)/(1+b^2) \in \mathbb{R} \textrm{ or } i \mathbb{R}$, and hence we have 
either $b\in \mathbb{R}$ or $|b|=1$; the two branches merge for $b=1$, in which case $\sigma=0$.
For real $b$, $\bhbar$ are $\tilde{\bhbar}$ are purely imaginary. 
For $|b|=1$, then $\bhbar, \tilde{\bhbar} \in \mathbb{R}$.
If we analytically continue to more general values of $b$, then both $\bhbar, \tilde{\bhbar}$ are complex. For $S^1\times S^2$, we have $\sigma \in i \mathbb{R}$ and 
\begin{align}
(S^1\times S^2)_q: \quad
\bhbar=-\tilde{\bhbar}=\frac{4\pi i }{\sigma}\in \mathbb{R} \;.
\label{real_S1S2}
\end{align}
We will come back to the reality properties of $\bhbar, \tilde{\bhbar}$ when in the discussion of state-integral models
in the next section.

%%%%%%%%%%%%%%%%%%%%%%%%%%%%%%%%%%%%%%%%%%%%%%%%%%%
\subsection{Co-dimension 4 Defects }  \label{sec : co-dimension 4}
Co-dimesion 4 defects can be realized as in eq.~\eqref{codim_4_brane}, and we claimed there that such defects are labeled by 
\begin{align}
R\; :\; \textrm{unitary representations of $SU(N)$}\;.
\end{align}
This is easy to see, again by 
compactifying  the system along the M-theory circle (3rd direction in eq.~\eqref{codim_4_brane}). The defect is then 
described as a Wilson loop operator in 5d $\mathcal{N}=2$ SYM (we will come back to this viewpoint in Sec.~\ref{sec.5dSYM}). These defects are labelled by $R$.

These co-dimension 4 defects are mutually BPS with the co-dimension 2 defects and we  consider co-dimension 4 defect in a representation $R$ in the presence of co-dimension 2 defect of  $\rho$.  
We consider a Wilson loop $\hat{W}_{R} (\CK)$
in $SL(N)$ CS theory on a knot complement $M=\hat{M}\backslash K$  along a knot $\CK$ in a representation $R$:  
\begin{align}
\langle \hat{W}_{R} ( \CK)\rangle (\frakM) = \displaystyle\int_{\rm b.c.} \! [\mathcal{D}\mathcal{A}] \, e^{i S_{\rm CS}[\mathcal{A}]}\textrm{Tr}_R P  \exp \left(-\oint_{\CK} \mathcal{A}\right)\;, \label{path-integral representation of wilson loop VEVs}
\end{align}
with a boundary condition fixing the boundary holonomy around  knot $\CK$ as in eqs.~\eqref{co-dimension 2 defects as b.c. in CS theory-1} and \eqref{co-dimension 2 defects as b.c. in CS theory-2}.
Since the complex CS theory is topological, the Wilson line depends only on the isotopy class of $\mathcal{K}$
inside $M$.

This defect $\mathcal{K}$ will be a loop operator in 3d $\CN=2$ $T_{N}[\hat{M}\backslash K]$ theory.
The correspondence \eqref{3d3d_codim_4} again has incarnations as statements on the partition functions of 
$(S^3/\mathbb{Z}_k)_b$ and $(S^1\times S^2)_q$:
\begin{align}
\begin{split}
&\textrm{($S^3/\mathbb{Z}_k)_b$-partition function of $T_{N}[\hat{M}\backslash K]$ with line operator labelled by $R$} 
\\
&\qquad= \textrm{$SL(N)_{k \in \mathbb{Z}_{>0}, \sigma =k \frac{1-b^2}{1+b^2} \in \mathbb{R} \,\textrm{or}\, i \mathbb{R}} $ CS partition function on $\hat{M}$} \\
& \qquad \qquad \qquad\textrm{with a Wilson line of 
representation $R$ along $\mathcal{K}$ }  \;,
\\
&\textrm{$( S^2\times S^1)_q$-partition function 
of $T_{N}[\hat{M}\backslash K]$ with line operator labelled by $R$} 
\\
&\qquad= \textrm{$SL(N)_{k=0, \sigma \in i \mathbb{R} }$ CS partition function on $\hat{M}$} \\
& \qquad \qquad\qquad \textrm{with a Wilson line of 
representation $R$ along $\mathcal{K}$  } \;.  
\end{split}
\label{3d3d_codim_4-2}
\end{align}

On $(S^1\times S^2)_q$ and $(S^3/\mathbb{Z}_k)_b$ there are two supersymmetric cycles compatible with the supercharge used in localization: considering these 3-manifold as $S^1$ bundle over $S^2$, these cycles wrap the fiber $S^1$ located at the north/south poles of the base $S^2$.  
These two choices are represented on the Chern-Simons side by the
exchange of $\bhbar$ and $\tilde{\bhbar}$.\footnote{Such an exchange is an example of the 
 recently-found temperature reflection symmetry \cite{Basar:2014mha}.}

This concludes the discussion of the supersymmetric defects, and we now turn to the 
explicit computations of the partition functions.

%%%%%%%%%%%%%%%%%%%%%%%%%%%%%%%%%%%%%%%%%
\section{From State Integral Model}\label{sec.state_integral}
%%%%%%%%%%%%%%%%%%%%%%%%%%%%%%%%%%%%%%

\subsection{Generalities on State-Integral Models}\label{sec.state_integral_general}

Let us here describe state-integral models for the $SL(N)$ CS theory, based on an ideal triangulation of $M$. The models give finite dimensional integral expression for the CS partition function \eqref{Z_CS_bc}. 
There are known constructions in the literature for the case of $\rho=\textrm{maximal}$, 
see \cite{Dimofte:2011gm,Dimofte:2013iv,Dimofte:2014zga}. We will describe the construction slightly more generally,
to make contact with the discussion of non-maximal co-dimension 2 defects
 in Sec.~\ref{sec : cluster partition function}.

%--------------------------------------------------------------------
\paragraph{Octahedron Decomposition}
The construction of the state-integral model starts with 
an ideal triangulation  $\mathcal{T}$ of $M=\hat{M}\backslash K$ (with $k$ ideal tetrahedra)
\begin{align}
\CT\;:\; \quad  M=\left(\bigcup_{i=1}^{k} \Delta_i\right)/\sim\;,
\label{tetrahedron_decomp}
\end{align}
where $\sim$ means that we glue the tetrahedra by identifying the faces and edges.
The ideal triangulation is not unique and the integral expression for the state-integral model depends on the choice of it. However,
we can show that the resulting invariant after integral is independent on the choice, 
and hence it computes a topological invariant of the manifold. We next associate a set of `octahedra' $\Diamond_i^{a}$ to each ideal tetrahedron $\Delta_i$:
\begin{align}
\Delta_i \rightsquigarrow \left(\bigcup_{a=1}^{\sharp_i} \Diamond^{a}_i\right)/\sim\; , 
\label{octahedron_decomp}
\end{align}
where $\sim$ here means identification of the vertices of octahedra (as we will see in examples, eq.~\eqref{octahedron_decomp} is not really a decomposition of tetrahedron into octahedra, and is more a rule for 
associating a set of octahedra, and hence the symbol $\rightsquigarrow$ instead of $=$). See Fig.~\ref{fig:single-octahedron} for 
a figure of a single octahedron.
\begin{figure}[htbp]
\begin{center}
   \includegraphics[width=.3\textwidth]{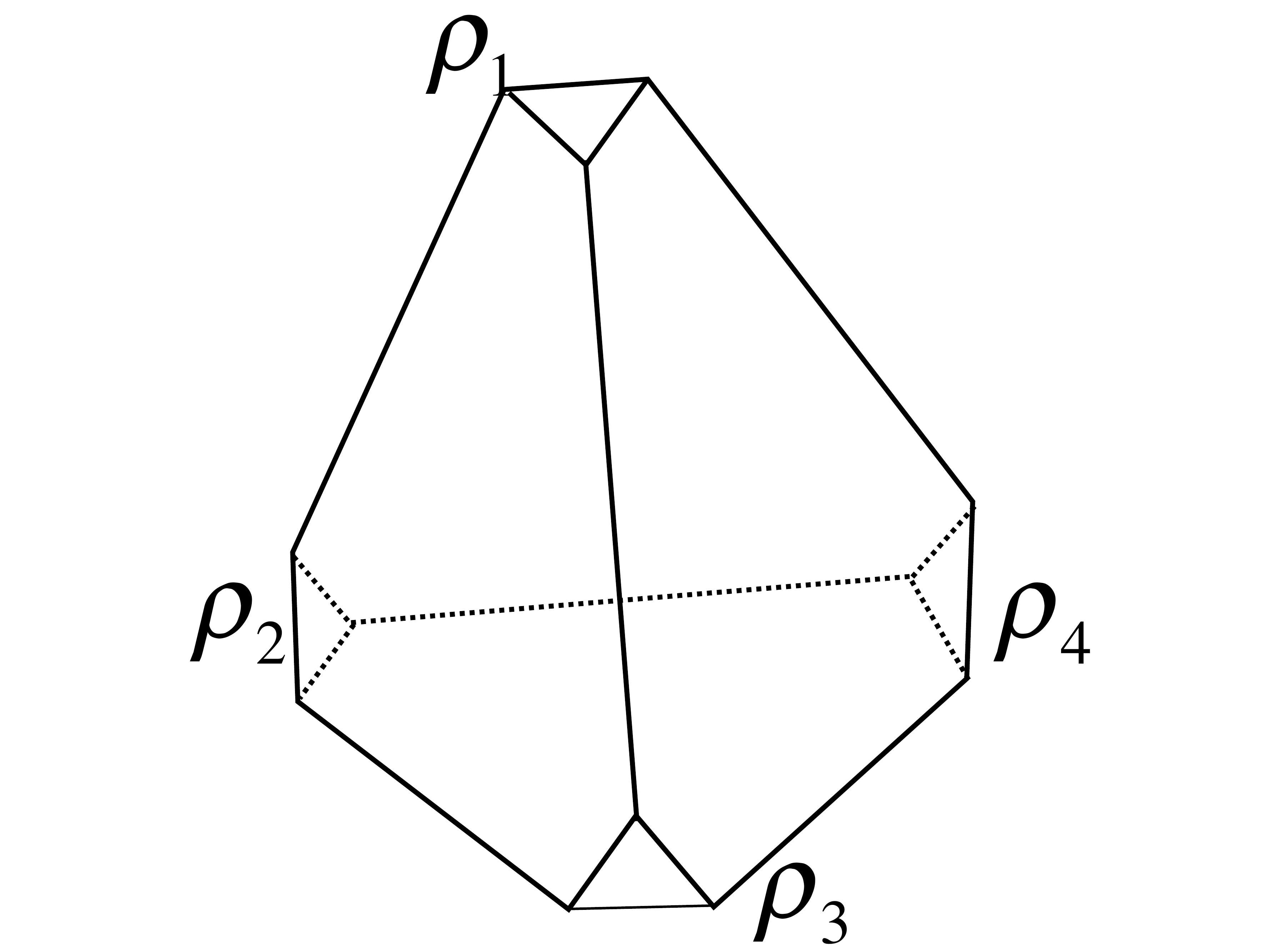}
   \end{center}
   \caption{The co-dimension 2 defect passes through the small neighborhood of a vertex of an ideal tetrahedron.
   In general the co-dimension 2 defects passing through four vertices are labeled by different $\rho$s. The octahedron decomposition 
   should be determined for a given choice of $\rho_{1,\ldots, 4}$. 
  There is no general known rule in the literature except when all $\rho$s are maximal, however we will discuss the non-maximal cases in the next section,
  where we identify octahedron structures (Fig.~\ref{quiver[21]-network})}
    \label{fig:rho-tetrahedron}
\end{figure}
The precise rule \eqref{octahedron_decomp} of how to associate octahedra to an ideal tetrahedron depends crucially on the choice of the co-dimension 2 defects.
 The co-dimension 2 defect in the ideal triangulation corresponds to loop(s) passing though the small neighborhood of the vertices of ideal tetrahedra, and hence each tetrahedron in general could have 
four different co-dimension 2 defects $\rho_1, \rho_2, \rho_3, \rho_4$ passing through its four vertices (Fig.~\ref{fig:rho-tetrahedron}).  
The completely general rule for the octahedron decomposition \eqref{octahedron_decomp} is not known at present, 
but we will discuss some examples (where $\rho$ is maximal and non-maximal)
in the rest of this paper.
Of course, by combining eqs.~\eqref{tetrahedron_decomp} and \eqref{octahedron_decomp}, we have a rule for 
associating octahedrons to the 3-manifold $M$:
\begin{align}
M\rightsquigarrow \left(\bigcup_{i=1}^k \bigcup_{a=1}^{\sharp_i} \Diamond^{a}_i\right)/\sim\; .
\label{combined_decomp}
\end{align}
We will denote the total number of octahedra by $\sharp_{\rm total}:=\sum_{i=1}^k \sharp_i$. The octahedron decomposition gives an algebraic ways to construct  the moduli space of flat connections:
\begin{align}
\begin{split}
&\CM_N (\hat{M}\backslash K, \rho) 
\\
&\quad:= \textrm{\{flat $SL(N)$ connections on $M$ satisfying  b.c. in  eq.~\eqref{co-dimension 2 defects as b.c. in CS theory-1}\}} \; 
\\
&\quad \,\,= \{ e^{Z_\eta''}+e^{-Z_\eta}-1=0\;, \; C_I (\{ Z, Z', Z''\}) \big|_{\hbar=0}=0\; \} \subset P(\partial \Diamond)^{\sharp_{\rm total} }\;,
\end{split} \label{Algebraic eqn for flat connections}
\end{align}
where $\eta$ is labeling for  $\sharp_\textrm{total}$ octahedrons,
and $P(\partial \Diamond)$  denoted a phase space associated to a octahedron (see Fig.~\ref{fig:single-octahedron})
\begin{align}
&P(\partial \Diamond) = \left\{ Z,Z',Z'': Z+Z'+Z''=i \pi + \frac{\hbar}{2} \right\} \simeq (\mathbb{R}\times S^1)^2\;,  
\label{P_Diamond}
\end{align}
with symplectic structure
\begin{align}
& \Omega = \frac{i}{\bhbar}   dZ''\wedge dZ + \frac{i}{\tilde{\bhbar}} d\bar{Z}'' \wedge d \bar{Z}\;.
\label{Omega}
\end{align}
\begin{figure}[htbp]
\begin{center}
   \includegraphics[width=.24\textwidth]{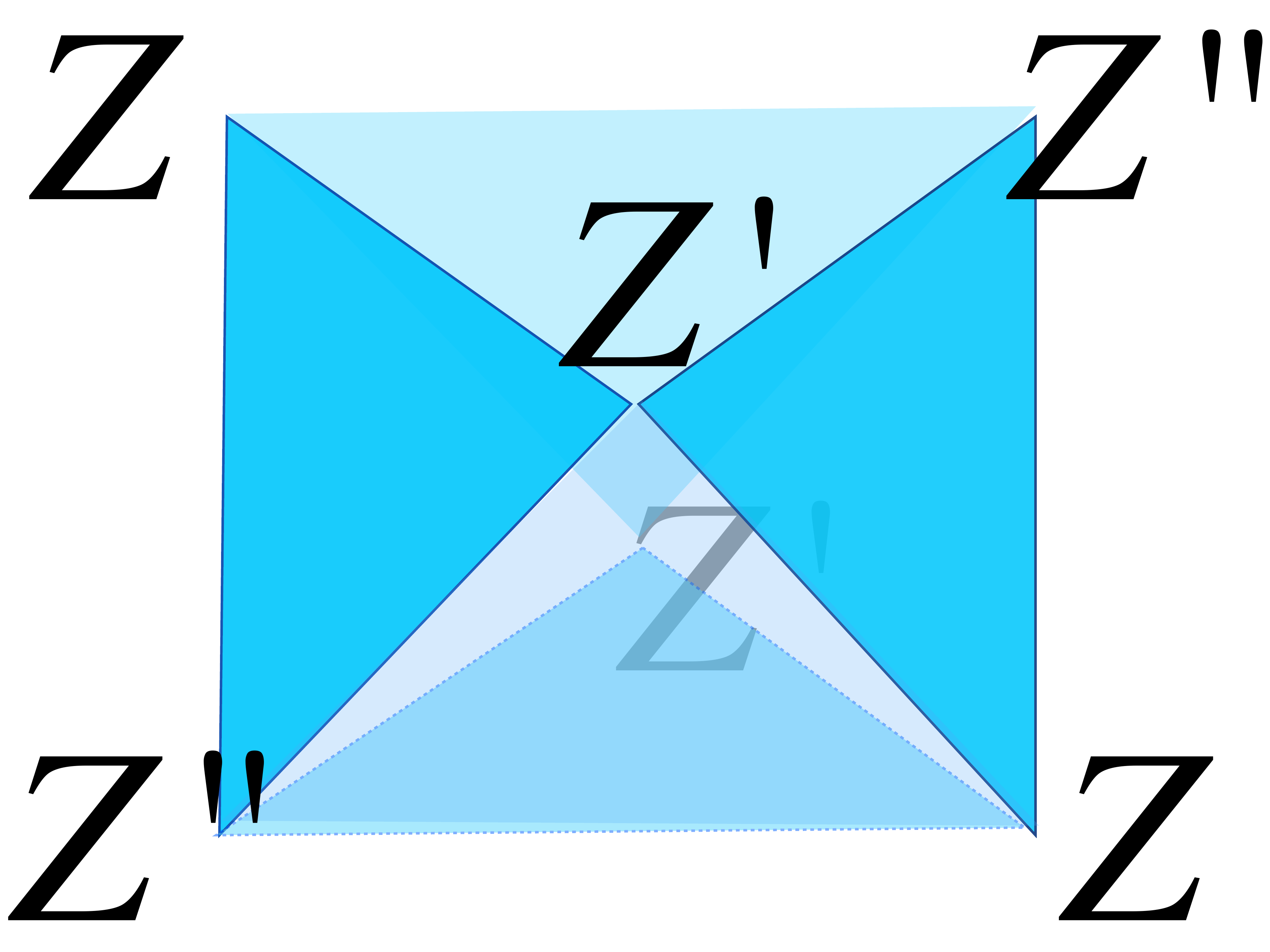}
   \end{center}
   \caption{For a single octahedron, we have the wavefunction $\psi_{\hbar, \tilde{\hbar}}(Z, \bar{Z})$ inside the phase space $P(\partial \Diamond)$.
  The phase space is constructed from three variables $Z, Z', Z''$, satisfying the constraint in eq.~\eqref{P_Diamond}.}
  \label{fig:single-octahedron}
\end{figure}
Geometrically, $(Z,Z',Z'')$ are vertex variables  assigned to each pair of vertices of octahedron, see Fig.~\ref{fig:single-octahedron}.  Imaginary part of these vertex variables are angle variables : $Z \sim Z+2 \pi i $. 
The  $C_I$ are variables associated with internal vertices in the octahedron decomposition,
\begin{align}
C_I = \textrm{(sum over vertex variables meeting at the $I$-th internal vertex)} - 2\pi i -\bhbar \;.
\end{align}
Number of independent  $C_I$ are always less than $\sharp_{\textrm{total}}$ and let the number be $\sharp_{\textrm{total}}- n_c$. Then, the dimension of the moduli space is $n_c$. 

For $N=2$, the octahedrons in the $N$-decomposition is one-to-one with tetrahedrons in the ideal triangulation and the octahedron vertex variables can be identified with edge variables of tetrahedrons. The edge variable measures the (complexified version of) dihedral angle between two faces meeting at the edge. In the identification, the construction in eq.~\eqref{Algebraic eqn for flat connections} has a geometrical meaning. These algebraic equations are firstly studied in the study of hyperbolic structure of 3-manifold. One way to construct hyperbolic structure is gluing hyperbolic tetrahedron smoothly. Edge variables of an ideal tetrahedron  in a hyperbolic space $\mathbb{H}^3$  satisfy the algebraic relation $e^{-Z}+e^{Z''}-1=0$ and $e^{Z+Z'+Z''}=-1$ and the internal edge conditions $C_I=0$ is the requirement for no conical singularity in the gluing. Thus, with a proper range of dihedral angle, solution for the algebraic equation give hyperbolic metric on 3-manifold. Then, using the relation between hyperbolic metric and $SL(2)$ flat connections, the solutions  give flat connections on the 3-manifold.  The algebraic variety in eq.~\eqref{Algebraic eqn for flat connections} further can be thought as a Lagrangian subvariety of $P_N(\partial M, \rho)$ defined by the following symplectic quotient
\begin{align}
P_N(\partial M, \rho) := P(\partial \Diamond)^{\sharp_{\rm total} }/\! /\{C_I=0\}\;. \label{boundary phase for M}
\end{align}
The symplectic quotient makes sense since actually all $\{ C_I \}$ are mutually commute, $\{C_I, C_J \}_{P.B}=0$ for any pair of $(I,J)$. In Sec. \ref{sec.codim_4_state}, we will explicitly construct the flat-connections (or equivalently its holonomies along cycles) on a knot-complement from the solution of algebraic equations.

\paragraph{Limitation of the Decomposition} 
Before explaining how to construct the  CS partition function from octahedron decomposition, as a cautionary remark, let us point out that 
there is an important limitation of  octahedron decomposition: only a sub-sector  of flat-connections can be obtained. For $\rho$=maximal case, for example, the $N$-decomposition only captures  fully irreducible flat-connections, i.e.  centralizer  of whose holonomies is trivial (see \cite{Chung:2014qpa} for details).
Due to this limitation, the wave functions of  the state-integral models based on octahedron decomposition cannot be glued to form a wave-function for the glued manifold,
and in particular Dehn's filling and Higgsing (the latter will be discussed in section.~\ref{sec: Higgsing}) cannot be done consistently on the  wave functions.  The limitation is related with the limitation of Abelian description of $T_N[M]$ theory, which will be discussed further in section.~\ref{sec : Necessity of Non-abelian}. 
Note that the same limitation exists for the cluster partition function in  section~\ref{sec : cluster partition function}. 

%--------------------------------------------------------------------
\paragraph{State-integral as an Overlap of Wavefunctions}
We can now write down the expression for the partition function of our state-integral model by quantizing eq.~\eqref{Algebraic eqn for flat connections}.
Schematically, the partition function for a knot complement $\hat{M}\backslash K$ \eqref{Z_CS_bc} is given by
\begin{align}
&Z^\textrm{state integral}_{\hat{M}\backslash K} ( \frakX_\a ) =\langle \frakX_\a \big{|} \hat{M}\backslash K \big{\rangle}= \big{\langle} \frakX_\a , C_I=0\big{|}    \Diamond^{\otimes \sharp_{\rm total}}  \big{\rangle}\;.\label{state-integral from ideal triangulation}
\end{align}

Let us explain the symbols here step by step. The partition function is written as an overlap of two states.
One of the two states is $\big{|}    \Diamond^{\otimes \sharp_{\rm total}}  \big{\rangle}: =
\otimes_{\eta=1}^{\sharp_{\rm total}}(|\Diamond_\eta \rangle)$.
This is a state in the Hilbert space $\otimes_{\eta=1}^{\sharp_{\rm total} }\CH (\partial \Diamond_\eta)$,
defined by a direct product of states $|\Diamond_{\eta}\rangle$, which in turn is a state in a Hilbert space $\CH^{(k,\sigma)} (\partial \Diamond_{\eta})$ for each $\eta$ (recall $k$ and $\sigma$ are the real and imaginary parts of the complexified level). 
Here, the Hilbert space  $\CH^{(k,\sigma)} (\partial \Diamond)$ is obtained by quantizing the phase space $P(\partial \Diamond)$ associated to a octahedron. The state is a quantization of the algebraic relation $e^{-Z}+e^{Z''}-1=0$, which define a Lagrangian sub-variety in the phase space, and satisfy the following operator equations
\begin{align}
(e^{-\hat{Z}}+e^{\hat{Z}''}-1) |\Diamond\rangle=(e^{-\hat{\bar{Z}}}+e^{\hat{\bar{Z}}''}-1) |\Diamond\rangle=0\;, \label{quantum lagrangian for octahderon}
\end{align}
where $\hat{Z},\hat{Z}', \hat{Z}''$ are quantized operators for vertex variables and $\hat{\bar{Z}},\hat{\bar{Z}}', \hat{\bar{Z}}''$ are their Hermitian conjugation.
Quantization of the phase space depends on the CS levels $(k,\sigma)$,
and we in particular need to impose different quantization conditions on position variables.
For the considerations of this paper, we need the following cases (see \cite{Dimofte:2014zga} for details):
\begin{align}
\begin{split}
&\textrm{position basis of $\CH^{(k=0, \sigma \in i \mathbb{R})}$ :}  \\
&\qquad \qquad \textrm{$\left\{ \Big|Z;\Pi_Z \Big\rangle:= \Big| Z= \frac{\bhbar}2 m + i \theta   , \bar{Z}  = \frac{\bhbar}2 m - i \theta \Big\rangle:   m \in \mathbb{Z}\;, \theta \sim \theta +2 \pi \in \mathbb{R} \right\} $ } \;, 
\\
&\textrm{position basis of $\CH^{(k \in 
\mathbb{Z}_{>0}, \sigma:=k\frac{1-b^2}{1+b^2}, |b|=1)}$ : } \\
& \qquad \qquad \textrm{$\left\{ \Big|Z;\Pi_Z \Big\rangle:=\Big|Z =\frac{2\pi}k ( b \mu + i \nu ) ,\bar{Z} =\frac{2\pi}k  ( b^{-1} \mu - i \nu ) \Big\rangle  : \mu \in \mathbb{R}\;, \nu \in \mathbb{Z}_k    \right\} $ } \;. 
\end{split}
\label{Quantization for several (k,s)}
\end{align}
Let us now comment on how to understand the quantization. First, when we consider $S^1\times S^2$, namely $k=0$ and$ \sigma \in i \mathbb{R}$, 
$\bhbar=-\tilde{\bhbar}$ is real (recall eq.~\eqref{real_S1S2}), 
and hence we learn from eq.~\eqref{Omega} that $\textrm{Re}[Z]$ is canonically conjugate to $\textrm{Im}[Z'']$.
The quantization of the $\textrm{Re}[Z]$ can then be understood from the
periodicity of the conjugate variable $\textrm{Im}[Z'']$. 
When we consider $(S^3/\mathbb{Z}_k)_b$, we choose $|b|=1$.\footnote{
This value is natural for the $SU(2)\times U(1)$ isometry-preserving squashing of $S^3$ considered in \cite{Imamura:2011wg},
which changes the relative size of the Hopf fibre and the base $S^3$ (in the Hopf fibration of $S^3$)
by the factor of $\ell^{-1}=(b+b^{-1})/2$. If we consider the $U(1)\times U(1)$ isometry-preserving squashing of
 \cite{Hama:2010av}, $b$ is geometrically real, $b\in \mathbb{R}$ requires analytic continuations on $b$.
The partition functions for the two squashings, although geometrically different, gives the same answer in the overlapping range, $0<\ell \leq 1$.
}
One simplification for this case is that we have 
\begin{align}
\bhbar^*=-\tilde{\bhbar} \;,
\label{hh_reality}
\end{align}
and the symplectic structure,
when expressed in terms of real part $\mu$
and imaginary part $m$ (as in eq.~\eqref{Quantization for several (k,s)}),
takes a simple form:
\begin{align}
\Omega= \frac{2\pi}{k^2} ( d\mu''\wedge d\mu + d\nu''\wedge d\nu) \;.
\end{align}
Since the imaginary part $m, m''$ has period $k$ and are canonical conjugates with each other,
we learn that $m$ should take values in $\mathbb{Z}_k:=\mathbb{Z}/k\mathbb{Z}$,
as stated in eq.~\eqref{Quantization for several (k,s)}.

If we consider the $(S^3/\mathbb{Z}_k)_b$ in eq.~\eqref{Quantization for several (k,s)},
with $k=1$,
$\nu$ in $Z, Z''$ is frozen to take a fixed value and $q=e^{2\pi i b^2}, \tilde{q}=e^{2\pi i b^{-2}}$ and $\bar{Z}=b^{-2} Z, \bar{Z}''=b^{-2} Z''$.
For  $k=1$ with $b$ real,  the quantization can be understood as an analytic continuation of $|b|=1$ case. In this case, only real parts of $Z,Z''$ can be varied and it is more like quantization of  $``SL(N,\mathbb{R})"$ theory with $X:=Z,P:=Z''$ and single positive real quantum parameter $\hbar_{\mathbb{R}}:=2\pi b^2$:
\begin{align}
&\langle X|e^{\hat{X}} =   \langle X |e^{X}\;, \; \langle X|e^{\frac{1}{i \hbar_{\mathbb{R}}} \hat{P}} = \langle X+1|\;, \;\langle X|P\rangle = e^{\frac{1}{\bhbar} Z Z'' +\frac{1}{\tilde{\bhbar}} \bar{Z}\bar{Z}''} =e^{\frac{1}{i \hbar_{\mathbb{R}}} XP}\;. \label{SL(2,R) like rep}
\end{align}

To fully characterize a position basis, choosing positions $\{\frakX_i \}$ is not enough but also need to specify its conjugate momentums $\{ \frakP_i\}$. The choice $\Pi:=(\frakX_i ;\frakP_i)$ of position/momentum variables is called a polarization and the corresponding position basis is denoted by 
\begin{align}
|\frakX_i ;\Pi\rangle \, .
\end{align}
We sometimes suppress the polarization choice $\Pi$ when it is obvious in the context. Its conjugate ket-state is
\begin{align}
|\frakX_i \rangle^\dagger  = \langle \bar{\frakX}_i |\;.
\end{align}
In general, we define a position basis $\langle \vec{\frakX} ,\Pi|$ in a polarization in $\Pi= (\frakX_i ; \frakP_i)$ as follows\footnote{These conditions does not fix the overall normalization of position basis. We may impose the orthonormality of the basis. Even then, however, overall phase factor cannot be fixed. Throughout this paper (except $k=0$ case), we are sloppy in the overall normalization. For $k=0$, there is a canonical choice for the normalization of the partition function, the superconformal index; the normalization factor is the supersymmetric version of the Casimir energy.}
\begin{align}
\begin{split}
&\langle \vec{\frakX};\Pi| e^{\hat{\frakX}_i}  = \langle \vec{\frakX};\Pi | e^{\frakX_i}\;, \quad  \langle \vec{\frakX} ;\Pi| e^{\hat{\bar{\frakX}}_i}  = \langle \vec{\frakX};\Pi | e^{\bar{\frakX}_i}\;,  
\\
&\langle \vec{\frakX} ;\Pi| e^{\sum_i \epsilon^i \hat{\frakP}_i}  = e^{\sum_i \epsilon^i \bhbar \,\partial_{\frakX_i}} \langle \vec{\frakX};\Pi |\;, \quad  \langle \vec{\frakX} ;\Pi| e^{\sum_i \epsilon^i \hat{\bar{\frakP}}_i}  =  e^{\sum_i \epsilon^i \tilde{\bhbar} \, \partial_{\bar{\frakX}_i}} \langle \vec{\frakX};\Pi |\;. \label{position basis}
\end{split}
\end{align}
In the polarization $\Pi$, $\frakX_i$ and $\frakP_i$ are position and momentum variables where the holomorphic symplectic form is written as 
\begin{align}
\Omega = \sum_i   \frac{i}{\bhbar} d\frakP_i \wedge d \frakX_i +  \frac{i}{\tilde{\bhbar}} d \bar{\frakP}_i \wedge d \bar{\frakX}_i\;. 
\label{Omega_complex}
\end{align}
$(\hat{\frakX}, \hat{\bar{\frakX}}, \hat{\frakP}, \hat{\bar{\frakP}})$ are quantum operators obtained by quantitating $\frakX, \bar{\frakX}, \frakP, \bar{\frakP}$ respectively. 

As an example, one possible polarization choice for an octahedron's  phase space \eqref{P_Diamond} is 
\begin{align}
\Pi_Z=(Z;Z'')\;. 
\end{align}
 In the polarization $\Pi_Z:=(Z,Z'')$, the octahedron's wave-function $|\Diamond\rangle$ is given by
 a version of  the quantum dilogarithm function \cite{Dimofte:2011ju} (see App.~\ref{app.qdilog})
\begin{align}
\langle Z;\Pi_Z| \Diamond\rangle :=\psi_{\hbar,\tilde{\hbar}}(Z, \bar{Z}):=
\prod_{r=0}^{\infty} \frac{1-q^{r+1}e^{-Z}}{1- \tilde{q}^{-r} e^{-\bar{Z}}} \;,   
\label{octheron's CS wave ftn}
\end{align}
with $q:=e^{\bhbar}=e^{\hbar}, \tilde{q}:=e^{\tilde{\bhbar}}=e^{\tilde{\hbar}}$. 
One can check that the wave function satisfy the operator equations in eq.~\eqref{quantum lagrangian for octahderon}.
The expression \eqref{octheron's CS wave ftn} is valid only for $|q|<1$ and $|\tilde{q}|>1$, and for general $q$ the expression requires analytic continuation.
For $S^3_b$ ($k=1$ in our notation), the quantum dilogarithm function $\psi_{\hbar, \tilde{\hbar}}(z, \bar{z})$
reduces to the Faddeev's non-compact quantum dilogarithm function \cite{FaddeevKashaevQuantum}. 

We can also choose $\Pi_{Z'}:=(Z';Z)$ or $\Pi_{Z''}:=(Z'';Z')$. The three choices are   related to each other by cycle permutation of  vertices $Z\rightarrow Z'\rightarrow Z''$, and  the corresponding wavefunction are all the same (up to an ovarall factor):%\cite{Dimofte:2011ju,Dimofte:2014zga}:
\begin{align}
 \langle \frakX;\Pi_Z |\Diamond\rangle =  \langle \frakX;\Pi_{Z'} |\Diamond\rangle= \langle \frakX;\Pi_{Z''} |\Diamond\rangle \;,
\end{align}
as is guaranteed from the property of the quantum dilogarithm function $\psi_{\hbar, \tilde{\hbar}}(z, \bar{z})$. 

In eq.~\eqref{position basis} we have treated $\hat{\frakX}$ and $\hat{\bar{\frakX}}$ as independent degrees of freedom,
however we wish to impose the following Hermiticity constraint:
\begin{align}
\hat{\frakX}^\dagger = \hat{\bar{\frakX}} \;, \quad \hat{\frakP}^\dagger = \hat{\bar{\frakP}} \;.
\end{align}
We need to make sure that this Hermiticity constraint is compatible with the symplectic structure \eqref{Omega_complex},
leading to the constraint \eqref{hh_reality}. Fortunately, that condition is satisfied for both of the cases considered in eq.~\eqref{Quantization for several (k,s)}.
The inner-product on the Hilbert space, from the compatibility between the Hermiticity and eq.~\eqref{position basis}, 
is uniquely determined up to an overall normalization as
\begin{align}
\langle \vec{\frakX} |\vec{\frakX}'\rangle = \delta (\vec{\frakX} -\vec{ \frakX'})\;,
\end{align}
and the completeness relation is\footnote{The delta function (integral) in the inner-product (completeness relation) should be understood as a Kronecker delta  (summation) for discrete variables. }
\begin{align}
\mathbb{I} = \int  d\vec{\frakX} \, | \vec{\frakX} \rangle \langle \vec{\frakX}|\;. \label{Completeness relation}
\end{align}

Coming back to eq.~\eqref{state-integral from ideal triangulation},
the variables $\{\frakX_\a\}$ denote the position variables 
in the boundary phase space \eqref{boundary phase for M} in a general choice of the polarization. 
The typical choice is to take $\{\frakX_\a=\frakM_\a\}$, where $\frakM_\a$ is the meridian variables \eqref{co-dimension 2 defects as b.c. in CS theory-1}.
The meridian variables $\frakM_\a$, as well as its canonical conjugate, the longitude  $\frakL_\a$, can be expressed linear combination of octahedra' vertex variables $Z, Z', Z''$ which commute with all $C_I$s.
We can also choose canonically conjugate variables $\{\Gamma_I\}$ of $C_I$, satisfying
the canonical commutation relations $\{ C_I, \Gamma_J \}_{P.B.} = -\hbar \, \delta_{IJ}$.
We then have a choice of polarization
\begin{align}
\vec{\frakX} =  ( \frakX_\a , C_I)\;, \quad  \vec{\frakP} =  ( \frakP_\a, \Gamma_I )\;.
\end{align}
and the state $| \frakX_\a, C_I=0 \rangle$ is a state defined in this polarization, with constraints $C_I=0$ imposed. 
In this way, we can consistently reduce the Hilbert-space for $P(\partial \Diamond)^{\sharp_{\textrm{totla}}}$ to the Hilbert space for $P_N(\partial M,\rho)$. This procedure is quantum version  of the symplectic quotient in eq.~\eqref{boundary phase for M}.  
Note that since we are setting $C_I=0$, the state is actually independent of the 
choice of the $\Gamma_I$; a change of the polarization $\Gamma_I \to \Gamma_I + \sum_J c_{IJ} C_J$
adds a Gaussian factor for $C_I$s to the wave function, 
which however is trivial due to the constraints $C_I=0$.

%--------------------------------------------------------------------
\paragraph{Integral Expression}

We can rewrite our partition function \eqref{state-integral from ideal triangulation}
into a more concrete expression, by 
inserting the completeness relation \eqref{Completeness relation} 
\begin{align}
Z^\textrm{state ntegral}_{\hat{M}\backslash K}(\frakX_\a ) &= \big{\langle } \frakX_\a |\hat{M}\backslash K \rangle =\big{\langle} \frakX_\a, C_I=0 \big{|}    \Diamond^{\otimes \sharp_{\rm total}}  \big{\rangle} \nn
\\
&=  \int d\vec{Z}\; \big{\langle} \frakX_\a, C_I=0  \big{|} \vec{Z}; (\Pi_Z)^{\otimes \sharp_{\rm total} }  \big{\rangle}  \big{\langle} \vec{Z}; (\Pi_Z)^{\otimes \sharp_{\rm total} } \big{|} \Diamond^{\otimes \sharp_{\rm total} } \big{\rangle}
\nn
\\
&= \int d\vec{Z} \;\big{\langle} \frakX_\a, C_I=0  \big{|} \vec{Z}; (\Pi_Z)^{\otimes \sharp_{\rm total} } \big{\rangle}\prod_{\eta=1}^{\sharp_{\rm total} } \psi_{\hbar, \tilde{\hbar}} (Z_\eta)
 \;. \label{results for general state-integral}
\end{align}
The matrix element $\langle \frakX_\a,C_I |\vec{Z}\rangle$ determines 
the change of the polarization. For our cases, this can be represented by an
$Sp(2\sharp_{\rm total} ,\mathbb{Z})$ canonical transformation plus affine constant shifts,
namely an element of the affine symplectic group $ISp(2\sharp_{\rm total} ,\mathbb{Z})$:%
\begin{align}
\Pi_{\frakX,C}&:=
\left(\begin{array}{c} \vec{\frakX}   \\ \vec{C}   \\\hline \vec{\frakP} \\  \vec{\G} \end{array}\right) = \left(\begin{array}{c|c}A & B \\\hline C &  D\end{array}\right) \left(\begin{array}{c} \vec{Z} \\\hline \vec{Z}'' \end{array}\right)  -
\left( i\pi+\frac{\hbar}2 \right) 
\left(\begin{array}{c} \nu \\\hline  \nu_p \end{array}\right)\nn
\\
&=: g \cdot  \left(\begin{array}{c} \vec{Z} \\ \hline \vec{Z}'' \end{array}\right) - \left(\begin{array}{c} \vec{\g} \\ \hline \vec{\d} \end{array}\right)
\label{Sp(2k,Z)+affine shifts}\;,
\end{align}
where we obtain the constant part with $\nu, \nu_p \in \mathbb{Z}$ when we use the 
relation in eq.~\eqref{P_Diamond}, namely $Z+ Z'+Z''=\bhbar/2+i \pi $.

Due to the differences \eqref{Quantization for several (k,s)} in the quantization conditions, 
details of state-integral models depends on CS levels $(k, \sigma)$. However, the expressions written in eqs.~\eqref{state-integral from ideal triangulation}, \eqref{octheron's CS wave ftn} and \eqref{results for general state-integral}, are true in general. 

%%%%%%%
\paragraph{State-integral Model for $k=0$} 

To make things concrete, let us specialize to the $S^1 \times S^2$ case of
eq.~\eqref{Quantization for several (k,s)}, with $k=0$ and $\bhbar=-\tilde{\bhbar}$ real. 
In this case, following eq.~\eqref{Quantization for several (k,s)} 
let us first represent the variables $\vec{\frakX}$ in terms of their real and imaginary parts to as
$\vec{\frakX}= \frac{\bhbar}2 \vec{m} + \log \vec{u}$ (with $|\vec{u}|=1$).
The wave-function  \eqref{octheron's CS wave ftn} can then be written as  
\begin{align}
\CI_{\Diamond} (m,u)&:= \langle m,u;\Pi_Z |\Diamond\rangle 
= \prod_{r=0}^{\infty} \frac{1-q^{r-\frac{m}{2}+1}u^{-1}}{1- q^{r -\frac{m}{2}} u}\;. \label{Octahedron's CS index}
\end{align}
For $k=0$, we use the letter $\CI$ instead of $Z$ for partition function since it corresponds to the superconformal index.
\begin{align}
\CI (m_\a, u_\a) = Z^{k=0}(\frakX_\a)|_{\frakX_\a =\frac{\hbar}2 m_\a + \log u _\a }.
\end{align}
We call this the index in `fugacity' position basis, since $u$ plays the role of the fugacity in the 
definition of the superconformal index (denoted $u_a$ in eq.~\eqref{index_as_trace}).

It is useful to  introduce the `charge' position basis $\langle m, e;\Pi|$, by taking the 
powers of the fugacities $u$:
\begin{align}
\langle m,u| :=\sum_e \langle m,e |u^{e} \;,
\label{ue_conversion}
\end{align}
which leads to 
\begin{align}
\CI_{\Diamond}(m,u) =\sum_{e \in \mathbb{Z}} \CI_{\Diamond}^{\rm c}(m,e) u^e\;, 
\quad \CI^{\rm c}_{\Diamond}(m,e):=\langle m,e;\Pi_Z|\Diamond \rangle \;.
 \label{Octahedron's CS index in charge basis}
\end{align}
The action of quantum position/momentum operators in the fugacity basis  can be read off from eq.~\eqref{position basis}:\footnote{We sometimes use subscript $+/-$ for holomorphic/anti-holomorphic variables or operators: $\hat{\frakX}_+ := \hat{\frakX},\hat{\frakX}_- := \hat{\bar{\frakX}}$.}
\begin{align}
&\langle m,u ;\Pi| \, e^{\hat{\frakX}_\pm  }= \langle m,u;\Pi| \, q^{\frac{m}{2}} u^{\pm 1}\;, \quad \quad \langle m,u;\Pi| \, e^{\hat{\frakP}_\pm} =e^{\frac{\hbar}2 u \partial_u } \langle m \pm 1 , u;\Pi| \;, 
\end{align}
which simplifies in the charge basis
\begin{align}
&\langle m,e ;\Pi| \, e^{\hat{\frakX}_\pm }= \langle m,e \mp 1; \Pi|\, q^{\frac{m}{2}} \;, \quad\quad \langle m,e;\Pi| \, e^{\hat{\frakP}_\pm} = \langle m\pm 1 , e;\Pi|\, q^{\frac{e}{2}}\;,
\label{(X,P) on charge basis}
\end{align}
and the completeness relations in the two basis are given by
\begin{align}
\mathbb{I} =\sum_{m,e\in\mathbb{Z}} | m,e\rangle \langle m,e| = \sum_{m \in \mathbb{Z}}\oint \frac{du}{2\pi i u}| m, u\rangle \langle m, u| \;.
\end{align}

One advantage of the charge basis, as is clear from eq.~\eqref{(X,P) on charge basis}, is that the symmetry between $m$ and $e$ is manifest, and 
consequently
the basis has following simple transformation property under the  $Sp(2\sharp_{\rm total} , \mathbb{Z})$+(affine shift) in eq.~\eqref{Sp(2k,Z)+affine shifts}:
\begin{align}
&\big{\langle} \vec{m}, \vec{e} ; \Pi_{\frakX,C} \big{|}  =  \big{\langle} (\vec{m}, \vec{e})
\cdot (g^t)^{-1} ;\Pi_Z^{\otimes \sharp_{\rm total}} \big{|} e^{\vec{e}\cdot \vec{\g}- \vec{m}\cdot \vec{\d} } \;, \label{basis change under polarization change}
\end{align}
where $\vec{m} = (m_\a,m_I)_{I=1\ldots \sharp_{\rm total} - \sharp_{C}}^{\a=1,\ldots, \sharp_{C}}$ and $\vec{e} = ( e_\a, e_I)_{I=1\ldots\sharp_{\rm total}  - \sharp_{C}}^{\a=1,\ldots, \sharp_{C}-1}$,
and $\sharp_{C}$ denotes the number of constraints $C_I=0$.

From eq.~\eqref{basis change under polarization change}, we find the index of $M\backslash \hat{K}$ to be
\begin{align}
\CI^{\rm c}_{M\backslash \hat{K}}(m_\a, e_\a) & = \langle m_\a, m_I; e_\a, e_I |\Diamond^{\otimes \sharp_{\rm total}}\rangle \big|_{C_I = \frac{\bhbar}{2} m_I + \log u_I  =0}  \;.
\end{align}
The imaginary part of the constraints $C_I=0$ reads $u_I=1$, and 
as we can see from eq.~\eqref{ue_conversion} this amounts to sum over the $e_I$s with equal weights:
\begin{align} 
\CI^{\rm c}_{M\backslash \hat{K}}(m_\a, e_\a) 
 &= \sum_{e_I  \in \mathbb{Z}}\langle m_\a, m_I=0; e_\a, e_I ;\Pi_{L,C}|\Diamond^{\otimes \sharp_{\rm total}}  \rangle
 \nn\\
 &= \sum_{e_I \in \mathbb{Z}} e^{\vec{e}\cdot \vec{\g} - \vec{m}\cdot \vec{\d}} (\CI^{c}_{\Diamond})^{\otimes \sharp_{\rm total}} \big{(}g^{-1} \cdot (m_\a, m_I=0, e_\a, e_I) \big{)} \;.
\label{index using state-integral from ideal triangulation}
\end{align}
This is the explicit expression of our index.

\paragraph{Abelian Description of $T_N[\hat{M}\backslash K]$}

The result eqs.~\eqref{results for general state-integral}
and \eqref{Sp(2k,Z)+affine shifts} have clear counterparts in 
3d $\mathcal{N}=2$ theories, and this is sufficient to give the 
Abelian description of the 3d $\mathcal{N}=2$ theory $T[M]$,
which we briefly comment here (see \cite{Dimofte:2011ju,Dimofte:2011py,Terashima:2013fg} for details).
First, in  eq.~\eqref{results for general state-integral}
we have a product of quantum dilogarithm functions inside the integrand.
Each of this factor represents a 3d $\mathcal{N}=2$ chiral multiplet. 
Second, the $Sp(2n, \mathbb{Z})$ transformation is interpreted as the 
$Sp(2n, \mathbb{Z})$ transformation for Abelian 3d $\mathcal{N}=2$ theories, 
defined from the diagonal/off-diagonal Chern-Simons terms \cite{Witten:2003ya,Dimofte:2011ju}.
We then have an integral over the parameters $\vec{Z}$, representing the 
Abelian gauge symmetries, and how the parameters $\vec{Z}$ appears in the 
arguments of the quantum dilogarithm function determines the
gauge charges of the corresponding $\mathcal{N}=2$ chiral multiplets.
We  also have the delta-function constraints. This means to include superpotential 
terms, breaking the symmetries; this superpotential in general contains 
fields not appearing  in the Lagrangian (monopole operators), and 
exactly which operator appears in the superpotential is determined by
an $Sp(2n, \mathbb{Z})$ matrix $g$ in eq.~\eqref{Sp(2k,Z)+affine shifts}.

%%%%%%%%%%%%%%%%%%%%%%%%%%%%%%%%%%%%
\subsection{Co-dimension 2 Defects}

Let us restrict to the case where $\rho=\textrm{maximal}$.
In this case, the rule for associating octahedra was worked out in \cite{Dimofte:2013iv},
by lifting to 3d the Fock-Goncharov construction on 2d Riemann surfaces \cite{2003math.....11149F}. 
We use an `$N$-decomposition' of the 3-manifold,
which can be obtained by replacing each ideal tetrahedron of an ordinary ideal triangulation
by a pyramid of $N(N^2-1)/6$ octahedra, $\{ \Diamond_{(a,b,c,d)} \}$ with $a,b,c,d=0,\ldots, N-2$ satisfying  $a+b+c+d=N-2$.\footnote{See figure 3 in \cite{Gang:2014ema} for the $N$-decomposition for $M$=(figure eight knot complement) with $N=4$.}
\begin{figure}[htbp]
\begin{center}
   \includegraphics[width=.52\textwidth]{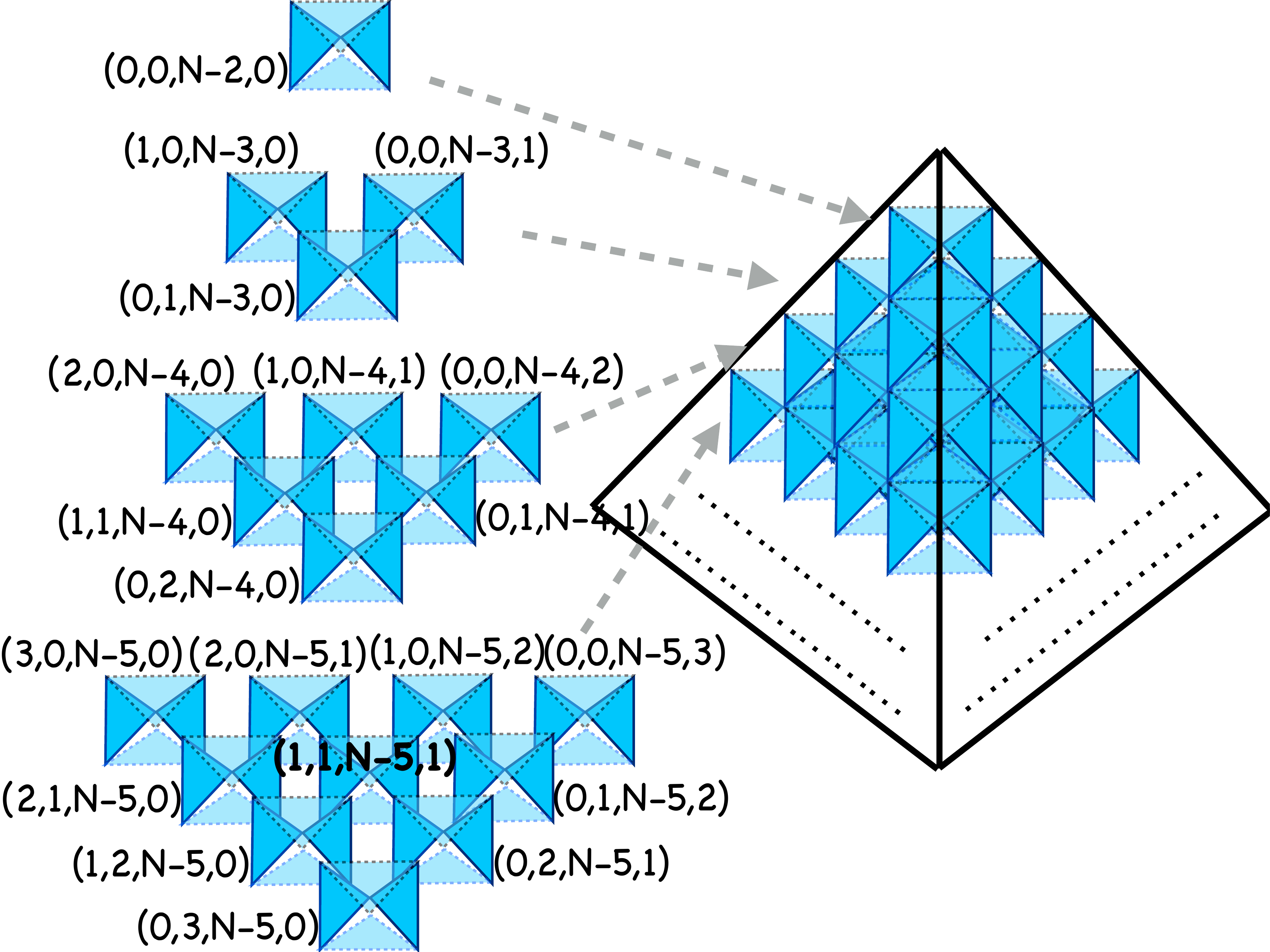}
   \end{center}
   \caption{$N$-decomposition of a single tetrahedron.  The $m$-th layer has $m(m+1)/2$ octahedra. Octahedrons are labelled by four non-negative integers $({a,b,c,d})$ satisfying $a+b+c+d=N-2$.}    \label{fig:N-decomposition}
\end{figure}
The number of octahedra per ideal tetrahedron is
\begin{align}
 \frac{1}{6} N(N^2-1)\;.
\label{total_N}
\end{align}
and grows as $O(N^3)$ as $N$ becomes large.
%

%---------------------------------------
\paragraph{Example : $M=S^3\backslash \mathbf{4}_1$  with $N=2$} 

\begin{figure}[htbp]
\centering\includegraphics[scale=0.15]{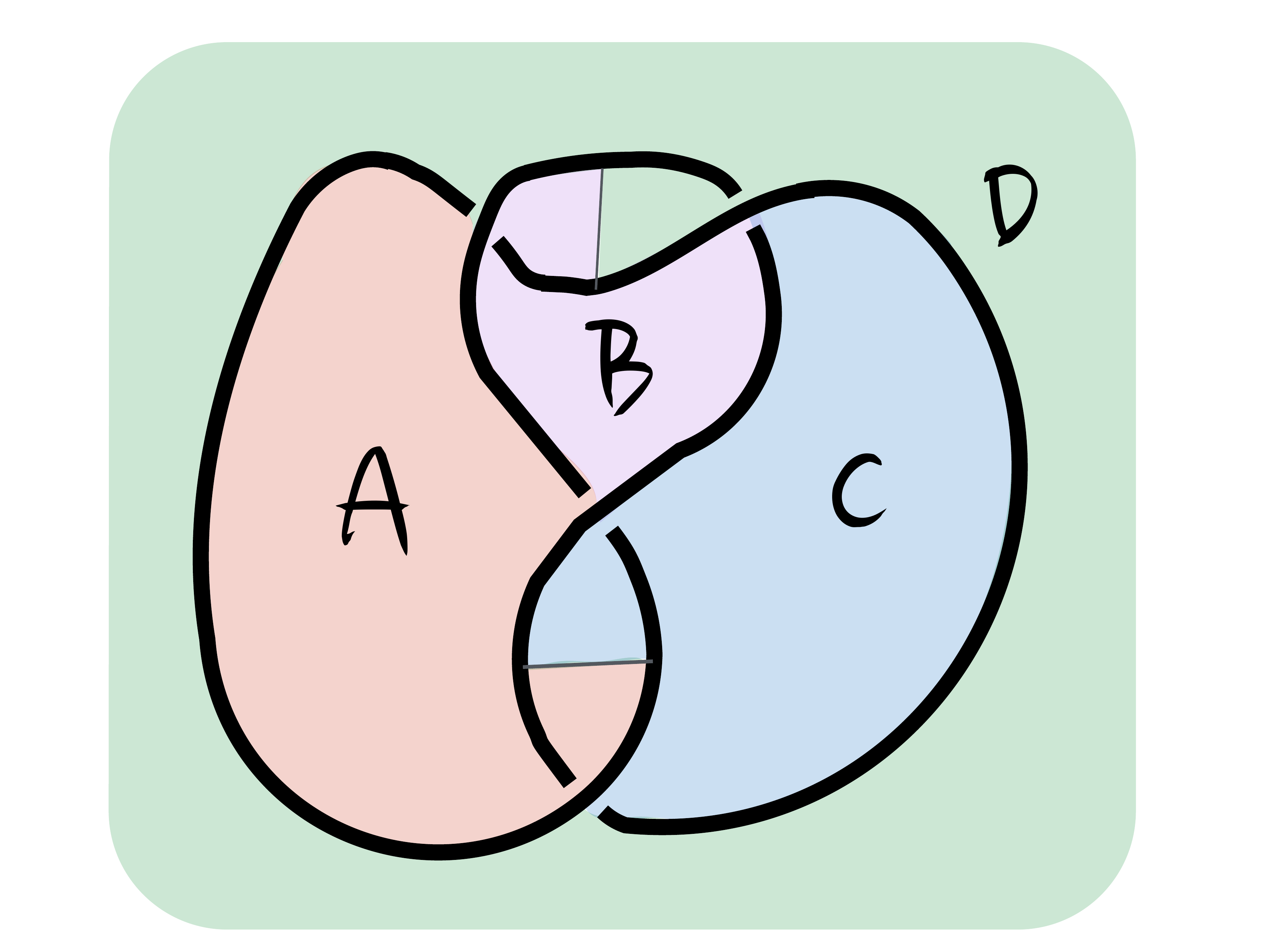}
\centering\includegraphics[scale=0.15]{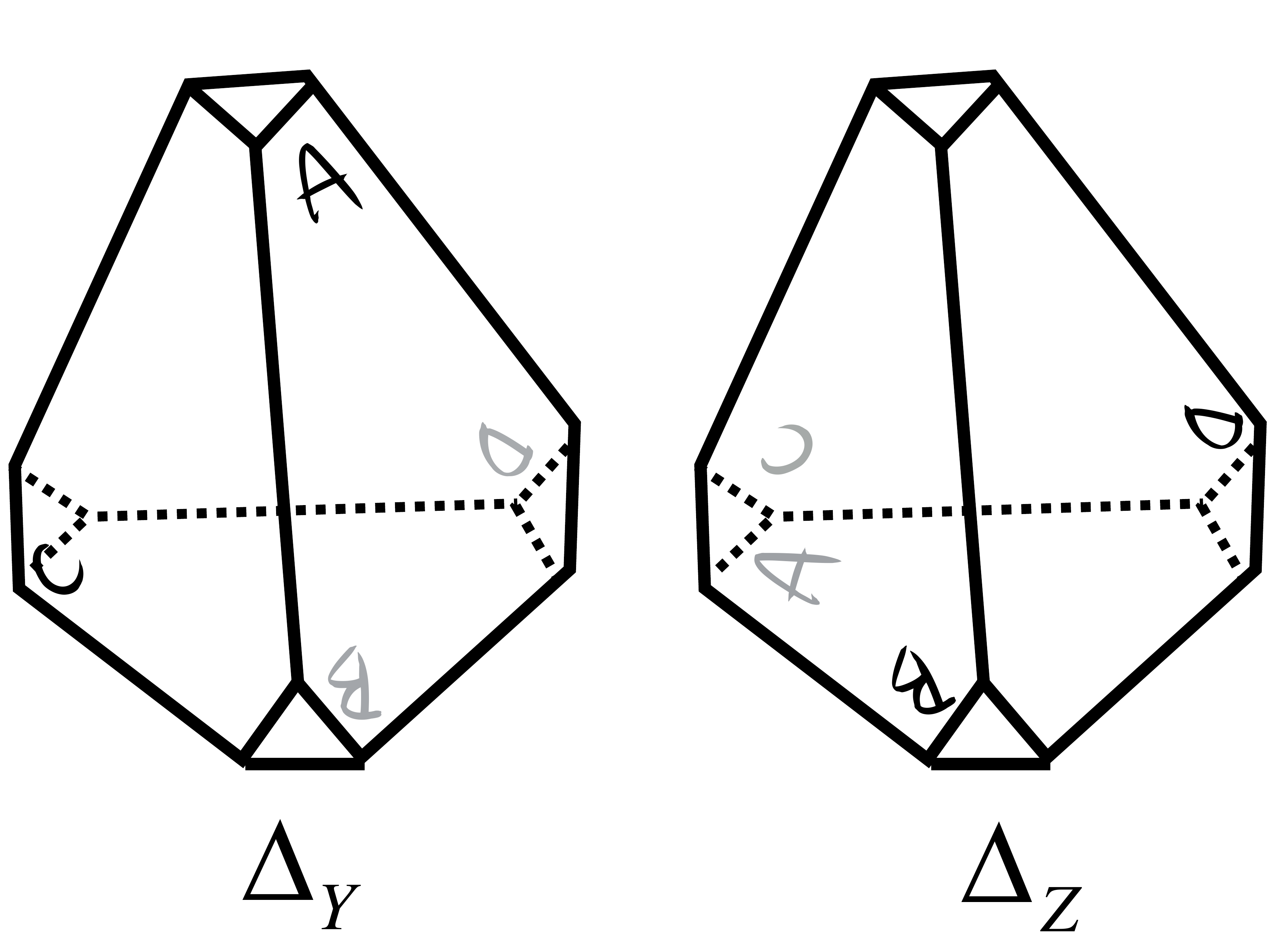}
\caption{The figure-eight knot $\bm{4}_1$ (left). We consider its knot complement in $S^3$, namely $S^3\backslash \bm{4}_1$. An ideal triangulation for the knot complement is drawn (right).}
\label{fig.4_1}
\end{figure}

Let us present an example of our procedure for the figure-eight knot complement $S^3\backslash \mathbf{4}_1$
(Fig.~\ref{fig.4_1}).
While these computations are not completely new, we present this example (and more for $N=3$ next), since
the results will be necessary for comparison with later sections.
Explicit expression for all internal vertex variables for $N$-decomposition of $S^3\backslash \mathbf{4}_1$ are given in eq.~\eqref{internal vertices for figure-eight}. 
The 3-manifold can be obtained by gluing two tetrahedra (which we call $\Delta_Y$ and $\Delta_Z$)
with following gluing datum
\begin{align}
&C= Y'+2Y+Z'+2Z- 2\pi i- \hbar\;, \nn
\\
& \frakL = Z-Z''\;, \quad \frakM=Z-Y'' \;. \label{gluing for figure-eight}
\end{align}
Under the polarization transformation
\begin{align}
\Pi_{\frakL,C} = \left(\begin{array}{c} \frakL \\  C \\ \hline \frakM \\  \Gamma \end{array}\right) = \left(\begin{array}{cccc}0 & 1 & 0 & -1 \\1 & 1 & -1 & -1 \\ 0 & 1 & -1 & 0 \\0 & 0 &1 & 0\end{array}\right) \left(\begin{array}{c} Y \\  Z \\ \hline Y'' \\  Z''\end{array}\right)= g \cdot \Pi_{Y,Z} \;,
\end{align}
the charge basis transforms as (see eq.~\eqref{basis change under polarization change})
\begin{align}
&\langle (m_\eta, m_c ,e_\eta, e_c);\Pi_{\frakL,C}|  \nn
\\
&=\langle(e_c+m_c-m_\eta, e_c+e_\eta,  e_c, e_c+e_\eta-m_\eta); \Pi_{Y,Z}| \;.\label{LC basis in figure-eight with N=2}
\end{align}
Finally, we obtain the $S^1\times S^2$ partition function:
\begin{align}
\CI^c_{S^3\backslash \mathbf{4}_1;N=2}  (m_\eta,  e_\eta) &= \sum_{e_c} \langle m_\eta, m_c=0 ,e_\eta, e_c;\Pi_{\frakL,C}|  \Diamond^{\otimes 2}\rangle \nn
\\
& = \sum_{e_c} \CI^{c}_{\Diamond}(e_c-m_\eta,e_c) \CI^c_{\Diamond} (e_c+e_\eta,e_c+e_\eta-m_\eta)\;.
\end{align}

\paragraph{Example : $M=S^3\backslash \mathbf{4}_1$  with $N=3$} 
The octahedral decomposition of $S^3\backslash \mathbf{4}_1$ with $N=3$ is studied in sec 7.4 of \cite{Dimofte:2013iv}.
For later comparison, let us rewrite the gluing equations using the labeling in fig.\ref{Labellings}:
\begin{align}
\begin{split}
&C_1=Y''_{0,0,1,0}+Y''_{1,0,0,0}+Y'_{1,0,0,0}+Z''_{0,0,1,0}+Z''_{1,0,0,0}+Z'_{0,0,1,0}-2\pi i -\hbar \;,
\\
&C_2=Y''_{0,0,0,1}+Y''_{0,1,0,0}+Y'_{0,0,0,1}+Z''_{0,0,0,1}+Z''_{0,1,0,0}+Z'_{0,1,0,0} -2\pi i -\hbar\;,
\\
&C_3=Y_{0,1,0,0}+Y_{1,0,0,0}+Z_{0,1,0,0}+Z_{1,0,0,0}+Y'_{0,1,0,0}+Z'_{1,0,0,0} -2\pi i -\hbar\;,
\\
&C_4=Y_{0,0,0,1}+Y_{0,0,1,0}+Z_{0,0,0,1}+Z_{0,0,1,0}+Y'_{0,0,1,0}+Z'_{0,0,0,1}-2\pi i -\hbar \;,
\\
&C_5=Y_{0,1,0,0}+Z_{1,0,0,0}+Y''_{0,0,1,0}+Y'_{1,0,0,0}+Z''_{0,0,0,1}+Z'_{0,1,0,0}-2\pi i -\hbar \;,
\\
&C_6=Y_{0,0,1,0}+Z_{0,0,0,1}+Y''_{0,1,0,0}+Y'_{0,0,0,1}+Z''_{1,0,0,0}+Z'_{0,0,1,0}-2\pi i -\hbar \;,
\\
&C_7=Y_{1,0,0,0}+Z_{0,0,1,0}+Y''_{0,0,0,1}+Y'_{0,1,0,0}+Z''_{0,1,0,0}+Z'_{0,0,0,1}-2\pi i -\hbar \; ,
\\
&C_8 =Y_{0,0,0,1}+Z_{0,1,0,0}+Y''_{1,0,0,0}+Y'_{0,0,1,0}+Z''_{0,0,1,0}+Z'_{1,0,0,0} -2\pi i -\hbar\;,
\\
& \frakL_1 =Z_{0,0,0,1}+Z_{1,0,0,0}-Z''_{0,0,1,0}-Z''_{0,1,0,0}\;, 
\\
&\frakL_2= Z'_{0,0,0,1}+Z'_{1,0,0,0}-Z'_{0,0,1,0}-Z'_{0,1,0,0}\;,
\\
&\frakM_1 =\frac{1}{2} \left(-Y_{0,0,0,1}-Y_{0,0,1,0}-Y_{0,1,0,0}-Y_{1,0,0,0}+Z_{0,0,0,1}+Z_{0,0,1,0}+Z_{0,1,0,0}+Z_{1,0,0,0}\right)\; ,
\\
&\frakM_2=\frac{1}{2} \left(-Y_{0,0,1,0}-Y_{0,1,0,0}+Z_{0,0,0,1}+Z_{1,0,0,0}\right)\;.
\end{split}
\label{octahedron's gluing for 4_1 with N=3}
\end{align}
From these gluing equations, 
the index \eqref{index using state-integral from ideal triangulation} for $T_{N=3}[S^3\backslash \mathbf{4}_1]$ can be written as 
\begin{align}
&\CI^c_{S^3\backslash \mathbf{4}_1;N=3}  (m_1, m_2, e_1, e_2) \nn
\\
& = \sum_{e_3, e_4,e_5,e_6,e_7, e_8} (-q^{\frac{1}{2}})^{e_3+e_4-e_5+m_1+m_2} \nn\\
& \quad \times \CI_{\Diamond}^{c}(e_7, e_3+e_6-e_7) \, \CI_{\Diamond}^{c}(-e_1+e_7, -e_2-e_3-e_7+e_8)  \nn
\\
&\quad \times \CI_{\Diamond}^{c}(-e_2-e_4-e_6+e_8, -e_1+e_7-e_8) \, \CI_{\Diamond}^{c}(-e_2-e_3+e_5-e_7-m_1, -e_1+e_6) \nn
\\
&\quad  \times \CI_{\Diamond}^{c}(-e_4-e_7-m_2, e_4+e_8+m_2) \, \CI_{\Diamond}^{c}(e_6-e_8-m_2, e_4-e_5-e_6+e_7+m_2) \nn
\\
&\quad \times \CI_{\Diamond}^{c}(-e_1+e_6-e_8-m_2, -e_2-e_4+e_5-e_6) \nn
\\
& \quad \times \CI_{\Diamond}^{c}(-e_3+e_5-e_6+e_8-m_1+m_2, e_3-e_5-e_8+m_1-m_2)\;.
\end{align}

%%%%%%%%%%%%%%%%%%%%%%%%%%%%%%%
\subsection{Co-dimension 4 Defects}\label{sec.codim_4_state}

Let us next consider co-dimension 4 defects in the state-integral model. As in Sec.~\ref{sec : co-dimension 4},
we consider a Wilson loop $W_{R} (\CK)$ in $SL(N)$ CS theory on a knot complement $M=\hat{M}\backslash K$  along a knot  $\CK$ in a representation $R$ (see eq.~\eqref{path-integral representation of wilson loop VEVs}).
In this paper, we choose to be the representation of $SL(N)$, obtained by naturally complexifying a
finite-dimensional representation of $SU(N)$.

We can also consider anti-holomorphic Wilson line operator by replacing $\mathcal{A}$ by $\overline{\mathcal{A}}$
in the exponent.

Recall that we have two `knots', one being the original knot $K$ defining the knots complement, 
and another the newly-added defect (knot) represented by $\mathcal{K}$. Note that the $K$ and $\mathcal{K}$ play different roles here, $K$ representing the co-dimension 2 defect and 
$\CK$ the co-dimension 4 (Fig.~\ref{fig:K_CK}).

In this section, we focus on the case when the co-dimension 2 defect along a knot $K \subset \hat{M}$ is maximal. %

%------------------------------------------------------------------------------------------------------
\paragraph{State-integral Model with Loop Operators}

What we wish to achieve here is to generalize the state-integral models, discussed in previous subsections, 
by 
including co-dimension 4 defects.

The basic idea is simple: we insert the Wilson line operator 
\begin{align}
\hat{W}_R(\CK)_\pm=\hat{W}_R(\CK)_\pm(\{\hat{Z}_\pm,\hat{Z}'_\pm, \hat{Z}''_\pm \})
\end{align}
into the partition function of the state-integral model \eqref{state-integral from ideal triangulation}:
\begin{align}
&\langle \hat{W}_R (\CK ) \rangle_\pm (\frakX_\a)=\big{\langle} \frakX_\a, C_I=0 \big{|}   \hat{W}_R (\CK )_\pm \big{|} \Diamond^{\otimes \sharp_{\rm total}}  \big{\rangle}\;.  \label{state-integral model for co-dimension 4 defect}
\end{align}
Here $+$/$-$ sign represents the choice of the  holomorphic/anti-holomorphic Wilson loop.

The remaining problem is to obtain the operator $\hat{W}_R(\CK)$.
Classically, 
the Wilson loop operator $W_R $ can be computed using `3d snake' (see App.~\ref{sec : snakes} as well as sec 4.3 of \cite{Dimofte:2013iv}),
and when represented in terms of vertex variables $\vec{Z}, \vec{Z}'$ and $\vec{Z}''$, we have:
\begin{align}
W_R (\CK)(\{ Z , Z' , Z'' \}) &= \sum_{k}  s_k\exp \left( \sum_{\eta=1}^{\sharp_{\rm total}} p^{(k)}_\eta Z_\eta+q^{(k)}_{\eta} Z'_\eta+r^{(k)}_\eta Z''_\eta \right)\;,
\label{W_R}
\end{align}
with $s_k , N p^{(k)}_\eta, Nq^{(k)}_\eta, N r^{(k)}_\eta \in \mathbb{Z}$. Note that 
the parameters here satisfy constraints
\begin{align}
Z_\eta+Z_\eta'+Z_\eta'' - i \pi =0\;, \quad  e^{Z_\eta''}+e^{-Z_\eta}-1=0\;, \quad C_I (\{ Z, Z', Z''\}) \big|_{\hbar=\tilde{\hbar}=0}=0\;, \label{classical ambiguity of WR}
\end{align}
and eq.~\eqref{W_R} are well-defined only up to these constraints. Note that 
eq.~\eqref{classical ambiguity of WR} implies\footnote{For comparison with literature, $Z'$ and $Z''$ are sometimes exchanged in the literature on hyperbolic geometry.}
\begin{align}
e^{Z_\eta'}=\frac{1}{1-e^{Z_\eta}} \;, \quad
e^{Z_\eta''}=1-e^{-Z_\eta} \;.
\label{ZZZ}
\end{align}

In the classical limit, $q\rightarrow 1$ ($\tilde{q}\rightarrow 1$), we expect that the   operator  $\hat{W}_R(\CK)_+$ ($\hat{W}_R(\CK)_-$) in eq.~\eqref{state-integral model for co-dimension 4 defect} will be equal to the classical expression $W_R(\CK)$. 
\begin{align}
\hat{W}_R(\CK)_{+}|_{q=1} = \hat{W}_R(\CK)_{-}|_{\tilde{q}=1} = W_R (\CK)\;. 
\end{align}
All we need to do is then to quantize the classical expression in eq.~\eqref{W_R}.

Quantization turns out to be highly non-trivial, however. First, in the quantization
procedure there are always ordering ambiguities. 
Second, the classical expression \eqref{W_R} is defined only up to the non-linear constraints \eqref{classical ambiguity of WR},
whose quantization is not automatic.
Third, the classical expression of $W_R(\CK)$ 
depends only on the homotopy class $\gamma$ of the knot $\CK$ inside the 3-manifold $\hatM\backslash K$,
and not on the full isotopy class of the knot $\CK$\footnote{In our notation, $\CK$  denotes a knot (defined by ambient isotopic equivalence) and $\gamma$ denotes a generator in $\pi_1(M)$ (defined by homotopy equivalence). Isotopic equivalence implies homotopy equivalence, but not the other way around. Classically, only homotopy equivalence class of knot is relevant. Indeed, the skein relation does not distinguish between an under-crossing and over-crossing
of a knot
in the classical limit $q=1$. };
by contrast, quantum mechanically we expect that two knots in different isotopy classes, even when the two are in the same homotopy class, 
will give different answers.

In this paper, we specialize to the 
case of knots which originates from knots of the 2d surface.
We can then quantize the loop operators following the procedure which we will explain later in Sec.~\ref{sec.codim_4_cluster}.

One disadvantage of this approach is that some of the loop operators do not come from 2d loop operators, 
and hence cannot be dealt with this method. We also have to assume that there is an underlying 2d surface for our 3-manifold $M$. The most typical case for this is when $M$ is a mapping torus of a 2d surface (see eq.~\eqref{mapping_torus}), as we will encounter many times in the rest of this paper.

One should keep in mind, however, that the restriction on the geometry is actually relatively mild, since
we can realize an arbitrary knot complement in $S^3$, using the formalism of \cite{Terashima:2013fg}.

Also, while such a description of the loop operators covers only a limit class,
we can then appeal to the skein relations of loop operators (see \cite{Coman:2015lna,Tachikawa:2015iba} for recent discussion in 2d),
from which we can recover even broader class of loop operators.
We leave the full exploration of this topic for future work \cite{InProgress2}.

%%%%%%%%%%%%%%%%%%%%%%%%%%%%%%%%%%%%%%%%%
\paragraph{Example: Figure-eight Knot Complement}

\begin{figure}[htbp]
\begin{center}
   \includegraphics[width=.4\textwidth]{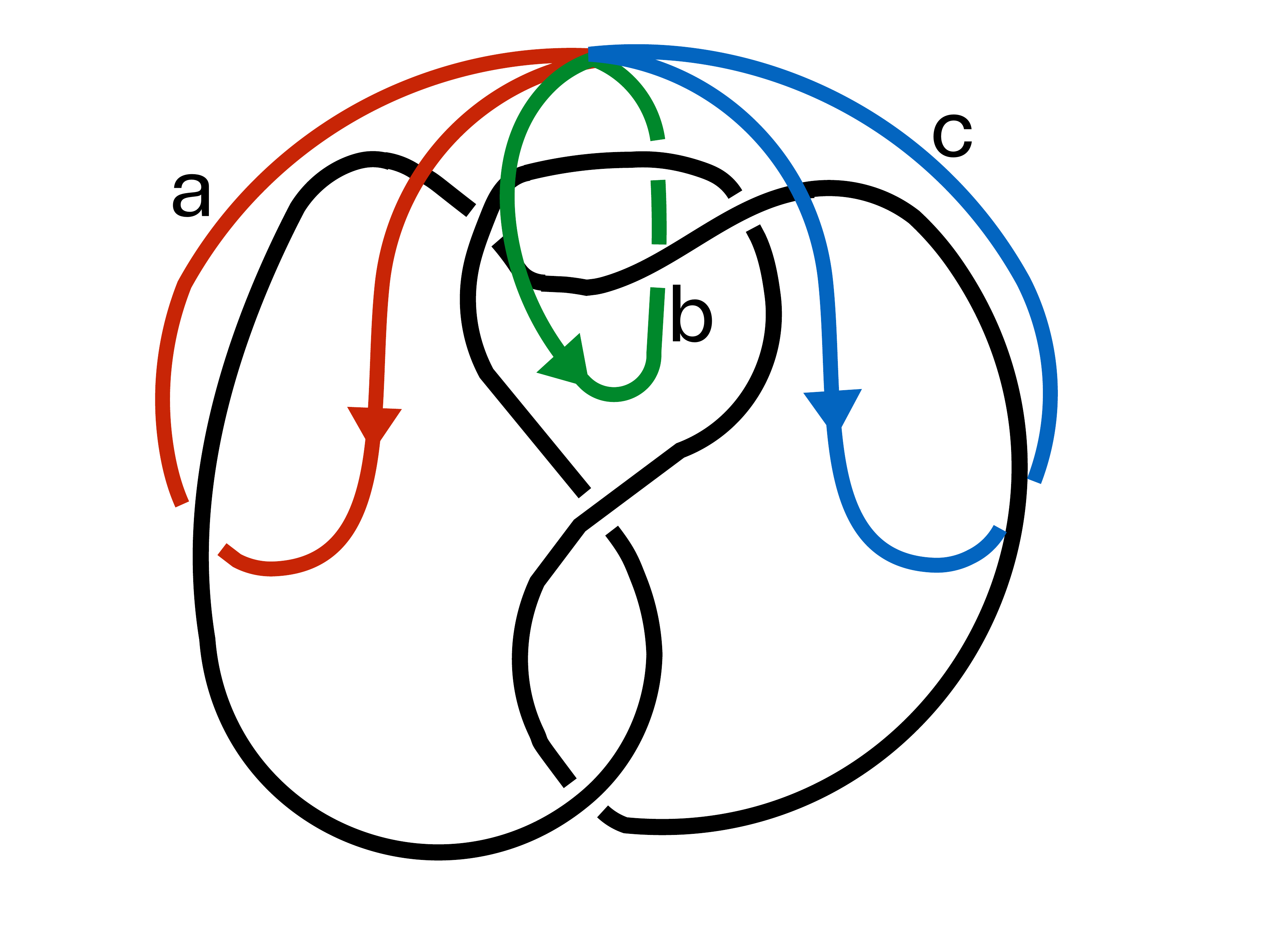}
      \includegraphics[width=.4\textwidth]{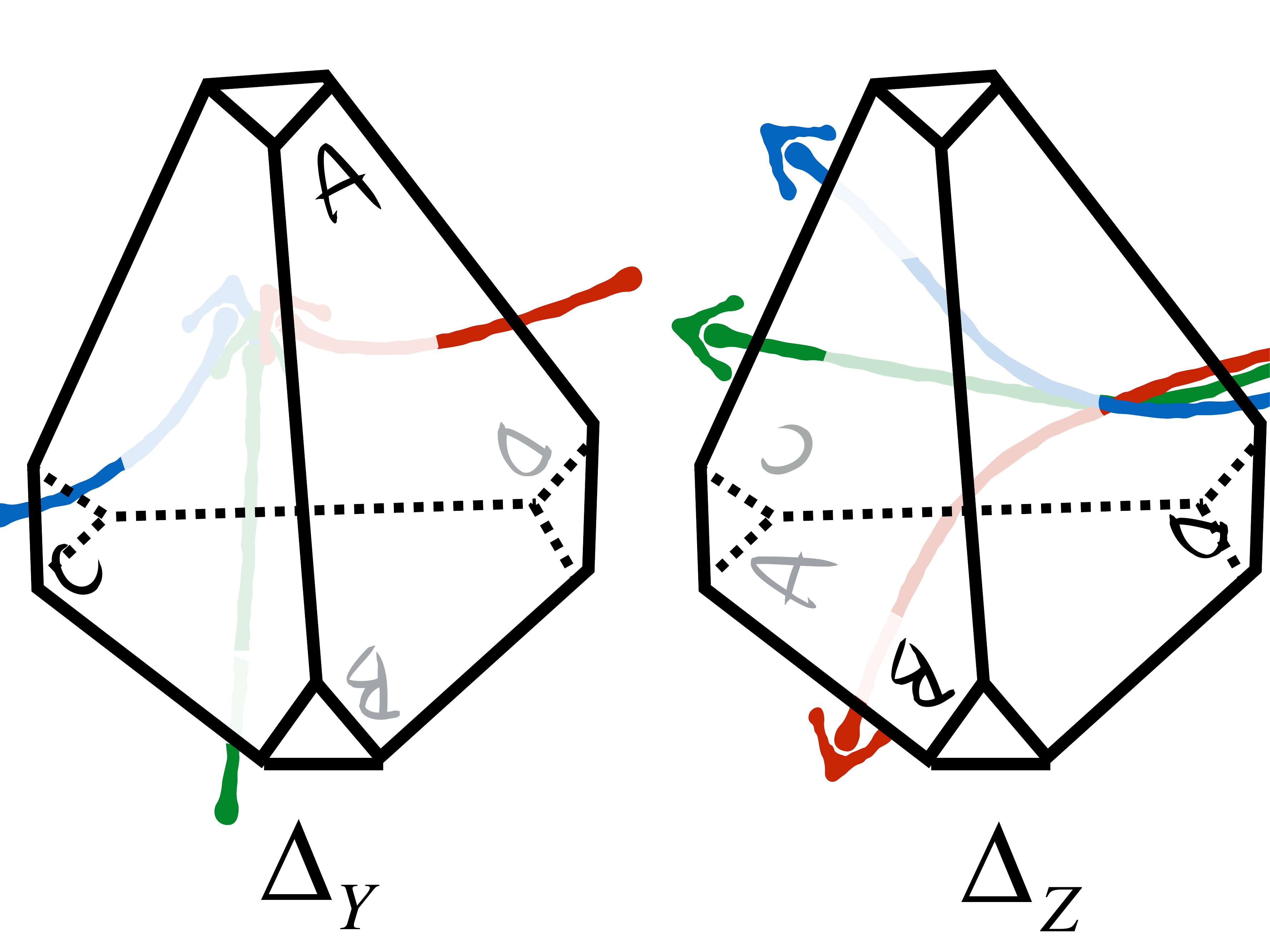}
   \end{center}
   \caption{Three generators $a, b, c$ of $\pi_1 (S^3 \backslash \mathbf{4}_1)$. The paths go through the faces of ideal tetrahedra. \newline \hspace{\linewidth}
    $a$ : (face D in $\Delta_Z$) $\rightarrow$ (A in $\Delta_Z$) $\rightarrow$ (A in $\Delta_Y$) $\rightarrow$ (D in $\Delta_Y$)$\rightarrow$(D in $\Delta_Z$),
    \newline \hspace{\linewidth}
     $b$ : (D in $\Delta_Z$) $\rightarrow$ (B in $\Delta_Z$) $\rightarrow$ (B in $\Delta_Y$) $\rightarrow$ (D in $\Delta_Y$)$\rightarrow$(D in $\Delta_Z$),
     \newline \hspace{\linewidth}
      $c$ : (D in $\Delta_Z$) $\rightarrow$ (C in $\Delta_Z$) $\rightarrow$ (C in $\Delta_Y$) $\rightarrow$ (D in $\Delta_Y$)$\rightarrow$(D in $\Delta_Z$). }
    \label{fig:figure 8 knot}
\end{figure}

Let us again study the example of the figure eight knot complement (Fig.~\ref{fig.4_1}).
The figure eight is often denoted by $\mathbf{4}_1$ (with $\mathbf{4}$ denoting the number of minimal crossings of its 2d projection), 
so its complement is $M= S^3 \backslash \mathbf{4}_1$.
The fundamental group of  $M= S^3 \backslash \mathbf{4}_1$ is 
generated by three generators $(a,b,c)$ depicted in Fig.~\ref{fig:figure 8 knot}:
\begin{align}
\pi_1 (M) = \langle a, b,c|\, ac^{-1}ba^{-1}c= bc^{-1}b^{-1}a=1\rangle\;. \label{fundamental group of figure eight knot}
\end{align}
  We can therefore consider co-dimension 4 defects 
along either $a, b$ or $c$.

The moduli space of $SL(N)$ flat connections can be written as
\begin{align}
\mathcal{M}_{\textrm{$SL(N)$ flat}} (S^3\backslash \mathbf{4}_1)
&=\textrm{Hom}(\pi_1(M), SL(N)) / \sim \nn \\
&= \{ A,B,C \in SL(N) :  A C^{-1} B A^{-1} C= B C^{-1} B^{-1}A=1 \}/ \sim\;,
\end{align}
where the equivalence relation is defined by  conjugation of the $SL(N)$.  The $SL(N)$ elements $A,B,C$ represent holonomy of flat connections around the cycles $a,b,c$ respectively.
\begin{align}
A = \textrm{Hol}(a):= P\,  e^{-\oint_a \mathcal{A}  }\;, \quad B = \textrm{Hol}(b):= P\,  e^{- \oint_b \mathcal{A}  }
\;, \quad C = \textrm{Hol}(c):= P\,  e^{-\oint_c \mathcal{A}  }\;.
\end{align} 
 
For a hyperbolic 3-manifold $M$, there are $SL(N)$ flat connections $\CA_{N}^{\textrm{geom}}$ and $\CA_{N}^{\textrm{conj}}$  which can be constructed using the hyperbolic structure on $M$:
\begin{align}
\mathcal{A}^{\textrm{conj}}_{N}:=\left(\mathcal{A}_{N}^{\textrm{geom}}\right)^\dagger :=\left([N]\cdot (\omega + i e)\right)^\dagger \;. \label{Flat connections from hyperbolic structure}
\end{align}

Let us first consider the complete hyperbolic structures.
This means we consider flat connections which satisfy the boundary condition in eq.~\eqref{co-dimension 2 defects as b.c. in CS theory-2}  with $\rho=\textrm{maximal}$, with unipotent monodromies on the boundary.
Here $\omega$ and $e$ denotes spin-connections and dreibein on $M$ and they form a $PGL(2)$ flat connection $\omega \pm i e$, and the connection can be promoted to an $SL(N)$  flat connection via the $N$-dimensional irreducible representation $[N]$ of $SL(2)$. 
For the figure-eight knot complement, gauge holonomies for $\mathcal{A}^{\textrm{geom}}_{N=2}$ around generators of the fundamental group are given by
\begin{align}
\textrm{Hol}(a)=\left(
\begin{array}{cc}
 1 & 0 \\
 e^{-\frac{i \pi}{3}  } & 1 \\
\end{array}
\right)\;, \quad  \textrm{Hol}(b)=\left(\begin{array}{cc}1& -1 \\ e^{- \frac{i\pi}3} & e^{\frac{i\pi}3}\end{array}\right)\;, \quad \textrm{Hol}(c)=\left(\begin{array}{cc}\sqrt{3} \, e^{- \frac{i\pi} 6} & e^{ \frac{2i \pi}{3}} \\ e^{ -\frac{i \pi}{3}} & e^{ \frac{i \pi}{3}}\end{array}\right)\;.
\label{Hol_abc_complete}
\end{align}
The complex length $\ell_{\mathbb{C}}$ of a 1-cycle $\gamma$ can be defined  as
\begin{align}
\textrm{Hol}(\gamma) (\CA^{\textrm{geom}}_{N=2})  \sim \left(\begin{array}{cc}e^{\frac{1}2 \ell_\mathbb{C} (\gamma)} & * \\0 & e^{-\frac{1}2 \ell_\mathbb{C} (\gamma)}\end{array}\right) \label{complex length}
\end{align}
up to a sign, and the absolute value of its real part $\ell(\gamma):= |\textrm{Re}\{ \ell_{\mathbb{C}}\}|$ of the complex length $\ell_{\mathbb{C}}$ is the hyperbolic length of the cycle, the length computed in a unit hyperbolic metric.
For example, listing the complex length for  the first few cycles with shortest (yet non-vanishing) length:
\begin{align}
\begin{split}
&\ell_{\mathbb{C}}(b) = 1.087070+1.722768 i\;,\; \ell_{\mathbb{C}}(a^{-1}b) = 1.087070-1.722768 i\;, \;
\\
&\ell_{\mathbb{C}}(ab) =\ell_{\mathbb{C}} (b^2 c^{-1})=1.662886+2.392124 i\;, \\
&\ell_{\mathbb{C}}(a^2 c^{-1}) =\ell_{\mathbb{C}} (c a^{-1}c)=1.662886-2.392124 i\;, \ldots \;.
\end{split}
\end{align}
%
%---------------------------------------------------------------------------------------------------

We can reproduce these results for $N=2$
from the the `3d snake' rule of App.~\ref{sec : snakes},
which gives the holonomies in terms of the octahedron variables.
The result of the computation, described in App.~\ref{sec : snakes}, gives
\begin{align}
\begin{split}
&\textrm{Hol} (a)= \left(
\begin{array}{cc}
 \sqrt{\frac{y''}{z}} & 0 \\
 \frac{-1+y''}{\sqrt{y'' z}} & \sqrt{\frac{z}{y''}} \\
\end{array}
\right) \;,
\\
&\textrm{Hol}(b)= \left(
\begin{array}{cc}
 \sqrt{\frac{z'}{y'}} &- \sqrt{\frac{z'}{y'}} \\
 \sqrt{y' z'} & \, -\sqrt{\frac{y'}{z'}}\left(z'-1\right) \\
\end{array}
\right) \;,
\\
&\textrm{Hol} (c)= \left(
\begin{array}{cc}
 \frac{y+z''-1}{\sqrt{y z''}} & \frac{1-y}{\sqrt{y z''}} \\
 \frac{z''-1}{\sqrt{y z''}} & \frac{1}{\sqrt{y z''}} \\
\end{array}
\right) \;.
\end{split}
\label{Hol_abc_y}
\end{align}
We can verify that 
these holonomies give a representation of $\pi_1(S^3\backslash \mathbf{4}_1)$ modulo the classical equations in eq.~\eqref{classical ambiguity of WR}.
Note that the variables $y, y', y''$ and $z, z', z''$ satisfy the gluing constraints of eq.~\eqref{gluing for figure-eight}
\begin{align}
y^2 y' z^2 z' =1 \;,\quad
e^{\frakL}=z z''^{-1} \;, \quad e^{\frakM}=z y''^{-1} \;,
\label{fig8_gluing_1}
\end{align}
as well as (recall eq.~\eqref{ZZZ})
\begin{align}
y'=\frac{1}{1-y} \;, \quad y''=1-y^{-1}\;, \quad z'=\frac{1}{1-z} \;, \quad z''=1-z^{-1} \;.
\label{fig8_gluing_2}
\end{align}

For $\frakL=\frakM=0$, these equations can be solved by
$y=y'=y''=z=z'=z''=e^{-\frac{\pi i}{3}}$,\footnote{This solution corresponds to the geometric flat connection $\CA_{N=2}^{\textrm{geom}}$. Another solution can be obtained by replacing $e^{-\frac{\pi i}{3}}$ by $e^{\frac{\pi i}{3}}$. The latter corresponds to the  conjugate flat connection $\CA_{N=2}^{\textrm{conj}}$. In general, the solution of gluing equations  with $\textrm{Im}[Z]\in (-\pi,0) +2 \pi  \mathbb{Z}$ for all vertex variables $Z$  corresponds  to $\CA_{N=2}^{\textrm{geom}}$. For non-hyperbolic case, there is no solution satisfying the angle conditions. For hyperbolic cases, the solution is unique if it exists and the existence  depends on the ideal triangulation $\CT$. } and the holonomies
eq.~\eqref{Hol_abc_y} reduce to eq.~\eqref{Hol_abc_complete}.

For the $\gamma=c^{-1} a$ and $b^{-1}$, we compute the Wilson line in the fundamental representation to be
\begin{align}
\begin{split}
&\textrm{Tr} \left[ \textrm{Hol}(c^{-1} a) \right]=\textrm{Tr}\left[\textrm{Hol} (c)^{-1}\, \textrm{Hol} (a)\right]= e^{Y''}+e^{-Y''}+e^{-Y''+Z+Z''}\;,
\\
&\textrm{Tr}\left[ \textrm{Hol}(b^{-1}) \right]=\textrm{Tr}\left[\textrm{Hol} (b)^{-1}\right]= e^{-Y+Z''}+e^{Y''-Z}+e^{-Y-Z}\;,
\end{split}
\end{align}
where we used eqs.~\eqref{fig8_gluing_1} and \eqref{fig8_gluing_2}.
We will argue in Sec.~\ref{sec.codim_4_cluster} that  the Wilson loop along the unknot\footnote{\label{foot.unknot}
By ``the unknot in the homotopy class $\gamma$'' we mean a knot inside $M=\hatM\backslash K$
which is $\gamma$ as a homotopy class and which is an unknot inside the ambient manifold $\hatM$,
namely an unknot when the co-dimension $2$ defect $K$ is removed.} $\CK_{c^{-1}a}$ and $\CK_{b}$ in this homotopy class  should be quantized as
\begin{align}
\begin{split}
& \hat{W}_{\Box}(\CK_{c^{-1}a})_+  =  \hat{W}_{\Box}(\CK_{a^{-1} c})_+ \simeq q^{\frac{1}{4}} \left( e^{\hat{Y}''} + e^{-\hat{Y}''}+ e^{\hat{Z}}e^{-\hat{Y}''+\hat{Z}''} \right) \;,
\\
& \hat{W}_{\Box}(\CK_{b^{-1}})_+  =  \hat{W}_{\Box}(\CK_{b})_+  \simeq q^{-\frac{1}{4}}\left( e^{-\hat{Y}}e^{\hat{Z}''} + e^{-\hat{Z}}e^{\hat{Y}''}+ e^{-\hat{Y}-\hat{Z}} \right)\;,
\end{split}
\label{Quantum loop for cIa}
\end{align}
where the equivalence relation $\simeq$ between 3d loop operators is defined by 
\begin{align}
\hat{O}\simeq \hat{O}' \quad\textrm{ if } \quad\langle C_I=0| \hat{O}|\Diamond^{\otimes L}\rangle  = \langle C_I=0| \hat{O}'|\Diamond^{\otimes L}\rangle \;. \label{equivalence of 3d loops}
\end{align}
Note that for our partition functions only the equivalence class matters. We only give explicit quantization for holomorphic Wilson loops, since quantization for the anti-holomorphic case is similar by replacing $(q,e^{\hat{Y}},e^{\hat{Z}})$ by $(\tilde{q}, e^{\hat{\bar{Y}}},e^{\hat{\bar{Z}}})$.
The partition function of the state-integral model with a holomorphic Wilson loop \eqref{state-integral model for co-dimension 4 defect} is then given by 
\begin{align}
\begin{split}
& \big{\langle}  \hat{W}_\Box (\CK_{c^{-1} a} ) \big{\rangle}_+ (m_\eta , \eta)
\\
&=\big{\langle} \frakL=\frac{\hbar}2 m_\eta+\log \eta, C= \frac{\hbar}2 m_c+\log u_c =0 \big{|}   \hat{W}_\Box (\CK_{c^{-1} a} )\big{|} \Diamond^{\otimes 2}  \big{\rangle} 
\\
&=\sum_{e_\eta,e_c} \eta^{e_\eta} \big{ \langle} m_\eta, m_c=0,e_\eta , e_c;\Pi_{\frakL,C} \big{|} \, q^{\frac{1}4}  ( e^{\hat{\frakP}_{1}} +e^{-\hat{\frakP}_{1}} + e^{ \hat{\frakX}_{2}} e^{ -\hat{\frakP}_{1}+\hat{\frakP}_{2}} )   \big{|}\Diamond^{\otimes 2} \big{\rangle} 
\\
& =\sum_{e_\eta,e_c}  \eta^{e_\eta}\big{ \langle}(e_c-m_\eta, e_c+e_\eta,  e_c, e_c+e_\eta-m_\eta); \Pi_{Y,Z}  \big{|} \, q^{\frac{1}4}  ( e^{\hat{\frakP}_{1}} +e^{-\hat{\frakP}_{1}} + e^{ \hat{\frakX}_{2}} e^{ -\hat{\frakP}_{1}+\hat{\frakP}_{2}} ) \big{|}\Diamond^{\otimes 2} \big{\rangle}
\\
&= q^{\frac{1}4}\, \sum_{e_\eta,e_c}   \eta^{e_\eta} \bigg{[} \CI^{c}_{\Diamond}(e_c-m_\eta+1,e_c)\, \CI^c_{\Diamond} (e_c+e_\eta,e_c+e_\eta-m_\eta)\, q^{\frac{e_c}{2} }
\\
&\qquad \qquad \qquad + \CI^{c}_{\Diamond}(e_c-m_\eta-1,e_c)\, \CI^c_{\Diamond} (e_c+e_\eta,e_c+e_\eta-m_\eta)\, q^{-\frac{e_c}{2} }\;.  
\\
& \qquad \qquad \qquad  +\CI^{c}_{\Diamond}(e_c-m_\eta-1,e_c)\, \CI^c_{\Diamond} (e_c+e_\eta+1,e_c+e_\eta-m_\eta-1)q^{\frac{e_c+2e_\eta-m_\eta-1}2}\; \bigg{]}.  
\end{split}
\end{align}
Here we use eq.~\eqref{LC basis in figure-eight with N=2}. In the computation, $(\hat{\frakX}_i,\hat{\frakP}_i)_{i=1,2}$ denote position/momentum operators in $\Pi_{Y,Z}$ and its action on the charge basis can be obtained using eq.~\eqref{(X,P) on charge basis}. Repeating the similar calculation, we have
\begin{align}
\begin{split}
& \big{\langle}  \hat{W}_\Box (\CK_{b} ) \big{\rangle}_+(m_\eta, \eta) =
\\
&= q^{-\frac{1}4}\, \sum_{e_\eta,e_c}   \eta^{e_\eta} \bigg{[} \CI^{c}_{\Diamond}(e_c-m_\eta,e_c+1)\, \CI^c_{\Diamond} (e_c+e_\eta+1,e_c+e_\eta-m_\eta)\, q^{\frac{e_\eta}{2} }
\\
&\qquad \qquad \qquad + \CI^{c}_{\Diamond}(e_c-m_\eta+1,e_c)\, \CI^c_{\Diamond} (e_c+e_\eta,e_c+e_\eta-m_\eta+1)\, q^{-\frac{e_\eta}{2} }\;.  
\\
& \qquad \qquad \qquad  +\CI^{c}_{\Diamond}(e_c-m_\eta,e_c+1)\, \CI^c_{\Diamond} (e_c+e_\eta,e_c+e_\eta-m_\eta+1)q^{\frac{-e_\eta-2e_c+m_\eta}2}\; \bigg{]}.  
\end{split}
\end{align}
For later use, we list several first orders in the $q$-expansion
\begin{align}
\begin{split}
&\big{\langle} \hat{W}_\Box (\CK_{c^{-1}a} ) \big{\rangle}_+ (m_\eta=0, \eta) 
\\
&=  - \left(\eta+\frac{1}{\eta}\right)q^{\frac{3}{4}}-3\, q^{\frac{5}{4}}-\left(\eta+\frac{1}{\eta}\right)q^{\frac{7}{4}}+\left(-1+\eta^2+\frac{1}{\eta^{2}}\right)q^{\frac{9}{4}} +3\left(\eta+\frac{1}{\eta}\right) q^{\frac{11}{4}}+\ldots  \;.
\\
&\big{\langle} \hat{W}_\Box (\CK_{b} ) \big{\rangle}_+ (m_\eta=0, \eta) 
\\
&=q^{-\frac{1}4} -3\, q^{\frac{3}4} +\left( -\frac{3}\eta - 3\eta \right) - 6\, q^{\frac{7}4} +\left(-\frac{2}\eta - 2\eta\right)q^{\frac{9}4}+\left(- 1 +\frac{1}{\eta^2}+\eta^2 \right)q^{\frac{11}4}+\dots
\end{split}
 \label{Simple wilson loop in figure eight}
\end{align}
Later in Sec.~\ref{sec.TSUN_codim_4} we will compare our answer 
with an independent computation from  $T[SU(N)]$ theory.

%%%%%%%%%%%%%%%%%%%%%%%%%%%%%%%%%%%%%%%%
\section{From Cluster Partition Function} \label{sec : cluster partition function}
%%%%%%%%%%%%%%%%%%%%%%%%%%%%%%%%%%%%%%%%%

We next come to one of the central materials of this paper, the discussion of cluster partition functions.
We first present the general expression of our cluster partition function, and then explain how that is related to the
discussion of 3-manifolds. We will finally work out explicit examples.

%%%%%%%%%%%%%%%%%%%%%%%%%%%%%%%%%%%%%%%%%%%%%
\subsection{General Formula}

Let us first summarize our results for the cluster partition function,
building on and generalizing the result of \cite{Terashima:2013fg}.
Since the derivation is technically involved, we present the derivation in the App.~\ref{app.cluster_derivation}.

We shall present our result for the case of the $(S^3)_b$ partition function which corresponds to  $k=1$,
in order to make contact with the results of \cite{Terashima:2013fg}. The formula however
easily generalizes to other cases in eq.~\eqref{Quantization for several (k,s)}.

We use the formula for the `trace' of the cluster partition function (an expression
before giving a trace is also given in App.~\ref{app.cluster_derivation}; we here give only the minimal results needed for the computations of examples in this paper).
This is determined from a quiver (an oriented graph) $Q$, represented by an 
anti-symmetric matrix $(Q_{i,j})$
changing from $Q(0):=Q$ to $Q(L)$. We denote the number of vertices of $Q(t)$ by $|Q|$,
which is actually independent of $t$.
This change of the quiver is prescribed by the so-called 
mutations of the quiver at vertices $\bm{m}=(m_0, \cdots, m_{L-1})$ of $Q$,
as well as a sequence of permutations $\bm{\sigma}=(\sigma_0, \ldots, \sigma_{L-1})$.\footnote{
There is an unfortunate conflict of notations where: $\sigma$ was also used for the Chern-Simons level in eq.~\eqref{complex CS action}. The $\sigma$ for the permutation is either written in bold ($\bm{\sigma}$) or has an index (as in $\sigma_i$).
We hope that no confusion arises from this.
}
Then the assumptions we make is that after all the mutations and permutations, the quiver $Q(L)$ comes back to $Q(0)$  (see App.~\ref{app.cluster_derivation} for details of 
definitions).

Our trace of the cluster partition function is a trace of an operator in a quantum mechanical Hilbert space constructed from $Q, \bm{m}$ and $\bm{\sigma}$.
Its Fourier transform has an integral expression:%
\begin{align}
&\mathrm{F.T.}\left[\textrm{Tr}_{Q,\bm{m}, \bm{\sigma}} \right](\bfrakM) =  \int \left[ \prod_{t=0}^{L-1} d\vec{u}(t) dZ(t) dZ''(t) \right]  \prod_{t=0}^{L-1}\psi_{\hbar} \big{(}Z(t) \big{)} e^{-\frac{1}{4\pi i b^2} Z(t) Z''(t)} \; \nn
\\
&  \qquad \qquad  \qquad   \times \delta \left(\hat{C}_{Q,\bm{m}, \bm{\sigma}}\cdot \vec{U}- \vec{V}\right)
\prod_{t=0}^{L-1}\,  \delta \! \left( Z(t)+Z''(t)-2\sum_{i=1}^{|Q|}　Q_{m_t, i}(t)u_i(t) \right)\nn \\
&  \qquad \qquad  \qquad   \times
\prod_{\alpha=1}^{n_c} \delta\left(\sum_{i=1}^{|Q|} c_i^{\alpha}(0) u_i(0)\right) 
\;.
\label{trace_result_1}
\end{align}

Let us explain the notation in eq.~\eqref{trace_result_1}. First, this expression is a function of a set of parameters $\bfrakM=\{ \bfrakM_{\a} \}$,
where $\a$ runs over the set of conserved quantities of the quantum mechanics commuting with `time evolution' generated by $(\bm{m},\bm{\sigma})$. The cluster partition function $\textrm{Tr}_{Q,\bm{m},\bm{\sigma}}$ is a function on the  conserved quantities $\{ \frakL_\a \}$ and the above expression  is its Fourier transformation  \eqref{F.T of cluster ptn}. 
In practice, the number of such $\bfrakM_{\a}$ is given by $n_c$, the number of central elements in the cluster algebra commuting $\hat{\varphi}$ constructed from $(\bm{m},\bm{\sigma})$. 

The notation $\frakL$ comes from the fact that these variables correspond to longitude variables $\frakL$ in the case of our favorite example: $(Q,\bm{m},\bm{\sigma})$ associated to mapping torus $(\Sigma_{1,1}\times S^1)_{\varphi= \bm{L}\bm{R}}$. Note that while the notation $\bfrakM$ is reminiscent of the meridian variable $\frakM$, the two are not the same, $\bfrakM$ can be identified as $-\frakM$ up to an ambiguity of the cluster partition function. The ambiguity   will be studied in sec.~\ref{sec: cluster partition function  for 3-manifold} (see eq.~\eqref{ambiguity in cluster ptn}). The ambiguity shifts $\bfrakM$ by some linear combinations for longitude variables.
\begin{align}
\bfrakM \sim  \bfrakM + (\textrm{some linear combination of $\frakL$})\;. \label{framing ambiguity in bfrakM}
\end{align}
Note that (as we have seen around eq.~\eqref{Dehn_filling}) the concept of the longitude/meridian depends on the choice of the ambient manifold $\hatM$, and 
does not have an intrinsic meaning for a given manifold $M=\hatM\backslash K$. We call meridian/longitude for the our favorite example case viewing the  mapping torus as a knot complement whose ambient manifold is $S^3$. If we view the knot complement as knot complement on a torus bundle $(T^2\times S^1)_{\bm{LR}}$, $\frakL$ should be interpreted as `meridian' variables. Viewing the torus bundle as the ambient space $\hat{M}$, the ambiguity in eq.~\eqref{framing ambiguity in bfrakM} is nothing but the framing ambiguity, ambiguity in the choice of longitude.\footnote{
In $S^3$ or more generally a homology $3$-sphere, we can fix this ambiguity by 
imposing the condition that the total intersection number of the longitude with the knot itself is
zero. Such a canonical choice does not exist, however, for a general 3-manifold.}

\begin{figure}[htbp]
\begin{center}
\centering{\includegraphics[scale=0.3]{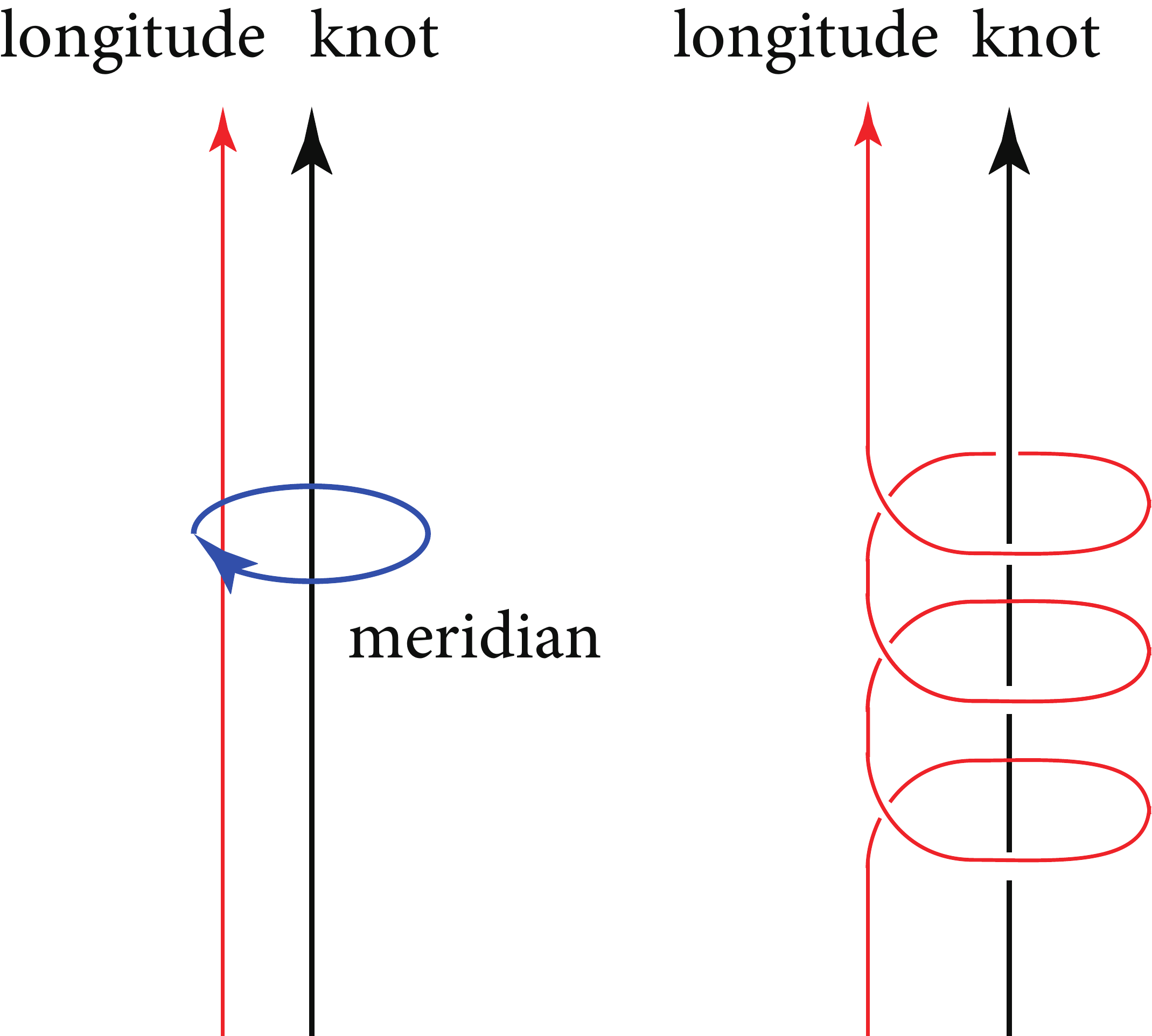}}
\end{center}
\caption{
The framing ambiguity for the longitude; we can add integer multiples of the meridian to the longitude.
When the ambient 3-manifold for the figure eight knot is chosen to be $(T^2\times S^1)_{\bm{LR}}$,
and not $S^3$, the role of the longitude and the meridian is reversed
and hence the framing ambiguity is the ambiguity of the meridian as in eq.~\eqref{framing ambiguity in bfrakM},
the $\bfrakM$ there is meant to be the meridian when the ambient 3-manifold is chosen to be $S^3$.
}
\label{framing_ambiguity}
\end{figure}

The integral is over a set of parameters 
$\vec{u}(t)=\{ u_i(t) \}$, $Z(t)$ and $Z''(t)$\footnote{Our convention of $Z(t), Z'(t), Z''(t)$ here is different from that in \cite{Terashima:2013fg}, to better match 
the notations in the state-integral models.
}, 
with $t$ running over time ($t=0, \cdots, L-1$)
and $i$ running over all the vertices of the quiver $Q$
($i=1, \cdots, |Q|$).
The integrand contains a product of a special function $\psi_{\hbar}(x)$, the quantum dilogarithm function
defined in App.~\ref{app.qdilog}.

The arguments of the delta functions in the second line of eq.~\eqref{trace_result_1} are given by
\begin{align}
&\hat{C}_{Q,\bm{m},\bm{\sigma}} \cdot \vec{U}:=
\left(\begin{array}{cccc}I & 0 & 0 & -\sigma_{L-1}^{-1} \cdot\hat{P}_{m_{L-1}} \\  -\sigma^{-1}_{0}\cdot\hat{P}_{m_0} & I & 0 & \vdots  \\ \vdots & \ddots & \vdots & \vdots \\0 &  \cdots & -\sigma^{-1}_{L-2}\cdot \hat{P}_{m_{L-2}}  & I\end{array}\right) 
\left(\begin{array}{c}\vec{u}(0) \\ \vec{u}(1) \\ \vdots \\ \vec{u}(L-1)\end{array}\right)  \; , \nn
\\
&\vec{V}:=  \left(\begin{array}{c}- \frac{1}2 \sum_{i=1}^{|Q|}  \left(\sum_{\a=1}^{|\textrm{Ker}(Q)|} c_i^\a(0) \bfrakM_\a\right)  \hat{e}_i - \frac{1}2  Z(0) \hat{e}_{m_0} \\ - \frac{1}2 Z(1) \, \hat{e}_{m_1}\\ \vdots \\ - \frac{1}2 Z(L-1)\, \hat{e}_{m_{L-1}}\end{array}\right)\;.
\label{trace_result_2}
\end{align}
These delta functions impose linear constraints among the integration variables.
Here $\hat{P}_{m_k}$ is a linear transformation acting on $\vec{u}(t)=u_i(t)$ (for each fixed $t$, see eq.~\eqref{hatP_def}):
\begin{align}
\hat{P}_{m_k} u_i(t)= 
\begin{cases}
-u_i(t)  & (i=m_t) \\
u_i(t)+ [Q_{i, m_t}(t)]_{+} u_{m_t}(t) & (i\ne m_t)
\end{cases} \ .
\end{align}
We take $e_k$ to be a row vector  $e_k= (\overset{1}{\check{0}}, 0, \cdots, \overset{k}{\check{1}},0,\cdots, \overset{|Q|}{\check{0}})$. Also, $\bm{\sigma}$ inside 
$\hat{C}_{Q,\bm{m},\bm{\sigma}}$ acts linearly on the $u_i(t)$s by changing the subscripts, namely
$\sigma\cdot u_i(t)=u_{\sigma(i)}(t)$. Note that $\hat{C}_{Q, \bm{m}, \bm{\sigma}}$ is of size $|Q|\times |Q|$,
and $\vec{U}, \vec{V}$ of size $|Q| \times L$. 

There is one subtlety in the integral expressions:
the integral is naively divergent since there are flat directions 
in the integral variables such that the integrand is kept invariant,
and we need to mod out such flat directions.
We find the flat directions are given by 
\begin{align}
\delta^{(\alpha)} u_i (t) = c^{\alpha}_i (t)\;, \quad \alpha =1, \ldots, n_c\;, \label{flat-directions}
\end{align}
where $c_i^\alpha (0)$ are vectors spanning the kernel of $Q(0)$ and $c_i^{\alpha}(t)$ is defined recursively by
\begin{align}
c_i^\alpha (t+1) :=\sigma_t^{-1} \cdot \hat{P}_{m_t} \big{ (} c_i^{\alpha}(t)\big{)} \;.
\label{c_alpha_def}
\end{align} 
Since 
we have (see App.~\ref{app.kernel} for explicit proof)
\begin{align}
\sigma_t^{-1}\cdot \hat{P}_{m_t}(\textrm{Ker}\, Q(t)) = \textrm{Ker}(Q(t+1)) \;,
\label{kernel_transf}
\end{align} 
it follows that $c^\alpha(t)  \in \textrm{Ker}(Q(t))$ and the $\delta$-functions $\delta (Z+Z''-\sum_i 2 Q_{m_t,i} u_i)$ is invariant under the flat directions. For other $\delta$-functions $\delta (C \cdot \vec{V}-\vec{U})$, the invariance under the flat directions is manifest except for $\delta \big{(} u(0)- \sigma_{L-1}^{-1}\cdot \hat{P}_{m_{L-1}}(u (L-1)) \big{)}$, whose invariance is also guaranteed from the condition $c^{\alpha}(L)=c^{\alpha}(0)$ (see eq.~\eqref{conditions on (Q,m,sigma) 2} in
App.~\ref{app.cluster_derivation}).
To kill the flat directions, we impose additional $\delta$-functions in the integration; this is exactly what appears in the last line of  eq.~\eqref{trace_result_1}.
%

%%%%%%%%%%%%%%%%%%%%%%%%%%%%%%%%%%%
\subsection{Applications to 3-manifolds} \label{sec: cluster partition function  for 3-manifold}

Having obtained a general formula for the cluster partition function,
the remaining task is to choose $(Q, \bm{m}, \bm{\sigma})$ appropriately, namely to fit the 3-manifold problem,
thereby establishing the link between cluster algebras and 3-manifolds.
Our presentation is a generalization of \cite{Terashima:2011qi,Terashima:2011xe,Nagao:2011aa,Terashima:2013fg}, which we follow closely.\footnote{See \cite{FST1} (\cite{Nagao:2011aa}) for mathematical discussion of cluster algebras in the context of 2-manifolds (3-manifolds).}

\paragraph{Flat Connections and Quivers}
Let us consider the moduli space $\mathcal{M}_{N}(\Sigma,\vec{\rho})$ of $SL(N)$ flat connection
on a Riemann surface $\Sigma$, with the specified holonomy at each of the punctures $p_a$ of type $\rho_a$. We consider a 2d surface $\Sigma_{g,h}$ of genus $g$ with $h$ punctures $\{p_a\}_{a=1}^h$, and we assume $\chi(\Sigma)=2-2g-h<0$.
We can use the same boundary conditions as in 
eqs.~\eqref{co-dimension 2 defects as b.c. in CS theory-1} and \eqref{co-dimension 2 defects as b.c. in CS theory-2},
although here the holonomy is meant to be the holonomy along a puncture in a 2d surface, not along a knot in 3-manifold.
This choice is again labeled by an embedding $\rho_a$ of $SU(2)$ into $SU(N)$,
for each puncture.
\begin{align}
\begin{split}
&\textrm{$\CM_{N}(\Sigma_{g,h}, \vec{\rho}):=$  $\Big\{ $moduli space of $SL(N)$ flat-connections on $\Sigma$}\nn \\
&\qquad\qquad\textrm{with a boundary condition around each puncture $p_a$}\nn\\
&\qquad\qquad\textrm{which is determined by  $(\rho_a, \frakL^{(a)}_{\a})$ as in eq.~\eqref{co-dimension 2 defects as b.c. in CS theory-1}$\Big\}$ }  \;.
\end{split}
\end{align}
More concretely ($P_a:=SL(N)$ holonomy around puncture $p_a$),
 \begin{align}
 &\CM_{N}(\Sigma_{g,h}, \vec{\rho}) = \textrm{Hom}\bigg{[}\pi_1 (\Sigma_{1,1})\rightarrow SL(N): \textrm{with fixed conjugacy class $P_a$ } \nn 
 \\
 &\qquad\qquad  \qquad \qquad \quad  \textrm{ of the form }\eqref{co-dimension 2 defects as b.c. in CS theory-1} \bigg{]}/\sim\;,
 \end{align}
where $\sim$ denotes an equivalence relation defined  by conjugation action of $SL(N)$. The image of $\gamma\in \pi_1 (\Sigma_{g,h})$ under a homomorphism can be thought of as an $SL(N)$ monodromy along $\gamma$ of the flat-connection determined by the homomorphism.
Let us first count the dimension of $\CM_{N} (\Sigma_{g,h}, \vec{\rho}) $ for general $\vec{\rho}$:
\begin{align}
\textrm{dim}_{\mathbb{C}}\CM_{N} (\Sigma_{g,h}, \vec{\rho})  &=  (2g+h)(N^2-1)-(N^2-1)- \sum_{a}\textrm{dim }\mathbb{L}^{(\rho_a)}-(N^2-1)  \nn
\\
&=(2g+h-2) (N^2-1)-  \sum_a \textrm{dim }\mathbb{L}^{(\rho_a)}\;.
\label{Moduli_flat_dim}
\end{align}
Note that the dimension is always even (the middle-dimensional real slice, the moduli space of $SL(N, \mathbb{R})$ flat connection, is already a K\"{a}hler manifold). Let us explain the counting in eq.~\eqref{Moduli_flat_dim}.
The fundamental group for $\Sigma_{g,h}$ is given by $(2g+h)$ generators with one relation.  In the counting, $(2g+h)(N^2-1)$ comes form $(2g+h)$ $SL(N)$ matrices which are image of the generators under the homomorphism. $-(N^2-1)$ comes from one matrix relation  among $(2g+h)$ $SL(N)$ matrices and $-\sum_a \textrm{dim }\mathbb{L}^{(\rho_a)}$  comes from the constraint  fixing the conjugacy class of $P_a$. The last term $-(N^2-1)$ comes from the quotient by $SL(N)$ ($/\sim$).\footnote{This counting is valid only around  generic points in the moduli space where all $SL(N)$ is broken by holonomies,  namely when the centralizer of images of $(2g+h)$ generators is trivial (the flat connection is called irreducible in this case).}

The moduli space $\mathcal{M}_{N}(\Sigma_{g,h}, \vec{\rho})$
is a K\"ahler manifold, with a canonical holomorphic symplectic structured given by 
\begin{align}
\Omega = \frac{1}{\bhbar} \oint_{\Sigma} \delta  \CA  \wedge \delta \CA + \frac{1}{\tilde{\bhbar}} \oint_{\Sigma} \delta  \bar{\CA}  \wedge \delta  \bar{\CA}\;. 
\end{align}
Moreover, the moduli space allows for a nice set of coordinate charts, parametrized by local coordinates $\{Y_i, \bar{Y}_i\}_{i=1}^{\textrm{dim}_\mathbb{C} \CM_N}$.
Namely, the moduli space is a cluster $\mathcal{X}$-variety, meaning that
on each patch we have a constant bilinear form determined from a quiver $Q$, 
\begin{align}
\{ Y_i, Y_j\}_{P.B.} =   \bhbar \, Q_{ji}\;, \quad \{ \bar{Y}_i, \bar{Y}_i\}_{P.B.} =  \tilde{\bhbar} \, Q_{ji}\;.
\end{align}
and the coordinate transformation between different patches are given by the 
transformation rules of the so-called $y$-variables of the cluster algebra (see eq.~\eqref{classical mutation}). While this general story is expected to be true for any $N$ and $\vec{\rho}$,
it is a non-trivial problem to work out the explicit cluster coordinates on the moduli space.
In the literature, the known constructions are primarily for the case where 
all the $\rho_a$ are the maximal punctures \cite{2003math.....11149F}. Naming after its inventor, the local coordinates for maximal punctures  are called Fock-Goncharov (FG) coordinates and the corresponding  quiver $Q$ is called FG quiver. 
The quiver is determined from an ideal triangulation of the 2d surface $\Sigma_{g,h}$. For more general punctures, there are very few papers (see e.g.\ \cite{Xie:2012dw}).
The quiver $Q$ associated to $\CM_{N,\vec{\rho}}(\Sigma_{g,h})$  is expected to have the following properties
\begin{align}
\begin{split}
|Q|& =\textrm{dim}_{\mathbb{C}}\CM_{N} (\Sigma_{g,h},\vec{\rho})  +\sum_{a=1}^h \textrm{rank}(H_{\rho_a})   
\\
&=(2g+h-2) (N^2-1)-  \sum_{a=1}^h \left( \textrm{dim }\mathbb{L}^{(\rho_a)}-\ell(\rho_a)\right)-h\;,  
\\
|\textrm{Ker}(Q)|&= \sum_{a=1}^h \textrm{rank}(H_{\rho_a}) = \sum_{a=1}^h \ell(\rho_a) -h\;.
\end{split} \label{Quiver-counting}
\end{align}
Here $|Q|$ the size of the square matrix and $|\textrm{Ker}(Q)|$ is the dimension of kernel of the matrix $Q$. Central elements of the cluster algebra $\CA_Q$ can be identified as distinct eigenvalues, $\frakL^{(a)}_\a$, of fixed holonomies around punctures. 
This is why we identify $|\textrm{Ker}(Q)|$ as $\sum_{a=1}^h \textrm{rank}(H_{\rho_a})$. Note that $\textrm{rank}(H_\rho)$ counts  independent parameters of fixed boundary holonomy of type $\rho$, see eq.~\eqref{co-dimension 2 defects as b.c. in CS theory-1}.
  More physically, given $\Sigma$ as well as $\vec{\rho}$ at punctures,
we can consider compactifications of 6d $(2,0)$ theory on $\Sigma$, giving rise to 
4d $\mathcal{N}=2$ theories $T_N[\Sigma]$. The moduli space 
$\mathcal{M}_{N}(\Sigma,\vec{\rho})$
is then the Coulomb branch of the 4d $\mathcal{N}=2$ theory compactified on $S^1$,
and the coordinates (cluster $y$-variables) are the identified with the 
VEV of the IR line operators therein \cite{Gaiotto:2009hg}.

\paragraph{Mapping Class Group and Mutations} 
We next describe the geometric meaning of $\bm{m}$ (and $\bm{\sigma}$):
they describe the action of the mapping class group.

The mapping class group (MCG) of a Riemann surface $\Sigma_{g,h}$ induces a sequence of flips on ideal triangulation on the Riemann surface.   More physically the MCG  corresponds to  an action of the generalized S-duality group of the 4d $\mathcal{N}=2$ theory $T[\Sigma]$ \cite{Gaiotto:2009we}.
On the  cluster  coordinates, flips of the ideal triangulation can be represented as a sequence of mutations and permutations. Classically,  a mutation $\mu_k$ on the $k$-th node in the quiver $Q$ induces the following transformation on the cluster coordinates $y_i:=e^{Y_i}$
\begin{align}
\mu_k\; :\; y_{i} \rightarrow y_{i} \, y_{k}^{\textrm{max}(Q_{ik},0)} \left(1 +y_{k}^{-1}\right)^{Q_{ik}}\;. \label{classical mutation}
\end{align}
By quantizing the moduli space $\mathcal{M}_{N}(\Sigma,\vec{\rho})$, we obtain a Hilbert space $\CH^{(k,\sigma)} (\Sigma,\vec{\rho})$ which depends on the quantum parameters $(h, \tilde{\hbar})$ (or equivalently $(k,\sigma)$, see eq.~\eqref{Quantization for several (k,s)}. An element $\varphi \in $ MCG$(\Sigma)$ is promoted to a linear operator $\hat{\varphi}$ acting on the Hilbert space after the quantization and it  gives a projective representation of MCG$(\Sigma)$.
Let us more explain about the projectivity of the representation. The quantized operators depends on fixed central elements $\frakL^{(a)}_\a$, which are related to fixed holonomy $P_a$ around the $a$-th puncture:
\begin{align}
\hat{\varphi} (\frakL^{(a)}_\a)  \quad \textrm{or} \quad \hat{\varphi}([P_a])\;.
\end{align}
Here $[P_a]$ denotes $SL(N)$ conjugacy class of $P_a$. Let $\{ \varphi_n\}$ be generators of 
the MCG
\begin{align}
\textrm{MCG} =\big{\langle} \{\varphi_n\} : \textrm{relations among generators of the form  $\prod_{i} (\varphi_{n_i})^{c_i}=1$}   \big{\rangle}\;.
\end{align}
Then, we propose that  the projectivity of the representation take following form in general
\begin{align}
\prod_{i} (\hat{\varphi}_{n_i})^{c_i}= \textrm{exp} \bigg{(}  \sum_{a=1}^{h} \frac{n_a} \CN \left(\frac{\textrm{Tr}(\log P_a)^2}{2\hbar}+\frac{\textrm{Tr}(\log P^\dagger_a)^2}{2\tilde{\hbar}}\right) \bigg{)} \;, \quad \{n_a\}\in \mathbb{Z}^{h}\;, \label{projectivity of MCG reps}
\end{align}
with a proper positive integer $\CN$ which might depend only on $N$.
In App.~\ref{app.rho_simple}, we will explicitly confirm the projectivity for $\Sigma_{0,1}$ with $N=3$ and $\rho=\textrm{`simple'}$. We leave the proof of the proposal for general case as future work. 

The quantum operator $\hat{\varphi}$ can be written as product of  quantum mutation operators $\hat{\mu}_k$ \eqref{mu_k} (quantization of \eqref{classical mutation}) and permutation operators $\hat{\sigma}_t$.

\paragraph{Mapping Tori}

\begin{figure}[htbp]
\begin{center}
\includegraphics[scale=0.2]{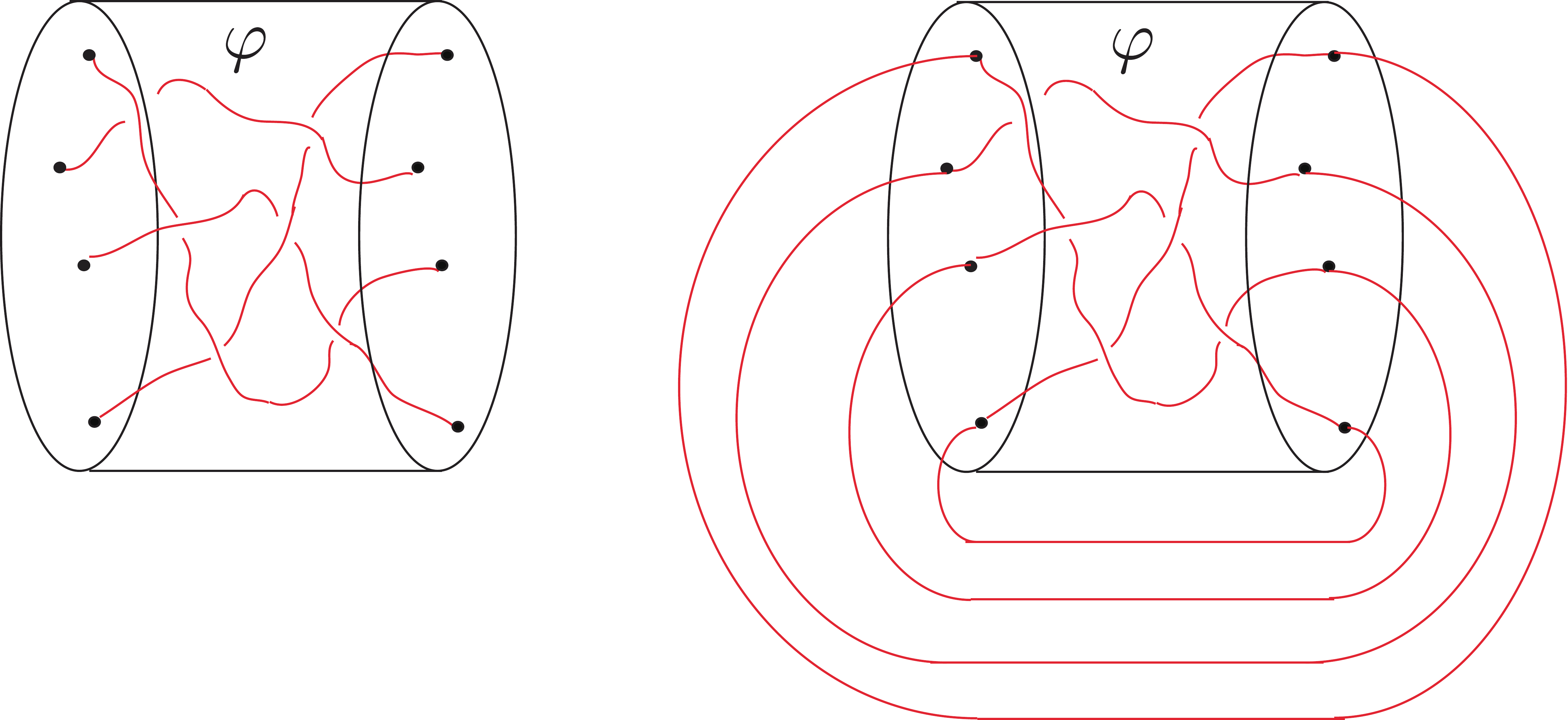}
\end{center}
\caption{An example of a mapping cylinder $(\Sigma_{g=0, h=4} \times [0,1])_{\varphi}$ (left) and 
a mapping torus $(\Sigma_{g=0, h=4} \times S^1)_{\varphi}$ (right).
The mapping class $\varphi$ in this example permutes the four punctures, generating non-trivial braids in between;
after the identification of the boundaries, we obtain a mapping torus. In this example
we have a  two-component link, and hence only two out of the four holonomies around the four punctures of the sphere
are independent.
}
\label{time_evolution}
\end{figure}

We now know the quiver $Q$ as well as a mutation sequence $\bm{m}$ associated with 
a change of the ideal triangulation. We can translate this into a 3-manifold by 
re-interpreting a flip as an ideal tetrahedron. This way, the time evolution of a quiver 
is translated into a 3-manifold cobordism between two 2-manifolds (Fig.~\ref{time_evolution}).
This is already a 3-manifold geometrically.
However,
if we wish to make more explicit the connection with the choice of a 3-manifold,
in particular with the defects inside a closed 3-manifold $\hat{M}$,
we need to close off the 
two boundary components.

In the formulation in terms of cluster partition function, there are two different methods to obtain
3-manifolds \cite{Terashima:2013fg}. The first is to attach a handlebody, or rather its generalization with knot-like defect, called a tanglebody in \cite{Terashima:2013fg}. This gives rise to 
an arbitrary knot in $S^3$, represented in the so-called plat representation.

In another method, we simply identify the two boundaries. We then obtain 
the mapping torus, a torus bundle $M$ over the Riemann surface, twisted by a mapping class group element $\varphi$:
\begin{align}
\begin{split}
&M= \left( \Sigma\times S^1 \right)_{\varphi}:  =  \{ (x,t)\in \Sigma \times [0,1] \}/\sim 
\\
&\textrm{where the equivalence relation is given by }(x,0)\sim (\varphi(x),1)\;.
\end{split}
\label{mapping_torus}
\end{align}
Here, the choice of $\varphi$ determines the mutation sequence $\bm{m}$;
$\varphi$ maps one triangulation to another, which can be equally represented by a sequence of 
change of the ideal triangulation. This in turn could be realized by a series of quiver mutations.
The permutations $\bm{\sigma}$ is then chosen such that the quiver comes back to itself after $L$ mutations.
At the level of the cluster partition function, identifying the two boundary Riemann surfaces means to 
take a trace $\textrm{Tr}_{Q,\bm{m}, \bm{\sigma}}$ of the cluster partition function.\footnote{
As is clear from this discussion, we in general need to use only one permutation, at the last step $t=L-1$.
However, for practical computations it is useful to have flexibility and allow for permutation after each mutation,
which was why we had a sequence of permutations in the discussion of the previous subsection.
} Since the phase space $\CM_N(\Sigma,\vec{\rho})$ can be thought as phase space of $SL(N)$ CS theory on $\mathbb{R}_{\textrm{time}}\times \Sigma$, the corresponding  cluster partition function will give CS partition function on the mapping torus with fixed conjugacy class of holonomy along punctures. 
\begin{align}
\begin{split}
\textrm{Tr}_{Q,\bf{m},\bm{\sigma}}(\frakL^{(a)}_\a) &= \textrm{Tr}(\hat{\varphi})(P_a)
\\
&=\{ \textrm{CS partition function on $M$ with fixed conjugacy class} 
\\
&\quad \quad \textrm{of  holonomies $[P_a]$ around $a$-th puncture} \} \ ;.
\end{split}
\end{align}
Due to the projectivity \eqref{projectivity of MCG reps}, the cluster partition function is only defined up to the $\mathbb{Z}^h$ which is phase factor (recall $h$ is the number of punctures of the 2d surface $\Sigma$). 
\begin{align}
\begin{split}
&\textrm{Tr}(\hat{\varphi})(P_a) \textrm{ is defined up to a phase factor of the form }  
\\
&\textrm{exp} \bigg{(}  \sum_{a=1}^{h} \frac{n_a}{\CN}\left(\frac{\textrm{Tr}(\log P_a)^2}{2\hbar}+\frac{\textrm{Tr}(\log P^\dagger_a)^2}{2\tilde{\hbar}}\right) \bigg{)} \;, \quad \{n_a\}\in \mathbb{Z}^{h} \;.\label{ambiguity in cluster ptn}
\end{split}
\end{align}
The ambiguity is a version of the well-known framing ambiguity in the CS partition function.  

Note that not all knot complements can be expressed as mapping tori of 2-manifolds,
and the class of the 3-manifolds we discuss here is not the most general
(such a 3-manifold is a complement of the so-called fibered knot).
However, mapping tori provide excellent examples for practical computations,
and include interesting examples. For example, the mapping torus of the once-punctured torus bundle
contains the complement of the so-called figure-eight knot in $S^3$, and we will study this example extensively.
If we consider twice-punctured torus bundles, we could obtain all the torus knots and the so-called two-bridge knots in 
$S^3$, as well as more general knots in lens spaces \cite{Cattabriga}.
In more physical terms, notice also that the 3d $\CN=2$ theories originating from mapping tori 
can be thought of as duality domain walls of 4d $\mathcal{N}=2$ theories, 
which we will comment further in Sec.~\ref{sec.TSUN}.

%%%%%%%%%%%%%%%%%%%%%%%%%%%%%%%%%%%%%%%%%%%%%%%%%%%%
\subsection{Relation with State-Integral Models}\label{subsec.relation}

In the previous subsections, we have already written down rules for writing down cluster partition functions;
we can now directly proceed to the computations.
Before coming to examples, however, it is useful to
rewrite the expression of the cluster partition function
in a different form,
such that the connection with the 
state-integral models in Sec.~\ref{sec.state_integral} becomes clearer.
This not only helps to reproduce the results of Sec.~\ref{sec.state_integral}
from the cluster partition function, but also to 
explore new state-integral models hitherto unknown in the literature.

The rewriting is actually rather simple:
since the delta functions \eqref{trace_result_1} are linear, we can easily solve the constraints and
 integrate them out.
Indeed, the number of integration variables, as well as the number of constraints, are given by
\begin{align}
\begin{split}
&\sharp \textrm{ of integration variables : } \overbrace{L |(Q)|}^{\vec{u}(t)} + \overbrace{2L}^{Z(t), Z''(t)} \;, 
\\
&\sharp \textrm{ of constraints : } \overbrace{L |Q|}^{\delta(C\cdot \vec{U}-\vec{V})}+ \overbrace{L}^{\delta(Z(t)+Z''(t)-2 \sum Q_{m_t,i} u_i(t) )} + \overbrace{n_c}^{\delta(\sum_i c_i^{\alpha} u_i(0))}
\;,
\end{split}
\end{align}
leading to $L-n_c$ remaining integration variables. Here, let us choose to integrate out only the $\vec{u}(t)$s.
The partition function then can be written in the form
\begin{align}
&\mathrm{F.T.}\left[\textrm{Tr}_{Q,\bm{m}, \bm{\sigma}} \right](\bfrakM) = \int \prod_{t=0}^{L-1} \left[ dZ(t)dZ''(t) \right] \bigg{[} \prod_{t=0}^{L-1}\psi_{\hbar} \big{(}Z(t) \big{)} e^{- \frac{1}{4 \pi i b^2} Z\cdot Z''} \bigg{]} \delta \left(\cdots\right)\;.
\label{trace_result_3}
\end{align}
Here the $\delta$-functions give $L+n_c$ linear constraints on $2L$ integration variables $\{ Z(t),Z''(t)\}$. 

We claim that the constraints in the  $\delta$-functions in the cluster partition function \eqref{trace_result_3} can be written in the following form:
\begin{align}
\begin{split}
&\bfrakM_{\a} -\sum_{t=0}^{L-1}  A^x_{\a} (t) Z(t)  =0\;,  
\\
&C_I := \sum_{t=0}^{L-1}A^{c}_I (t) Z(t) +B^{c}_I (t)Z''(t)=0 \;,
\\
&\frakL_\a:=\sum_{t=0}^{L-1} C^p_\a (t)Z(t) +D^p_\a Z''(t) =0\;,
\label{delta-fuctions-to-gluing}
\end{split}
\end{align}
with $\a = 1, \ldots,n_c\;, \; I=1,\ldots, L-n_c$.
We argue moreover that the integer matrices $(A^x, A^c, B^c, C^p,D^p)$ 
\begin{align*}
&(A^x)_{\a,t}:=A_\a(t)\;, \;(A^c)_{I,t}:=A^c_I(t)\;, \; (B^c)_{I,t}:=B^{c}_{I}(t)\;, \\
& (C^p)_{\a,t}:=C^{p}_{\a}(t)\;,\; (D^p)_{\a,t}:=D^p_\a(t) \nn
\end{align*}
satisfy
\begin{align}
\begin{split}
&A^x \cdot (B^c)^T =\mathbf{0},\; A^x \cdot  (D^p)^T = \mathbb{I}\;, \quad A^C\cdot (B^c)^T - (B^c)^T \cdot A^c = \mathbf{0} \;,
\\
&A^c\cdot (D^p)^{T}- B^c\cdot (C^p)^T=\mathbf{0}\;, \quad C^p\cdot (D^p)^T- D^P \cdot (C^P)^T =\mathbf{0} \;. \label{A,B,C,D matrices}
\end{split}
\end{align}
In fact, the constraints eqs.~\eqref{delta-fuctions-to-gluing} and \eqref{A,B,C,D matrices} are a part of the following assumption:
there exists a $Sp(2L,\mathbb{Q})$ matrix $g_{\textrm{cluster}}$ and 
a set of coordinates $\Gamma_I$ (or equivalently matrices $(C^\g, D^\g)$) such that 
\begin{align}
&g_{\textrm{cluster}} :=\left(\begin{array}{cc}A^x_{n_c \times L} & \mathbf{0}_{n_c \times L} \\ A^c_{(L-n_c)\times L} & B^c_{(L-n_c)\times L} \\ C^p_{n_c\times L} & D^p_{n_c\times L} \\ C^\g_{(L-n_c)\times L} &  D^\g_{(L-n_c)\times L}\end{array}\right)  \textrm{ satisfying} \nn
\\
&\left(
\begin{array}{c}
\bfrakM_{\a} \\
C_I \\
\frakL_{\a} \\
\G_I
\end{array}
\right)
=
g_{\rm cluster} \cdot
\left(
\begin{array}{c}
Z(t) \\
Z''(t) 
\end{array}
\right)\;.
\label{g_transf}
\end{align}
While we do not have a general proof of the aforementioned statements applicable to general quiver mutations,
we will find that this assumption is satisfied for all the examples discussed in this paper,
and is consistent with the results from the state-integral models (in fact, a similar condition was implicitly assumed in the discussion of the state-integral models
in Sec.~\ref{sec.state_integral}).

Once we accept this assumption,
we can interpret the linear transformation \eqref{g_transf} as a 
change of the polarization in the quantization, from the polarization $(Z(t); Z''(t))$ to 
$(\bfrakM_\a, C_I; \frakL_\a, \G_I)$. 
 More concretely, we will prove in App.~\ref{app.delta_proof} that 
\begin{align}
\langle \bfrakM_\a ,C_I =0 | Z(t)\rangle = 
\int \prod_{t=0}^{L-1}dZ''(t) \,  e^{- \frac{1}{4 \pi i b^2}\sum_t Z(t) Z''(t)}  \delta (\textrm{eq.}\eqref{delta-fuctions-to-gluing})\;,
\label{XC_lemma}
\end{align}
up to an overall constant factor. Here we use language of state-integral model explained in Sec.~\ref{sec.state_integral}: $\langle \bfrakM_\a, C_I|$ denote a position basis of $\CH(\partial \Diamond)^{\otimes L}$ in the polarization where positions are $\{ \bfrakM=A^x\cdot Z, C=A^c\cdot Z+B^c\cdot Z'' \}$ and  its conjugate momentums are $\{\frakL = C^p \cdot Z+ D^p\cdot Z'',\Gamma:=C^\g \cdot Z+D^{\g}\cdot Z''\}$ while $\langle Z(t)|$ is a position basis in the polarization where $Z(t)$ and $Z''(t)$ are positions and momentums respectively. When $C_I=0$, the position basis $\langle \bfrakM, C |$ is independent on the choice of $\Gamma_I$.  
Using eq.~\eqref{XC_lemma},
we can then rewrite our cluster partition function  \eqref{trace_result_3} \eqref{delta-fuctions-to-gluing} into the following form:
\begin{align}
\mathrm{F.T.}\left[\textrm{Tr}_{Q,\bm{m}, \bm{\sigma}} \right](\bfrakM) &=
\int \left[ \prod_{t=0}^{L-1} dZ(t) \right] \langle \bfrakM_\a ,C_I=0 | Z(t)\rangle \langle Z(t) |\Diamond^{\otimes L}\rangle \nn \\
& = \langle \bfrakM_\a, C_I=0| \Diamond^{\otimes L}\rangle \;, 
\end{align}
where in the second line we used the completeness relation \eqref{Completeness relation},
and we used the octahedra's wave-function \eqref{octheron's CS wave ftn} :
\begin{align}
\langle Z(t) |\Diamond^{\otimes L}\rangle = \prod_{t=0}^{L-1} \psi_{\hbar}\left(Z(t)\right) \ .
\end{align}
By performing Fourier transformation again, we finally have
\begin{align}
\textrm{Tr}_{Q,\bm{m}, \bm{\sigma}}(\frakL)  = \langle \frakL_\a, C_I=0| \Diamond^{\otimes L}\rangle \;, 
\label{trace_as_overlap}
\end{align}
where the $\langle \frakL_\a, C_I|$ is position basis in a polarization where $(\frakL, C_I)$ are positions and $(-\bfrakM, \Gamma_I)$ are momentums.
We therefore came to the conclusion that the cluster partition function takes exactly the same form as 
the partition function of state-integral models, and for precise comparison all we need to do is to compare the delta function constraints (gluing equations in 3-manifold examples). Once we obtain the octahedral gluing equations from the cluster partition function which is derived for only $k=1$, the cluster partition function for other $k$ can be obtained using the  state-integral models for other $k$. Conversely, given a cluster partition function we can recover the 
gluing equations, and hence the octahedron structures of the ideal triangulation. In fact, the latter point can be made somewhat more manifest by making a connection between octahedra and the `mutation network' of \cite{Terashima:2013fg}, which we now turn to.

%------------------------------------------------------------------------------------------------------------------
\paragraph{Mutation Networks versus Octahedra}

For a general cluster partition function (including those not coming from 3-manifolds), 
a useful method to encode the data of quiver mutations
is to use the formalism of the mutation network introduced in \cite{Terashima:2013fg}.

\begin{figure}[htbp]
\centering\includegraphics[scale=0.65]{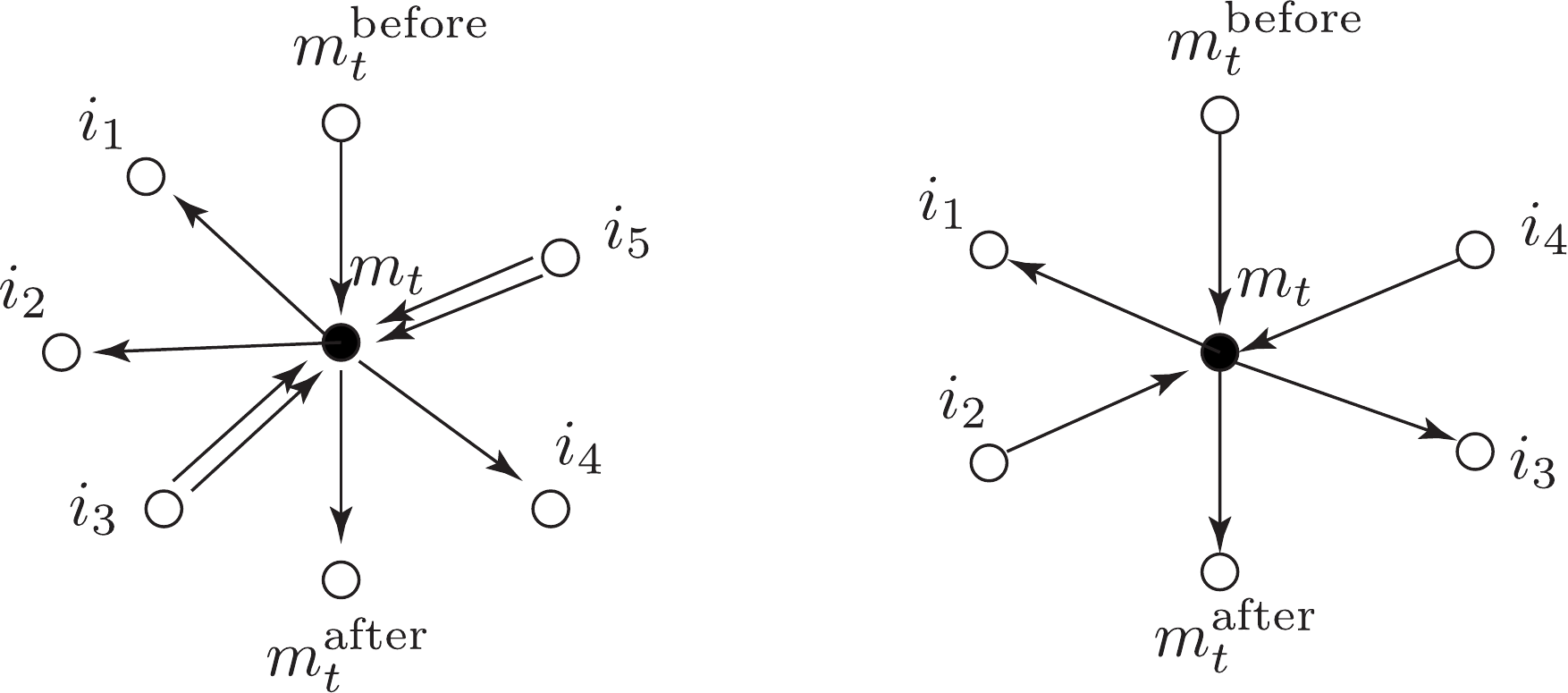}
\label{fig.mutation_network}
\caption{A mutation network represents combinatorial structure of the quiver mutations.
A mutation at vertex $m_t$ is represented graphically as shown here, here
the black vertex representing the mutation is connected (possibly with multiplicity) to the 
vertices $i_1, i_2, \ldots, $ affected by the mutation, and in particular the white vertex for the
mutated vertex $m_t$ itself is duplicated, one for before the notation, and another for after the mutation.
For applications to 3-manifold, the mutation network looks as on the right locally around a black vertex,
and it plays the same role
as the octahedron decomposition. In this sense, a mutation net work is more general than the octahedron decomposition.}
\end{figure}

Let us quickly summarize the concept of the mutation network (see \cite{Terashima:2013fg}).
The mutation network is a graph consisting of 
black vertices and white vertices,
with 
\begin{enumerate}[leftmargin=*]
\item Black vertices represent mutations of the quiver; we have a vertex for each mutation $m$.
\item White vertices represent the vertices of the quiver. Each time a mutation is performed we
add a new vertex, representing the vertex after the mutation. This means that (if we are interested in the 
trace of the cluster partition function) the total number of white vertices 
is given by $|Q|+L$.

\item Suppose a mutation is performed at the vertex $m$.
We prepare two vertices $m^{\rm before}$ as well as $m^{\rm after}$, 
representing the vertex $m$ before and after the mutation.
Then, the black vertex representing the mutation is connected to the white vertex representing $m^{\rm before}$
and $m^{\rm after}$, as well as to all the white vertices 
whose corresponding quiver vertices are connected to the vertex $m$ (Fig.~\ref{fig.mutation_network}).

\end{enumerate}

\begin{figure}[htbp]
\centering\includegraphics[scale=0.8]{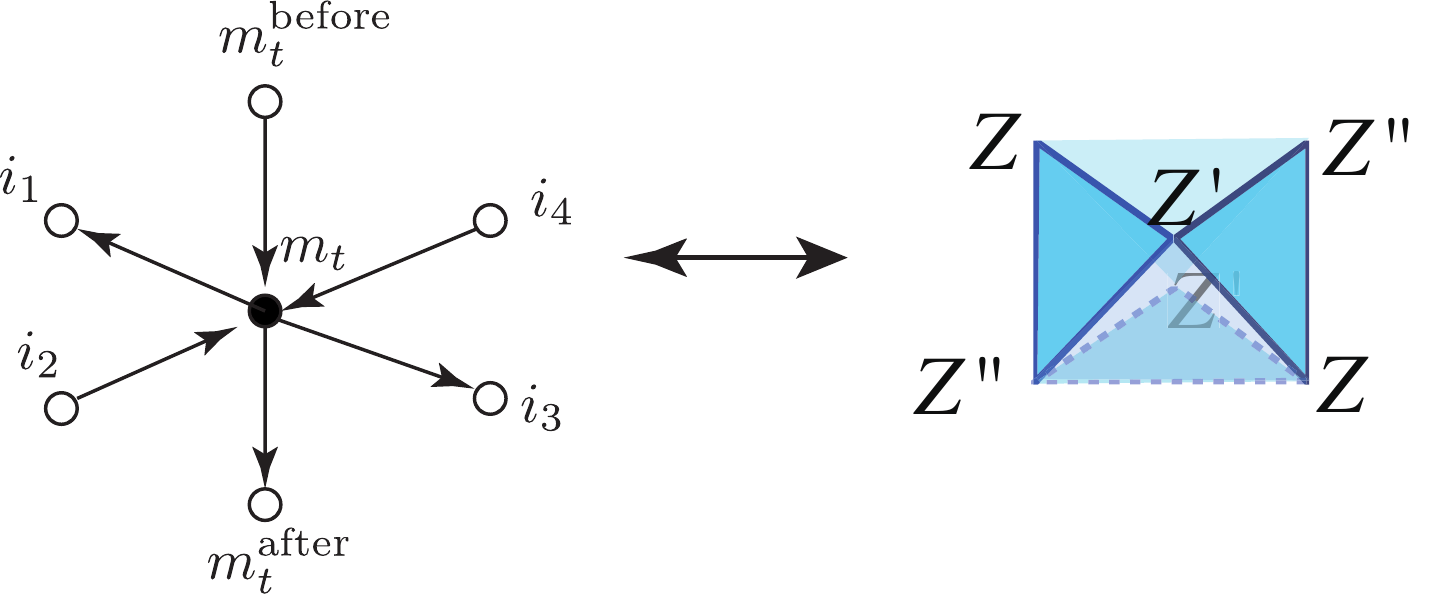}
\caption{For the 3-manifold cases, there is a one-to-one correspondence
between a mutation network and an octahedron;
a black vertex (mutation) of the mutation network represents an octahedron,
and the white vertices connected to it represents the vertices of the octahedron.
This makes it possible to write down octahedron structures for non-maximal punctures.
}
\label{fig:network_octahedron}
\end{figure}

Let us now specialize to the case
of this subsection, namely a mutation sequence coming from the 
ideal triangulations of the Riemann surface.
What is remarkable in this case is that 
we always mutation at the vertex with two lines coming in and two lines coming out (see examples in the next subsection).\footnote{
If this pattern continues, this will be a strong constraint on the possible mutation sequence for 
quiver sequences coming from general $\rho$.
Interestingly, this means that the mutation of the quiver can be thought of as a
Seiberg duality \cite{Seiberg:1994pq} of the 4d $\mathcal{N}=1$ quiver gauge theory defined in \cite{Xie:2012mr,Franco:2012mm},
and we can associate {\it 4d} superconformal indices (or its lens space generalization) following \cite{Yamazaki:2012cp,Terashima:2012cx,Yamazaki:2013nra}.
}
In this case, the mutation network always looks as in the right of Fig.~\ref{fig.mutation_network},
namely the black vertex is connected to six vertices.
We can identify this with an octahedron, which has six edges;
a mutation (a black vertex) corresponds to an octahedron, 
and a vertex (a white vertex) corresponds to an edge of the octahedron (Fig.~\ref{fig:network_octahedron}).
This means that the given a quiver mutation,
we can unambiguously write down the mutation network,
and consequently a octahedron-type decomposition.
This is a powerful machinery to write down 
octahedron decompositions, 
even in cases hitherto unknown in the literature,
for example for the example of the simple punctures to be discussed in the next subsection.

%------------------------------------------------------------------------------------------------------------------
\paragraph{Summary}

Comparing our expression of the cluster partition function \eqref{trace_as_overlap} with that for the state-integral model \eqref{state-integral from ideal triangulation}, we immediately find the following correspondence:
\begin{align*}
\begin{array}{|c|c|}  \hline 
 \textrm{cluster partition function } & \textrm{state-integral model} \\\hline \hline
 \textrm{mutation network} & \textrm{octahedron decomposition} \\
\hline
 \textrm{mutation} & \textrm{octahedron} \\
\hline
 \textrm{quiver vertex affected by a mutation} & \textrm{vertex of a octahedron} \\
\hline n_c & \textrm{complex dimension of boundary phase space} 
\\\hline    Z(t), Z''(t), & \multirow{2}{*}{\textrm{octahedron's vertex variables}} \\
  Z'(t):=i \pi + \frac{\hbar}2 - Z(t)-Z''(t)  & 
  \\\hline    
  \frakL_\a/  \bfrakM_\a  &  \textrm{boundary position/momentum }
\\\hline  
\textrm{delta function constraint} & \textrm{gluing equations of octahedra} \\
  C_I=0 &  \textrm{at an internal vertex} 
\\\hline
\end{array} 
\end{align*}

We can also associate a 3d $\mathcal{N}=2$ theory  (``cluster $\mathcal{N}=2$ theories'')
$T[Q, \bm{m}, \bm{\sigma}]$ with Abelian gauge groups: the procedure is exactly the same
as in Sec.~\ref{sec.state_integral_general}.
The dictionary for the 3d $\mathcal{N}=2$ theory is given as follows:
\begin{align*}
\begin{array}{|c|c|}  \hline  \textrm{cluster partition function } & \textrm{3d $\mathcal{N}=2$ theory} \\\hline \hline
 \textrm{mutation network} & \textrm{a graph indicating matter and symmetry} \\
\hline
 \textrm{mutation} & \textrm{chiral multiplet} \\
\hline
 \textrm{quiver vertex affected} & \textrm{symmetry under which} \\
  \textrm{by a mutation} & \textrm{a chiral multiplet is charged} \\
\hline n_c& 
 \textrm{rank of global symmetries} 
\\\hline Z(t),Z''(t),  & \textrm{loop operators for symmetries}\\
   Z'(t):=i \pi + \frac{\hbar}2 - Z(t)-Z''(t)  &  \textrm{of a free chiral multiplet}
\\\hline   \frakL_\a / \bfrakM_\a &  \textrm{flavor Wilson/vortex loops}\\
\hline
 \textrm{delta function constraint}  &  \textrm{superpotential $W$ constraint}\\
 C_I=0 & \textrm{R-charge($W$)=2}
\\\hline
\end{array} 
\end{align*}

\subsection{Inclusion of Co-dimsion 4 Defects}\label{sec.codim_4_cluster}

 In our cluster partition function formalism, we can include co-dimension 4 defects
by inserting a loop operator in the 2-manifold direction.
The loop operator is classically a holonomy along a 1-cycle of the 2-manifold, which we can quantize systematically
 using the Fock-Goncharov coordinate. The equivalence between the FG quantization and quantization using Skein-relation is demonstrated in \cite{Coman:2015lna}.

Given an ideal triangulation on the 2-manifold, there are rules for reading off the 
holonomies for the 1-cycle, as we will explain in App.~\ref{sec : snakes}.
This will give rise to classical expression for the Wilson loop operator in terms of FG coordinates $Y_i$
\begin{align}
W_R (\CK)=\textrm{Tr}_R\, \textrm{Hol}(\gamma:=[\CK])=\sum_{k} c_{k} \, e^{\sum_{i=1}^{|Q|}a^{(k)}_{i} Y_{i}}  \;,
\end{align}
where $\gamma$ is an element of $\pi_{1}(\Sigma)$. Unlike in 3d case, these $Y_i$ variables are only constrained by linear equations, which identify some linear combination of $Y_i$ with eigenvalues of holonomies around punctures, and allows for a simple quantization rule:
\begin{align}
\hat{W}_R(\CK_\gamma)= \sum_{k} \hat{c}_k\, e^{\sum_{i=1}^{|Q|} a^{(k)}_i\sfY_i}
\end{align}
where $\sfY$s are quantized FG coordinates and $\CK_\gamma$ is the unknot in the homotopy class $\gamma$ (recall footnote \ref{foot.unknot}). 
Also, $\hat{c}_k$ is a quantization of $c_k$, replacing the integer $c_k$ by in general a Laurent polynomial in $q^{\half}$, 
symmetric under the exchange $q\to q^{-1}$ ({\it cf.} \cite{Gaiotto:2010be}).

Since we consider mapping tori (and hence the `time' $t$ is periodic), we
can choose to insert the Wilson line into the cluster partition function at $t=0$.\footnote{If we choose to insert Wilson lines at several different times,
we could discuss correlators of Wilson lines.}
Generalizing the computation of cluster partition function in Sec.~\ref{sec : cluster partition function} with the insertion of  loop operators, we wish to compute
\begin{align}
\begin{split}
\textrm{F.T.}[\langle W_R ( \CK_\gamma) \rangle_{Q,\bm{m},\bm{\sigma}}] (\bfrakM)  
&=\displaystyle\int \prod d \frakL_\a   \, e^{ \frac{1}\hbar \sum_{\alpha} \frakL_\a \cdot \bfrakM_\a}\textrm{Tr} \big{[}   \hat{W}_R (\CK_\gamma) \hat{\varphi}_{Q,\bm{m}, \bm{\sigma}} \big{]}(\frakL_\a)  \;,
\\
& =\displaystyle\sum_{k} c_k \, \textrm{F.T.}[\langle e^{\sum_i a_i^{(k)} \sfY_i}\rangle_{Q,\bm{m},\bm{\sigma}}](\bfrakM) \;.
\label{state-integral model for co-dimension 4 defect for mapping torus}
\end{split}
\end{align}
The computation of eq.~\eqref{state-integral model for co-dimension 4 defect for mapping torus}
is similar to the case without the Wilson lines: we insert the complete set in between the operators,
converting the expression into integrals. We then carry out some obvious integrals.
After the computation (see App.~\ref{app.cluster_with_Wilson}), we have
\begin{align}
\begin{split}
&\textrm{F.T.}[\langle e^{\sum_i a_i \sfY_i}\rangle_{Q,\bm{m},\bm{\sigma}}](\bfrakM) 
\\
&= \int  \big{(}  \prod_{t=0}^{L-1} d\vec{u}(t) dZ(t) dZ''(t)   \big{)}  \exp \bigg{(}-\frac{1}2\sum_{i=1}^{|Q|} a_i Q_{i m_0} Z(0)-\sum_{i,j=1}^{|Q|} a_i Q_{ij} u_j(0) \bigg{)}
\\
& \times  \prod_{t=0}^{L-1} \psi_{\hbar} \big{(}Z(t) \big{)} e^{-\frac{1}{2\hbar} Z(t) Z''(t)}    \delta \left(\hat{C}_{Q,\bm{m}, \bm{\sigma}}\cdot \vec{U}- \vec{V}_a\right)
\\
& \times
\prod_{t=0}^{L-1}\,  \delta \! \left( Z(t)+Z''(t)-2\sum_{i=1}^{|Q|}　Q_{m_t, i}(t)u_i(t) \right) 
\prod_{\alpha=1}^{|\textrm{Ker}(Q)|} \delta\left(\sum_{i=1}^{|Q|} c_i^{\alpha} u_i(0)\right) 
\;.
\end{split}
\end{align}
with 
\begin{align}
&\vec{V}_a := \vec{V} -\frac{\hbar}2  \left(\begin{array}{c}  \vec{a} \\ \mathbf{0}_L \\ \ldots \\ \mathbf{0}_L\end{array}\right)\;.
\label{Z_cluster_with_Wilson}
\end{align}
The matrix $\hat{C}_{Q,\bm{m},\bm{\sigma}}$ and vectors $\vec{U}$  and $\vec{V}$ are defined in eq.~\eqref{trace_result_2}. Let us change the above expression suitable to compare with 3d Wilson loop in eq.~\eqref{state-integral model for co-dimension 4 defect}.
First we  shift the dummy integration variables $(Z,Z'',u_i)$ properly in order to  cancel the effect of the Wilson loop in $\delta$-function : $(Z,Z'',u_i)\rightarrow (Z+\delta Z, Z''+\delta Z'', u_i+\delta u_i)$
\begin{align}
\begin{split}
&\hat{C}_{Q,\bm{m},\bm{\sigma}}\cdot \delta \vec{U}=- \frac{1}2 \left(\begin{array}{c} \delta Z(0) \hat{e}_{m_0} +\hbar \vec{a}\\ \delta Z(1)\hat{e}_{m_1} \\ \vdots \\ \delta Z(L-1) \hat{e}_{m_{L-1}} \end{array}\right) \;, \;\quad \textrm{with}\quad \delta \vec{U}:=\left(\begin{array}{c} \delta \vec{u}(0) \\ \delta \vec{u}(1)\\ \ldots \\ \delta \vec{u}(L-1)\end{array}\right) \;,
\\
& \delta Z(t)+\delta Z''(t)-2\sum_{i=1}^{|Q|}Q_{m_t,i }(t)\delta u_i(t)=0\;, \quad  t=0,\ldots, L-1\;,
\\
&\sum_{i=1}^{|Q|} c_i^\a \delta u_i (0) =0\;. \label{shifts in dummy variables}
\end{split}
\end{align}
The shifts should depends only on $\vec{a}$. 
In general, we do not know the existence  of the solutions; however,
in several examples we found that such solutions exist. Assuming existence of the solution, we now have
\begin{align}
\begin{split}
&\textrm{F.T.}[\langle e^{\sum_i a_i \sfY_i}\rangle_{Q,\bm{m},\bm{\sigma}}](\bfrakM)   
\\
&=\int  \left[ \prod_{t=0}^{L-1}dZ(t) dZ''(t)  \right] \prod_{t=0}^{L-1} \psi_{\hbar}\big{(}Z(t)+\hbar W_p(t)\big{)} e^{W_x(t) Z(t) } q^{  \Delta (t)}e^{- \frac{1}{2\hbar} Z(t)\cdot Z''(t)} \delta \big{(}\textrm{eq.~}\eqref{delta-fuctions-to-gluing}\big{)}\;,  
\\
&= \big{\langle} \bfrakM,C_I=0 \big{|}  q^{\sum_{t=0}^{L-1}  \Delta (t)} e^{\sum_t W_x (t) \hat{Z}(t) }e^{\sum_t W_p(t)  \hat{Z}''(t) }\big{|}\Diamond^{\otimes L}\big{\rangle}\;,
\end{split}
\end{align}
where 
\begin{align}
\begin{split}
&  W_p(t) :=\frac{1}{\hbar} \delta Z(t)\;, 
\\
& \hbar \Delta(t)+ W_x (t) Z(t) :=- \frac{1}{2\hbar} Z(t)\delta Z''(t) - \frac{1}{2\hbar} \delta Z(t) Z''(t) - \frac{1}{2\hbar}\delta Z(t) \delta Z''(t) 
\\
&\qquad  -\frac{\delta_{t,0}}2 \sum_{i=1}^{|Q|} a_i Q_{im_0}\big{(}Z(0)+\delta Z(0)\big{)}-\delta_{t,0}\sum_{i,j=1}^{|Q|}a_i Q_{ij} \big{(}u_j(0)+\delta u_j(0)\big{)}\bigg{|}_{\textrm{$\delta$-functions in eq.~\eqref{trace_result_1}}} \label{2d loops to 3d loops-1} \;.
\end{split}
\end{align}
Thus final expression for cluster partition function with insertion of a 2d Wilson loop is exactly same as the partition function of the 3d state-integral model with a Wilson line insertion \eqref{state-integral model for co-dimension 4 defect}, 
under the identification
\begin{align}\begin{split}
&\textrm{2d loop operator } \sum c_k \, e^{ a_i^{(k)}  \sfY_i}  
\\
&\Longleftrightarrow \; \textrm{3d loop operator }  \sum_k c_k\, q^{\sum_t \Delta(t)} e^{\sum_t W_x (t)  \hat{Z}(t)}  e^{\sum_t W_p (t)  \hat{Z}''(t)}\;. \label{2d loops to 3d loops-2}
\end{split}
\end{align}

%%%%%%%%%%%%%%%%%%%%%%%%%%%%%%%%%%%%%%%%%%%%%%%%%%%%%%%%
\subsection{Examples}\label{subsec.examples}
%%%%%%%%%%%%%%%%%%%%%%%%%%%%%%%%%%%%%%%%%%%%%%%%%%%%%%

We will now come to the analysis of concrete examples. For concreteness we will below concentrate on the case when 
$\Sigma$ is a once-punctured torus or a four times punctured sphere. In both cases, the mapping class group is 
(or contains in the latter case) $PSL(2, \mathbb{Z})$,
generated by two elements
\begin{align}
\bm{L} =\left(\begin{array}{cc}
1 & 0 \\
1 & 1 \\
\end{array}\right) \;,
\quad
\bm{R}= \left(\begin{array}{cc}
1& 1 \\
0 & 1 \\
\end{array} \right) \;.
\end{align}
We primarily consider the case where $\varphi=\bm{LR}$ ($\bm{L}:=\bm{ST}^{-1}\bm{S}^{-1}, \bm{R}=\bm{T}$).
Interestingly,  for once-punctured torus case the resulting 3-manifold coincides with the complement of the figure eight knot $\bm{4}_1$ inside $S^3$:
\begin{align}
\left(\Sigma_{1,1}\times S^1 \right)_{\varphi = \bm{LR}} = (\textrm{Figure-eight knot complement on $S^3$}) \;. \label{4_1 as mapping tori}
\end{align}
Note that this is a concrete example where knots in different closed 3-manifolds $\hatM$ generate the same 3-manifold with boundary;
one 3-manifold is $(T^2\times S^1)_{\varphi = \bm{LR}}$, which is not hyperbolic and rather is the so-called solvmanifold;
another is $S^3$, which is again not hyperbolic. The two closed 3-manifolds are related by $(0,1)$-Dehn surgeries.
We can also consider other $(p,q)$-surgeries, and the resulting closed 3-manifold $M_{p,q}$, with a
knot inside it, again generates the same cusped 3-manifold $S^3\backslash \bm{4}_1$, with
the choice of polarization on the boundary torus induced by the $(p,q)$-Dehn surgery (recall eq.~\eqref{Dehn_filling}).
In general, Thurston's hyperbolic Dehn surgery theorem states that 
for a given hyperbolic cusped 3-manifold, its Dehn fillings are hyperbolic except for a finite values of $(p,q)$ (note here $p$ and $q$ are taken to be coprime).\footnote{For the figure eight knot, 
it is known that such exceptional Dehn surgeries are 
$(p,q)=(1,0), (0, 1), (\pm 1, 1), (\pm 2, 1), (\pm 3, 1), (\pm 4, 1)$ \cite[section 4]{ThurstonLecture}.
}

%%%%%%%%%%%%%%%%%%%%%%%%%%%%%%%%%%%%%%%%%%%%%%%%%%%%%%%%
\subsubsection{Co-dimension 2 Defects: \texorpdfstring{$\rho=\textrm{maximal}$}{rho=maximal}}
%%%%%%%%%%%%%%%%%%%%%%%%%%%%%%%%%%%%%%%%%%%%%%%%%%%%%%

Let us first start with the case of $\rho=\textrm{maximal}$. 

%%%%%%%%%%%%%%%%%%%%%%%%%%%%%%%%%%%%%%
\paragraph{Ex 1. $(\Sigma_{1,1}\times S^1)_{\varphi}$ with General $\varphi$ and $N$} %
We will give cluster partition function  datum $(Q,\bm{m},\bm{\sigma})$ for  mapping torus $(\Sigma_{1,1}\times S^1)_{\varphi}$ with general  $N$. 
\begin{figure}[htbp]
\centering\includegraphics[scale=0.2]{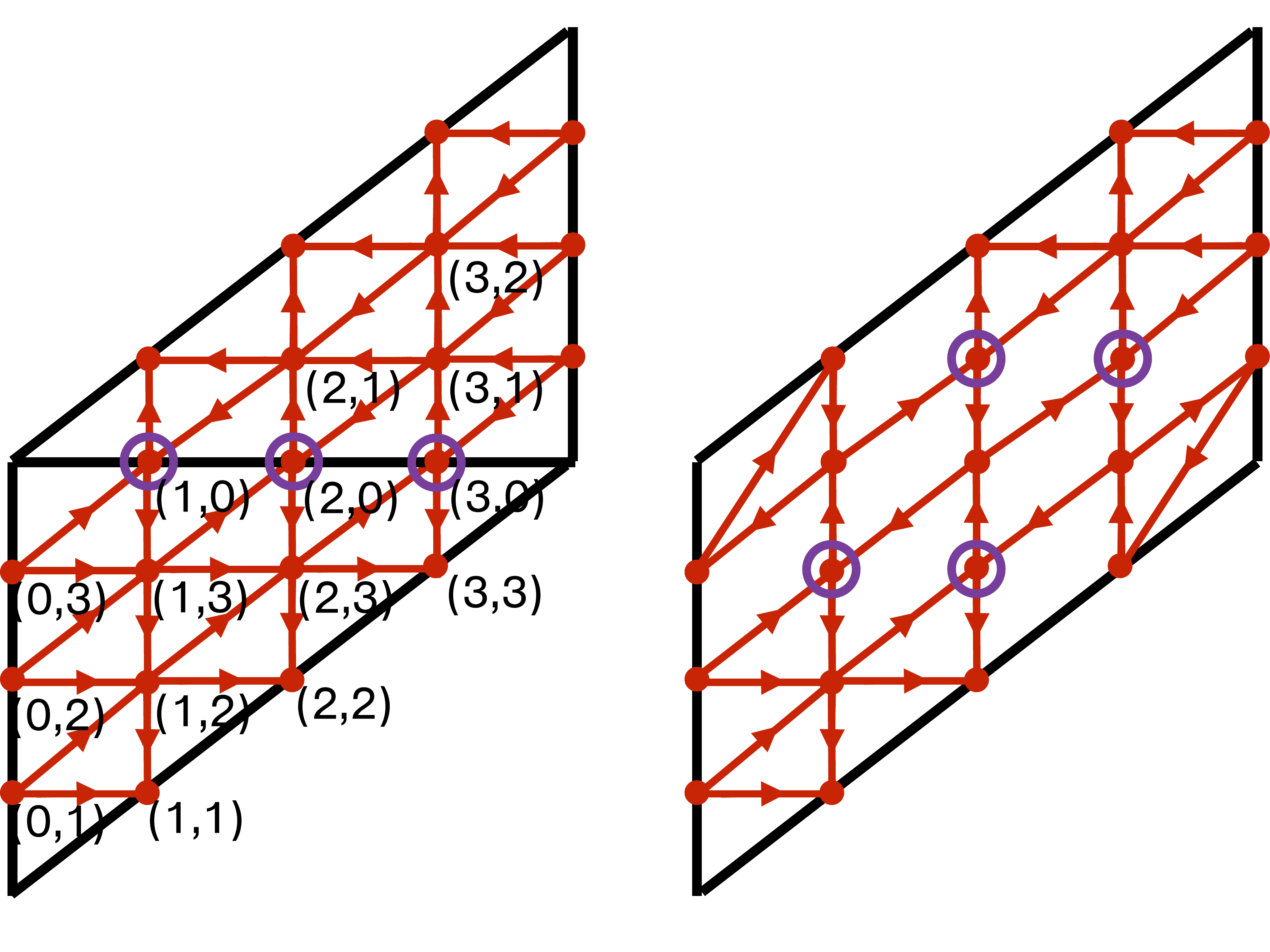}
\centering\includegraphics[scale=0.2]{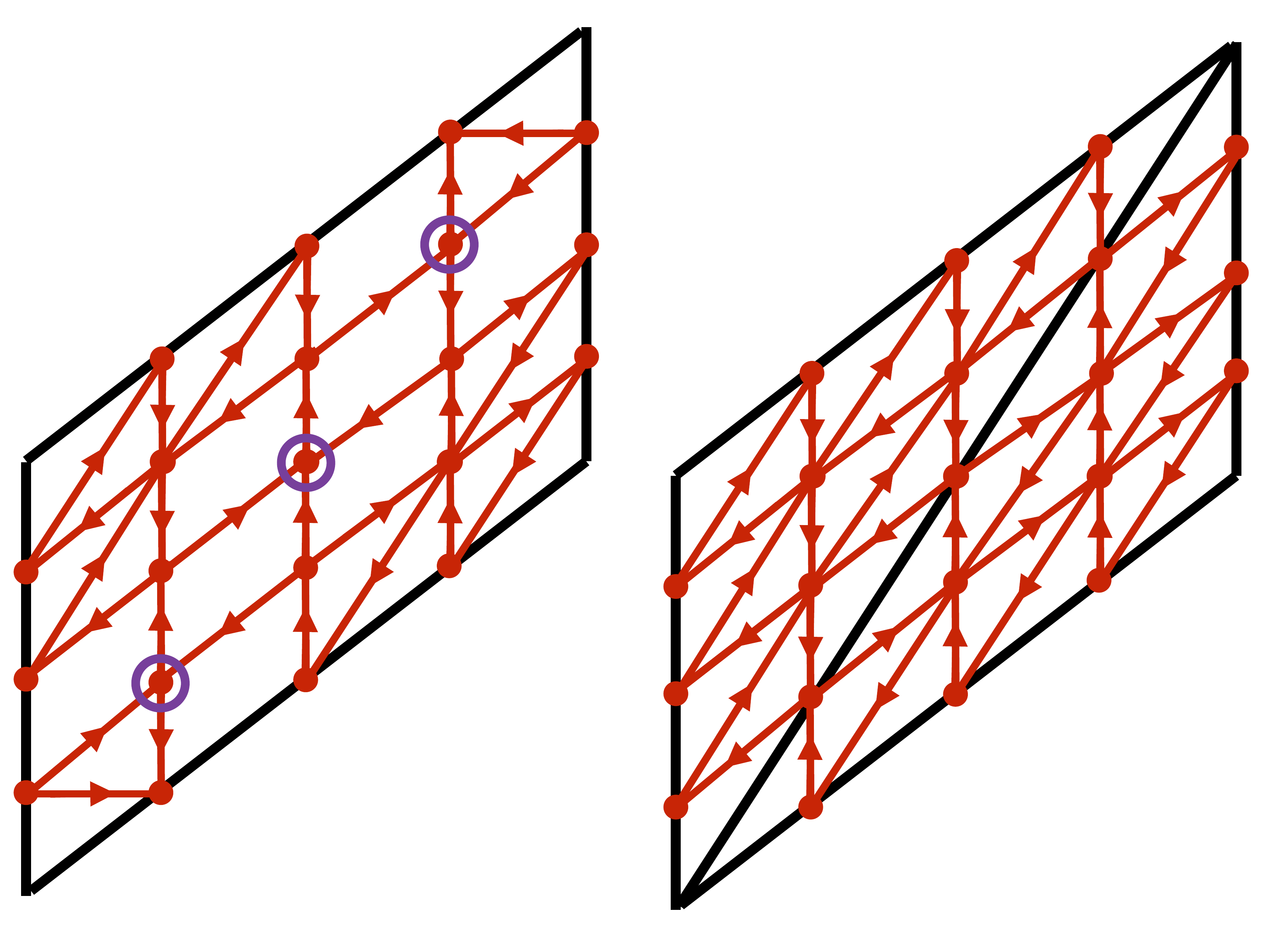}
\caption{The Fock-Goncharov quiver (red lines) associated with an ideal triangulation of $\Sigma_{1,1}$ with $N=4$. The deal triangulation with two triangles is drawn in black (first from the left). A puncture is located on vertices of two triangles.  A flip of the ideal triangulation,  which corresponds to $\bm{L} \in SL(2,\mathbb{Z})$, causes a sequence of $3+4+3=10$ mutations (on vertices with violet circles).}
\label{N=4_quiver}
\end{figure}

The Fock-Goncharov (FG) quiver $Q$ can be obtained using a tessellation for each triangle of a triangulation of $\Sigma_{1,1}$. In the tessellation, we introduce $(N-1)$ nodes in each edges of triangles and fill nodes inside the triangulation in a natural way.   The quiver with $N=4$ is depicted in Fig.~\ref{N=4_quiver}.  
The quiver contains $(N^2-1)$ nodes, which according to their positions will be labelled by an element in $\left(\mathbb{Z}^2
\backslash \{0,0\} \right) / (N\mathbb{Z} \times N\mathbb{Z})$:
\begin{align}\begin{split}
&\textrm{nodes (vertices) of the FG quiver } = \{ (a,b) \in  \mathbb{Z}^2\backslash {(0,0)}\}  \;,\\
& (a,b) \sim (a+N \mathbb{Z},b +N \mathbb{Z}) \;. 
\end{split}
\end{align}
There are $N-1$ central elements in the algebra $\CA_Q$ defined from
the FG quiver:
\begin{align}
|\textrm{Ker}(Q)| = N-1\;.
\end{align}
All  central elements are commute with  mapping class group elements which will be constructed below. Thus, we have
\begin{align}
n_c= |\textrm{Ker}(Q)|=N-1\;.
\end{align}
In the FG quiver, a mapping class group element $\bm{L}$ can be realized as 
\begin{align}
\bm{L} & = \left( \prod_{p=0}^{N-2}\prod_{(a,b)\in \Box_L(p)}\mu_{(a,b)} \right) \sigma_L\nn\\
&:= \left(\prod_{(a,b)\in \Box_L(0)}\mu_{(a,b)} \prod_{(a,b)\in \Box_L (1)}\mu_{(a,b)} \cdots \prod_{(a,b)\in \Box_L(N-2)}\mu_{(a,b)} \right) \sigma_L\;,
\end{align}
where $\Box_L (p)$ denotes a subset of $\mathbb{Z}^2\backslash \{ (0,0)\}$ with $(p+1)(N-p-1)$ entries
\begin{align}
\begin{split}
\Box_L (p) &:= \{ \bm{m}_{L,(p, r,s)}:=(1,N-p)+r(1,0)+s(1,2): \\
& \qquad \qquad\qquad\qquad \qquad0 \leq r \leq N-2-p, \; 0  \leq s \leq p \}\;.
\end{split}
\end{align}
For example, for $N=4$ we have $p=0, 1, 2$ and 
\begin{align}
\begin{split}
\Box_L (0) &:= \{ \bm{m}_{L,(p, r,s)}:=(1,0)+r(1,0) : \;
0 \leq r \leq 2 \}\;,
\\
\Box_L (1) &:= \{ \bm{m}_{L,(p, r,s)}:=(1,-1)+r(1,0)+s(1,2): \;
 r=0,1 \;, s=0,1 \}\;,
\\
\Box_L (2) &:= \{ \bm{m}_{L,(p, r,s)}:=(1,-2)+s(1,2):  \; 0  \leq s \leq 2 \}\;,
\end{split}
\end{align}
which coincides with the circled vertices in Fig.~\ref{N=4_quiver}.
For a given $p$, the ordering of mutations $\mu_{(a,b)\in \Box_L(p)}$ is irrelevant since they all mutually commute.
Thus, the sequence of mutations for $L$ is
\begin{align}
\begin{split}
&\bm{m}_L = (\{ \vec{m}_{L,(p, r,s)} \}_{0 \leq r \leq N-2-p, 0 \leq s \leq p, 0 \leq p \leq  N-2}\;, 
\\
&\textrm{with a partial ordering}  \quad (p,r,s) < (p',r',s')   \textrm{ if }  p<p'\;.
\end{split}
\end{align}
The permutation $\sigma_L$ is given by 
\begin{align}
\sigma_L : (a,b)\longrightarrow (a, a+b) \;.
\end{align}

Similarly, $\bm{R}\in SL(2,\mathbb{Z})$ can be realized as
\begin{align}
\bm{R}&= \left(\prod_{p=0}^{N-2}\prod_{(a,b)\in \Box_R(p)}\mu_{(a,b)} \right)  \sigma_R\nn \\
&:= \left(\prod_{(a,b)\in \Box_R(0)}\mu_{(a,b)} \prod_{(a,b)\in \Box_R (1)}\mu_{(a,b)} \ldots \prod_{(a,b)\in \Box_R(N-2)}\mu_{(a,b)} \right) \sigma_R\;, 
\end{align}
with 
\begin{align}
\begin{split}
&\Box_R (p) =  \{ \bm{m}_{R,(p, r,s)}:=(N-p,N-1-p)-r(0,1)+s(2,1) : \\
& \qquad \qquad\qquad\qquad \qquad 0 \leq r \leq N-2-p, \; 0  \leq s \leq p \}\;,  
\\
& \sigma_R : (a,b)\rightarrow (a+b, b)\;.
\end{split}
\end{align}
Thus, 
\begin{align}
\begin{split}
&\bm{m}_R =(\{ \vec{m}_{R,(p, r,s)} \}_{0 \leq r \leq N-2-p, 0 \leq s \leq p, 0 \leq p \leq  N-2}\;, 
\\
&\textrm{with a partial ordering}  \quad (p,r,s) < (p',r',s')   \textrm{ if }  p<p'\;.
\end{split}
\end{align}
The total number of mutations for a single flip is
\begin{align}
|\bm{m}_L| = |\bm{m}_R| = \sum_{p=0}^{N-2} (p+1)(N-p-1) = \frac{1}{6} N (N^2-1)\;. 
\label{total_N_cluster}
\end{align}
which coincides with eq.~\eqref{total_N} with $k=1$ ({\it i.e.} for a single tetrahedron).

For a pseudo-Anosov map $\varphi \in SL(2,\mathbb{Z})$ (i.e, $|\textrm{Tr}(\varphi)>2|$), the mapping torus 
$(\Sigma_{1,1}\times S^1)_{\varphi}$ is a hyperbolic 3-manifold. As an element of $SL(2, \mathbb{Z})$, pseudo-Anosov map $\varphi$ can be always decomposed into $\bm{L}$ or $\bm{R}$ up to conjugation
\begin{align}
\varphi  = \varphi_1 \varphi_2 \ldots \varphi_\sharp\;, \qquad  \textrm{with \quad $\varphi_i= \bm{L}$ or $\bm{R}$} \;.
\end{align}
The $SL(N)$ CS partition function  on the  corresponding mapping torus can be realized as cluster partition function  with the 
following datum
\begin{align}
\begin{split}
&\bm{m}= \{ \bm{m}_{\varphi_1} ,\bm{m}_{\varphi_2},\ldots, \bm{m}_{\sharp}\}\;, 
\\
&\bm{\sigma} = \{ \overset{0}{\check{\mathbb{I}}}, \mathbb{I}, \cdots, \overset{\frac{1}{6} N(N^2-1)-1}{\check{\sigma_{\varphi_1}}},\mathbb{I},\cdots, \overset{\frac{k}{6} N(N^2-1)-1}{\check{\sigma_{\varphi_k}}}, \mathbb{I},\cdots,    \overset{\frac{\sharp}{6} N(N^2-1)-1}{\check{\sigma_{\varphi_\sharp}}} \}\;, \label{cluster datum for mapping torus}
\end{split}
\end{align}
with the Fock-Goncharov quiver $Q$. Using these, it is straightforward to write down the cluster partition function,  and  get the final answer  \eqref{trace_result_3} with linear constraint on $Z(t)$ and $Z''(t)$ with $t=0, \ldots, \frac{\sharp}{6} N(N^2-1)-1$. As we will see below in 
several examples, one can check that the linear constraints can be written as in the form of eq.~\eqref{delta-fuctions-to-gluing}. There is a pictorial way to  understand these linear constraints: they are equivalent to octahedra's gluing equations in the $N$-decomposition of the mapping torus. To see the equivalence, it is better to use following labelling for $t$
\begin{align}
\begin{split}
&\{ Z(t),Z''(t) \}_{t=0,1,\ldots, \frac{\sharp}6 N(N^2-1)} \\
&\qquad\rightsquigarrow \left\{Z^{(k)}( p,r,s), \;Z''^{(k)}(p,r,s) \right\}^{k=1,\ldots, \sharp}_{0 \leq r \leq N-2-p, \, 0 \leq s \leq p,\,  0 \leq p \leq  N-2} \;. \label{mutation label-1}
\end{split}
\end{align}
Topologically, the mapping torus can be decomposed into $\sharp$ ideal tetrahedra $\{\Delta_k\}$
\begin{align}
\left(\Sigma_{1,1}\times S^1\right)_{\varphi} = \left(\bigcup_{k=1}^{\sharp} \Delta_k\right)/\sim\;.
\end{align}
\begin{figure}[htbp]
\centering\includegraphics[scale=0.2]{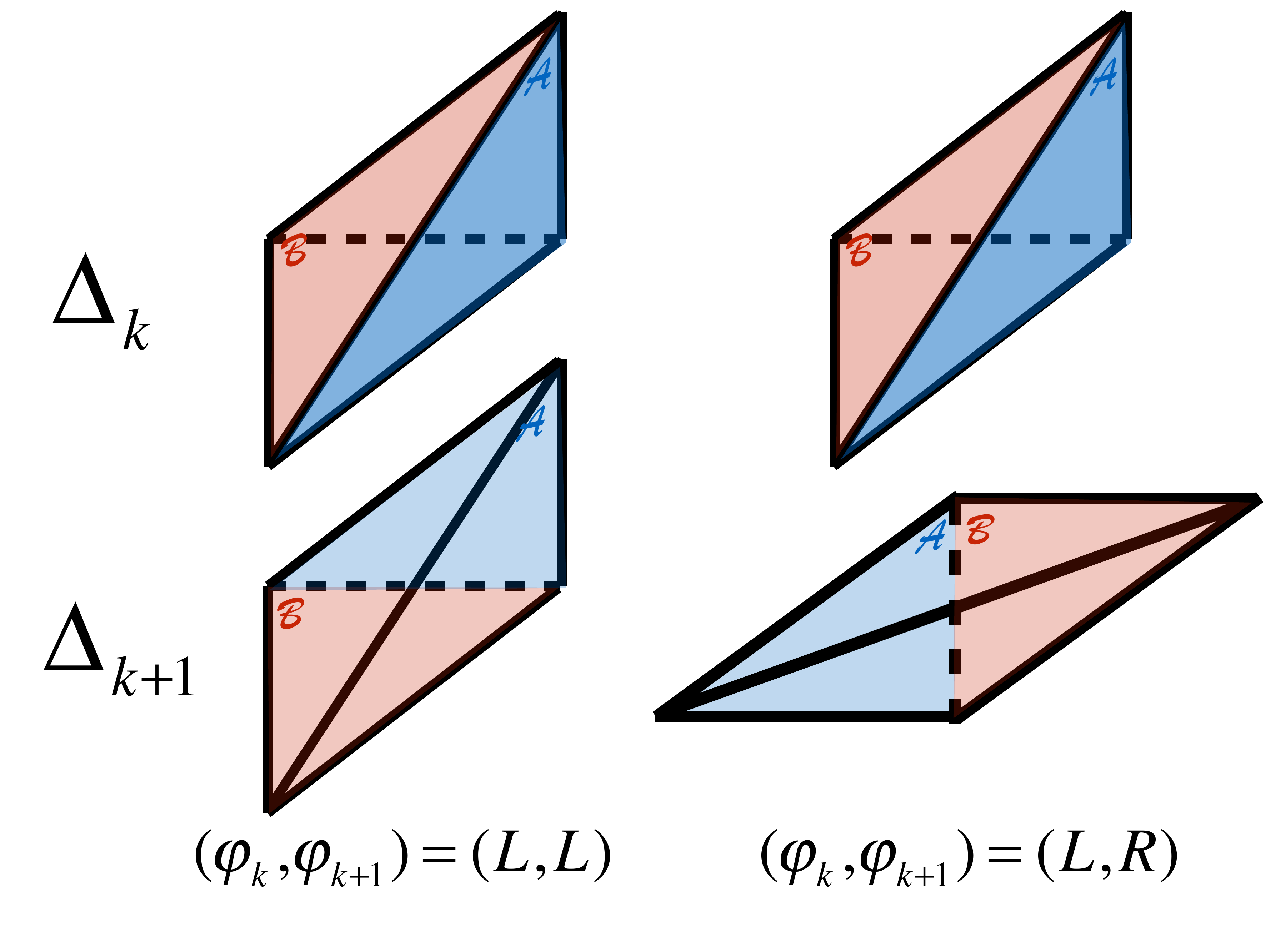}
\centering\includegraphics[scale=0.2]{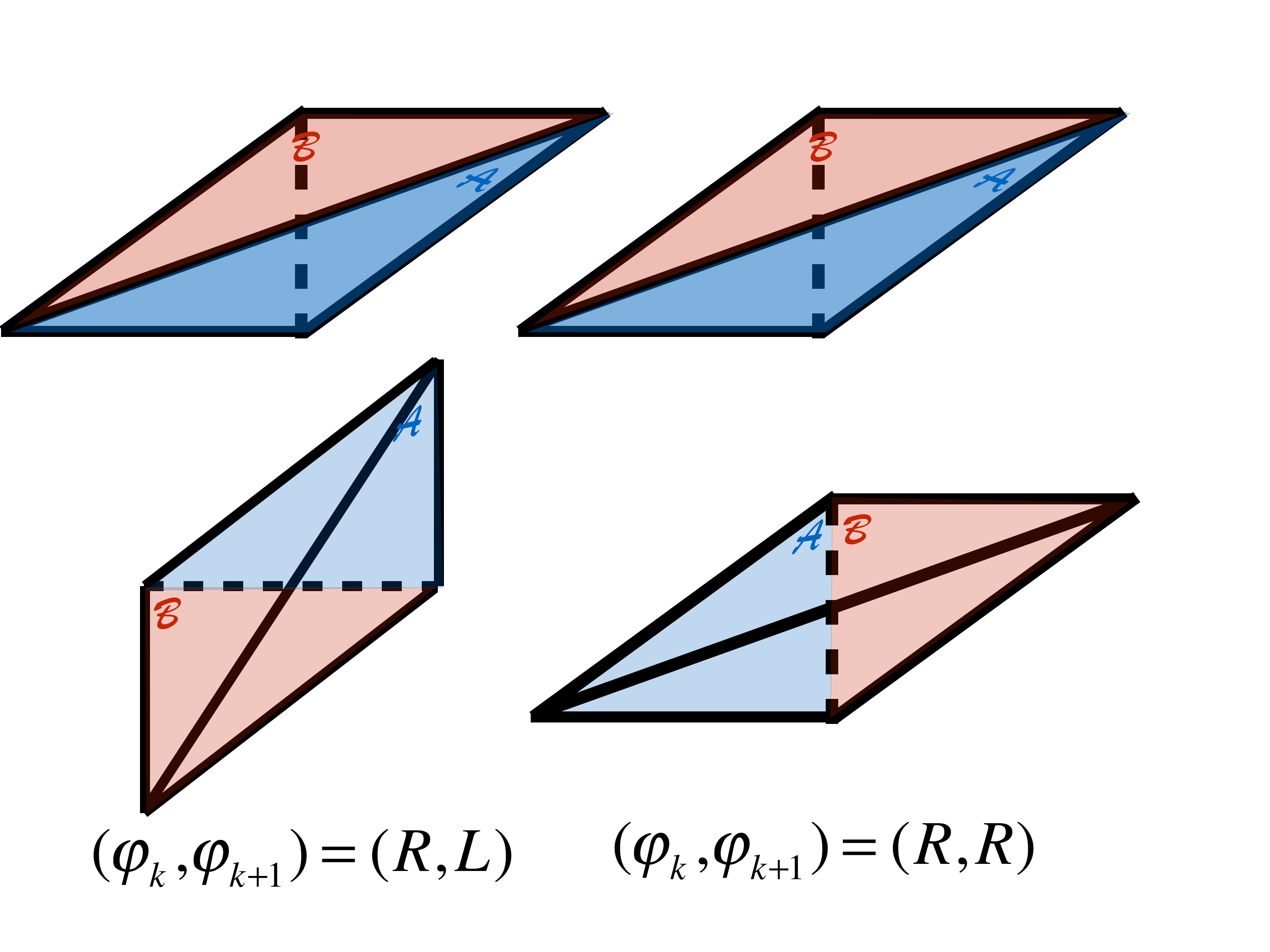}
\caption{An ideal triangulation of $(\Sigma_{1,1}\times S^1)_\varphi$ with $\varphi= \varphi_1 \ldots \varphi_{\sharp}$. The two faces in the top of  $\Delta_k$ are glued to the two faces in the bottom of $\Delta_{k+1}$. We glue them according to the sequence $(\varphi_k , \varphi_{k+1})$, namely the two faces with same color are glued together ({\it cf.} \cite{Terashima:2011xe}).}
\label{Mapping-torus-triangluation}
\end{figure}
Thus, $N$-decomposition of the mapping torus introduce $\frac{\sharp}{6} N(N^2-1)$ octahedra whose vertex variables are labelled by 
\begin{align}
(Z,Z',Z'')^{(k)}_{\a,\b,\g,\d} \;:\; \a+\b+\g+\d=N-2\;\; (\a,\b,\g,\d \geq 0)\;.
\end{align}
\begin{figure}[htbp]
\centering\includegraphics[scale=0.20]{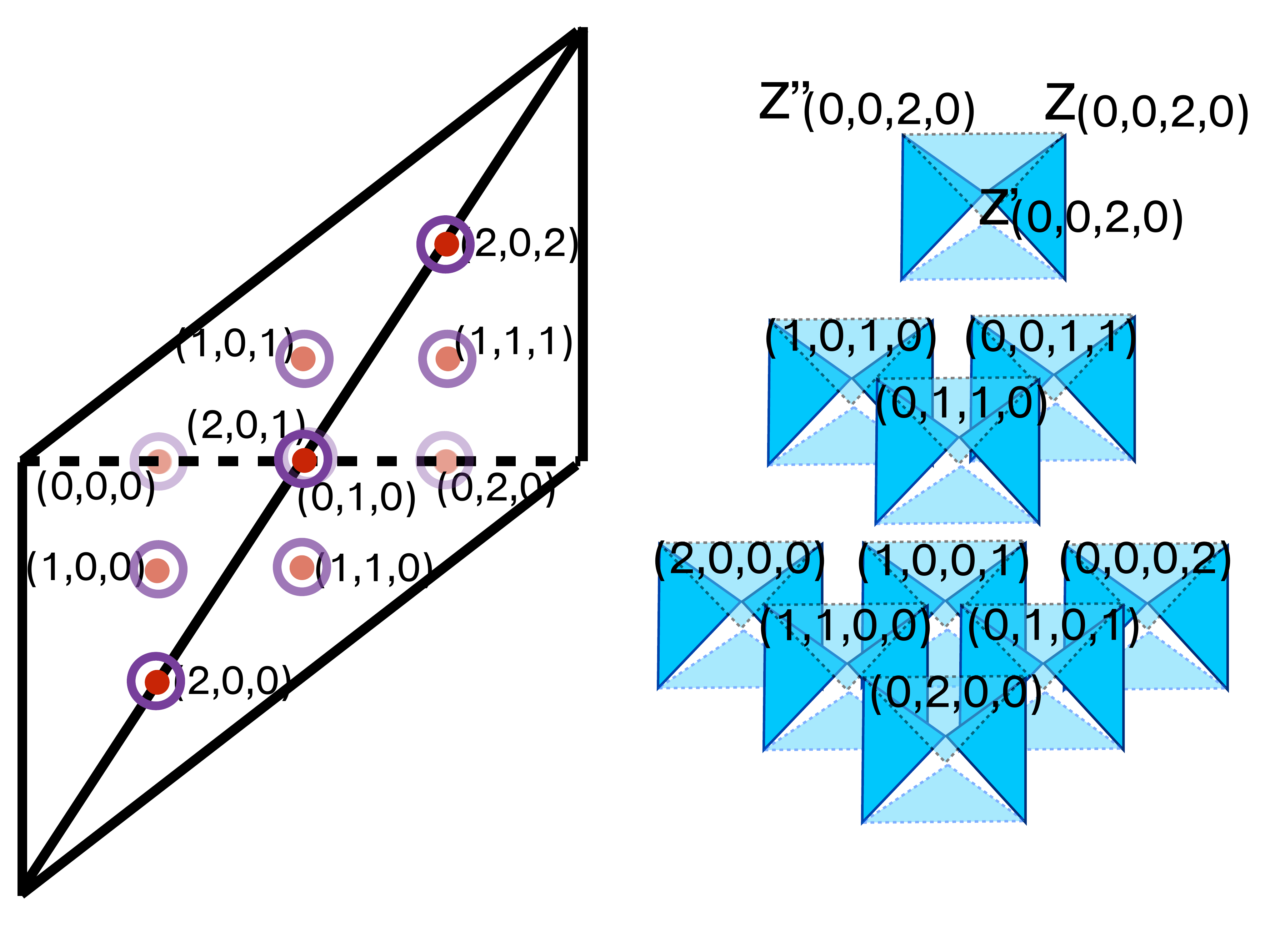}
\centering\includegraphics[scale=0.20]{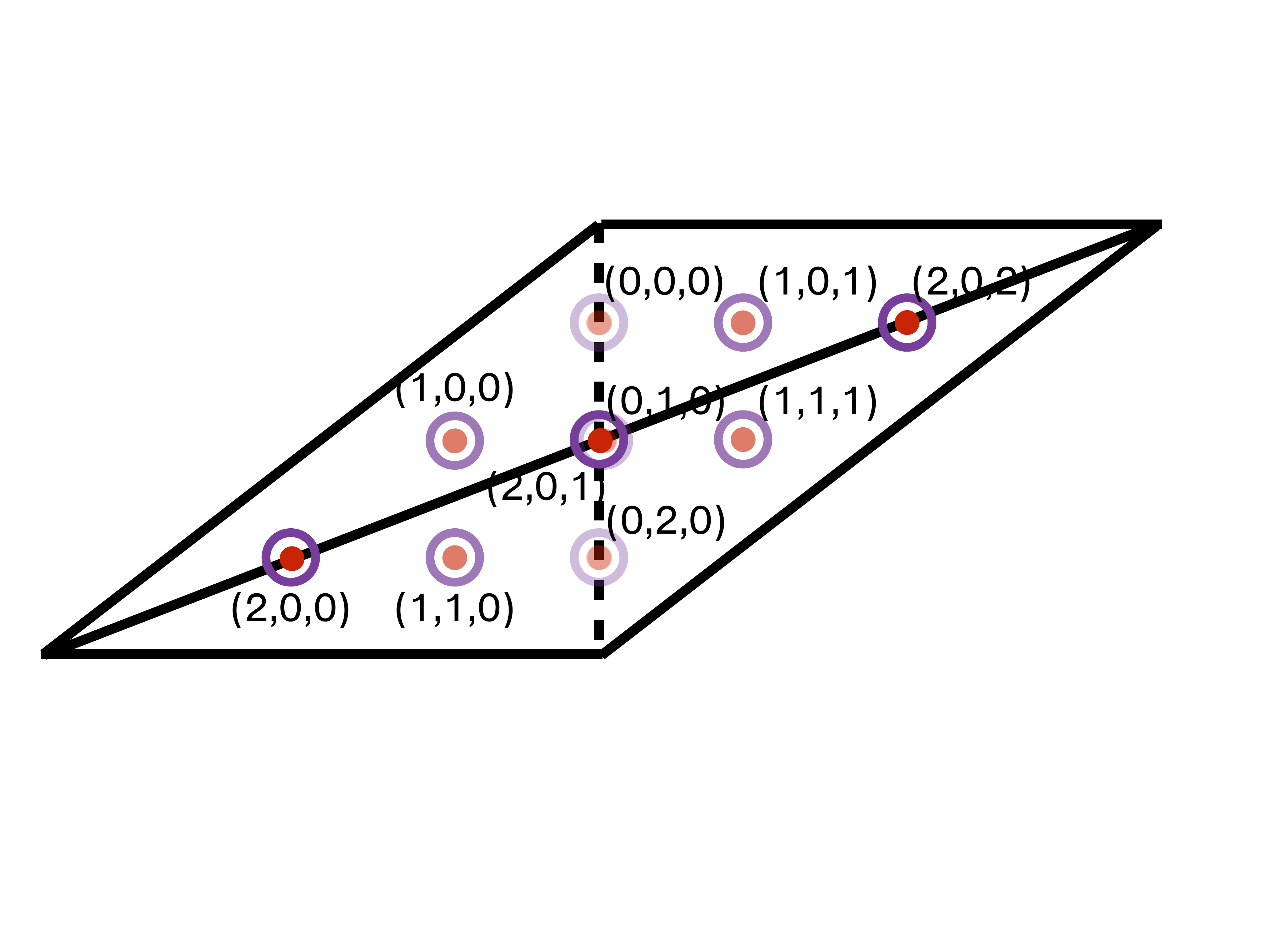}
\caption{Left: Labeling $(p,r,s)$ of mutations in a single flip corresponding to $\bm{L}\in SL(2,\mathbb{Z})$ of FG quiver with $N=4$. Right: Labeling $(Z,Z',Z'')_{(\a,\b,\g,\d)}$ of octahedra in the $N=4$-decomposition of a single tetrahedron. Right: Labeling $(p,r,s)$ of mutations in a single flip corresponding to $\bm{R}\in SL(2,\mathbb{Z})$.  }
\label{Labellings}
\end{figure}
Our convention for labelling octahedra is depicted in  Fig.~\ref{Labellings} using the example for $N=4$.
Then the linear constraints in the cluster partition function are equivalent to gluing equations for the $N$-decomposition under the following identification:
\begin{align}
&(Z,Z',Z'')^{(k)}_{\alpha, \beta, \gamma, \delta} = \left( 
Z^{(k)}(p,r,s) , i \pi + \frac{\hbar}2 -Z^{(k)}(p,r,s)- Z''^{(k)} (p,r,s) , Z''^{(k)}  (p,r,s)\right)  \nn
\\
&\textrm{with }  \quad p = \b+\g, r= \d,s=\g \;. \label{mutation label-2}
\end{align}
The $N$-decomposition for the mapping torus can be drawn  by replacing each tetrahedron in the Fig.~\ref{Mapping-torus-triangluation} by a pyramid of $N(N^2-1)/6$ octahedra with labelling depicted in Fig.~\ref{Labellings}. 
It is notationally rather cumbersome to 
explicitly write down  the full gluing equations for general $\varphi$.
We therefore here write down gluing equations for the internal vertices  in the $N$-decomposition with $\varphi =\bm{LR}$.

Depending on its location in ideal tetrahedra, the internal vertices can be divided into 3 classes: `interior' type  located inside ideal tetrahedra and `edge' type  on the edges and `face' type on the faces. In total, there are $\frac{1}3 N(N^2-1)$ internal vertices:
\begin{align}
\begin{split}
\frac{1}3 N(N^2-1)&= \frac{(N-1)(N-2)(N-3)}3 +2 (N-1)(N-2) +2(N-1) \nn
\\
&= \sharp(\textrm{`interior' vertices})+\sharp (\textrm{`face' vertices})+ \sharp(\textrm{`edge' vertices})\;.
\end{split}
\end{align}
Consequently, the gluing equations also come in three different types. The first is the ``interior type''
\begin{align}
\begin{split}
& Z''^{(k)}_{\a,\b,\g,d}+Z'^{(k)}_{\a,\b-1,\g,\d+1}+Z^{(k)}_{\a-1,\b,\g,\d+1}+Z''^{(k)}_{\a-1,\b-1,\g+1,\d+1} \\
& \qquad \qquad +Z'^{(k)}_{\a-1,\b,\g+1,\d}+Z^{(k)}_{\a,\b-1,\g+1,\d} = 2\pi i  +\hbar  \;,
\end{split}
\end{align}
with $k\in \{ 1,2\},\;1 \leq \a,\b\leq N-2 ,\; 0 \leq \g,\d \leq N-3,\;\a+\b+\g+\d=N-2$.
The second is the ``edge type''
\begin{align}
\begin{split}
&Z'^{(1)}_{0,d,N-2-d,0}+Z''^{(1)}_{N-2-d,d,0,0}+Z^{(1)}_{0,0,N-2-d,d}+Z''^{(2)}_{0,0,N-2-d,d} \\
&\qquad \qquad+Z^{(2)}_{N-2-d,d,0,0}+Z'^{(2)}_{N-2-d,0,0,d} = 2\pi i +\hbar  \;,
\\
& Z^{(1)}_{d,0,N-2-d,0}+Z^{(1)}_{0,d,0,N-2-d}+Z'^{(1)}_{d,0,0,N-2-d}+Z^{(2)}_{0,d,0,N-2-d} \\
& \qquad \qquad+Z^{(2)}_{d,0,N-2-d,0}+Z'^{(2)}_{0,d,N-2-d,0} = 2\pi i +\hbar  \;,
\end{split}
\end{align}
with $0 \leq d \leq N-2$.
Finally, we have the gluing equations of the ``face type'':
\begin{align}
\begin{split}
&Z'^{(1)}_{0,r-s,N-2-r,s}+Z^{(1)}_{0,r-s,N-1-r,s-1} +Z''^{(1)}_{0,r-s+1,N-2-r,s-1}+Z''^{(2)}_{N-2-r,r-s,0,s} 
\\
&\qquad \qquad Z^{(2)}_{N-2-r+1,r-s,0,s-1} +Z'^{(2)}_{N-2-r,r-s+1,0,s-1} = 2 \pi i +\hbar\;,  
\\
& Z'^{(1)}_{r-s+1,s-1,N-2-r,0}+Z^{(1)}_{r-s,s,N-2-r,0} +Z''^{(1)}_{r-s,s-1,N-1-r,0}+Z''^{(2)}_{r-s+1,0,N-2-r,s-1} 
\\
&\qquad \qquad Z^{(2)}_{r-s,0,N-2-r,s}+Z'^{(2)}_{r-s,0,N-r-1,s-1} = 2 \pi i +\hbar\;,  
\\
&Z'^{(1)}_{N-2-r,0,s,r-s}+Z^{(1)}_{N-2-r,0,s-1,r-s+1} +Z''^{(1)}_{N-r-1,0,s-1,r-s}+Z''^{(2)}_{0,N-1-r,s-1,r-s} 
\\
&\qquad \qquad Z^{(2)}_{0,N-2-r,s,r-s}+ Z'^{(2)}_{0,N-2-r,s-1,r-s+1} = 2 \pi i +\hbar\;,  
\\
&Z'^{(1)}_{s-1,N-1-r,0,r-s}+Z^{(1)}_{s,N-2-r,0,r-s} +Z''^{(1)}_{s-1,N-2-r,0,r-s+1}+Z''^{(2)}_{s-1,N-2-r,r-s+1,0} 
\\
&\qquad \qquad Z^{(2)}_{s,N-2-r,r-s,0} + Z'^{(2)}_{s,N-2-r,r-s,0} = 2 \pi i +\hbar\;,  \label{internal vertices for figure-eight}
\end{split}
\end{align}
with
$1\leq s\leq r\leq N-2$.
Out of these equations, only $\frac{1}3 N(N^2-1)-(N-1)$ of them are  independent. These equations are equivalent  to linear constraints $\{C_I=0\}$ appearing in the cluster partition function  \eqref{delta-fuctions-to-gluing} (this can be checked explicitly for a given $N$).
The remaining $2(N-1)$ linear constraints, $\bfrakM_\a = A^x_\alpha  \cdot Z$ and $\frakL_\a :=A^p_\a \cdot Z+B^p_\a \cdot Z'' =0$, 
pick up choice of polarizations in octahedron's gluing equations. The $\{\frakL_\a\}$ will be identified with longitude variables $\{\mathfrak{L}_\a\}$, which corresponds to puncture variables in 2d. Longitude variables corresponds eigenvalues of longitude holonomy 
\begin{align}
\textrm{distinct eigenvalues of Hol($\mathfrak{l}$)} =  \{e^{\mathfrak{L}_\a}\}\;.
\end{align}
The matrix elements of $\textrm{Hol} (\mathfrak{l})$ (\textrm{Hol}(P)) can be expressed as rational functions on octahedron's vertex variables (cluster $y$-variables) using 3d (2d) snakes. Unlike longitude variables, it is non-trivial to find the precise identification for the $\{\bfrakM_\a\}$ variables, and these variables 
are in general a linear combination of the longitude and meridian variables.

\paragraph{Ex 1-1. $(\Sigma_{1,1}\times S^1)_{\varphi}$ with $\varphi=\bm{LR}$ and $N=2$}  The quiver is given by
\begin{align}
&\textrm{vertices } =\{(1,0),(0,1),(1,1) \}\;,  \nn
\\
&Q=\left(\begin{array}{ccc}0 & -2 & 2 \\ 2 & 0 & -2 \\ -2 & 2 & 0\end{array}\right)\;.
\end{align}
The mapping class group $SL(2,\mathbb{Z})$  of the once-punctured torus is realized as 
\begin{align}
\begin{split}
&\bm{L}= \mu_{(1,0)} \sigma_{L}\;, \quad \bm{R}= \mu_{(0,1)} \sigma_R\;, \;\textrm{with} 
\\
&\sigma_R : (0,1) \longleftrightarrow (1,1)\;, \quad \sigma_L : (1,0) \longleftrightarrow (1,1)\;.
\end{split}
\end{align}
Using eq.~\eqref{4_1 as mapping tori}, the $SL(2)$ CS partition function on $S^3\backslash \mathbf{4}_1$ can be realized as a cluster partition function $\textrm{Tr}_{Q, \bm{m},\bm{\sigma}}$ with the following data: 
\begin{align}
\begin{split}
&\bm{m}=\{m_0= (1,0), m_1=(0,1) \}  \;, 
\\
&\bm{\sigma} =\{ \sigma_0 = \sigma_L, \sigma_1= \sigma_R\}\;.
\end{split}
\end{align}
The kernel of $Q$ is spanned by $c=(1,1,1)^T$, and hence
the central element is given by
\begin{align}
\sum_{i=1}^{3}c_{i}Y_i=Y_{(1,0)}+Y_{(0,1)}+Y_{(1,1)}\;.
\end{align}
We can then straightforwardly write down the expression for the cluster partition function from
the results \eqref{trace_result_1} and \eqref{trace_result_2},
and the delta function constraints are given by:
\begin{align}
&C_{Q,\bm{m},\sigma}\cdot \bigg{(}u_{(1,0)}(0), u_{(0,1)}(0),u_{(1,1)}(0), u_{(1,0)}(1), u_{(0,1)}(1),u_{(1,1)}(1) \bigg{)}^T  - \vec{V}=\mathbf{0}\;, \nn
\\  
&Z(0)+Z''(0) +4 u_{(0,1)}(0)-4u_{(1,1)}(0) = Z(1)+Z''(1)-4u_{(1,0)}(1)+4u_{(1,1)}(1)=0\;,  
\end{align}
where
\begin{align}
\begin{split}
&\hat{C}_{Q, \bm{m},\sigma} = \left(\begin{array}{cc}I  & -\sigma_1^{-1}\cdot \hat{P}_{(0,1)}(1) \\ -\sigma_0^{-1} \cdot \hat{P}_{(1,0)}(0) & I \end{array}\right) = \left(\begin{array}{cccccc}1 & 0 & 0& -1 & 0 & 0 \\0 & 1 & 0 & 0 & 0 & -1 \\0 & 0 & 1 & -2 & 1 & 0 \\ 0 & 0 & -1 & 1 & 0 & 0 \\0 & -1 & 0 & 0 & 1 & 0 \\1 & 0 & -2 & 0 & 0 & 1\end{array}\right)\;, 
\\
& \vec{V} =- \frac{1}{2}  \bigg{(}  \bfrakM+ Z(0),  \bfrakM,   \bfrakM, 0, Z(1),   0 \bigg{)}^T\;. 
\end{split}
\end{align}
These constraints imply
\begin{align}
\begin{split}
&\bfrakM=Z(0)-Z(1)\;, \quad \frakL:= -Z(0)+Z''(0) =0\;, \\
& C := Z(0)-Z''(0)+Z(1)-Z''(1)=0\;.
\end{split}
\label{41_constraint_N2}
\end{align}
This is compatible with  eq.~\eqref{delta-fuctions-to-gluing}.
These equations gives the gluing equations   \eqref{gluing for figure-eight} for the ideal triangulation for the mapping torus with the identification
\begin{align}
\begin{split}
&\left( Z(0),Z'(0):= \pi i + \frac{\hbar}2 - Z(0)-Z''(0), Z''(0) \right)  \; \Longleftrightarrow  \; \big{(} Y , Y', Y'' \big{)} \;, 
\\
&\left( Z(1),Z'(1):= \pi i + \frac{\hbar}2 - Z(1)-Z''(1), Z''(1)  \right)  \; \Longleftrightarrow \; \big{(} Z , Z', Z''  \big{)}\;, 
\\
&(\bfrakM,\frakL) \; \Longleftrightarrow \; ( -\frakM+\frakL,\frakL)\;.  
\end{split}
\label{Z0Z1 to YZ}
\end{align}
Thus using eq.~\eqref{trace_as_overlap} we see that the cluster partition function is same as the partition function of the 
state-integral model in a polarization $\Pi=(\frakL, \frakM-\frakL)$
\begin{align}
\textrm{Tr}_{Q,\bm{m},\bm{\sigma}} (\frakL) = \big{\langle} \frakL, C_I=0 \big{|}  \Diamond^{\otimes 2}\big{\rangle}\;.
\end{align}
Such a change of the polarization in the cusped boundary is expected since
our partition functions have framing ambiguities as in eq.~\eqref{ambiguity in cluster ptn}.

\begin{figure}[htbp]
\begin{center}
   \includegraphics[width=.4\textwidth]{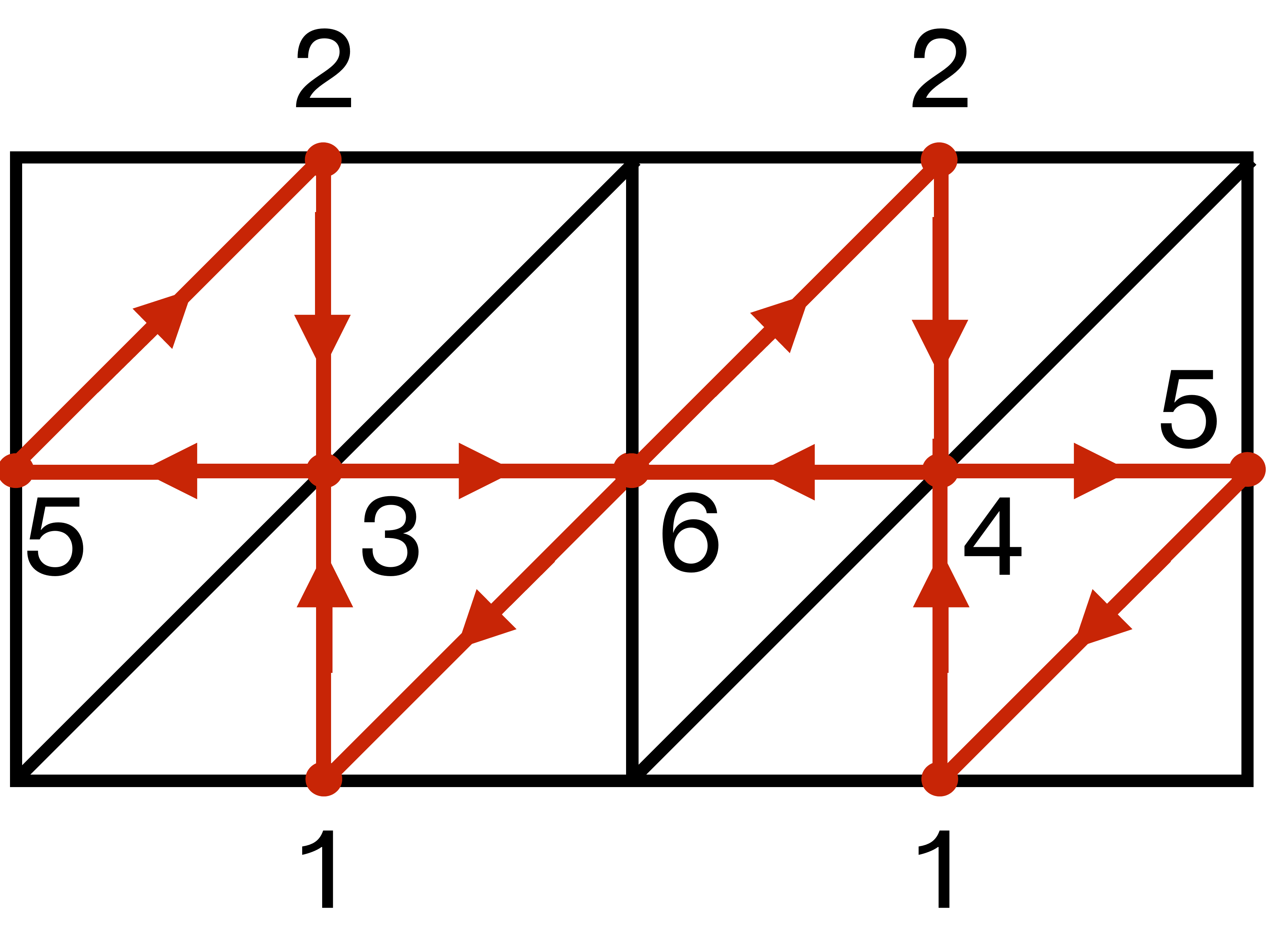}
   \end{center}
   \caption{Quiver for $\CM_{N=2}(\Sigma_{0,4}, \vec{\rho}=[1,1]^{\otimes 4})$. $\Sigma_{0,4}$ can be decomposed into four ideal triangles.} .
    \label{fig:sigma04}
\end{figure}

\paragraph{Ex 1-2. $(\Sigma_{1,1}\times S^1)_{\varphi}$ with $\varphi=\bm{LR}$ and $N=3$}  The quiver is given by
\begin{align}
\begin{split}
&\textrm{Vertices}=\{(1, 0), (2, 0),(0, 1), (1, 1), (2, 1), (0, 2), (1, 2), (2, 2)\} \;, 
\\
&Q= \left(
\begin{array}{cccccccc}
 0 & 0 & 0 & 1 & -1 & -1 & 1 & 0 \\
 0 & 0 & -1 & 0 & 1 & 0 & -1 & 1 \\
 0 & 1 & 0 & -1 & -1 & 0 & 1 & 0 \\
 -1 & 0 & 1 & 0 & 1 & 0 & -1 & 0 \\
 1 & -1 & 1 & -1 & 0 & -1 & 0 & 1 \\
 1 & 0 & 0 & 0 & 1 & 0 & -1 & -1 \\
 -1 & 1 & -1 & 1 & 0 & 1 & 0 & -1 \\
 0 & -1 & 0 & 0 & -1 & 1 & 1 & 0 \\
\end{array}
\right)\;.
\end{split}
\end{align}
The mutations $\bm{m}$ and and the permutations $\bm{\sigma}$ are
\begin{align}
\begin{split}
&\bm{m} = (\{1, 0\}, \{2, 0\}, \{1, 2\}, \{2, 1\}, \{0, 2\}, \{0, 1\}, \{2, 1\}, \{1, 2\})\;,
\\
&\bm{\sigma} =\{\mathbb{I},\mathbb{I},\mathbb{I},\sigma_L, \mathbb{I},\mathbb{I},\mathbb{I},\sigma_R \}\;.
\end{split}
\end{align}
We choose two central elements to be 
\begin{align}
\begin{split}
&\sum_{i=1}^{8}c^{(1)}_{i}Y_i= -Y_{(1,0)}-Y_{(2,0)}-Y_{(0,1)}-Y_{(1,1)}-Y_{(2,1)}-Y_{(0,2)}-Y_{(1,2)}-Y_{(2,2)}\;, 
\\
&\sum_{i=1}^{8}c^{(2)}_{i}Y_i= -Y_{(2,1)}-Y_{(1,2)}\;.
\end{split}
\end{align}
Applying these $(Q,\bm{m},\bm{\sigma})$ to eqs.~\eqref{trace_result_1} and \eqref{trace_result_2}, we have following constraints 
\begin{align}
\begin{split}
&C_I\big{(}\textrm{in }eq.~\eqref{octahedron's gluing for 4_1 with N=3}\big{)}=0\;,\;\;  I=1,\ldots, 8\;, 
\\
&\frakL_\a \big{(}\textrm{in eq.}~\eqref{octahedron's gluing for 4_1 with N=3}\big{)}=0\;, \;\bfrakM_\a = -\frakM_\a  (\textrm{in }eq.~\eqref{octahedron's gluing for 4_1 with N=3})\;, \;\; \a=1,2\;.
\end{split}
\end{align}
These constraints are  same as expected from eq.~ \eqref{delta-fuctions-to-gluing}. Here we use eqs.~\eqref{mutation label-1} and \eqref{mutation label-2}, and express $\{Z^*(t)\}$ by $\{Y^{*}_{\a,\b,\g,\d}:=Z^{*(1)}_{\a,\b,\g,\d} ,Z^{*}_{\a,\b,\g,\d}:=Z^{*(2)}_{\a,\b,\g,\d}\}$ where $*$ is $'$ or $''$ or nothing. Thus, we see that the cluster partition function is same as the partition function of the 
state-integral model in a polarization $\Pi=(\frakL_1,\frakL_2, \frakM_1,\frakM_2)$
\begin{align}
\textrm{Tr}_{Q,\bm{m},\bm{\sigma}} (\frakL_1,\frakL_2) = \big{\langle} \frakL_1,\frakL_2, C_I=0 \big{|}  \Diamond^{\otimes 2}\big{\rangle}\;.
\end{align}
\paragraph{Ex 2. $(\Sigma_{0,4}\times S^1)_{\varphi}$ with $\varphi=\bm{LR}$ and $N=2$} Ideal triangulation and FG quiver for $\Sigma_{0,4}$ with $N=2$ is given in figure.~\ref{fig:sigma04}.
The skew-symmetric matrix $Q$ is 
\begin{align}
Q=\left(
\begin{array}{cccccc}
 0 & 0 & 1 & 1 & -1 & -1 \\
 0 & 0 & 1 & 1 & -1 & -1 \\
 -1 & -1 & 0 & 0 & 1 & 1 \\
 -1 & -1 & 0 & 0 & 1 & 1 \\
 1 & 1 & -1 & -1 & 0 & 0 \\
 1 & 1 & -1 & -1 & 0 & 0 \\
\end{array}
\right)\;.
\end{align}
Mapping class group of $\Sigma_{0,4}$ contains $SL(2,\mathbb{Z})$ and its two generators $\bm{L,R}$ can be written as
\begin{align}
\begin{split}
&\bm{L}=\mu_1 \mu_2 \sigma_L\;, \quad \bm{R}=\mu_5 \mu_6 \sigma_R\;, \quad \textrm{with}
\\
&\sigma_L=\left(
\begin{array}{cccccc}
 0 & 0 & 0 & 1 & 0 & 0 \\
 0 & 0 & 1 & 0 & 0 & 0 \\
 1 & 0 & 0 & 0 & 0 & 0 \\
 0 & 1 & 0 & 0 & 0 & 0 \\
 0 & 0 & 0 & 0 & 1 & 0 \\
 0 & 0 & 0 & 0 & 0 & 1 \\
\end{array}
\right)\;, \quad \sigma_R=\left(
\begin{array}{cccccc}
 1 & 0 & 0 & 0 & 0 & 0 \\
 0 & 1 & 0 & 0 & 0 & 0 \\
 0 & 0 & 0 & 0 & 0 & 1 \\
 0 & 0 & 0 & 0 & 1 & 0 \\
 0 & 0 & 1 & 0 & 0 & 0 \\
 0 & 0 & 0 & 1 & 0 & 0 \\
\end{array}
\right)\;.
\end{split}
\end{align}
Classical transformation on the y-variables for these generators are
\begin{align}
\begin{split}
&\bm{L} : (y_1, y_2, y_3, y_4, y_5)
\\
& \;\; \rightarrow \left\{\frac{y_1 y_2 y_3}{\left(y_1+1\right) \left(y_2+1\right)},\frac{y_1 y_2 y_4}{\left(y_1+1\right) \left(y_2+1\right)},\frac{1}{y_2},\frac{1}{y_1},\left(y_1+1\right) \left(y_2+1\right) y_5,\left(y_1+1\right) \left(y_2+1\right) y_6\right\} \;,
\\
&\bm{R} : (y_1, y_2, y_3, y_4, y_5) 
\\
& \;\; \rightarrow \left\{\frac{y_1 y_5 y_6}{\left(y_5+1\right) \left(y_6+1\right)},\frac{y_2 y_5 y_6}{\left(y_5+1\right) \left(y_6+1\right)},\frac{1}{y_5},\frac{1}{y_6},y_4 \left(y_5+1\right) \left(y_6+1\right),y_3 \left(y_5+1\right) \left(y_6+1\right)\right\}\;.
\end{split}
\end{align}
Note that only two out of four central elements are invariant under  $\varphi=\bm{LR}$. We choose two central elements as
\begin{align}
\begin{split}
 &\sum_{i=1}^6 c^{(1)}_i Y_i :=\frac{1}3( Y_1 +2Y_2+Y_3+2Y_4+2Y_5+Y_6)\;, 
 \\
 &\sum_{i=1}^6 c^{(2)}_i Y_i := \frac{1}3( 2Y_1 +Y_2+2Y_3+Y_4+Y_5+2Y_6)\;.
 \end{split}
\end{align}
Integrating out $\{\vec{u}_i(t)\}$ in the delta-function constraints in cluster partition function \eqref{trace_result_1}, we have
\begin{align}
\begin{split}
&\bfrakM_{1}= Z(0)-Z(2)\;, \quad \bfrakM_{2}=Z(1)-Z(3)\;, 
\\
&\frakL_1 := Z''(0)-Z(1)=0\;, \quad \frakL_2:=Z''(1)-Z(0)=0\;,
\\
&C_1 := Z(1)+Z(3)-Z''(0)-Z''(2)=0\;, \quad C_2 = Z(0)+Z(2)-Z''(1)-Z''(3)=0\;.
\end{split}
\end{align}
which have  expected structure in eq.~\eqref{delta-fuctions-to-gluing}.

%%%%%%%%%%%%%%%%%%%%%%%%%%%%%%%%%%%%%%%%%%%%%%%%%%%%%%
\subsubsection{Co-dimension 2 Defects: \texorpdfstring{$\rho=\textrm{non-maxiaml}$}{rho=simple}}
\label{sec.Xe_rule}
\begin{figure}[htbp]
\begin{center}
   \includegraphics[width=.31\textwidth]{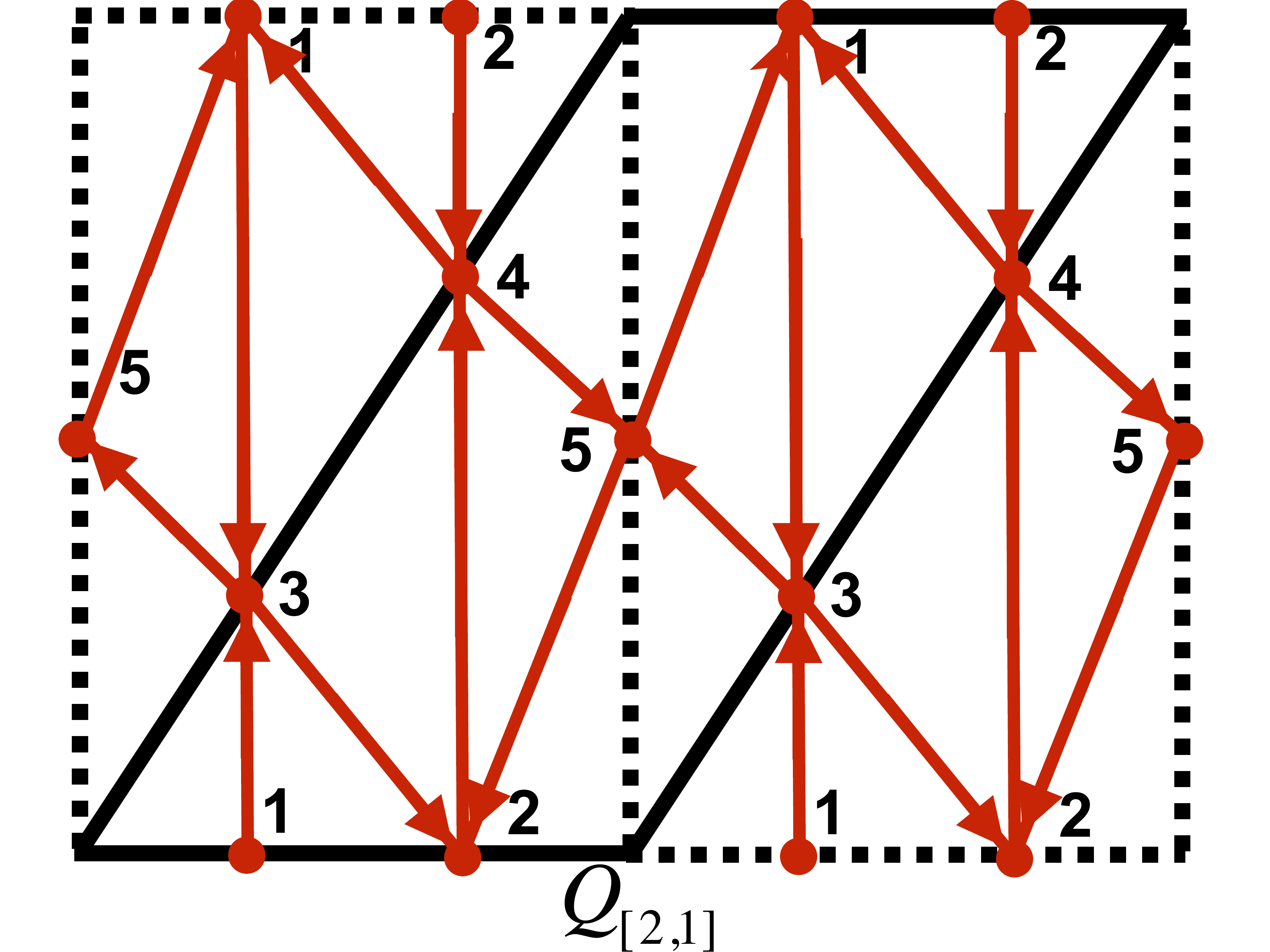}
   \;
      \includegraphics[width=.31\textwidth]{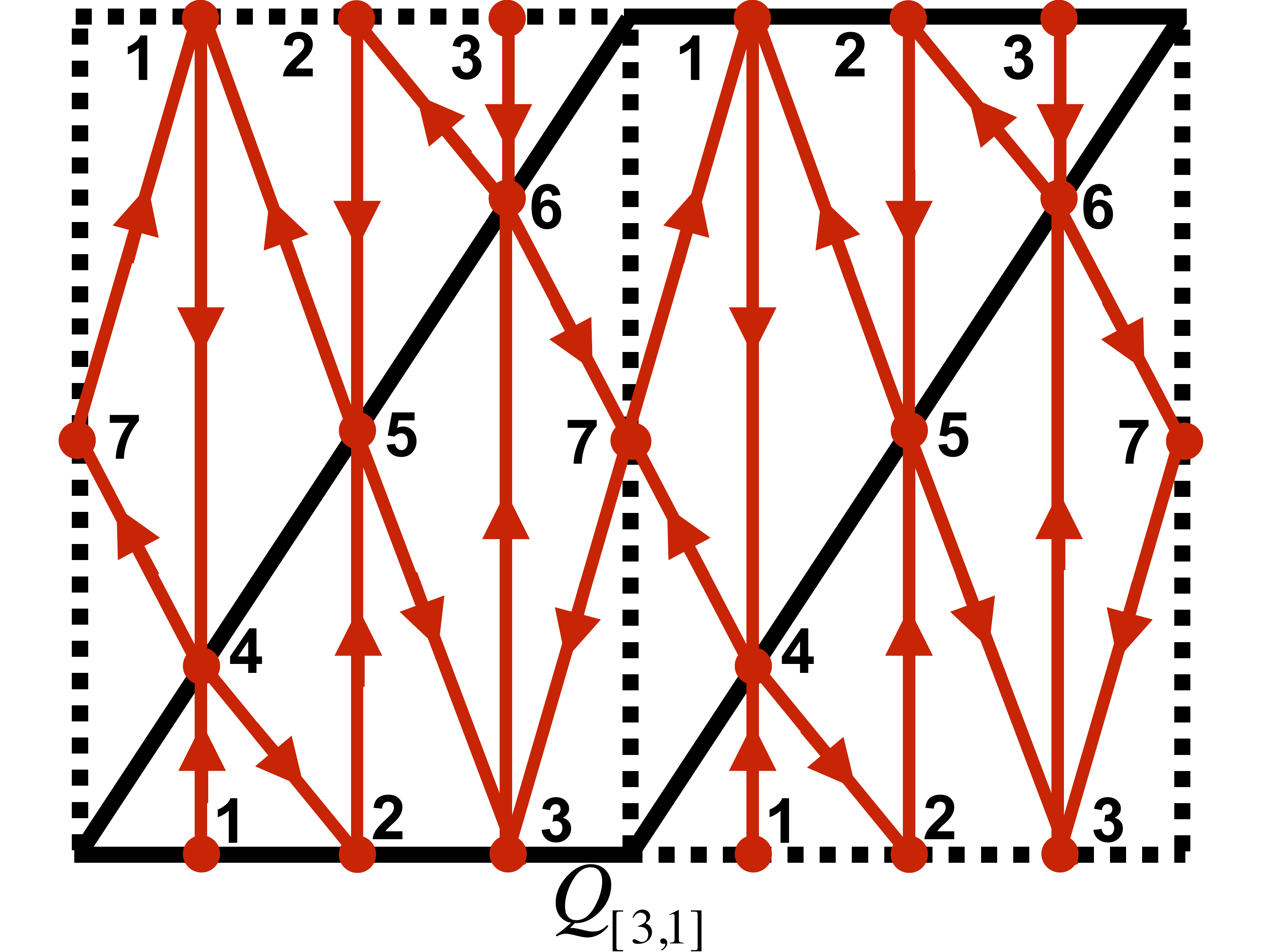}
      \;
        \includegraphics[width=.31\textwidth]{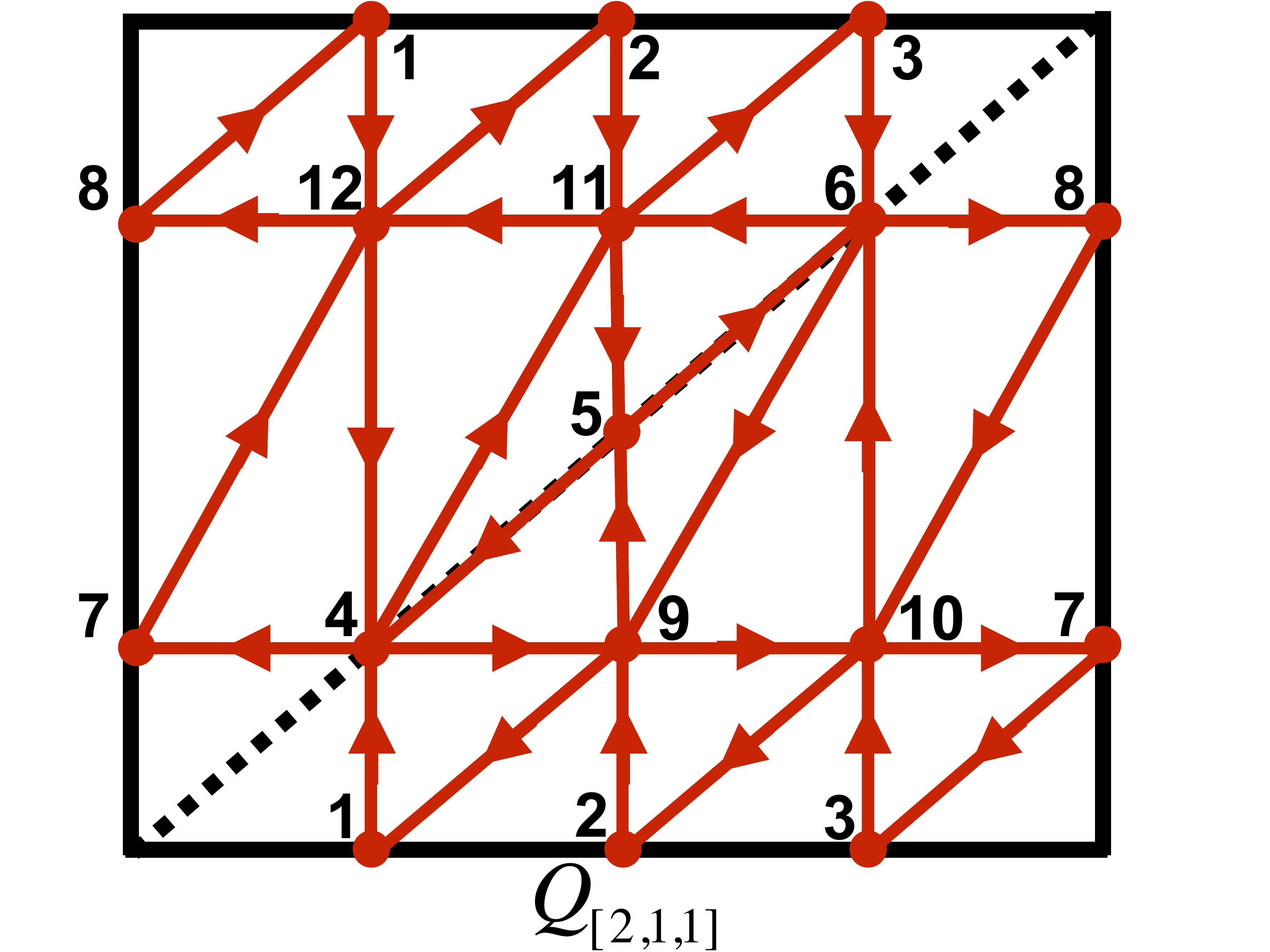}
   \end{center}
   \caption{Quivers for $\CM_{N=3}(\Sigma_{1,1}, \textrm{simple}),\CM_{N=4}(\Sigma_{1,1}, \textrm{simple})$ and $\CM_{N=4}(\Sigma_{1,1}, [2,1,1])$ (from left to right). Fundamental region of torus is chosen as the region surrounded by  black  lines.
  For $\rho$=simple case, generalization to arbitrary $N$ is obvious, there are $2N+1$ nodes in the quiver.}
    \label{fig:non-maximal-once-punctured-quiver}
\end{figure}

Let us next turn to non-maximal $\rho$. We proceed to work on simple but yet non-trivial examples: $\Sigma=\Sigma_{1,1}$ (once-punctured torus) with $N=3$ and $4$. 
We study the quiver for  $N=3$ with $\rho=[2,1]$ (simple) and for $N=4$ with $\rho =[3,1]$ (simple) and $\rho=[2,1,1]$.
In these cases, the proposed  quivers are drawn  in Fig.~\ref{fig:non-maximal-once-punctured-quiver}.

Notice that our quiver breaks the symmetry between the three edges of the quiver, which was present for $\rho$=maximal case.\footnote{This 
is an interesting feature, and means that some of the considerations for the maximal case
requires certain modifications.
For example, it looks like that the the specification of the quiver requires not just an ideal triangulation, but 
in addition an ordering of the vertices of each ideal triangle. We hope to explore this issue 
further in the future, and here are content in working out an example of the once-punctured torus.}
There are several indications that this gives the correct quiver for the case at hand.
First, we can indeed find a sequence of mutations realizing the flips, and satisfying the 
relations of the full mapping class group $SL(2, \mathbb{Z})$. In fact,
if we choose a random quiver this is almost never the case.
Second, the quiver gives the correct dimensionality \eqref{Quiver-counting} for the
moduli space of flat connections.
Third, the our quiver is consistent with
the proposal of \cite{Xie:2012dw}
(motivated by generalized s-rule in the 5-brane configuration),
as well as the mathematical work of \cite{FominP} (in particular its Figure 9).

\paragraph{Ex 3. $N=3$  with $\rho=[2,1]$ }
The quiver is drawn in Fig.~\ref{fig:non-maximal-once-punctured-quiver} and the skew-symmetric matrix $Q$ is
\begin{align}
Q = \left(\begin{array}{ccccc}0 & 0 & 2 & -1 & -1 \\0 & 0 & -1 & 2 & -1 \\ -2 & 1 & 0 & 0 & 1 \\1 & -2 & 0 & 0 & 1 \\1& 1 & -1 & -1 & 0\end{array}\right)\;.
\end{align}
Here we have $|\textrm{Ker}(Q)|=1$ and a central element in the cluster algebra is 
\begin{align}
\sum_{i=1}^5 c_i Y^i:=Y_1+Y_2+Y_3+Y_4+Y_5 \;. \label{central element for Q_{2,1}}
\end{align}
Two generators $\bm{S}, \bm{T}$ of the mapping class group $SL(2,\mathbb{Z})$ can be represented as
\begin{align}
\bm{S} =\mu_5 \sigma_S\;, \quad \bm{T}= \mu_3 \mu_4 \sigma_T \;.
\end{align}
The permutation $\sigma_S$ and $\sigma_T$ are given by 
\begin{align}
&\sigma_S = \left(
\begin{array}{ccccc}
 0 & 0 & 0 & 1 & 0 \\
 0 & 0 & 1 & 0 & 0 \\
 1 & 0 & 0 & 0 & 0 \\
 0 & 1 & 0 & 0 & 0 \\
 0 & 0 & 0 & 0 & 1 \\
\end{array}
\right)\;, \quad 
 \sigma_T = \left(
\begin{array}{ccccc}
 0 & 0 & 1 & 0 & 0 \\
 0 & 0 & 0 & 1 & 0 \\
 1 & 0 & 0 & 0 & 0 \\
 0 & 1 & 0 & 0 & 0 \\
 0 & 0 & 0 & 0 & 1 \\
\end{array}
\right) \;.
\end{align}
Classical transformation of $\bm{S}$ and $\bm{T}$ are
\begin{align}
\begin{split}
&\bm{S} : (y_1, y_2, y_3, y_4, y_5) \rightarrow \big{(}\mu_5(y_1),\mu_5(y_2),\mu_5(y_3),\mu_5(y_4),\mu_5(y_5)\big{)}\cdot  \sigma_S
\\
& \qquad =\left\{y_3 \left(y_5+1\right),y_4 \left(y_5+1\right),\frac{y_2 y_5}{y_5+1},\frac{y_1 y_5}{y_5+1},\frac{1}{y_5}\right\}\;,
\\
&\bm{T} : (y_1, y_2, y_3, y_4, y_5) \rightarrow \left( \mu_2\mu_1(y_1),\mu_2\mu_1(y_2),\mu_2\mu_1(y_3),\mu_2\mu_1(y_4),\mu_2\mu_1(y_5)\right) \cdot \sigma_T
\\
& \qquad =\left\{\frac{1}{y_3},\frac{1}{y_4},\frac{y_1 \left(y_3+1\right){}^2 y_4}{y_4+1},\frac{y_2 y_3 \left(y_4+1\right){}^2}{y_3+1},\frac{y_3 y_4 y_5}{\left(y_3+1\right) \left(y_4+1\right)}\right\}\;.
\end{split}
\end{align}
One can check that these generators form $SL(2,\mathbb{Z})$:
\begin{align}
\bm{S}\bm{S}\bm{S}\bm{S}\;, \; \bm{T}\bm{S}\bm{T}\bm{S}\bm{T}\bm{S} : \;(y_1,y_2, y_3, y_4, y_5) \rightarrow   \;(y_1,y_2, y_3, y_4, y_5) \;.
\end{align}
For mapping torus $M= (\Sigma_{1,1}\times S^1)_{\varphi}$ with $\rho=\textrm{simple}$, the corresponding cluster partition function  datum $\{\bm{m}, \bm{\sigma}\}$ can be obtained by decomposing $\varphi \in SL(2,\mathbb{Z})$ into products of $\bm{S},\bm{T}$ and its inverses.  For figure-eight knot complement, $\varphi=\bm{ST}^{-1}\bm{S}^{-1}\bm{T}$   and using $\hat{\mu}_k^2=$(identity)
\begin{align}
\begin{split}
&\varphi=\bm{ST}^{-1}\bm{S}^{-1} \bm{T} = \hat{\mu_5} \hat{\sigma}_S \hat{\sigma}^{-1}_T \hat{\mu}_4 \hat{\mu}_3  \hat{\sigma}_S^{-1} \hat{\mu}_5 \hat{\mu}_3 \hat{\mu}_4 \hat{\sigma}_T 
\\
&  \Longrightarrow \bm{m}= \{ 5, 4, 3, 5, 3,4\} \;,
\\ 
& \quad \;\; \;\; \bm{\sigma} =  \{ \sigma_S \sigma_T^{-1}, \mathbb{I}, \sigma_S^{-1}, \mathbb{I}, \mathbb{I}, \sigma_T \}\;.\
\end{split}
\end{align}
The $\delta$-functions in the cluster partition function  are given by  
\begin{align}
\begin{split}
&\bfrakM= Z(0)-Z(3)\;, \; \frakL:=Z(1)-Z(2)+Z(3)+Z''(0)-Z''(1) =0\;, \; 
\\
&C_1 :=  Z(4)+Z(5)-Z''(1)-Z''(2)=0\;, \\
&C_2:=-2Z(1)+2Z(2)+Z''(1)-Z''(2)=0\;, \;\;  
\\
&C_3 := -Z(0)-Z(3)+Z(4)+Z(5)-Z''(0)-Z''(3)=0\;,
\\
&C_4:=Z(0)+Z(1)+Z(2)+Z(3)-Z(5)-2Z''(4)=0\;, 
\\
&C_5 = -Z(4)+Z(5)+2Z''(4)-2Z''(5)=0\;.
\end{split}
\end{align}
From these gluing equations, the cluster partition function with $k=0$ (superconformal index) can be written as \eqref{trace_as_overlap}
\begin{align}
\begin{split}
&\CI_{S^3\backslash \bm{4}_1, \textrm{simple}}(m_\eta,\eta) :=\textrm{Tr}_{Q,\bf{m},\bm{\sigma}} (\CL)|_{\frakL=\frac{\hbar}2 m_\eta +  \log \eta} =\langle \frakL, ,C_I=0\big{|} \Diamond^{\otimes 5}\rangle|_{\frakL=\frac{\hbar}2 m_\eta +  \log \eta}\;.
\end{split}
\end{align}
Here the $\langle \frakL ,C_I \big{|}$ is a position basis in a polarization $\Pi=(\frakL, C_I; -\bfrakM, \Gamma_I)$  with a choice of $\G_I$ conjugate to $C_I$. 
In charge basis, the index can be computed using the general formula \eqref{index using state-integral from ideal triangulation}.
\begin{align}
\begin{split}
&\CI^c_{S^3\backslash \bm{4}_1, \textrm{simple}}(m_\eta,e_\eta) = \sum_{e_i \in \mathbb{Z}^5} \CI_\Diamond^c\left(e_2,e_2-e_3+\frac{e_4}{2}+\frac{e_5}{2}\right)\CI_\Diamond^c\left(e_3,-e_2+e_3+\frac{e_4}{2}+\frac{e_5}{2}\right) 
\\
&\qquad \qquad \times \CI_\Diamond^c\left(e_4,e_1+\frac{e_2}{2}+\frac{e_3}{2}-\frac{e_5}{2}+\frac{e_\eta}{2}\right) \CI_\Diamond^c\left(e_5,e_1+\frac{e_2}{2}+\frac{e_3}{2}-\frac{e_4}{2}+\frac{e_\eta}{2}\right)\\
&\qquad \qquad  \times \CI_\Diamond^c\left(e_1,-e_1+\frac{e_4}{2}+\frac{e_5}{2}-e_\eta+m_\eta\right) 
\CI_\Diamond^c\left(e_1+e_\eta,-e_1+\frac{e_4}{2}+\frac{e_5}{2}-m_\eta\right) \;.
\end{split}
\end{align}
The octahedron's index $\CI_\Diamond^c$ in charge basis is defined in eq.~\eqref{Octahedron's CS index in charge basis} and for non integer $(m,e) \notin \mathbb{Z}^2$ the index defined to be zero. 
For example, listing first several order of the index in fugacity basis:
\begin{align}
\begin{split}
\CI_{S^3\backslash \bm{4}_1, \textrm{simple}} (0, \eta)&:= 1+ \left(2\eta+\frac{2}{\eta} \right) q^{\frac{3}{2}}+\left(8+2 \eta^2 + \frac{2}{\eta^2}\right) q^2+\left(6 \eta+\frac{6}\eta \right)q^{\frac{5}{2}} 
\\
& \qquad\qquad +\left(2- 3 \eta^2- \frac{3}{\eta^2}\right)q^3+\ldots  \;,
\\
\CI_{S^3\backslash \bm{4}_1, \textrm{simple}} (1, \eta)&:=  \eta \left(\frac{1}{\eta^2}+\frac{1}{\eta} + \eta +\eta^2 \right) q+\eta \left(6 +3\eta +  \frac{3}{\eta}\right) q^2 \\
&\qquad \qquad+\eta \left(-6 - \frac{1}{\eta^3}- \frac{3}{\eta^2}-\frac{5}{\eta}-5 \eta-3\eta^2 -\eta^3\right)q^3+\ldots \;. \label{N=3 4_1 simple from cluster}
\end{split}
\end{align}

In App.~\ref{app.rho_simple} we repeat the same computation without relying on the general machinery of the cluster partition function, and more directly from the analysis of the Hilbert space associated with the cluster algebra mutations.
More interestingly the index can be reproduced from index computation using a non-Abelian description of the $T_{N=3}[S^3\backslash \bf{4}_1, \textrm{simple}]$ in Sec.~\ref{TSU3_index_glue}. %

\begin{figure}[htbp]
\begin{center}
\includegraphics[width=.36\linewidth]{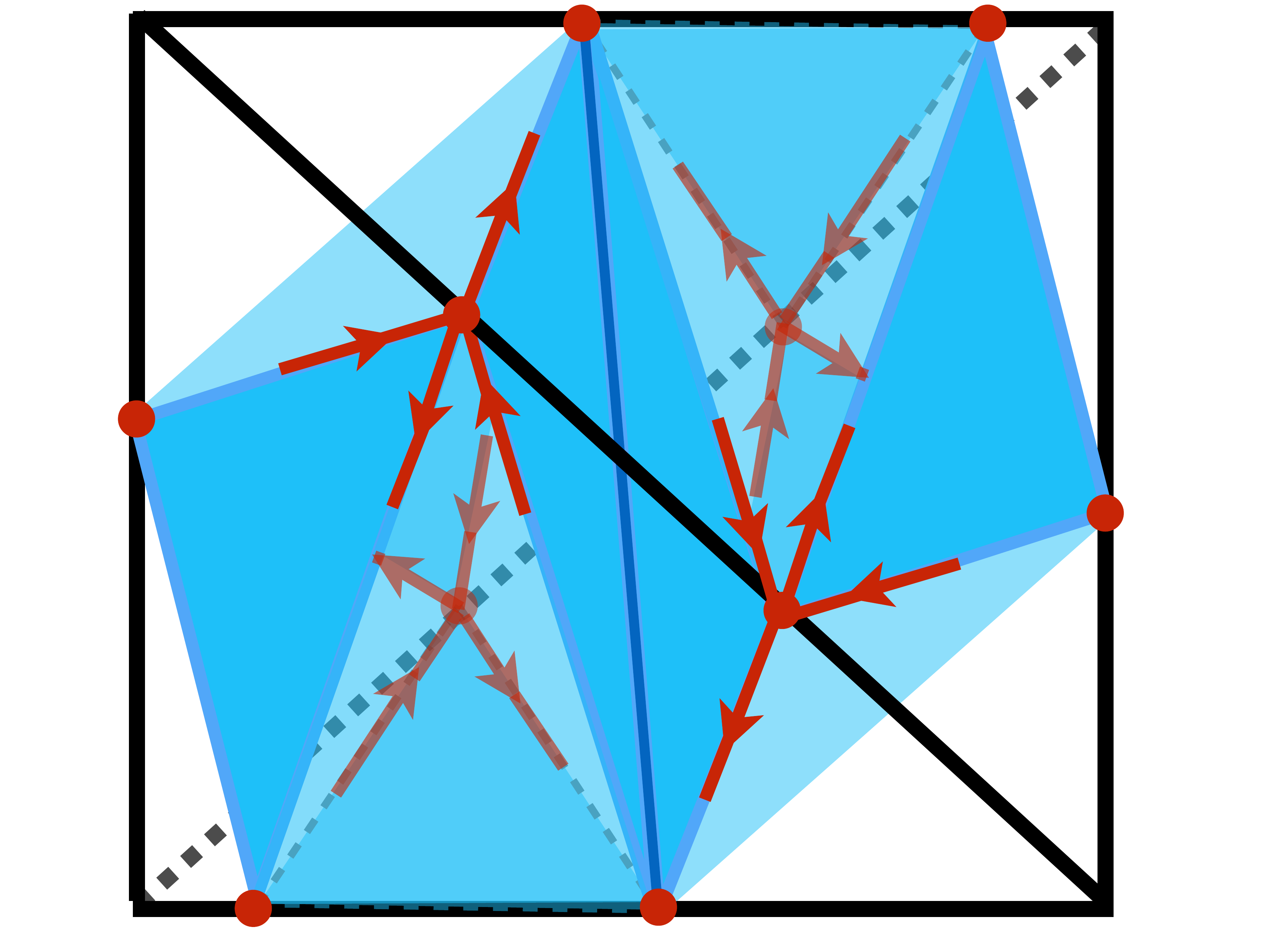}
\includegraphics[width=.36\linewidth]{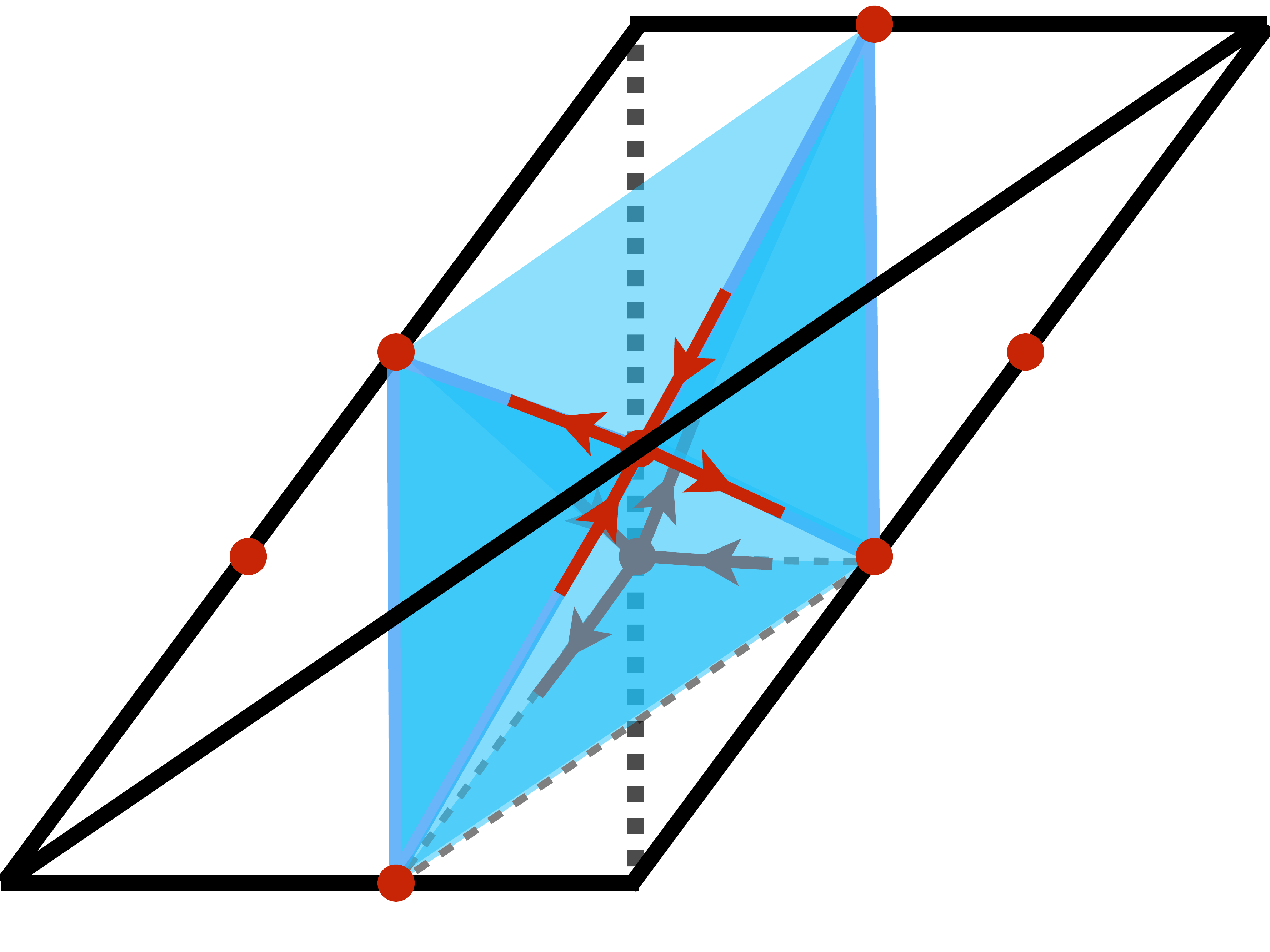}
\end{center}
\caption{Octahedron structure for the simple puncture cases for a single ideal tetrahedron, 
as determined from the connection between the mutation network and the octahedron decomposition (Fig.~\ref{fig:network_octahedron}). Interestingly, there are at least two different patterns; left (right) figure corresponds to $\bm{S}$ ($\bm{T}$), containing two (one) octahedron(s). This is in contrast to the case of the maximal puncture,
where we always use the same $N$-triangulation octahedron pattern for an ideal tetrahedron.
}
\label{quiver[21]-network}
\end{figure}

Following the comment around Fig.~\ref{fig:network_octahedron}, we can write down the octahedron 
structure in this case (Fig.~\ref{quiver[21]-network}). %An ideal tetrahedron in this case contains two octahedra,
%as expected from two triangulations.
It would be an interesting problem to see if such a octahedron decomposition defines
a state-integral model, whose partition function is independent of the choice of a ideal triangulation.

\paragraph{Ex 4. $N=4$  with $\rho=[3,1]$ } The quiver is drawn in Fig.~\ref{fig:non-maximal-once-punctured-quiver} and the anti-symmetric matrix $Q_{ij}$ is
\begin{align}
\left(
\begin{array}{ccccccc}
 0 & 0 & 0 & 2 & -1 & 0 & -1 \\
 0 & 0 & 0 & -1 & 2 & -1 & 0 \\
 0 & 0 & 0 & 0 & -1 & 2 & -1 \\
 -2 & 1 & 0 & 0 & 0 & 0 & 1 \\
 1 & -2 & 1 & 0 & 0 & 0 & 0 \\
 0 & 1 & -2 & 0 & 0 & 0 & 1 \\
 1 & 0 & 1 & -1 & 0 & -1 & 0 \\
\end{array}
\right) \;.
\end{align}
We have $|\textrm{Ker}(Q)|=1$ and a central element in the cluster algebra is 
\begin{align}
\sum_{i=1}^5 c_i Y^i:=Y_1+Y_2+Y_3+Y_4+Y_5+Y_6+Y_7\;. 
\end{align}
Two generators $\bm{S}, \bm{T}$  can be represented as\footnote{Interestingly, $\bm{S}$ can be represented as a single permutation, with no mutations.}
\begin{align}
\bm{S}= \sigma_S \;, \quad \bm{T}= \mu_7 \mu_4 \mu_6 \mu_1 \mu_5 \mu_3 \mu_5 \sigma_T \;,
\end{align}
where permutations matrices are 
\begin{align}
\sigma_S=
\left(
\begin{array}{ccccccc}
 0 & 0 & 1 & 0 & 0 & 0 & 0 \\
 0 & 1 & 0 & 0 & 0 & 0 & 0 \\
 1 & 0 & 0 & 0 & 0 & 0 & 0 \\
 0 & 0 & 0 & 0 & 0 & 1 & 0 \\
 0 & 0 & 0 & 0 & 1 & 0 & 0 \\
 0 & 0 & 0 & 1 & 0 & 0 & 0 \\
 0 & 0 & 0 & 0 & 0 & 0 & 1 \\
\end{array}
\right), \quad
\sigma_T=\left(
\begin{array}{ccccccc}
 1 & 0 & 0 & 0 & 0 & 0 & 0 \\
 0 & 0 & 0 & 0 & 1 & 0 & 0 \\
 0 & 0 & 0 & 0 & 0 & 0 & 1 \\
 0 & 0 & 0 & 1 & 0 & 0 & 0 \\
 0 & 0 & 1 & 0 & 0 & 0 & 0 \\
 0 & 0 & 0 & 0 & 0 & 1 & 0 \\
 0 & 1 & 0 & 0 & 0 & 0 & 0 \\
\end{array}
\right) \;.
\end{align}
The full expression of classical transformation for $\varphi=\bm{T}$ is rather complicated but we  checked that
\begin{align}
\bm{S}\bm{S}\;, \; \bm{T}\bm{S}\bm{T}\bm{S}\bm{T}\bm{S} : \;(y_1,y_2, y_3, y_4, y_5,y_6,y_7) \rightarrow   \;(y_1,y_2, y_3, y_4, y_5,y_6,y_7) \;.
\end{align}

\paragraph{Ex 6. $N=4$  with $\rho=[2,1,1]$} In the case, we only did minimal consistency check, namely reproducing expected size of quiver, $|Q|$, and the number of central elements in $\CA_Q$. From the general counting rule in eq.~\eqref{Quiver-counting},
\begin{align}
\begin{split}
&|Q_{[2,1,1]}|= 1\times (4^2-1)-\left(\textrm{dim }\mathbb{L}^{([2,1,1])}- \ell([2,1,1])\right)-1= 15 - (5-3)-1=12\;,
\\
&|\textrm{Ker}(Q^{[2,1,1]})| = \ell([2,1,1])-1= 2\;.
\end{split} \nn
\end{align}
%

%%%%%%%%%%%%%%%%%%%%%%%%%%%%%%%%%%%%%%%%%%%
\subsubsection{Co-dimension 4 Defects}\label{sec. ex of codim_4_cluster}
In this subsection, we give a concrete example for uplifting 2d loop operators to 3d loops. Using the uplift, the problem of quantization of some class of 3d loop operator is mapped to the problem of quantization of 2d loops, which has been more studied.  We explicitly work out the simplest example, $(\Sigma_{1,1}\times S^1)_{\varphi=\bm{LR}}$ with $N=2$. Generalization to arbitrary mapping torus with general $N$ and $\rho$=maximal is straight-forward.  

\paragraph{Ex 7. Co-dimension 4 Defects in $S^3\backslash \mathbf{4}_1$}  
Using the two 1-cycles $(\gamma_x,\gamma_y) \in \pi_1(\Sigma_{1,1})$, 
\begin{figure}[htbp]
\begin{center}
   \includegraphics[width=.3\textwidth]{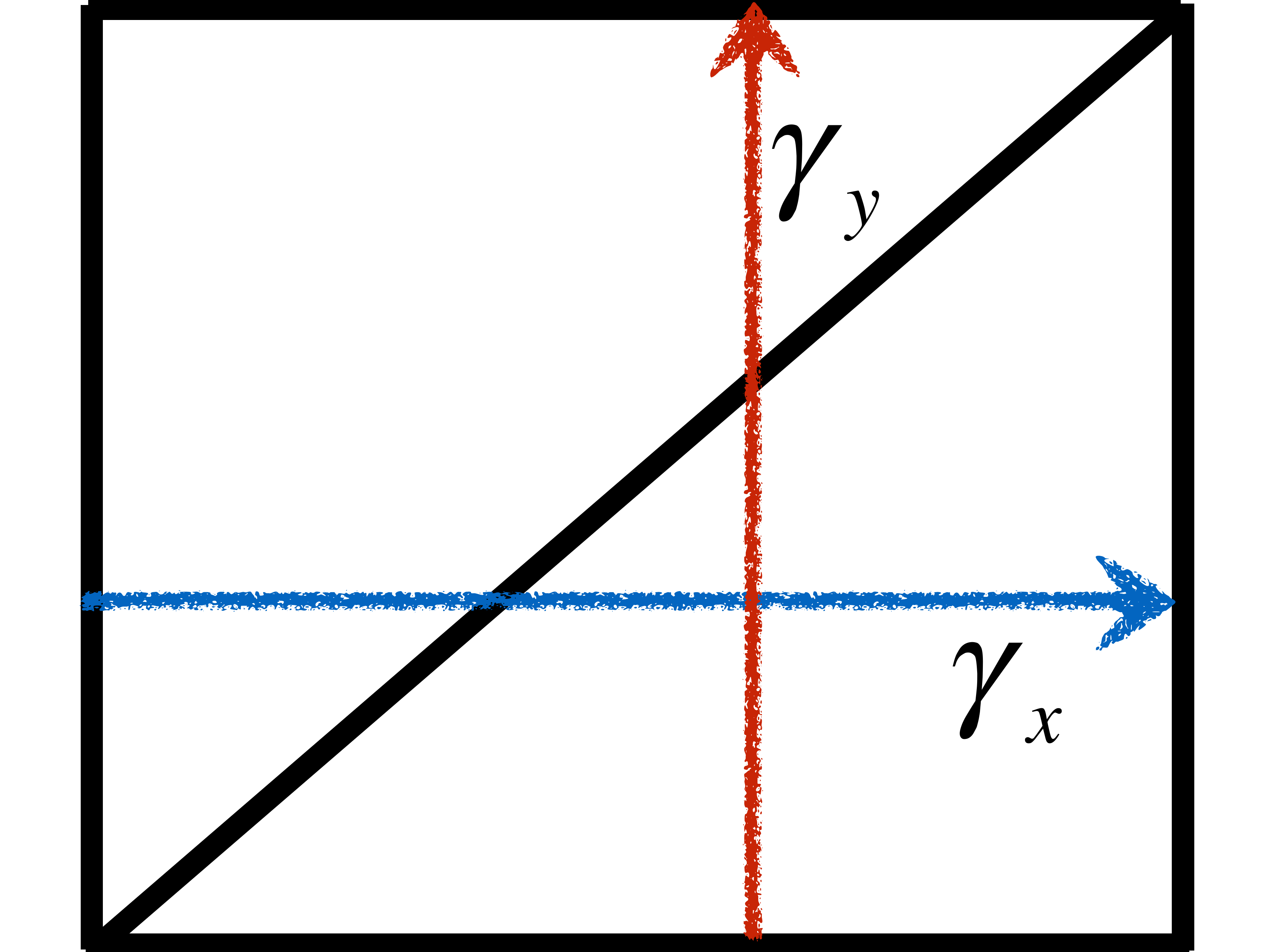}
   \end{center}
   \caption{Two 1-cycles in ideally-triangulated once-punctured torus.} 
    \label{fig:two-cycles in torus}
\end{figure}
the fundamental group
$\pi_1 (S^3\backslash \mathbf{4}_1) = \pi_1 \big{(} (\Sigma_{1,1}\times S^1)_{\varphi = \bm{LR}} \big{)}$ can be represented as %
\begin{align}
\pi_1 (S^3\backslash \mathbf{4}_1)= \langle \gamma_x, \gamma_y, \gamma_m | \gamma_m^{-1} \gamma_x \gamma_m =  \gamma_x \gamma_y\;, \gamma_m^{-1} \gamma_y \gamma_m = \gamma_y \gamma_x\gamma_y \rangle \label{fundamental group of figure eight knot 2}\;.
\end{align}
Here $\gamma_m$ can be understood as a 1-cycle along the $S^1$ in $(\Sigma_{1,1} \times S^1)_\varphi $ and it generates $\varphi$-transformation on two 1-cycles $(\gamma_x,\gamma_y)$ on $\Sigma_{1,1}$: 
\begin{align}
\gamma_m (\gamma_x, \gamma_y) \gamma_m^{-1} =  \varphi(\gamma_x,\gamma_y)\varphi^{-1} =\big{(}\varphi(\g_x),\varphi(\g_y)\big{)}\;.
\end{align}
Under $\bm{L}$ and $\bm{R}$ transformations, the generators of  $\pi_1 (S^3\backslash \mathbf{4}_1)$ transform as
\begin{align}
\bm{R}\;: \;(\gamma_x, \gamma_y) \rightarrow (\gamma_x ,\gamma_y\gamma_x)\;, \quad \bm{L}\;:\; (\gamma_x, \gamma_y) \rightarrow  (\gamma_x \gamma_y , \gamma_y )\;.
\end{align}
The cycle $\gamma_m$ can be identified as the meridian cycle in $\partial M$. The fundamental group has an automorphism defined by conjugation by $\gamma_m$:
\begin{align}
 \; (\gamma_x, \gamma_y,\gamma_m) \rightarrow  \gamma_m (\gamma_x, \gamma_y,\gamma_m) \gamma_m^{-1}\;. \label{Automorphism in fundamental group}
\end{align} 
This redundancy by the automorphism is reflected in the state-integral model \eqref{state-integral model for co-dimension 4 defect for mapping torus}  as 
the following invariance by conjugation:
\begin{align}
&\langle \hat{W}_R (\CK_\gamma) \rangle= \textrm{Tr} \big{[} \hat{W}_R (\CK_\gamma) \hat{\varphi} \big{]}= \textrm{Tr} \big{[}  \hat{\varphi} \hat{W}_R (\CK_{\varphi^{-1}(\gamma)} ) \big{]} = \textrm{Tr} \big{[}  \hat{W}_R (\CK_{\varphi^{-1}(\gamma)}) \hat{\varphi} \big{]} \nn
\\
&\qquad \quad \;\;\; =\langle \hat{W}_R (\CK_{\gamma_m^{-1}\gamma \gamma_m} ) \rangle \;.
\end{align}
The two sets of generators in eq.~\eqref{fundamental group of figure eight knot} and eq.~\eqref{fundamental group of figure eight knot 2}  of $\pi_1 (S^3\backslash \mathbf{4}_1)$  are related as
\begin{align}
\gamma_x = c^{-1}a \;, \quad \gamma_y = b^{-1} \;, \quad \textrm{and} \quad \gamma_m= c^{-1}\;. \label{relation between two generators of 41}
\end{align}
up to the automorphism.  Using the map \eqref{2d loops to 3d loops-2},  we have
\begin{align}
\begin{split}
&\exp \big{(} {a_{1} \sfY_{(1,0)}+a_{2} \sfY_{(0,1)}+a_{3} \sfY_{(1,1)}}  \big{)}
\\
&\Longleftrightarrow  q^{\frac{1}{2} (a_{1}^2+a_{2}^2+a_{3}^2-2a_{1}a_{2}-2a_{1}a_{3})+(a_1+a_2-a_3)\delta- \delta^2} 
\\
&\qquad\quad \times e^{(\delta-a_1) \hat{Z}(0)+(\delta+a_3-a_2)\hat{Z}(1)} e^{(2a_2-a_1-\delta) \hat{Z}''(0)+(a_3-a_2-\delta) \hat{Z}''(1)}
\\
&\qquad= e^{\delta \big{(}\hat{Z}(0)+\hat{Z}(1)-\hat{Z}''(0)-\hat{Z}''(1)\big{)}}  q^{\frac{1}{2} (a_{1}^2+a_{2}^2+a_{3}^2-2a_{1}a_{2}-2a_{1}a_{3})} 
\\
&\qquad\quad  \times e^{-a_1 \hat{Z}(0)+(a_3-a_2)\hat{Z}(1)} e^{(2a_2-a_1) \hat{Z}''(0)+(a_3-a_2) \hat{Z}''(1)} \;.
 \label{3d uplifting formula for 4_1}
\end{split}
\end{align}
Note that there is a 1-parameter ambiguity in the  uplifted operator parametrized by $\delta$, however they are all equivalent as a 3d loop operator
 under the gluing constraints \eqref{equivalence of 3d loops}. 
Simple example of  2d loop operators on $\Sigma_{1,1}$ for $N=2$ are
\begin{align}
\begin{split}
&\hat{W}_{\Box}(\CK_{\gamma_x}) = e^{\frac{1}2 \sfY_{(0,1)}+\frac{1}2\sfY_{(1,1)}} +e^{-\frac{1}2 \sfY_{(0,1)}- \frac{1}2 \sfY_{(1,1)}}+e^{\frac{1}2 \sfY_{(1,1)}-\frac{1}2 \sfY_{(0,1)}}\;,
\\
&\hat{W}_{\Box}(\CK_{\gamma_y}) = e^{\frac{1}2 \sfY_{(1,0)}+\frac{1}2\sfY_{(1,1)}} +e^{-\frac{1}2 \sfY_{(1,0)}- \frac{1}2 \sfY_{(1,1)}}+e^{-\frac{1}2 \sfY_{(1,1)}+\frac{1}2 \sfY_{(1,0)}}\;.
\end{split}
\end{align}
Applying the formula \eqref{3d uplifting formula for 4_1}, we have
\begin{align}
\begin{split}
& \hat{W}_{\Box}(\CK_{\gamma_x}) = q^{\frac{1}{4}} \left( e^{\hat{Z}''(0)} + e^{-\hat{Z}''(0)}+e^{\hat{Z}(1)}e^{-\hat{Z}''(0)+\hat{Z}''(1)} \right) \;,
\\
& \hat{W}_{\Box}(\CK_{\gamma_y}) = q^{-\frac{1}{4}} \left( e^{-\hat{Z}(0)}e^{\hat{Z}''(1)} + e^{-\hat{Z}(1)}e^{\hat{Z}''(0)}+ e^{-\hat{Z}(0)-\hat{Z}(1)} \right)\;.
\end{split}
\end{align}
These expressions are compatible with the expressions in eq.~\eqref{Quantum loop for cIa} via the map \eqref{relation between two generators of 41} and \eqref{Z0Z1 to YZ}.

%%%%%%%%%%%%%%%%%%%%%%%%%%%%%%%%%%%%%%%%%%%%%%%%%%%%%%%%
\section{From Domain Wall Theory \texorpdfstring{$T[SU(N)]$}{T[SU(N)]}} \label{sec.TSUN}
%%%%%%%%%%%%%%%%%%%%%%%%%%%%%%%%%%%%%%%%%%%%%%%%%%%%%%%%

\subsection{Necessity of Non-Abelian Description for \texorpdfstring{$T_N[M]$}{TN[M]}} \label{sec : Necessity of Non-abelian}

In previous sections we discussed state integral models and cluster partition functions,
and obtained their partition functions. As explained in sections \ref{sec.state_integral_general} and \ref{subsec.relation},
from the expression of the partition function we can recover the 
Abelian description of the associated 3d $\mathcal{N}=2$ theory $T_N[\hatM
\backslash K, \rho]$.
We also incorporated co-dimension 4 defects along $\mathcal{K}$.

There is one unsatisfactory aspect, however:
we really did not explain why, on the side of the 3d $\mathcal{N}=2$ theory,
the co-dimension 4 defects are labeled by a representation $R$ of $SU(N)$.
Note that in the Abelian descriptions, we could consider loops (Wilson loop, vortex loop, or its mixture)
for each Abelian gauge group, but there is no natural explanation 
for the origin of the discrete label $R$; there is simply no non-Abelian symmetries.

This strongly suggests that there should always be a non-Abelian descriptions of $T[M]$.
The Abelian descriptions presumably arises on the ``Coulomb branch''\footnote{As commented in footnote \ref{foot.3dN2}, in 3d $\mathcal{N}=2$ non-Abelian gauge theories there is no clear-cut distinction between Coulomb and Higgs branches, and hence the ``Coulomb branch'' is at best an approximate notion.} of the non-Abelian theory.
While the Abelian description is sufficient for the computation of the 
$S^3_b$ and $(S^1\times S^2)_q$ partition functions,
it will not be sufficient for the full understanding of loop operators and more generally
supersymmetric defects, 
as well as the discussion of the
quantum-corrected moduli space, for example.\footnote{The moduli space of vacua for a non-Abelian 3d $\mathcal{N}=2$ 
receives dramatic quantum corrections: instanton effects sometimes lift the Coulomb branch, and several different branches
merge, blurring the distinction between Coulomb and Higgs branches. \label{foot.3dN2}}
It is therefore an important problem in the 3d--3d correspondence to search for non-Abelian descriptions.

Fortunately, such a non-Abelian description of $T_N[M]$ is known in the literature, 
and it fact it was the proposal of \cite{Terashima:2011qi} in 2011 (see also \cite{Drukker:2010jp,Hosomichi:2010vh}).

The basic idea is as follows. Let us consider a 3-manifold $M$ with boundary $\partial M=\Sigma$.
From M5-brane compactifications, we expect the relations between the two associated theories, 
3d $\mathcal{N}=2$ theory $T_N[M]$, and 4d $\mathcal{N}=2$ theory (of the so-called class $\mathcal{S}$) $T_N[\Sigma]$ \cite{Gaiotto:2009we}. The natural expectation is that $T_N[M]$ is the boundary degrees of freedom for a certain 1/2-BPS boundary condition for $T_N[\Sigma]$. We can therefore analyze 
the boundary conditions of $T_N[\Sigma]$ theory,
and recover the $T_N[M]$ theory.

The analysis of such boundary conditions, however, in general is rather complicated,
partly because we often do not have a Lagrangian description of $T_N[\Sigma]$,
and partly because there are quantum corrections to the moduli space.\footnote{
For a generic choice of $\Sigma$, the boundary condition preserves only four supercharges out of the original eight supercharges of $T_N[\Sigma]$. In this sense, the situation is 
close to the analysis of 1/4 BPS boundary conditions of 4d $\mathcal{N}=4$, recently worked out in \cite{Hashimoto:2014vpa,Hashimoto:2014nwa}.
}

The situation simplifies for the case where $\Sigma$ is a torus $T^2$,
and hence $T_N[\Sigma]$ theory is the 4d $\mathcal{N}=4$ super Yang-Mills theory.
In this case, the relevant boundary conditions have been identified in the works of \cite{Gaiotto:2008sa,Gaiotto:2008ak},
giving rise to non-Abelian description of $T_N[M]$ involving the $T[SU(N)]$ theory (which we discussed already in a different context in Sec.~\ref{sec : co-dimension 2}).

We can also introduce a simple puncture to the torus\footnote{If we instead have a maximal puncture, the 
Lagrangian description for the $T_N[\Sigma_{g=1, h=1}]$ theory is not known for $N>2$.}, 
leading to the 4d $\mathcal{N}=2^*$ theory, namely the deformation of the 4d $\mathcal{N}=4$ theory by mass deformation of the $\mathcal{N}=2$ adjoint chiral multiplet. This leads to the 3d $\mathcal{N}=2$ deformation of the 3d $\mathcal{N}=4$ theory, by giving the real mass parameter to the axial $U(1)$ R-symmetry inside $SO(4)$ $\mathcal{N}=4$ R-symmetry \cite{Tong:2000ky}.

In the following we compute the partition function of the mapping torus of the 
once-punctured torus with a simple puncture, using 
such non-Abelian descriptions of $T_N[M]$.
The result will give a highly non-trivial cross-check for our understanding of the co-dimension 2 defect of type 
$\rho=\textrm{simple}$, 
while simultaneously checking the consistencies between Abelian and non-Abelian descriptions of $T_N[M]$.

%%%%%%%%%%%%%%%%%%%%%%%%%%%%%%%%%%%%%%%%%%%%%%%
\subsection{Co-dimension 2 Defects: \texorpdfstring{$\rho=\textrm{simple}$}{rho=simple} from \texorpdfstring{$T[SU(N)]$}{T[SUN(N)]} Theory}  \label{TSU3_index_glue}

Consider a mapping torus over an once-punctured torus with $\rho=\textrm{simple}$,
determined by an element $\varphi$ of $SL(2, \mathbb{Z})$.
The dual 3d $\mathcal{N}=2$ theory (which is a mass deformation of the 3d $\mathcal{N}=4$ theory),
which we denote by $\textrm{Tr}\big{(}T[SU(N), \varphi]\big{)}$,
can be obtained as follows (see \cite{Terashima:2011qi} for details). 

First, we identify the $\mathcal{N}=2$ mass-deformed $T[SU(N)]$ theory as $T[SU(N),\varphi=\bm{S}]$\footnote{
For simplicity we use the same name `$T[SU(N)]$ theory' both to $\mathcal{N}=4$ theory
and its $\mathcal{N}=2$ mass deformation.} and an empty theory with $SU(N)\times SU(N)$ flavor symmetry with background off-diagonal $\mathcal{N}=4$ Chern-Simons term as   $T[SU(N),\varphi=\bm{T}]$.
Second, for the theory $T[SU(N),\varphi_1\cdot \varphi_2]$, we then 
glue the two theories $T[SU(N),\varphi_1]$ and $T[SU(N),\varphi_2]$
by gauging the diagonal $SU(N)$ flavor symmetry.
By using the second rule recursively, we can define the 3d theory $T[SU(N),\varphi]$ for any $\varphi$,
and the S-duality of 4d $\mathcal{N}=4$ theory ensures that the resulting 3d theory is independent of the choice of decomposition. In the 3d--3d correspondence, the $T[SU(N),\varphi]$ theory corresponds to $SL(N)$ CS theory on mapping cylinder $\Sigma_{1,1}\times [0,1]$. The theory has $SU(N)_{\textrm{top}}\times SU(N)_{\textrm{bot}}\times U(1)_{\textrm{punct}}$  flavor symmetry.\footnote{'bot'/`top' means `bottom'/`top' of the 
mapping cylinder $\Sigma_{1,1} \times [0,1]$, since these two flavor symmetries are associated with the two boundaries of the mapping cylinder. `top' also represents `topological' since the flavor symmetry is often call the topological flavor symmetry.} The theory corresponding to mapping torus $(\Sigma_{1,1}\times S^1)_{\varphi}$ can be obtained by gluing two $SU(N)$ flavor symmetries by gauging diagonal $SU(N)$ subgroup of $T[SU(N),\varphi]$ theory.\footnote{
Note in general this involves gauging of an emergent $SU(N)$ symmetry which is not present in the Lagrangian (except for its Cartan).} Let denote the theory obtained  in this way as $\textrm{Tr}(T[SU(N),\varphi])$:
\begin{align}
\begin{split}
&\textrm{Tr}(T[SU(N),\varphi]) 
\\
&= \left(\textrm{the theory obtained by gauging diagonal $SU(N)$ of $T[SU(N),\varphi]$} \right) \;.
\end{split}
\end{align}
The mapping torus is a knot complement on a closed 3-manifold and the knot corresponds to a simple co-dimension 2 defect. Thus, we can identify 
\begin{align}
\textrm{Tr}(T[SU(N),\varphi]) = T_N[(\Sigma_{1,1}\times S^1)_\varphi, \textrm{simple}]\;.
\end{align}
 The  field theory on the left has $U(1)_\textrm{punct}$ symmetry, which can be identified with $H_{\rho=\textrm{simple}}$ of the theory on the right.

%%%%%%%%%%%%%%%%%%%%%%%%%%%%%%%%%%%%%%%
\paragraph{$T[SU(3)]$ Theory}  

Let us now describe our theory for $N=3$, and discuss their $S^1\times S^2$ partition functions.

Let us begin with the basic building block, namely the $T[SU(3)]$ theory.
The $T[SU(3)]$ theory is described by the following 3d $\mathcal{N}=4$ quiver,
where the square (circle) represents the flavor (gauge) symmetry:

\begin{align}
\scalebox{0.8}{
\begin{tikzpicture}[
roundnode/.style={circle, draw=green!60, fill=green!5, very thick, minimum size=7mm},
squarednode/.style={rectangle, draw=red!60, fill=red!5, very thick, minimum size=5mm},
]
%Nodes
\node[roundnode](first)                              {$U(1)_{\rm gauge}$};
\node[roundnode](second)       [right=of first] {$U(2)_{\rm gauge}$};
\node[squarednode](third)       [right=of second] {$SU(3)_{\rm bot}$};
%Lines
\draw(first.east) -- (second.west);
\draw (second.east) -- (third.west);
\end{tikzpicture}
} 
\end{align}

As we mentioned already, we are interested in an $\mathcal{N}=2$ mass deformation of the theory,
namely we turn on all the real mass/ FI parameters consistent with $\mathcal{N}=2$ supersymmetry.
The theory is $U(1)\times U(2)$ gauge theory which has  $SU(3)_\textrm{top} \times SU(3)_{\rm bot} \times U(1)_{\rm punct}$ flavor symmetry In terms of $\CN=2$ chiral superfields, the charge assignment for flavor/R-symmetries can be summarized as %
\begin{align}
\left.\begin{array}{c||c|c|c|c|c|c} & U(1)_{\rm gauge} & U(2)_{\rm gauge} & SU(3)_{\rm bot} & U(1)_{\rm punct} & SU(3)_{\rm top}  & U(1)_R
 \\ \hline
 \hline
\Phi_1  & 1 & \overline{\mathbf{2}} & \mathbf{1} & \frac{1}{2}&  \mathbf{1} & \frac{1}{2}
\\  \hline
\Phi_2 & -1 & \mathbf{2} &  \mathbf{1} &  \frac{1}{2} &  \mathbf{1} & \frac{1}{2}
\\ \hline
\Phi_3 & 0  & \mathbf{2} & \overline{\mathbf{3}} &   \frac{1}{2} &  \mathbf{1} & \frac{1}{2}
\\ \hline
\Phi_4 & 0 & \overline{\mathbf{2}} & \mathbf{3} &  \frac{1}{2} & \mathbf{1} & \frac{1}{2}
\\ \hline
\varphi_1 & 0 & \mathbf{1} & \mathbf{1} & -1 & \mathbf{1}& 1
\\ \hline
\varphi_2 & 0 & \textrm{adj} & \mathbf{1} & -1 & \mathbf{1}& 1
\end{array}\right.
\end{align}
Here
$(\Phi_1 , \Phi_2^\dagger)$ and $(\Phi_3, \Phi_4^\dagger)$ form $\mathcal{N}=4$ bi-fundamental hypermultiplets, and $\varphi_1$ and $\varphi_2$ form $\CN=2$ adjoint chiral multiplet inside an  $\CN=4$ vector multiplet. 
The $SU(3)_{\textrm{top}}$ is a quantum/emergent symmetry in the IR, and is not present in the classical Lagrangian,
except for its Cartan $U(1)^2$: these correspond to two topological $U(1)_J$ symmetries coupled to $U(1)$ factors in $U(1)_{\rm gauge}\times U(2)_{\rm gauge}$. 
In $S^1\times S^2$ partition function, we have a 
continuous parameter (fugacity) as well as a discrete parameter (magnetic flux) for 
each flavor symmetry.
We denote them by
\begin{align}
\left.\begin{array}{c||c|c|c|c|c} 
& U(1)_{\rm gauge} &U(2)_{\rm gauge} &SU(3)_{\rm bot} &U(1)_{\rm punct} &SU(3)_{\rm top}  \\
\hline\hline
\textrm{fugacity} &\zeta &(z_1, z_2)& (v_1,v_2) &\bareta &(w_1, w_2) \\
\hline
 \textrm{magnetic flux} & \sigma &  (s_1, s_2) &  (n_1,n_2) & m_\eta & (m_1, m_2)
 \end{array}\right.
\label{fugacity_notation}
\end{align}
Here $(v_1, v_2, v_3)$ and $(w_1, w_2)$ are fugacities for the following Cartan element of $SU(3)_{\rm bot}$ and $SU(3)_{\rm top}$, respectively:
\begin{align}
\begin{split}
&(v_1, v_2)\longleftrightarrow  \textrm{diag}{(H_1, H_2,  -H_1- H_2)} \in \mathfrak{su}(3)_{\rm bot} \;,
\\
&(w_1,w_2)\longleftrightarrow  \textrm{diag}{(H_1, H_2, -H_1- H_2)} \in \mathfrak{su}(3)_{\rm top} \;.
\end{split}
\end{align}

%%%%
\paragraph{Computation of Index}

We can now compute the index for the $T[SU(N), \varphi]$ theory. The index in the notation \eqref{fugacity_notation}
is a function
\begin{align}
\CI_{T[SU(N)]}(m_1,m_2, w_1, w_2|n_1, n_2, v_1, v_2; m_\eta, \bareta)  \ ,
\end{align}
whose explicit formula for $N=3$ we write down in detail in App.~\ref{app.TSU3}.
Given the index for $T[SU(N)]$ theory, 
the indices of theories $T[SU(N), \varphi]$ for general choices of $\varphi$ can be computed with the help of the following two facts,
which are index versions of the two recursive rules we described previously.

First, a multiplication of two elements $\varphi_1, \varphi_2$ leads to 
the gauging of the diagonal $SU(N)$ flavor symmetry of the corresponding 
two theories $T[SU(N), \varphi_{1, 2}]$. Written for $N=3$, this means %
\begin{align}
&\CI_{T[SU(3) ,\varphi_1 \varphi_2]}(m_1, m_2, w_1, w_2|n_1,n_2, v_1, v_2; m_\eta, \bareta) \nonumber
\\
& =\sum_{(p_1,p_2)} \oint \frac{du_1 du_2}{(2\pi i u_1)(2\pi i u_2)} \Delta_3 (u_1, u_2,p_1, p_2;q) \CI_{T[SU(3) ,\varphi_1]}(m_1, m_2, w_1, w_2|p_1,p_2, u_1, u_2; m_\eta, \bareta)  \nn
\\
& \qquad \times \CI_{T[SU(3) ,\varphi_2]}(p_1, p_2, u_1, u_2|n_1,n_2, v_1, v_2; m_\eta, \bareta)\;,   
\end{align}
where $\Delta_3$ is the measure from $\CN=2$  $SU(3)$ vector multiplets 
given in eq.~\eqref{vector_measure}. This means the basic building blocks are the theories for the generators of 
$SL(2, \mathbb{Z})$, namely for $\varphi=\bm{S}, \bm{T}$. To complete the rule we need to give the indices for 
the theories $T[SU(3), \varphi=\bm{S}, \bm{T}]$.
For $\varphi=\bm{S}$, we have the $T[SU(3)]$ theory whose index is
written down in App.~\ref{app.TSU3}. If we have $\bm{T}$ generators in addition, we have
Chern-Simons terms. For example,
\begin{align}
\begin{split}
&\CI_{T[SU(3) , \bm{T}^{k_1} \bm{S} \bm{T}^{k_2}]}(m_1, m_2, w_1, w_2|n_1,n_2, v_1, v_2; m_\eta, \bareta) 
\\
& \qquad=\big{(} w_1(-1)^{m_1}\big{)}^{-k_1 (2m_1 +m_2)} \big{(}w_2(-1)^{m_2}\big{)}^{-k_1 (2m_2+m_1)}   \big{(}v_1(-1)^{n_1}\big{)}^{-k_2 (2n_1 +n_2)} 
\\
&\qquad \qquad  \times \big{(}v_2(-1)^{n_2}\big{)}^{-k_2 (2n_2+n_1)}   
\CI_{T[SU(3)]}(m_1, m_2, w_1, w_2|n_1,n_2, v_1, v_2; m_\eta, \bareta)  \;. 
\end{split}
\end{align}
The sign factors $(-1)^{m_i}$ and $(-1)^{n_i}$ come from  shifts of spin of states on $S^2$ in the presence of magnetic fluxes \cite[Appendix A]{Aharony:2013dha}. 
We can explicitly verify that the relations of the $SL(2, \mathbb{Z})$ are satisfied up to an 
overall shift of the partition function: 
\begin{align}
\begin{split}
&\CI_{T[SU(3), \bm{S}^4 \cdot \varphi]} =\CI_{T[SU(3), \varphi \cdot \bm{S}^4]} =\CI_{T[SU(3), \varphi]}  \;, 
\\
&\CI_{T[SU(3), (\bm{T}\bm{S}\bm{T}\bm{S}\bm{T}\bm{S}) \cdot \varphi]} = \CI_{T[SU(3),   \varphi \cdot (\bm{T}\bm{S}\bm{T}\bm{S}\bm{T}\bm{S})  ] } =(-\bareta)^{-\frac{3}2 m_\eta} \CI_{T[SU(3), \varphi]} \;,
\end{split}
\label{I_ST_eg}
\end{align}
The phase factor shift is related to framing ambiguity in knot theory as explained around eqs.~\eqref{projectivity of MCG reps} (see also \eqref{projectivity for N=3 simple})
under the parameter identification $\bareta=-\eta$ and 
\begin{align}
\frakL :=\frac{\hbar}2 m_\eta + \log \eta\;,
\end{align}
where the effect of the relative minus sign is to shift the definition of the central element 
$\frakL$ in the cluster partition function by a constant factor of $i \pi$.

Finally, to obtain the index for mapping tori, 
we need to gauge the diagonal of the remaining two $SU(3)$ flavor symmetries:
\begin{align}
&\CI_{\textrm{Tr}(T[SU(3),\varphi])}(m_\eta, \bareta) \nn
\\
&=\sum_{(p_1,p_2)} \oint \frac{du_1 du_2}{(2\pi i u_1)(2\pi i u_2)} \Delta_3 (u_1, u_2,p_1, p_2;q) \,\CI_{T[SU(3) ,\varphi]}(p_1, p_2, u_1, u_2|p_1,p_2, u_1, u_2; m_\eta, \bareta)  \;.
\end{align}
Here the measure $\Delta_3$ is from an $\CN=2$  $SU(3)$ vector multiplet,  and is
given by 
\begin{align}
\begin{split}
&\Delta_3 (u_1, u_2,p_1, p_2;q) 
\\
&:=\frac{1}{\textrm{sym($p_1 ,p_2$)}} q^{\frac{-|p_1-p_2|-|p_1+2p_2|-|2p_2+p_1|}{2}}(1-  q^{\frac{|p_1-p_2|}{2}} u_1u_2^{-1})  (1-  q^{\frac{|p_1-p_2|}{2}} u_2 u_1^{-1})   
\\
&\quad \quad \times (1-  q^{\frac{|2p_1+p_2|}{2}}u_1^2 u_2)(1-  q^{\frac{|2p_1+p_2|}{2}} u_1^{-2} u_2^{-1})\\
&\quad \quad \times (1-  q^{\frac{|2p_2+p_1|}{2}} u_1 u_2^2)(1-  q^{\frac{|2p_2+p_1|}{2}} u_2^{-2} u_1^{-1})\;.
\end{split}
\label{vector_measure}
\end{align}
Here the range of allowed $(p_1, p_2)$ is
\begin{align}
(p_1, p_2) \in \mathbb{Z}/3\;, \quad p_1-p_2 \in \mathbb{Z} \;, \quad p_1 \geq p_2 \geq -(p_1+p_2)\;,
\end{align}
and the symmetric factor is defined by 
\begin{align}
\textrm{sym}(p_1, p_2): = 
\begin{cases}
6 &  (p_1=p_2=0) \\ 
2 & (p_1 = p_2 >0 \;  \textrm{or}\; p_2= -(p_1+p_2) <0)\\
1  & (\textrm{otherwise})
\end{cases}
\;.
\end{align}

The resulting expression is a complicated expression involving many integrals.
However, we can expand the integrand in power series in the fugacity $q$,
and we obtain the expression for $\CI_{\textrm{Tr}(T[SU(3),\varphi])}(m_\eta, \eta)$ in power series expansion in $q$.
For example, if we specialize to $m_{\eta}=0$, we obtain ($\varphi=\bm{S}\bm{T}^{-1}\bm{S}^{-1}\bm{T}$)
\begin{align}
\begin{split}
\CI_{_{\textrm{Tr}(T[SU(3),\varphi])} }(m_\eta=0, \bareta)
&= 1- \left(2\bareta+\frac{2}{\bareta} \right) q^{\frac{3}{2}}+\left(8+2 \bareta^2 + \frac{2}{\bareta^2}\right) q^2-\left(6 \bareta+\frac{6}\bareta \right)q^{\frac{5}{2}} 
\\
& \qquad\qquad +\left(2- 3 \bareta^2- \frac{3}{\bareta^2}\right)q^3+\ldots  \;,
\\
\CI_{_{\textrm{Tr}(T[SU(3),\varphi])} }(m_\eta=1, \bareta)
& = \left(\frac{1}{\bareta^2}-\frac{1}{\bareta} - \bareta +\bareta^2 \right) q+ \left(6 -3\bareta -  \frac{3}{\bareta}\right) q^2 \\
&\qquad \qquad+ \left(-6 + \frac{1}{\bareta^3}- \frac{3}{\bareta^2}+\frac{5}{\bareta}+5 \bareta-3\bareta^2 +\bareta^3\right)q^3+\ldots \;. 
\label{N=3 4_1 simple from T[SU(3)]}
\end{split}
\end{align}
These results are consistent with 
eqs.~\eqref{N=3 4_1 simple from cluster} and \eqref{STSinvTinv}, again under the parameter identification
$\bareta =-\eta$.
As emphasized already, these results are simultaneously (1) consistency checks between
Abelian and non-Abelian description of $T_N[M]$ and (2) consistency checks of the simple-puncture result in Sec.~\ref{sec.Xe_rule}.

%%%%%%%%%%%%%%%%%%%%%%%%%%%%%%%%%%%%%%%%%%%%%
\subsection{Co-dimension 4 Defects} \label{sec.TSUN_codim_4}
Let us now come to the question of the co-dimension 4 defect, which we commented at the beginning of this section.
In the non-Abelian description of $T_N[M,\rho]$ theory, there are several $SU(N)$ gauge  groups and one natural choice of loop operator is  a Wilson loop in representation $R$ of one of $SU(N)$s. In the 3d--3d correspondence, a loop operator is mapped to a Wilson loop of representation $R$ in $SL(N)$ CS theory.   The choice of gauge group is mapped to choice of a knot $\CK$, trajectory of the Wilson loop. 
For simplicity we here focus on one of the simplest cases, the loop operators in the 
figure-eight knot complement.  Note that this is an example
where co-dimension 2 defect with $\rho=\textrm{simple}$ coexists with a 
co-dimension 4 defect. In the case, we can give concrete examples of the map and  verify the map by explicitly checking the 3d--3d correspondence for $k=0$. For the check, we use localization methods in the field theory computation and use state-integral model  \eqref{state-integral model for co-dimension 4 defect} in CS theory computation. 

%-----------------------------------------------------------------
\paragraph{Loop Operators in $T_{N=2}[S^3\backslash \mathbf{4}_1]$}
Let first recall that the $T[SU(2)]$ has $SU(2)_{\textrm{top}}\times SU(2)_{\textrm{bot}}$ flavor symmetry,
which we can represent graphically as
\begin{align}
\scalebox{0.8}{
\begin{tikzpicture}[
roundnode/.style={circle, draw=green!60, fill=green!5, very thick, minimum size=5mm},
squarednode/.style={rectangle, draw=red!60, fill=red!5, very thick, minimum size=5mm},
TSUNnode/.style={circle, draw=blue!60, fill=blue!5, very thick, minimum size=5mm},
]
%Nodes
\node[squarednode](first)                              {$SU(2)_{\textrm{top}}$};
\node[TSUNnode](second) [right=of first] {$T[SU(2)]$};
\node[squarednode](third) [right=of second] {$SU(2)_{\textrm{bot}}$};
%Lines
\draw(first.east) -- (second.west);
\draw (second.east) -- (third.west);
\end{tikzpicture}
}
\end{align}
The 3d $\mathcal{N}=2$ theory $T_{N=2} [S^3\backslash \mathbf{4}_1]=\textrm{Tr}(T[SU(2),\bm{LR}]) $ can be then constructed by gluing two $T[SU(2)]$ theories as follows:
\begin{align}
\textrm{
\scalebox{0.8}{\begin{tikzpicture}[
node distance=2.5cm,
roundnode/.style={circle, draw=green!60, fill=green!5, very thick, minimum size=5mm},
squarednode/.style={rectangle, draw=red!60, fill=red!5, very thick, minimum size=5mm},
TSUNnode/.style={circle, draw=blue!60, fill=blue!5, very thick, minimum size=5mm},
]
%Nodes
\node[roundnode](N){$SU(2)_1$};
\node[TSUNnode](W)[below left of=N]{$T[SU(2)]$};
\node[TSUNnode](E)[below right of=N]{$T[SU(2)]$};
\node[roundnode](S)[below right of=W] {$SU(2)_{-1}$};
%Lines
\draw(N) -- (W);
\draw(N) -- (E);
\draw(S) -- (E);
\draw(S) -- (W);
\end{tikzpicture}
}
}
\label{TSU2_glue}
\end{align}
In the gluing we gauge diagonal subgroups of $(SU(2)_+)_{\textrm{top}}\times (SU(2)_-)_{\textrm{bot}}$ and  $(SU(2)_+)_{\textrm{bot}}\times (SU(2)_-)_{\textrm{top}}$ by introducing  dynamical $SU(2)$ vector multiplets   with CS level $- 1$ and $+1$ respectively. We propose the following map for 3d--3d correspondence: 
\begin{align}
\begin{split}
&\textrm{ Wilson loop charged under $SU(2)_{-1}$ in $\textrm{Tr}(T[SU(2),\bm{LR}])$ theory}  
\\
&\qquad\qquad\Longleftrightarrow \textrm{ Wilson loop along $\CK_{\gamma=c^{-1}a}$ in $SL(2)$ CS theory on $S^3\backslash {\bf 4_1}$ }\;,
\\
&\textrm{ Wilson loop charged under $SU(2)_{+1}$ in $\textrm{Tr}(T[SU(2),\bm{LR}]$ theory}  
\\
&\qquad\qquad\Longleftrightarrow \textrm{ Wilson loop along $\CK_{\gamma=b}$ in $SL(2)$ CS theory on $S^3\backslash {\bf 4_1}$} \;.
\end{split}
\end{align}
This correspondence can be confirmed by computing the index of the 3d $\mathcal{N}=2$ theory
with the Wilson loop and comparing it to the computation in eq.~\eqref{Simple wilson loop in figure eight}.   The index can be computed as follows:
\begin{align}
&\langle W_\Box (SU(2)_{-1}) \rangle^{\textrm{Tr}(T[SU(2),\bm{LR}])}_\pm  \nn\\
&\, =  \sum_{2m_1,2m_2=0}^\infty \oint \frac{du_1}{2\pi i u_1} \frac{du_2}{2\pi i u_2} \Delta (m_1,u_1)\Delta (m_2, u_2)  (-1)^{\half}(q^{\frac{m_1}{2}} u_1^{\pm 1} +q^{-\frac{m_1}{2}} u_1^{\mp 1})  \nn
\\
&\,   \times  \big{(}u_1(-1)^{m_1}\big{)}^{-2m_1}\big{(}u_2(-1)^{m_2}\big{)}^{2m_2} \CI_{T[SU(2)]} (m_1,u_1|m_2,u_2;m_\eta, \bar\eta) \CI_{T[SU(2)]} (m_2,u_2|m_1,u_1;m_\eta, \bareta)\;,
\label{Wilson_TSU2}
\end{align}
and similarly
\begin{align}
&\langle W_\Box (SU(2)_{+1}) \rangle^{\textrm{Tr}(T[SU(2),\bm{LR}])}_\pm \nn\\
&\, =  \sum_{2m_1,2m_2=0}^\infty \oint \frac{du_1}{2\pi i u_1} \frac{du_2}{2\pi i u_2} \Delta (m_1,u_1)\Delta(m_2, u_2) (-1)^{-\half}  (q^{\frac{m_2}{2}} u_2^{\pm 1} +q^{-\frac{m_2}{2}} u_2^{\mp 1})  \nn
\\
&\,   \times  \big{(}u_1(-1)^{m_1}\big{)}^{-2m_1}\big{(}u_2(-1)^{m_2}\big{)}^{2m_2}  \CI_{T[SU(2)]} (m_1,u_1|m_2,u_2;m_\eta, \bar{\eta}) \CI_{T[SU(2)]} (m_2,u_2|m_1,u_1;m_\eta, \bar{\eta})\;.
\label{Wilson_TSU2-2}
\end{align}
Here $\CI_{T[SU(2)]}(m,u|m',u';m_\eta, \bareta)$ is the index for $T[SU(2)]$  theory; we refer  to \cite{Gang:2013sqa} for explicit formula. $(m,u)$,  $(m', u')$ and $(m_\eta, \bareta)$ are (monopole flux, fugacity) for  $SU(2)_{\textrm{top}}$, $SU(2)_{\textrm{bot}}$  and  $U(1)$  axial symmetry respectively of the $T[SU(2)]$ theory. 
The factor $\big{(}u_1(-1)^{m_1}\big{)}^{-2m_1}\big{(}u_2(-1)^{m_2}\big{)}^{-2m_2}$ comes from CS terms of level $+1$ and $-1$ for two $SU(2)$s and $\Delta(m,u)$ is the index contribution from $SU(2)$ $\CN=2$ vector multiplet
\begin{align}
\begin{split}
&\Delta(m,u):=\frac{1}{\textrm{sym}(m)}\left(q^{\frac{m}{2}} u-q^{-\frac{m}{2}}u^{-1})(q^{\frac{m}{2}}u^{-1}-q^{-\frac{m}{2}}u\right)\;, 
\\
&\textrm{sym}(m): = 
\begin{cases}
2 &  (m=0) \\ 
1 & (m>0)\\
\end{cases} \;.
\end{split}
\end{align}

The factor $(q^{\frac{m_1}{2}} u_1^{\pm 1} +q^{-\frac{m_1}{2}}u_1^{\mp 1})$ in eq.~\eqref{Wilson_TSU2} comes from classical action of the $SU(2)_{+1}$ fundamental Wilson loop.  
In the saddle point of localization, fields  in the $SU(2)_{+1}$ vector multiplet  are given by 
\begin{align}
\begin{split}
&i A = \frac{m_1}2 \left(\begin{array}{cc}1 & 0 \\0 & -1\end{array}\right) (\pm 1-\cos \theta) d\phi + \frac{\log u_1}{i \beta} \left(\begin{array}{cc}1 & 0 \\0 & -1\end{array}\right)  d\tau\;, 
\\
& \sigma = \frac{m_1}2   \left(\begin{array}{cc}1 & 0 \\0 & -1\end{array}\right) \;.
\end{split}
\end{align}
Two BPS trajectories $x_\pm^\mu (s)|_{s=0}^{\beta:=\log q}$ at north/south poles are
\begin{align}
\begin{split}
&+ \;: \;(\tau, \theta, \phi)(s)  =(s, 0, *)\;,
\\
&-\;:\; (\tau, \theta, \phi)(s)  =(-s, \pi, *)\;.
\end{split}
\end{align}
Here $\tau \in [0,\beta]$ is a coordinate for $S^1$ and $(\theta,\phi)$ are the standard spherical coordinates for $S^2$. Thus we have 
\begin{align}
\textrm{Tr}_\Box Pe^{\oint_\pm  (- A_\mu \frac{dx^\mu}{ds} +\sigma |\frac{dx}{ds}| ) ds }
\big|_{\textrm{saddle-point}} = q^{\frac{m_1}{2}} u_1^{\pm 1}+q^{-\frac{m_1}{2}} u_1^{\mp 1}\;.
\end{align}
The phase factor $(-1)^{\pm \half}$ in \eqref{Wilson_TSU2} is one subtle point: it reflects  an overall shift in spin of states on $S^2$ in the presence of the loops.    
Once we write down the expression \eqref{Wilson_TSU2}, we can evaluate the integral order by order in $q$. For example,
\begin{align}
\begin{split}
&\langle W_\Box (SU(2)_{-1}) \rangle^{\textrm{Tr}(T[SU(2),\bm{LR}])}_+ (m_\eta=0, \bareta) 
\\
&=  \left(\bareta+\frac{1}{\bareta}\right)q^{\frac{3}{4}}-3\, q^{\frac{5}{4}}+\left(\bareta+\frac{1}{\bareta}\right)q^{\frac{7}{4}}+\left(-1+\bareta^2+\frac{1}{\bareta^{2}}\right)q^{\frac{9}{4}} -3\left(\bareta+\frac{1}{\bareta}\right) q^{\frac{11}{4}}+\ldots  \;,
\\
&\langle W_\Box (SU(2)_{+1}) \rangle^{\textrm{Tr}(T[SU(2),\bm{LR}])}_+ (m_\eta=0, \bareta) 
\\
&=q^{-\frac{1}4} -3 q^{\frac{3}4} +\left( \frac{3}\bareta + 3\bareta \right)q^{\frac{5}4} - 6 q^{\frac{7}4} +\left(\frac{2}\bareta +2\bareta\right)q^{\frac{9}4}+\left(- 1 +\frac{1}{\bareta^2}+\bareta^2 \right)q^{\frac{11}4}+\ldots \;.
\end{split}
\end{align}
and we can verify that the result is consistent with our previous computation from the state integral model \eqref{Simple wilson loop in figure eight} (again under the parameter identification
$\bareta =-\eta$).

Once we identify  fundamental Wilson loops  in  the $SU(2)$ theory in 3d--3d correspondence, generalization to higher representation or higher $N$ is obvious. This approach will  provide simple way to quantize (and identify the maps in 3d--3d correspondence for) Wilson loops in higher representation which are not obvious in terms of IR Abelian variables (cluster coordinates).
We can also consider other loop operators, such as vortex loops. We still  keep the non-Abelian structure on
 vortex loops by defining them as non-Abelian $SL(2,\mathbb{Z})$-transformation on Wilson loops. This construction gives natural non-Abelian generalization of Abelian vortex loops studied in \cite{Kapustin:2012iw}. Non-abelian $SL(2,\mathbb{Z})$ action on 3d $\mathcal{N}=2$ theory with $SU(N)$ flavor theory can be generated by two operations: one is gluing $T[SU(N)]$ theory using the $SU(N)$ symmetry which corresponds to $\bm{S}$ and the other is adding  background CS term with level $-1$ for the $SU(N)$ flavor symmetry which corresponds to $\bm{T}$. Both operations preserve $SU(N)$ flavor symmetry and known to form $SL(2,\mathbb{Z})$. 
 The automorphism in eq.~\eqref{Automorphism in fundamental group} implies   (where $\varphi = \bm{LR}=\bm{ST}^{-1}\bm{S}^{-1}\bm{T}$)
\begin{align}
\begin{split}
&(\textrm{VEV of loop operator } \mathcal{O} \textrm{ charged under the $SU(2)_{\pm1}$)} \\
& \qquad \textrm{ = }  (\textrm{VEV of loop operator } \varphi (\mathcal{O} )\textrm{ charged under the $SU(2)_{\pm1}$} ) \;.
\end{split}
\end{align}
This property will also be useful in quantizing loop operators in mapping torus.

%%%%%%%%%%%%%%%%%%%%%%%%%%%%%%%%%%%%%%%%%%%%%%%%%%%
\section{From 5d \texorpdfstring{$\CN=2$}{N=2} SYM}\label{sec.5dSYM}

We now come to another non-Abelian description, namely the 5d $\CN=2$ SYM.
In \cite{Cordova:2013cea,Lee:2013ida} the partition function of 
5d $\mathcal{N}=2$ SYM with gauge group $G$ on $S^2\times M$
has been computed by localization. 
The result coincides with the
partition function of the complexified pure Chern-Simons theory
with gauge group $G_{\mathbb{C}}$, the complexification of $G$:
\begin{align}
Z_\textrm{5d $G$ $\mathcal{N}=2$ SYM}[S^2\times M]=Z_\textrm{3d $G_{\mathbb{C}}$ CS}[M] \ ,
\end{align}
with parameter identification \cite{Lee:2013ida}\footnote{Our $\sigma$ is $i \sigma$ in \cite{Lee:2013ida}.}
\begin{align}
k=0\; , \quad \sigma=\frac{8\pi^2 i r}{g} \; .
\end{align}

In this section we include supersymmetric 1/2-BPS Wilson line to this computation (following \cite{Lee:2013ida}), 
and show that the after localization the VEV of the Wilson in the 5d $\mathcal{N}=2$ 
SYM reproduces the VEV of the Wilson line in the complexified Chern-Simons theory.
This gives a direct derivation of eq.~\eqref{3d3d_codim_4}.

%%%%%%%%%%%%%%%%%%%%%%%%%%%%%%%%%%
\subsection{Localization of 5d \texorpdfstring{$\CN=2$}{N=2} SYM on \texorpdfstring{$S^2\times M$}{S2 x M}}

\paragraph{Conventions}

Let us first summarize our conventions for 5d $\CN=2$ SYM, following \cite{Lee:2013ida}.
We use the indices $M, N, \cdots$ ($A, B, \cdots$) for the spacetime (internal space)
indices which runs from $1$ to $5$, while $I, J, \cdots$ for $Sp(4)$ R-symmetry indices.
We use the following representations for the five-dimensional gamma matrices $\G^M$ for the spacetime, and $\hat\G^A$ for the internal space:
\begin{align}
\begin{split}
  \G^m & = \g^m \otimes {\bf 1}_2 \qquad (m=1,2) \ , \\
  \G^\m & = \g^3 \otimes \g^\m \qquad (\m=1,2,3) \ , \\
  \hat \G^\m & = \g^\m \otimes \g^3 \qquad (\m=1,2,3)  \ , \\
  \hat \G^i & = {\bf 1}_2 \otimes \g^{i-3} \qquad (i=4,5) \ ,
  \end{split}
  \label{gamma_convention} 
\end{align}
where $\g^m = ( \tau^1 , \tau^2)$, $\g^\mu = (\tau^1,\tau^2,\tau^3)$
and $\tau_i$ are Pauli matrices. The charge conjugation operator $C$ and
the $Sp(4)_R$ invariant tensor $\hat C^{IJ}$ are given by
\begin{align}
\begin{split}
  C & = \left( \tau^1 \right)^{ab} \otimes \e^{\dot a \dot b} 
 \ ,\\
  \hat C & = \e^{\a \b} \otimes \left( \tau^1 \right)^{\dot \a \dot \b}\ ,\
  \end{split}
\end{align}
where we hereafter denote five-dimensional spinor
indices ($Sp(4)$ R-symmetry indices) by $\CI = (a,\dot a)$ ($I=(\a,\dot \a)$). Each of these indices $a$, $\dot a$,
$\a$ and $\dot \a$ is raised and lowered by the antisymmetric tensor
$\e^{ab}$, $\e^{\dot a \dot a}$, $\e^{\a\b}$ and $\e^{\dot\a \dot \b}$ with $\e^{12}=-\e_{12}=1$.
Our convention for bilinear of 5-dimensional spinors is
\begin{align}
  \vare \l = - \vare_{\CI} C^{\CI \CJ} \l_{\CJ}\ ,
  \qquad \vare \G^M \l = - \vare_{\CI} (C\G^M)^{\CI \CJ} \l_{\CJ}\ ,
  \qquad \text{etc.}
\end{align}

\paragraph{Topological Twist}

When we place 5d $\CN=2$ SYM on 
$S^2\times M$,
one needs to partially topologically twist the theory along 
the curved three-manifold $M$  ({\it cf.} \cite{Marcus:1995mq,Blau:1996bx}).
Let us denote by $SO(3)_\text{twist}$ the diagonal subgroup of the $SO(3)$
local Lorentz group on $M$ and the $SO(3)_R$.
The leftover
$SO(2)_R$ is then identified as the $U(1)_R$ R-symmetry of the $SU(2|1)$
supersymmetry algebra on $S^2$.
Under the symmetry group $SO(3)_\text{twist} \times U(1)_R$, various fields can be decomposed as follows
\begin{align}
\begin{split}
  A^M \ & : \   {\bf 1}_{\pm 2} \oplus {\bf 3}_0 \equiv A^m \oplus A^\mu  \ ,  \\
  \l_I \ & : \ {\bf 1}_{\pm 1} \oplus {\bf 3}_{\pm 1} \equiv \left( \l, \bar \l \right) \oplus \left(
  \psi^\mu, \bar \psi^\mu \right) \ ,  \\
  \phi^A \ & : \ {\bf 1}_{\pm 2} \oplus {\bf 3}_{0} \equiv \varphi_{\pm} \oplus
 \phi^{\mu} \ ,
 \label{twist_matter_content}
 \end{split}
\end{align}
while the supersymmetry parameters can be decomposed as
\begin{align}
  \vare_I \ : \ {\bf 1}_{\pm 1} \oplus {\bf 3}_{\pm 1}\ .
  \label{ep_twist}
\end{align}
The $SU(2|1)$ supersymmetry of our interest (${\bf 1}_{\pm 1}$ in eq.~\eqref{ep_twist}) can be parameterized by the singlets $(\xi,\bar \xi)$
under the $SO(3)_\text{twist}$, which takes the following form
\begin{align}
  \big( \vare_{I} \big)_{a\dot a}= \frac i2 \e_{\dot a \a} \left( \xi_a \otimes\vare^+_{\dot \a}
  - (\g^3 \bar \xi)_a \otimes \vare^-_{\dot \a}\right)\ ,
  \label{4Q}
\end{align}
where $\xi$ and $\bar \xi$ satisfy the Killing spinor equation on the two-sphere
\begin{align}
  \nabla_m \xi = + \frac{1}{2r} \g_m \g^3 \xi \ , \qquad
  \nabla_m \bar \xi = - \frac{1}{2r} \g_m \g^3 \bar \xi\ ,
  \label{S2_spinor}
\end{align}
and
\begin{align}
  \vare^+ := \begin{pmatrix} 1 \\ 0 \end{pmatrix}\ , \qquad
  \vare^- := \begin{pmatrix} 0 \\ 1 \end{pmatrix}\ .
  \label{ep_pm}
\end{align}
%

%%%%%%%%%%%%%%%%%%%%%%%%%%%%%%%%%%
\subsection{Co-dimension 4 Defects as Wilson Lines}\label{codim_4_as_Wilson}

\paragraph{5d Wilson Line}

Let us consider a Wilson line on $S^2\times M$,
which spreads along a 1-cycle $\gamma$ on $M$ and is located at a specific point $p$ on $S^2$.
The choice of the (non-self-intersecting) 1-cycle $\gamma$ is arbitrary,
and $\gamma$ can be any knot inside $M$.

We consider the VEV of this Wilson line, in a representation $R$ of the 
gauge group $G$:
\begin{align}
\langle W_R \rangle = \left\langle \textrm{Tr}_R \, P \exp\left( \int_{ \{p \}\times \gamma} \left(-A_{\mu} \mp i \phi_{\mu}\right)  \right)  \right\rangle \ ,
\label{5d_W}
\end{align}
where $\gamma$ is an arbitrary closed path inside the 3-manifold $M$,
and $A_{\mu}$ and $\phi_{\mu}$ are the 1-forms on $M$ after the topological twist \eqref{twist_matter_content}.

To check the remaining supersymmetry of this Wilson line, let us first recall the supersymmetry variation of the fields $A_{\mu}$ and $\phi_{\mu}$:
\begin{align}
\begin{split}
\delta A_{\mu}&= i \varepsilon_I \hat{C}^{IJ} \Gamma_{\mu} \lambda_J   \ , \\
\delta \phi_{\mu}&= \varepsilon_I (\hat{C} \hat{\Gamma}_{\mu})^{IJ} \lambda_J  \ .
\end{split}
\end{align}
It then follows that 
\begin{align}
\delta \left(A_{\mu} \pm i \phi_{\mu}\right) =
i\varepsilon_I \left( \hat{C}^{IJ} \Gamma_{\mu}  \pm  (\hat{C} \Gamma_{\mu} )^{IJ} \right) \ \lambda_J \ ,
\end{align}
and hence preserves a fraction of the supersymmetry given by
\begin{align}
\varepsilon_I ( \hat{C}^{IJ} \Gamma_{\mu} \pm  (\hat{C} \hat{\Gamma}_{\mu})^{IJ})=0 \ .
\label{susy_fraction}
\end{align}
Writing $I=(\alpha, \dot{\alpha}), J=(\beta, \dot{\beta})$, and using the 
expressions of $C, \hat{C}, \Gamma_{\mu}, \hat{\Gamma}_{\mu}$ given in eqs.~\eqref{gamma_convention}, 
\eqref{susy_fraction} amounts to 
\begin{align}
(\varepsilon_{\alpha \dot{\alpha}})_{a \dot{a}} \,
 \epsilon^{\alpha \beta} \otimes (\tau^1)^{\dot{\alpha} \dot{\beta}}  \otimes (\tau^3)^{ab} \otimes (\tau^{\mu})^{\dot{a}\dot{b}}  \pm 
(\varepsilon_{\alpha \dot{\alpha}})^{b \dot{b}} \,
(\epsilon \gamma^{\mu})^{\alpha \beta} \otimes (\tau^1 \tau^3)^{\dot{\alpha} \dot{\beta}} 
 =0 \ .
 \label{tmp_1}
\end{align}

What is crucial for our purposes is whether or not the Wilson line preserve the same supersymmetry used for the localization computation.
Substituting the supercharge \eqref{4Q} into eqs.~\eqref{tmp_1}, \eqref{tmp_1}, one obtains
\begin{align}
\begin{split}
&    (\varepsilon^+ \tau^1)^{\dot{\beta}} \otimes (\xi \tau^3)^b \otimes (\tau^{\mu})^{\beta\dot{b}}
\pm     (\varepsilon^+ \tau^1 \tau^3)^{\dot{\beta}} \otimes \xi^b \otimes (\tau^{\mu})^{\beta\dot{b}} 
\\
&\qquad
-    (\varepsilon^- \tau^1)^{\dot{\beta}} \otimes ((\tau^3\bar\xi) \tau^3)^b \otimes (\tau^{\mu})^{\beta\dot{b}}
\mp     (\varepsilon^- \tau^1 \tau^3)^{\dot{\beta}} \otimes  (\tau^3\bar\xi)^b \otimes (\tau^{\mu})^{\beta\dot{b}}
\\
& =   (\varepsilon^-)^{\dot{\beta}} \otimes (\xi  (1\pm \tau^3))^b \otimes (\tau^{\mu})^{\beta\dot{b}}
+
    (\varepsilon^+)^{\dot{\beta}} \otimes (\bar\xi  (1\mp \tau^3))^b \otimes (\tau^{\mu})^{\beta\dot{b}} \; .
\end{split}
\end{align}
This means that the remaining supersymmetry should satisfy
\begin{align}
\xi(1\pm \tau^3)=0  \ , \quad
 \bar\xi (1\mp \tau_3)=0 \ .
 \label{xi_project}
 \end{align}
These conditions pick up two supercharges out of the four supercharges preserved on $S^2\times M$.
 
Recall that $\xi, \bar{\xi}$ are the 
Killing spinors on $S^2$ \eqref{S2_spinor}, 
and hence depends non-trivially on the position at $S^2$.
When we parametrize the $S^2$ by
\begin{align}
ds^2=r^2 (d\theta^2 + \sin^2\theta d\varphi^2) \ ,
\end{align}
then $\xi$ and $\bar\xi$ are explicitly written as
\begin{align}
\xi=e^{\frac{i \varphi}{2}} e^{-\frac{i\theta}{2} \tau^2} \epsilon_+ \ , \quad
\bar\xi=e^{-\frac{i \varphi}{2}} e^{\frac{i\theta}{2} \tau^2} \epsilon_- \ ,
\end{align}
where $\epsilon_{\pm}$ are given in eq.~\eqref{ep_pm}, and in particular
eigenstates of $\tau^3$.

Due to the presence of the factor $e^{-\frac{i\theta}{2} \tau^2}$, $\xi$ and $\bar\xi$ in general do not have a definite chirality. However, the situation is special for north pole ($\theta=0$) and south poles ($\theta=\pi$):
\begin{align}
\begin{split}
\tau^3 \xi_{NP}=\xi_{NP} \ ,  \quad \tau^3 \xi_{SP}=-\xi_{SP} \;, \\
\tau^3 \bar \xi_{NP}= - \bar \xi_{NP} \ ,  \quad \tau^3 \bar \xi_{SP}=\bar \xi_{SP} \;.
\end{split}
\label{NPSP}
\end{align}

Comparing eqs.~\eqref{xi_project} and \eqref{NPSP}, 
we learn that we can include holomorphic (or anti-holomorphic) Wilson lines
in the south (north) pole of $S^2$.

\paragraph{Localization}

Having established the presence of supersymmetry, we can now appeal to the supersymmetric localization computation. Since the Wilson line preserves the supercharges used for the localization,
the computation works in exactly the same manner, the only difference being that 
we have to evaluate the Wilson line at the saddle point locus.

As explained in \cite{Lee:2013ida},
at the saddle point, both $A_{\mu}$ and $\phi_{\mu}$, and hence its complex combination 
$\CA_{\mu}=A_{\mu}+i \phi_{\mu}$, are constant along the $S^2$ directions, and 
has a non-trivial profile only along $M$:
\begin{align}
\mathcal{A}_{\mu}(x^M) =\mathcal{A}_{\mu}(x^{\mu}) \ .
\end{align}
The action then reduces to the pure Chern-Simons action of the complexified Chern-Simons theory.

The Wilson line \eqref{5d_W} then reduces to the 
holomorphic (or anti-holomorphic) Wilson line of the complex Chern-Simons theory (up to a constant factor, the volume of $S^2$):
\begin{align}
\langle W_R\rangle=\Big\langle \textrm{Tr}_R \, P \exp\left(- \int_{\gamma} \mathcal{A}   \right) \Big \rangle_\textrm{$G_{\mathbb{C}}$ pure Chern-Simons}
\label{3d_W} \;,
\end{align}
where $R$ is the representation of $G_{\mathbb{C}}$, which is obtained by 
a natural complexification of the representation of $G$.

%%%%%%%%%%%%%%%%%%%%%%%
\subsection{Co-dimension 2 Defect: Higgsing and Refinement}\label{sec: Higgsing}
%%%%%%%%%%%%%%%%%%%%%%%%%%%%%%%%%%%%%%%%%%%%%%%%%%%%%

The discussion of the previous subsection raises a natural question:
could be perform similar localizations for co-dimension 2 defects \eqref{3d3d_codim_2}, 
directly from 5d $\CN=2$ SYM?

While this is a well-defined question, it is not too straightforward to cary out in detail
localization computations with co-dimension $2$ defects in 5d $\mathcal{N}=2$ SYM.
Instead we choose to take a different route, which turns out to be 
a rather useful shortcut.

%------------------------------------------------------------------------------------
\paragraph{Higgsing Prescription}

Our starting point was already explained in Sec.~\ref{sec : co-dimension 2},
namely  the expectation that the co-dimension 2 defect of type $\rho$
is described by coupling to $T_{\rho}[SU(N)]$ theory.
Formulated for 5d $\CN=2$ SYM, we have ({\it cf.} \cite{Bullimore:2014upa,Yonekura:2013mya})
\begin{align}
\begin{split}
&\textrm{5d $\CN=2$ $SU(N)$ SYM+ co-dimension 2 defect of type $\rho$} 
\\
&\textrm{=   5d $\CN=2$ $SU(N)$ SYM coupled to 3d $T_{\rho}[SU(N)]$ theory} 
 \;.
\end{split}
\label{codim2_as_TSUN}
\end{align}
We can then compactify the 5d $\CN=2$ SYM on a 3-manifold, which should keep intact the 
relation \eqref{codim2_as_TSUN}.

Now it becomes evident how to change the type of $\rho$: we start with say $\rho=\textrm{maximal}$,
and `remove' the $T[SU(N)]=T_{\rho=\textrm{maximal}}[SU(N)]$ theory from the theory $T_{N}[M;\rho=\textrm{maximal}]$,
and then glue together $T_{\rho}[SU(N)]$ theory. Very schematically, 
\begin{align}
T_{N}[M;\rho] \sim  T_{N}[M;\rho=\textrm{maximal}] -  T[SU(N)]  +  T_{\rho}[SU(N)] \;.
\end{align}

Of course, this prescription does not make sense unless we clarify what mean by ``removing'' $T[SU(N)]$ theory.
Fortunately, $T[SU(N)]$ has a very special property which we already mentioned in Sec.~\ref{sec.TSUN}:
$T[SU(N)]$ is a representation of the element $\bm{S}$ of the mapping class group $PSL(2, \mathbb{Z})$,
and hence it squares to a trivial theory under gluing, as expected from S-duality of 4d $\mathcal{N}=4$ SYM \cite{Gaiotto:2008ak}.%\DG{In eqref{I_ST_eg, S^4=I.} \footnote{As we have seen in {\it e.g.} eq.~\eqref{I_ST_eg},
 %when we glue two $T[SU(N)]$ theories
%the partition function gives a non-trivial overall shift (associated with the framing anomaly of the Chern-Simons theory).
%This is however only an overall shift and does not really affect the consideration of Higgsing here.
%} 
 Namely, to remove $T[SU(N)]$ theory
we just need to glue $T[SU(N)]$ theory. Hence, (again very schematically)\footnote{Our Higgsing proposal is reminiscent of the 
discussion of surface defects for superconformal indices for 4d class S theories \cite{Gaiotto:2012xa}.
There are differences, however, in that in the 4d case we glue a trinion (a bifundamental multiplet),
whereas in our 3d setup we glue an annulus. 
}
\begin{align}
T_{N}[M;\rho] \sim  T_{N}[M;\rho=\textrm{maximal}] 
+ \left( T[SU(N)]  + T_{\rho}[SU(N)] \right) \;.
\label{Higgsing_schematic}
\end{align}
 
\begin{figure}[htbp]
\begin{center}
   \includegraphics[width=.45\textwidth]{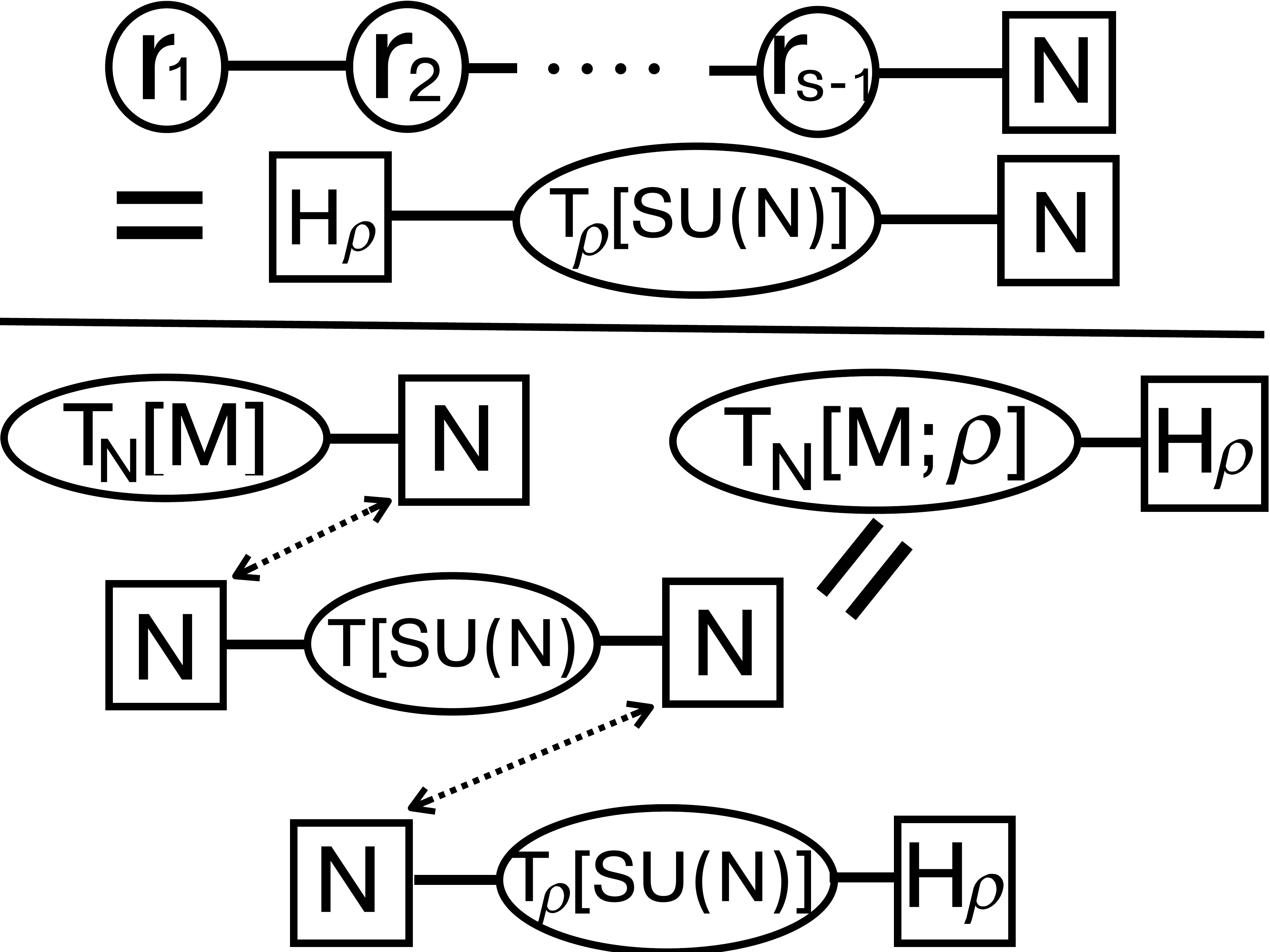}
      \includegraphics[width=.45\textwidth]{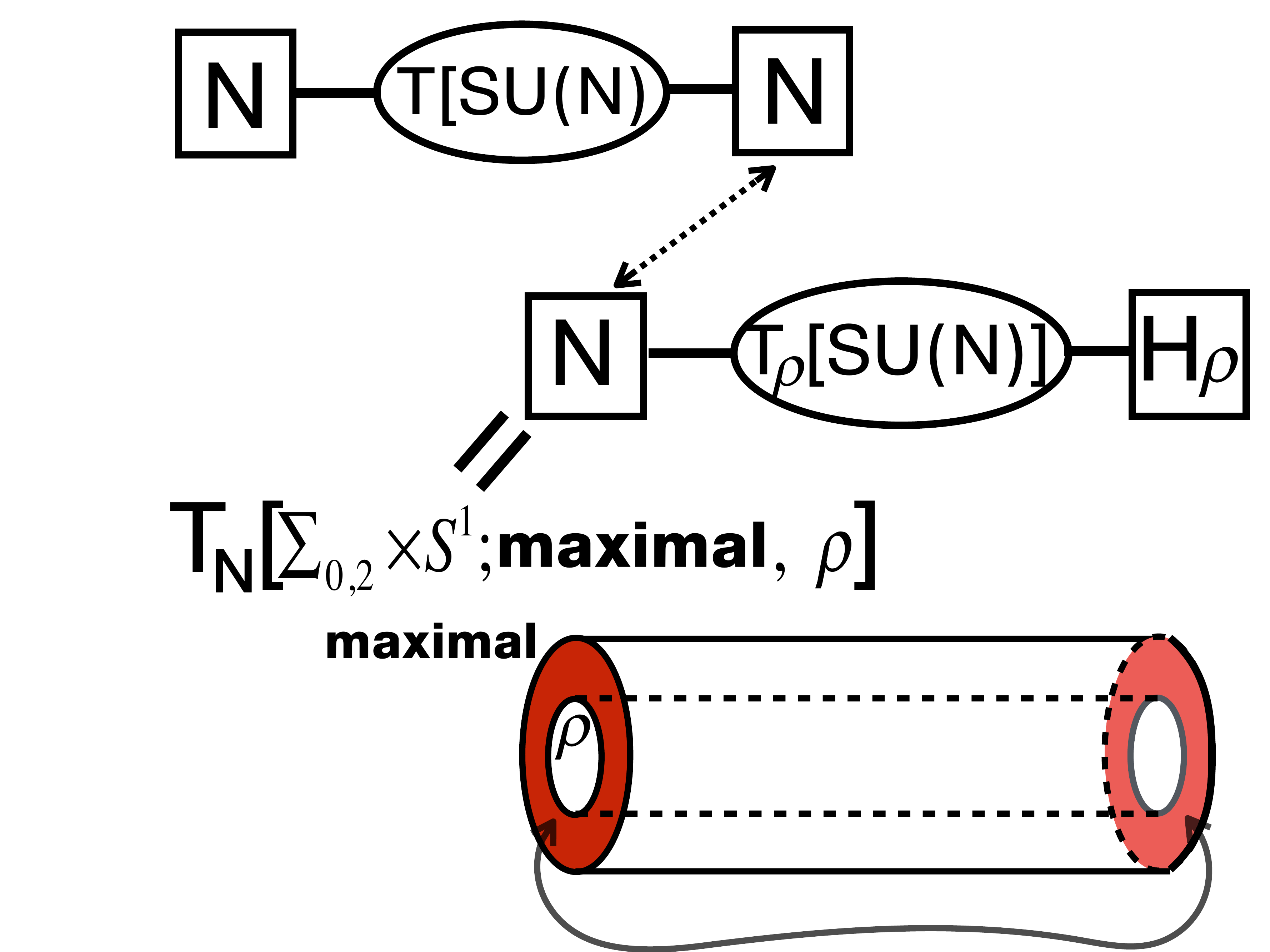}
   \end{center}
   \caption{Left above : quiver diagram for $T_{\rho}[SU(N)]$ with $\rho=[n_1, n_2, \ldots , n_s]$ theory.  Here $ r_{k}:= n_1+n_2\ldots +n_k$.  Circle vertices represent $\CN=4$ vector multiplets with gauge group $U(r_i)$ and the square vertices denotes the $(SU(N) \textrm{or } H_\rho)$ flavor symmetry. Lines connecting those symmetry groups represent  bi-fundamental hypermultiplet.  Topological symmetries $U(1)_J^{s-1}$ for each $U(1)$ factors of $U(r_i)$ gauge will be enhanced to $H_{\rho}$ at IR.  Second diagram is a simple representation of the $T_{\rho}[SU(N)]$.  Left below : Higgsing procedure which gives  $T_N[M,\rho]$ theories from the $T_N[M]$. Dotted arrow line represent gauging a diagonal $SU(N)$ flavor symmetry. Right : The theory used in the Higgsing procedure can be identified with $T_N[\Sigma_{0,2}\times S^1, \textrm{maximal},\rho]$}
    \label{fig:Higgsing-defect}
\end{figure}

To make this more precise, 
we start from $T_N[\hat{M}\backslash K]$ which has a $SU(N)_{\rm orig}$ flavor symmetry. 
Let us note that $T[SU(N)]$ has $SU(N)_1\times SU(N)_{2}$ symmetry\footnote{More precisely one of the $SU(N)$ flavor symmetries is $SU(N)/\mathbb{Z}_N$, and correspondingly there are two choices in eq.~\eqref{codim2_as_TSUN},
depending on whether you gauge $SU(N)$ or $SU(N)/\mathbb{Z}_N$.
This subtlety does not matter (up to an overall constant factor of $N!$)
for the partition functions considered in this paper.}, and similarly $T_{\rho}[SU(N)]$ theory has 
global symmetry $SU(N)_3\times H_{\rho}$, where $H_{\rho}$ is defined in eq.~\eqref{H_rho}.
We can then couple the two theories by gauging the diagonal $SU(N)$ symmetry of the 
$SU(N)_{\rm orig}$ and $SU(N)_1$, and similarly of $SU(N)_2$ and $SU(N)_3$,
where gauging makes the corresponding background $\CN=4$ vector multiplet dynamical.
The resulting theory has $H_{\rho}$ as the remaining flavor symmetry,
and is identified with $T_{N}[M, \rho]$, see  Fig.~\ref{fig:Higgsing-defect}. 

It is tempting to propose the following wavefunction
interpretation of the Higgsing procedure 
\eqref{Higgsing_schematic}.
First, it is known that 
 the 3d theory obtained by gluing $T[SU(N)]$ and $T_{\rho}[SU(N)]$ plays the role of an 
 overlap of two states: (see \cite{Nishioka:2011dq} for a similar proposal in the case of the $S^3_{b=1}$ partition function)
\begin{align}
&\langle \frakL^{\textrm{maximal}}  | \frakL^{\rho}  \rangle  \nn
\\
&\Longleftrightarrow
\quad \textrm{3d theory obtained by gluing $T[SU(N)] $ and $T_{\rho}[SU(N)]$}\;,
\label{TSUN_kernel}
\end{align}
where $\frakL^{\rho}$ and $\frakL^{\textrm{maximal}}$ are the mass parameters for the 
$H_{\rho=\textrm{maximal}}$ and $H_{\rho}$ flavor symmetries.
Then we propose that the Higssing procedure is simply an integral transformation,
with kernel given in eq.~\eqref{TSUN_kernel}: 
\begin{align}
\langle  \frakL^{\rho} | \hat{M}\backslash K  \rangle  =  \int  \left[ d\frakL^{\textrm{maximal}}  \right]   \langle  \frakL^{\rho}      | \frakL^{\textrm{maximal}}  \rangle   \langle\frakL^{\textrm{maximal}}| \hat{M} \backslash K  \rangle \;, \label{Higgsing-ptn}
\end{align}
This naturally can be interpreted as  the completeness relation
\begin{align}
\mathbb{I}= \int  \left[ d\frakL^{\textrm{maximal}}  \right]
 | \frakL^{\textrm{maximal}}  \rangle \langle \frakL^{\textrm{maximal}} |   \;.
\end{align}
Note that the partition function of the state-integral model does take the form of the overlap of two states \eqref{state-integral from ideal triangulation}. 

One should keep here in mind that there is a subtlety in this Higgsing procedure:
the 3d theory $T_{N=2}[\Sigma_{0,2}\times S^1, \textrm{maximal},\rho]$ is a  `bad' theory in the sense of \cite{Gaiotto:2008ak} (meaning that some operators decouple in the IR, and the UV R-symmetry does not coincide with the IR R-symmetry of the superconformal algebra)
and its supersymmetric partitions functions on curved backgrounds diverges.\footnote{Both $T[SU(N)]$ and $T_{\rho}[SU(N)]$ theories are good, but the problem happens when we combine them by gauging the diagonal $SU(N)$ flavor symmetry. For example, if we glue two $T[SU(N)]$ theories, the node for the newly-gauged
gauge symmetry $U(N)$ has $N_f=2(N-1)$ flavors.}
The divergence is actually expected from the 3d--3d correspondence,
since there is no flat connection (saddle point of CS theory) on $\Sigma_{0,2}\times S^1$ with different  fixed holonomies  along two  punctures in $\Sigma_{0,2}$  and there is one-to-one correspondence between \cite{Witten:2010cx}
\begin{align}
&\textrm{(saddle points of path integral)} \Longleftrightarrow (\textrm{convergent contours in path-integral})\; .
\end{align}
Although the `Higgsing' theory $T_{N}[\Sigma_{0,2}\times S^1]$ is a `bad' theory, by coupling the theory to a theory $T_{N}[M, \textrm{maximal}]$ we can obtain a `good' theory. Similarly by gluing $\Sigma_{0,2}\times S^1$ to $\hat{M}\backslash K$, we obtain $\hat{M}\backslash K$ which can allow flat connections with boundary longitude holonomy of type $\rho$.

%-----------------------------------------------------------------------------------------------------------------
\paragraph{Additional $U(1)_t$ and `Refined' CS Partition Function} 

Before closing this section, 
let us comment on  one implication of eq.~\eqref{Higgsing_schematic}.
The result \eqref{Higgsing_schematic} means that we can turn on an 
extra parameter, and therefore `refine' the complex CS theory. To describe this, note that while we are interested in 3d $\mathcal{N}=2$ theory $T[\hatM\backslash K, \rho]$,
the 3d theories $T_{\rho}[SU(N)]$ have 3d $\mathcal{N}=4$ supersymmetry.
From the viewpoint of 3d $\mathcal{N}=2$ theory, this means that 
we have an extra symmetry: this is the axial $U(1)_t$ symmetry,
which is a Cartan of $SO(4)_R$ $\CN=4$ 
symmetry which commute with the $\CN=2$ R-symmetry $SO(2)_R\subset SO(4)_R$.
The real mass (fugacity) parameter for this $U(1)_t$ symmetry appears non-trivially in the partition functions \eqref{Higgsing-ptn},
and gives a natural 1-parameter generalization (`refinement') of the complex CS partition function.
In the brane configuration of eq.~\eqref{codim_2_brane},
this $U(1)_t$ symmetry is the axial combination of $U(1)_{78}$ and $U(1)_{9\sharp}$;
the $U(1)_{9\sharp}$ rotation is the rotation of $\mathbb{R}^2$ of eq.~\eqref{brane_RTMR} ({\it cf.} \cite{Chung:2014qpa}). 

This refinement can also be understood by considering the BPS equations, which are obtained by topological twisting of 5d $\CN=2$ SYM theory on a 3-manifold:
\begin{align}
F_\CA := d\CA +\CA \wedge \CA=0\;, \quad D_\CA \varphi =0\;.
\end{align}
The 5d theory has five adjoint scalars and three of them become one-form after topological twisting and form a complex $SL(N)$ connection $\CA:=A_\mu +i \phi_\mu$. The remaining two scalars form a complex field $\varphi$. The $U(1)_{9\sharp}$ symmetry mentioned above
rotates the scalar $\varphi$, and could act non-trivially when $\varphi$ is non-zero. At a generic point in the moduli space of flat connections, $SL(N)$ gauge group is totally broken by holonomies, and the BPS equation implies $\varphi=0$.%\DG{Since $\varphi$ is a scalar, gauge transformation on zero is always zero.} (up to gauge transformation). 
At non-generic points of the moduli space, however,
there could be unbroken subgroup remaining (this is when the flat connection is reducible),
and $\varphi$ can have non-trivial VEVs along the unbroken direction. 
It is expected that the VEVs for the scalar parametrize the Higgs (or mixed) branch of the corresponding 3d  $T_{N}[\hat{M}\backslash K, \rho]$ theory ({\it cf.} \cite{Xie:2014pua} for similar discussion in 4d--2d).  For hyperbolic 3-manifold $\hat{M}\backslash K$, there is a special $SL(N)$ flat connection  $\CA^{\textrm{conj}}_\rho:=\rho^t (\CA_{N=2}^{\textrm{conj}})$  for each $\rho$ which can be   constructed from the hyperbolic structure. Here the $SL(2)$ flat connection $\CA^{\textrm{conj}}_{N=2} := \omega - i e$, where $\omega$ and $e$ are dreibeins and spin-connections of the hyperbolic metric.   
Since the holonomies of the flat connection commute with $H_{\rho^t}$, 
the VEV of $\varphi$ can take values in $H_{\rho^t}$.
In particular, the VEV of $\varphi$ can be non-zero except when $\rho=\textrm{maximal}$. Thus at the point $\CA^{\textrm{conj}}_\rho$ for  non-maximal $\rho$, we can introduce additional $U(1)_t$ which rotates $\varphi$.

It would be interesting to investigate this refinement further, and in particular to understand the connection with the categorification of the knots invariants and the refined topological strings ({\it cf.} \cite{Gukov:2004hz,Aganagic:2011sg,Witten:2011zz}).

%%%%%%%%%%%%%%%%%%%%%%%%%%%%%%%
\section{From Large \texorpdfstring{$N$}{N} Holography} \label{sec:sugra}
%%%%%%%%%%%%%%%%%%%%%%%%%%%%%%%%%%%%%%
In this section we study the holographic duals of co-dimension 2 and 4 defects. We compute gravity free energy with various defects which corresponds to $(S^3)_b$-free energy of 3d $T_{N}[M]$ theory with defects at conformal point with real mass $\frakM_\a=0$. Via  the 3d--3d correspondence \eqref{3d-3d relation} and \eqref{3d3d_codim_4-2}, the free-energy is  related to free energy of $SL(N)$ Chern-Simons theory on $M$ with a defect at quantized CS level $k=1$.

%%%%%%%%%%%%%%%%%%%%%%%%%%%%%%%%%%% 
\subsection{Supergravity Background }

Let us first begin with the $D=11$ supergravity background of
$N$ M5-branes wrapping a hyperbolic 3-manifold $\hat{M}$ \cite{Pernici:1984nw,Gauntlett:2000ng,Donos:2010ax,Gang:2014qla,Gang:2014ema}, and review its properties. M-theory on the background can be thought as gravity dual of 3d $T_{N}[\hat{M}]$ theory.

The metric takes the form of a warped product $AdS_4 \times \hat{M} \times \tilde{S}^4$:
\begin{align}
\dd s^2_{11} = l_{\rm P}^2 (2\pi N)^{\frac{2}{3}} (1+\sin^2\theta)^{\frac{1}{3}}\Bigg[ \dd s^2(AdS_4) + \dd s^2(H^3) 
+ (1+\sin^2\theta)^{\frac{2}{3}} \dd s^2(\tilde{S}^4)
\Bigg] \;,
\label{metric}
\end{align}
where $l_{\rm P}$ is the eleven-dimensional Planck constant.
The warp factors depend on $\theta$, which is one of the coordinates of the squashed 4-sphere $\tilde{S}^4$. The metric of the 4-sphere $\tilde{S}^4$ is given as follows:
\begin{align}
\dd s^2(\tilde{S}^4)  = (1+\sin^2\theta) \left[
\frac{1}2 \left(\dd\theta^2 + \frac{\sin^2\theta}{1+\sin^2\theta} \dd\phi^2 \right) + \frac{\cos^2\theta}{1+\sin^2\theta} \dd s^2(\tilde{S}^2)
\right] \;.
\label{squashed_S4}
\end{align}
The round two-sphere $\tilde{S}^2$ is fibered over $H^3$, {\it i.e.} $ds^2(\tilde{S}^2)=\sum_a(d\mu^a+\bar{\omega}^a_{\;\; b} \mu^b)^2$ with $\sum_{a=1}^3 (\mu^a)^2=1$ and $\bar{\omega}^a_{\;\;b}$ representing the spin connection of $H^3$. The remaining coordinates $\theta, \varphi$ cover the range
$0\le \theta \le \frac{\pi}{2}$ and $0 \le \varphi \le 2\pi$.
Although the metric \eqref{squashed_S4} looks non-illuminating at first, 
it is in fact the metric 
of an ellipsoid embedded in $\mathbb{R}^5$. One can easily check that, starting with 
\be
x_1^2 + x_2^2 +x_3^2 +2 x_4^2  +2 x_5^2   = 1 \;,
\ee
and the parametrization
\begin{align}
\begin{split}
x_1 &= \cos\theta \cos\vth \;, \ \quad
x_2 = \cos\theta \sin\vth\cos\vph \;, \\
x_3& = \cos\theta \sin\vth\sin\vph \;, \quad
x_4 = \frac{1}{\sqrt{2}}\sin\theta \cos\phi \;, \quad
x_5 = \frac{1}{\sqrt{2}}\sin\theta \sin\phi \;, 
\end{split}
\end{align}
we obtain eq.~\eqref{squashed_S4}, aside from the fibration structure over $H^3$.
The hyperbolic space $H^3$ is a special Lagrangian 3-cycle $\hatM$, which should be more precisely expressed as  $\hatM=H^3/\Gamma$, where $\Gamma$ is a torsionless discrete subgroup of $PSL(2, \mathbb{C})$. Then $\hatM$ is assigned a finite volume, and without orbifold singularities. We use eq.~\eqref{metric} only as a local form of the metric, and
replace $H^3$ with $\hatM=H^3/\Gamma$.
The rotational invariance associated with 
the Killing vector $\partial_\phi$ is dual to $U(1)$ R-symmetry of the dual ${\cal N}=2$ superconformal
field theory in $D=3$.

The fact that the above supergravity background describes $N$ M5-brane is confirmed through 
the 4-form flux quantization.
When restricted to the squashed 4-sphere, the flux is given as 
\be
G |_{\tilde{S}^4}= - N \pi l_{\rm P}^3\, \dd\left[
\frac{\cos^3\theta}{1+\sin^2\theta}
\right]
 \wedge \dd\phi \wedge \textrm{Vol} (\tilde S^2) \;. 
\label{gflux}
\ee
This result will be crucial later in the study of the probe M5-brane action dual to co-dimension 4-defects.

For the computation of $S^3$ partition function of the dual superconformal field theory
in three-dimensions, we need to consider the dimensional reduction of the theory from eleven down 
to four-dimensions \cite{Gang:2014qla,Gang:2014ema}. 
The 11-dimensional
supergravity action evaluates to 
\be
S_{\rm sugra} = \frac{N^3}{24\pi^3} {\rm Vol}(\hatM) \int_{AdS_4} \sqrt{g} (R+6) \;. 
\ee

The factor in front of the four-dimensional Einstein-Hilbert action gives the four-dimensional
Newton constant: if we follow the standard convention and use the result above, 
\be
\frac{1}{16\pi G_4} = \frac{N^3}{24\pi^3} \;, 
\ee
and accordingly via holography $S^3$-free energy  on the field theory is given by \cite{Emparan:1999pm}\footnote{Free energy is defined as $F=-\log |Z_{S^3}|$.}
\be
F_{S^3} (T_{N}[\hat{M}])= \frac{\pi}{2G_4} = \frac{N^3}{3\pi} \textrm{Vol}(\hatM) \;.
\label{F_H3}
\ee
The $N^3$ behavior is a manifestation of the famous $N^3$ 
scaling behavior of the $N$ M5-branes.
By replacing the round $S^3$ by squashed one $(S^3)_b$, the large $N$ free energy modified in the following universal way \cite{Martelli:2011fu,Martelli:2011fw}
independent of any details of the 3d $\mathcal{N}=2$ theory:
\begin{align}
F_{(S^3)_b}(T_{N}[\hat{M}]) =\frac{1}4 (b+b^{-1})^2 F_{S^3}  = \frac{N^3}{12\pi } (b+b^{-1})^2 \textrm{vol}(\hat{M})\;.
\end{align}
The gravity computation is reliable only for a closed hyperbolic $\hat{M}$.  But as noticed  in \cite{Gang:2014qla,Gang:2014ema}, the large $N$ formula also can be applicable to $T_{N}[\hat{M}\backslash K, \rho]$ when the $M=\hat{M}\backslash K$ is hyperbolic and $\rho$ is maximal:
\begin{align}
F_{(S^3)_b}(T_{N}[\hat{M}\backslash K,\textrm{maximal}]) = \frac{N^3}{12\pi } (b+b^{-1})^2 \textrm{vol}(\hat{M}\backslash K)\;. \label{F_H3 for knot complement}
\end{align}

Note that for non-hyperbolic 3-manifolds the volume $\textrm{Vol}(\hatM)$ is defined to be zero.
In this case, the result \eqref{F_H3} becomes trivial, and we need to 
analyze the subleading corrections of order $N^2$. Note that this is consistent with the fact that
the only known supergravity solution is for hyperbolic 3-manifolds. For non-hyperbolic 3-manifolds we expect that the 
associated 3d $\mathcal{N}=2$ theory is massive, and its IR theory will be trivial, apart from possible topological degrees of freedom.

%%%%%%%%%%%%%%%%%%%%%%%%%%%%%%%%%%%%%%%%%%%
\subsection{`Simple' Co-dimension 2 Defects}
 
\subsubsection{Single Probe M5}
We are now done with the review material and come to the discussion of supersymmetric defects.

Let us now consider putting supersymmetric defects into the $D=11$ geometry above.
The first example we take is M5-brane whose worldvolume expands the whole
$AdS_4$, and also a geodesic $\gamma$ in $H^3$ and a great circle in the two-sphere. 
We use probe approximation where
the backreaction to geometry is neglected; it would be interesting to 
construct fully back-reacted geometry.

This configuration was studied and its supersymmetry 
was verified in \cite{Bah:2014dsa}. This object is analogous to a puncture for
the case of M5-branes wrapping a Riemann surface. 

Here we will calculate the expectation value of the dual operator when the dual field
theory is put on $S^3$. The induced metric times M5-brane tension gives
\be
\Delta F_{S^3} = T_5 ({4\pi^2l_{\rm P}^6})(2\pi)\, \ell(\gamma)\, \textrm{Vol}(AdS_4) \;,
\label{sf}
\ee
where $\ell(\gamma)$ is the length of the geodesic $\gamma$,
and $T_5$ is the tension of the M5-brane and 
\begin{align}
T_5=\frac{1}{(2\pi)^5 l_{\rm P}^6} \;.
\end{align}
Using the regularized volume of $AdS_4$ \footnote{
To derive this, note that in the Euclidean signature the hyperbolic space
with constant curvature metric can be written as
\be
\dd s^2 = d\rho^2 + \sinh^2 \rho \, \dd \Omega_3^2 \;.
\ee
Here the boundary is the three-sphere with unit radius. Then the volume with regulator cutoff
$\rho_0$ is 
\begin{align}
\textrm{Vol}(AdS_4) &= \textrm{Vol}(S^3) \int^{\rho_0}_0 \sinh^3 \rho \, \dd \rho 
=2\pi^2\left( \frac{1}{3}\cosh^3 \rho_0 - \cosh \rho_0 +\frac{2}{3} \right) \;.
\end{align}
Extracting the finite piece, we obtain eq.~\eqref{vol_AdS}.
}
\be
\textrm{Vol}(AdS_4) |_{\rm reg} = \frac{4\pi^2}{3} \;,
\label{vol_AdS}
\ee
we obtain
\be
\Delta F_{S^3} = \frac{\ell(\gamma) N^2}{3}  \;.
\label{S_N2}
\ee
This single probe M5 brane can be identified as gravity dual of a `simple' co-dimension 2 defect. The difference $\Delta F_{S^3}$ in free energies measure the increase of the free energy by the defect at large $N$:
\begin{align}
\Delta F_{S^3} = F(T_{N}[\hat{M}\backslash K_\gamma, \textrm{simple}]) - F(T_{N}[\hat{M}]) \;.
\end{align}
Since this defect has $O(N^2)$ scaling, we can consistently neglect the backreaction and the probe approximation is well-justified.
This is in contrast with the previously-discussed case of the maximal puncture \eqref{F_H3 for knot complement},
which has $O(N^3)$ scaling and hence the defect would rather change the geometry $\hatM$ into $\hatM\backslash K$.
The similar scaling occurs in 4d--2d story, adding `simple' puncture increase anomaly coefficients by  $O(N^2)$ while `maximal' puncture increase the coefficients by $O(N^3)$ \cite{Gaiotto:2009gz}.

%%%%%%%%%%%%%%%%%%%%%%%%%%%%%%%%%%
\subsubsection{\texorpdfstring{Large $N$ of $T_N[(\Sigma_{1,1}\times S^1)_\varphi ,\textrm{simple}]$ }{
Large N of TN[(Sigma(1,1)x S1)(varphi, simple)]}}

Let us next study the large $N$ limit from the non-Abelian gauge theory description of $T_N[(\Sigma_{1,1}\times S^1)_\varphi, \textrm{simple}]$ theory (Sec.~\ref{sec.TSUN}).  The closed 3-manifold $\hat{M}_\varphi$ is a torus bundle $(T^2\times S^1)_\varphi$ which is not hyperbolic. So, we cannot use the gravity solution to predict the large $N$ behavior of the theory. 

The theory is build by gluing $T[SU(N)]$ theory, and as is explained in Sec.~\ref{app.TSU3},
its $(S^1\times S^2)_q$ partition function is complicated already for $N=3$. Fortunately, there is a dramatic simplification when we consider the $S^3_{b=1}$ partition function of the 
3d $\CN=4$ version of the $T[SU(N)]$ theory (namely when no real mass parameter for axial $U(1)_t$ symmetry 
is turned on). In this case, the partition function of the $T[SU(N)]$ theory 
takes a rather simple form \cite{Nishioka:2011dq,Benvenuti:2011ga,Gulotta:2011si}.
Denoting by $\vec{\mu}=(\mu_1, \ldots, \mu_N)$ and $\vec{\nu}=(\nu_1, \ldots, \mu_N)$ the real mass and FI parameters
for the $SU(N)\times SU(N)$ flavor symmetry (and hence $\sum_i \mu_i=\sum_i \nu_i=0$),
the $(S^3)_{b=1}$ partition function of $T[SU(N)]$ theory is given by
\begin{align}
Z_{T[SU(N)]}[\vec{\mu}, \vec{\nu}] = 
\frac{1}{N!}
\sum_{\sigma \in \mathfrak{S}_N } (-1)^{\sigma} \frac{e^{2\pi i\, \vec{\nu}\cdot \sigma(\vec{\mu})}}
{\Delta(\vec{\mu}) \Delta(\vec{\nu})} \;,
\label{NTY_formula}
\end{align}
where we denoted $\vec{\nu}\cdot \sigma(\vec{\mu})=\sum_{i=1}^N \nu_i \mu_{\sigma(i)}$,
the sum in eq.~\eqref{NTY_formula} is over the the symmetric group, and $\Delta(\vec{\mu})$ is the sinh
Vandermonde determinant:
\begin{align}
\Delta(\vec{\mu}):=
\prod_{i<j} \sinh\pi (\mu_i-\mu_j)  \ .
\end{align}

One natural question is the large $N$ behavior of eq.~\eqref{NTY_formula}.
Interestingly, at the conformal point (namely in the limit $\vec{\mu}, \vec{\nu}\to 0$)
eq.~\eqref{NTY_formula} has a free energy which scales as $N^2 \log N$ in the large $N$ limit \cite{Assel:2012cp}.
This, however, is not in contradiction with 
our holographic computations \eqref{F_H3} or \eqref{S_N2},
since the corresponding 3-manifold (mapping cylinder) is not a hyperbolic 3-manifold.

Let us turn to the 
mapping torus theory discussed in Sec.~\ref{TSU3_index_glue}.
The  mapping torus $(\Sigma_{1,1}\times S^1)_{\varphi}$, 
 admits a hyperbolic structure ({\it i.e.} $\varphi$ is pseudo-Anosov) if 
\begin{align}
|\textrm{Tr}(\varphi)|>2 \;.
\label{pseudo_Anosov}
\end{align}
The condition \eqref{pseudo_Anosov} can be satisfied, for example, by choosing $\varphi=\bm{S T}^k$ 
with $k\ge 3$. The $S^3_{b=1}$-partition function of the mapping torus theory is given by
\begin{align}
Z_{S^3}\big{[}T_N[(\Sigma_{1,1}\times S^1)_\varphi, \textrm{simple}]\big{]}&=\int d\vec{\mu}\,  \Delta(\vec{\mu})^2 \, \delta(\vec{\mu}- \vec{\nu}) \overbrace{Z_{T[SU(N)]}[\vec{\mu}, \vec{\nu}]}^{\bm{S}}  \overbrace{e^{k \pi i \vec{\mu}^2}}^{\bm{T^k}} \nn\\
& =\frac{1}{N!} \sum_{\sigma\in \mathfrak{S}_N} (-1)^{\sigma} \int d\vec{\mu} \,\, e^{2\pi i \left( \frac{k \vec{\mu}^2}{2} + \vec{\mu}\cdot \sigma(\vec{\mu}) \right)}
\;,
\label{I_wish_evaluate} 
\end{align}
where $d\vec{\mu}=\prod_{i=1}^N d\mu_i$ and $\delta(\vec{\mu}-\vec{\nu})=\prod_{i=1}^N \delta(\mu_i-\nu_i)$. We evaluate this integral in App.~\ref{app.deriv_TSUN}.
It turns out that the result is a rather simple:
\begin{align}
Z_{S^3_{b=1}}\big{[}T_N[(\Sigma_{1,1}\times S^1)_{\varphi=\bm{ST^k}}, \textrm{simple}]\big{]}= \frac{ 1}{c_k^{N^2} (-c_k^{-2};-c_k^{-2})_{N}}\;,
\label{Z_TSUN_largeN}
\end{align}
where $c_k$ is the largest eigenvalue of the $2\times 2$ matrix
$\varphi=\bm{ST}^k$ and the $q$-Pochhammer symbol $(a;q)_n$ is defined by
\begin{align}
(a;q)_n:=\prod_{k=0}^{n-1} (1-a q^k)\;.
\label{q_Pochhammer}
\end{align}
It then immediately follows that in the large $N$ limit we have
\begin{align}
F_{S^3_{b=1}}\big{[}T_N[(\Sigma_{1,1}\times S^1)_{\varphi=\bm{ST^k}}, \textrm{simple}]\big{]}\longrightarrow \frac{N^2}{2} \log c_k^2\;.
\label{TSUN_largeN}
\end{align}
Note that we again see the $O(N^2)$ behavior for a `simple' defect. 

%---------------------------------------------------------------------------------------------
\paragraph{Entropy of Pseudo-Anosov Map and Hyperbolic Volume}

The fact that the large the eigenvalue of $\varphi$ with the largest absolute value appears
in  eq.~\eqref{TSUN_largeN} is has interesting geometrical implications, which we now turn to.

The starting observation is that mathematically, $c_k^2$ is known to 
be the {\it dilation} of the pseudo-Anosov map, which in turn 
coincides with the entropy (topological entropy) $\textrm{Ent}(\varphi)$ of the  map $\varphi$ \cite{Travaux}:
\begin{align}
\log (c_k^2)= \log (\textrm{largest real eigenvalue of $\varphi$}) = \textrm{Ent}(\varphi)\;,
\end{align}
and hence 
\begin{align}
\lim_{N\rightarrow \infty }\frac{1}{N^2} F_{S^3}\big{[}T_N[(\Sigma_{1,1}\times S^1)_{\varphi=\bm{ST^3}}, \textrm{simple}]\big{]} =\frac{1}2 \textrm{Ent}(\varphi) \;.
\label{entropy_conjecture}
\end{align}
This topological entropy $\textrm{Ent}(\varphi)$ is associated with a mapping class $\varphi$, but can also be thought of as a quantity associated with the mapping torus $(\Sigma\times S^1)_{\varphi}$, since the latter is determined by the former.

Motivated by this computation for $\Sigma=\Sigma_{1,1}$ and $\varphi=\bm{ST}^k$, we conjecture that the equality \eqref{entropy_conjecture} holds
for $\Sigma$ and any pseudo-Anosov map $\varphi$ (such that $(\Sigma\times S^1)_{\varphi}$ is hyperbolic, see eq.~\eqref{pseudo_Anosov} for $\Sigma=\Sigma_{1,1}$).
It would be interesting to prove or disprove this conjecture.

To some readers it might not be obvious (apart from obvious mathematical interest)
why a gauge/string theorist might be interested in the conjecture \eqref{entropy_conjecture}.
The reason comes from the following interesting 
inequality between $\textrm{Ent}(\varphi)$, and the hyperbolic volume $\textrm{Vol}(\varphi)$,  of a mapping torus $(\Sigma \times S^1)_{\varphi}$:\cite[Theorem 1]{KojimaMcShane},
which states that the inequality
\begin{align}
\textrm{Ent}(\varphi) \geq \frac{1}{3 \pi |\chi(\Sigma_{1,1})|} \textrm{Vol}(\varphi) =\frac{1}{3 \pi} \textrm{Vol}(\varphi)\;, \label{Thm}
\end{align}
holds for any pseudo-Anosov $\varphi \in \textrm{MCG}(\Sigma)$. 

Interestingly, the mathematical quantities in both sides of the inequality \eqref{Thm} are related to co-dimension 2 defects: $\rm{Ent}(\varphi)$ is related to a `simple' defect and $\rm{Vol}(\varphi)$ is related to a `maximal' defect. 
More explicitly,
\begin{align}
\begin{split}
&\textrm{Ent}(\varphi)= \lim_{N\rightarrow \infty }\frac{2}{N^2} \times (\textrm{$S^3$ free energy of $N$ M5s on $\hat{M}_\varphi$ + `simple' defect on $K_\varphi$}) \;,
\\
&\frac{\textrm{Vol}(\varphi)}{3\pi }= \lim_{N\rightarrow \infty }\frac{1}{N^3} \times (\textrm{$S^3$ free energy of $N$ M5s on $\hat{M}_\varphi$ + `maximal' defect  on $K_\varphi$}) \;, \label{Ent,Vol with co-dimension 2 defect}
\end{split}
\end{align}
where a closed 3-manifold $\hat{M}_\varphi$ and a knot $K_\varphi$ in $\hat{M}$ are defined by following conditions
\begin{align}
 \hat{M}_\varphi :=(\Sigma_{1,0}\times S^1)_\varphi \;\textrm{ and }  (\Sigma_{1,1}\times S^1)_\varphi=\hat{M}_\varphi \backslash K_\varphi\;.
\end{align}
The second equation in eq.~\eqref{Ent,Vol with co-dimension 2 defect} follows from eq.~\eqref{F_H3 for knot complement}. 

It is therefore tempting to understand the inequality \eqref{Thm} physically as the condition (in the large $N$ limit)
that a defect with $\rho$ maximal (with $O(N)$ M5-branes) does not decay into 
a set of $O(N)$ M5-branes with $\rho$ simple, namely that the latter has more energy than the former.
We leave more detailed analysis as future work \cite{InProgress}.

%%%%%%%%%%%%%%%%%%%%%%%%%%%%%%%%%%
\subsection{Co-dimension 4 Defects}

%%%
\subsubsection{Fundamental Representation as M2-brane}
The M2-brane configuration which is dual to a Wilson loop operator in fundamental representation of 
3d CS theory on $H^3$, is also 
 a line operator for the theory dual to $AdS_4$. In addition to a loop in the boundary field theory, 
 the M2-brane is extended along the holographic direction and also a great circle within $S^2$. 
 This configuration is a co-dimension 4 defect within
the M5-brane $(2,0)$ field theory. 

Adopting the standard regularization scheme, the 
 holographic expectation  value of a circular Wilson loop with unit radius is given 
from the evaluation of M2-brane action, 
 \be
S = T_2 ( {2\pi l_{\rm P}^3} ) (2\pi)  \ell(\gamma)
 = N \, \ell(\gamma) \;.
 \ee
 Here $\ell(\gamma)$ is the length of a geodesic loop $\gamma$ in $H^3$, wrapped by the M2-brane.
We have used the tension of the M2-brane
\be
T_2 =  \frac{1}{(2\pi)^2 l_{\rm P}^3} \;.
\ee
The action computes the expectation values of the co-dimension 4 defects and it is related to  Wilson loop in $SL(N)$ CS theory with $k=1$ and $b=1$ (or equivalently $\sigma=0$) via the 3d--3d correspondence \eqref{3d3d_codim_4-2}:
\be
\log  \langle \hat{W}_{\Box}(\CK_\gamma) \rangle^{k=1,b=1}_{\textrm{norm}}  = N \ell(\gamma)  \quad \textrm{at large $N$}.
\ee
Here $\langle \mathcal{O}\rangle_{\textrm{norm}}$ denotes the normalized expectation value of an operator $\mathcal{O}$,
\begin{align}
\langle \mathcal{O} \rangle_{\textrm{norm}}:=\frac{\langle \mathcal{O} \rangle}{Z} =\frac{\langle \mathcal{O} \rangle}{\langle 1 \rangle} \;.
\end{align}
Again, as  $(S^3)_b$ free energy, the dependence on $b$ at large $N$ for the defect wrapping $AdS_2$ factor in $AdS_4$ can be   universally restored as \cite{Farquet:2014bda}:
\begin{align}
\begin{split}
&\log  \langle \hat{W}_{\Box}(\CK_\gamma)_+ \rangle^{k=1,b}_{\textrm{norm}} = \frac{1+b^2}{2}\log  \langle \hat{W}_{\Box}(\CK_\gamma)_+ \rangle^{k=1,b=1}_{\textrm{norm}} = \frac{1+b^2}{2} N \ell(\gamma) \;,
\\
&\log  \langle \hat{W}_{\Box}(\CK_\gamma)_- \rangle^{k=1,b}_{\textrm{norm}}  =  \frac{1+b^{-2}}{2}\log  \langle \hat{W}_{\Box}(\CK_\gamma)_+ \rangle^{k=1,b=1}_{\textrm{norm}} = \frac{1+b^{-2}}2 N \ell(\gamma) \;. \label{co-dimension 4 in gravity}
\end{split}
\end{align}
The subscript $\pm$ represents two supersymmetric cycle in $(S^3)_b$: $+$ for a cycle of length $b$ and $-$ for a cycle of length $b^{-1}$. They correspond to holomorphic/anti-holomorphic Wilson loop in $SL(N)$ CS theory.
%%%%%%%%%%%%%%%%%%%%%%%%%%%%%%%%%%%%%%%%%%%%%%%%
\subsubsection{Antisymmetric Representation as M5-brane}

The next object of our interest is the M5-brane which is in $AdS_2$ subspace of $AdS_4$,
a line defect $\gamma$ in $H^3$, and also occupying a three-dimensional sphere within the four-sphere transverse
to the source M5-branes. We expect that the probe M5-brane corresponds to a co-dimension 4 defect in  $R=A_K$, $K$-th anti-symmetric representation, with $K \sim O(N)$.\footnote{In the brane configuration of eq.~\eqref{codim_4_brane}, let us Euclideanize the $0$-direction and reduce the M5-branes to (Euclidean) D4-branes in type IIA theory. The number of Dirichlet-Neumann directions for the two D4-branes is $8$.
This implies that the zero-point ground-state energy of fundamental strings between the D4-branes (originating from M2-branes between the M5-branes in M-theory) 
is in the R sector in the NSR formalism. This behaves as a fermion and hence anti-symmetrizes the Chan-Paton indices, so that the fundamental strings are naturally anti-symmetrized.}
In the un-wrapped version the worldvolume of this probe M5-brane 
occupies $AdS_3\times S^3$ and already studied in \cite{Mori:2014tca}. We need to use the PST action to 
properly identify the solution and evaluate the on-shell action.

For supersymmetry and also 
for satisfying the brane equation of motion, it is essential to turn on the three-form
gauge field on M5-brane. Without it, the action is just induced worldvolume metric times tension,
as usual. The nontrivial configuration we need to be careful about is the three-sphere. Let
us identify this part of the worldvolume coordinates with $\vth,\vph$ from $\tilde S^2$, and $\phi$.
Then the angle $\theta$ is in general a function of these three coordinates. To be supersymmetric
it should be independent of the R-symmetry angle $\phi$, and without losing generality we assume it
is also independent of $\vph$. Now in terms of $u(\vth)\equiv (\sin\theta(\vth))^2$, 
the action before turning
on 3-form field (and also ignoring the contribution from background $G$-flux) is given by
\be
S = \frac{\ell(\gamma) N^2}{4} \int d\vth \sin \vth
\sqrt{u'^2 + \frac{8u(1-u)^2}{1+u}} \;.
\label{actionzero}
\ee
To incorporate the gauge field, we introduce
\be
F_3 = b'(\vth)\, \dd \phi \wedge\dd \vth \wedge\dd \vph \;.
\ee
The flux quantization on the brane requires
\be
\int  b'(\vth) \dd \vth = (2\pi)^3 K \, l^3_P \;,
\label{asquant}
\ee
where $K$ is an integer. Later, we will relate $K$ to the rank of anti-symmetric representation 
for the Wilson loop. The physical gauge field on the worldvolume should also include the pull-back
of the 3-form gauge potential in the background, {\it i.e.} $dC_3 = G_4$ and 
\begin{align}
H_3 &= F_3 - C_3 \nn\\
&= \left[
b' + {N\pi l^3_P}\left(\frac{(1-u)^{\frac{3}{2}}}{1+u}-1\right)\sin\vth
\right]
\dd \phi \wedge\dd \vth \wedge\dd \vph \;.
\end{align}
It is crucial for us to include an integration constant $-1$ here: it makes sure that
$C_3$ vanishes at $\theta=0$, where the 3-sphere part of M5-brane shrinks to zero size in 
our solution. 

To compute the contribution of $H_3$ in the PST action, we need to compute and multiply
$\sqrt{1+H^2_3}$ to eq.~\eqref{actionzero}. It gives us 
\be
S = \frac{\ell(\gamma) N^2}{4} \int d\vth\, \sin \vth \mathscr{L} \;,
\ee
where
\be
\mathscr{L}: = \sqrt{u'^2 + \frac{8u(1-u)^2}{1+u}+
\frac{4}{N^2\pi^2l^6_P}(1+u)\left(\frac{b'}{\sin\vth}+N\pi l^3_P\left(\frac{(1-u)^{\frac{3}{2}}}{1+u}-1\right)\right)^2} \;.
\label{actionone}
\ee
This action can be treated as a classical mechanical system with ``time'' $\vth$. We may first
take advantage of the gauge symmetry and derive the displacement field 
\be
D = \frac{4(1+u)}{(N\pi l^3_P)^2\mathscr{L}}  
\left[\frac{b'}{\sin\vth}+N\pi l^3_P\left(\frac{(1-u)^{\frac{3}{2}}}{1+u}-1\right)\right] \;,
\label{displace}
\ee
which is a constant. We can plug it back to the original action \eqref{actionone} and take 
a partial Legendre transformation:
\begin{align}
\mathscr{H} &= D b' - \mathscr{L} \sin\vth
\nn\\
&= - \sin\vth
\left(
\sqrt{u'^2 + \frac{8u(1-u)^2}{1+u}} \sqrt{1-\frac{d^2}{1+u}} 
+ \frac{2d(1-u)^{\frac{3}{2}}}{1+u}
\right)\;.
\end{align}
Here we introduced $d:=2/(N\pi l^3_PD)$.
One can check that the Euler-Lagrange equation from $\mathscr{H}$ is
interpreted as a Lagrangian of $u,u'$ is the same as the equation of motion derived 
from eq.~\eqref{actionone}.

Because of the explicit ``time'' dependence on $\vth$, the above action defies straightforward
integration. However it turns out that there is a relatively simple solution:
\be
u = 1 - \frac{d^2}{\cos^2\vth} \;.
\label{assol}
\ee
Our solution \eqref{assol} defines a 3-sphere through 
a {\it constant latitude} condition, $x_1=d$.\footnote{This is reminiscent of a similar result for 
4d $\mathcal{N}=4$ SYM \cite{Yamaguchi:2006tq,Gomis:2006sb} and 5d $\mathcal{N}=1$ SCFTs \cite{Assel:2012nf}.}

Now it is easy to compute the action. The worldvolume flux quantization \eqref{asquant}
gives 
\be
d = 1 - \frac{2K}{N} \,,
\ee
and the action evaluates to
\be
S = \frac{\ell(\gamma) N^2}{4}(1-d^2) = \ell(\gamma) N^2 \frac{K}{N}\left( 1-\frac{K}{N} \right) \;.
\label{codim4_action}
\ee
Note that this expression is consistent with the symmetry $K\to N-K$ of the 
$K$-th antisymmetric representations. Also, when $K$ is small, this reduces to 
$S = LN K$, which is $K$ times the action of the M2-brane computed previously.
This is to be expected since the M5-brane solution for $K \sim N$ can be thought of as the blow-up of the 
M2-branes when the flux charge $K$ is large.  Restoring $b$-dependence and relating to Wilson loop in $SL(N)$, as we did for probe M2-brane case, we expect that
\begin{align}
\begin{split}
&\log  \langle \hat{W}_{A_K}(\CK_\gamma)_+ \rangle^{k=1,b}_{\textrm{norm}} = \frac{1+b^2}{2} N^2 \kappa (1-\kappa) \ell(\gamma) \;,
\\
&\log  \langle \hat{W}_{A_K}(\CK_\gamma)_- \rangle^{k=1,b}_{\textrm{norm}}  = \frac{1+b^{-2}}{2} N^2 \kappa (1-\kappa) \ell(\gamma) \;, \label{co-dimension 4 in gravity-2}
\end{split} 
\end{align}
in the limit
\begin{align}
N \rightarrow \infty\;, K \rightarrow \infty \; \textrm{ with fixed } \kappa:=\frac{K}{N} \;.
\end{align}

We can also incorporate several different M5-branes, all occupying different latitudes on the sphere.
Let us assume that the $i$-th M5-brane
corresponds to the $K_i$-th anti-symmetric representation, and it wraps the cycle $\gamma_i$ in $\hatM$,
where $K_i$ is of order $O(N)$. We assume that $i$ runs over $i=1,\ldots, s$, where $s$ is of order $O(N^0)$.\footnote{
If we allow $s$ to be of order $O(N^1)$, then eq.~\eqref{codim4_action_general} could be of order $O(N^3)$
and the probe approximation breaks down.
}
Since these M5-branes preserve the same supersymmetry, we expect that there are no forces between them,
and the action, in the leading large $N$ limit, should be a sum of the contribution from each M5-brane:
\be
S =  N^2  \sum_{i=1}^s \ell(\gamma_i) \frac{K_i}{N}\left( 1-\frac{K_i}{N} \right) \;.
\label{codim4_action_general}
\ee

Let us consider the special case where the 1-cycles $\gamma_i$ inside the 3-manifold $\hatM$ are all the same, $\gamma_i=\gamma$.
We propose ({\it cf.} \cite{Gomis:2006sb}) that the M5-brane configuration for eq.~\eqref{codim4_action_general}
represents
a Wilson line in the Chern-Simons theory, in representation $R$ labeled by the partition
\begin{align}
\lambda=[\underbrace{1,\cdots, 1}_{K_1-K_2}, \underbrace{2,\cdots, 2}_{K_2-K_3}\ldots, \underbrace{s,\cdots, s}_{K_s}] \,,
\quad 
\lambda^t=[K_1, K_2 \cdots, K_s]
\;,
\label{lambda_general}
\end{align}
where without losing generality we assumed $K_i\ge K_{i+1}$.
Notice that this representation reduces to the anti-symmetric representation $A_K$ when
$s=1$, namely $K_1=K$, $K_{i\ge 2}=0$.
We will check the consistency of this proposal in the next subsection.\footnote{
In our leading supergravity approximation, we really do not distinguish between the 
representation \eqref{lambda_general} and the tensor product representation $\otimes_i A_{K_i}$.
The two differ by an exponentially suppressed contributions.
}

%%%%%%%%%%%%%%%%%%%%%%%%%%%%%%%%%%%%%%%%%%%%%
\subsection{Chern-Simons Perturbation}
%%%%%%%%%%%%%%%%%%%%%%%%

In this section, we  try to understand the above gravity computations in $SL(N)$ CS theory.  The gravity computation is only reliable at large $N$ but exact in $b$. In CS theory, $b$ is related to perturbative expansion parameter \eqref{3d-3d relation}. Here, we consider the case when $k=1$ and $b$ (and correspondingly $\sigma$) is real. In the case, as discussed around eq.~\eqref{SL(2,R) like rep}, we  use a real parameter $\hbar_{\mathbb{R}}:=2\pi b^2$ as perturbative expansion parameter.

\paragraph{Contour for $\sigma \in \mathbb{R}$}
The large $N$ free energy \eqref{F_H3 for knot complement} was reproduced in a highly non-trivial numerical way  using state-integral model in \cite{Gang:2014qla,Gang:2014ema}. %
What is remarkable in the comparison is that the large $N$ free energy is wholly reproduced by perturbative expansion of CS theory around a {\it single} flat connection $\CA^{\textrm{conj}}_N$ in eq.~\eqref{Flat connections from hyperbolic structure}. 
As discussed around eq.~\eqref{Contour for pure imaginary sigma},  we don't know correct path-integral cycle for real $\sigma$ unlike $\sigma = i \mathbb{R}$ case.  The non-trivial  consistency check in \cite{Gang:2014qla,Gang:2014ema} gives a possible candidate for the correct contour in the 3d--3d correspondence for $k=1$ and real $\sigma$ :
\begin{align}
\begin{split}
&\mathcal{C}_{ \sigma \in \mathbb{R}} =  \sum_{\a,\b} n_{(\a, \b)} \CJ^{(\a,\b)}\;, \quad \textrm{with}
\\
&n_{\a,\b} \neq 0 \textrm{ if and only if } (\a,\b)=(\textrm{conj},0) \textrm{ as $b\rightarrow 0$}\;,
\\
&n_{\a,\b} \neq 0 \textrm{ if and only if } (\a,\b)=(0,\textrm{conj})  \textrm{ as $b\rightarrow \infty$}\;.
\end{split}
\end{align}
$\CA^{\alpha=0}$ denotes the trivial flat connection. For $N=2$, there is some empirical supporting evidence for it using state-integral models \cite{Andersen:2011bt,Gang:2014ema}.

In weak coupling limit $\hbar_{\mathbb{R}} \rightarrow 0$, the holomorphic Wilson loop expectation value can be perturbatively expanded 
around saddle point configurations
\begin{align}
 \langle \hat{W}_{R} (\CK_{\gamma} ) \rangle= \sum_{(\alpha,\beta)}  n_{(\alpha,\b)} \exp \bigg{(} W^{(\alpha,\b)}_0 (\CK_\gamma,R)  +\ldots +W^{(\alpha,\b)}_n (\CK_\gamma,R) \hbar_{\mathbb{R}}^n +\ldots
  \bigg{)}\;,
\end{align}
where as before $(\alpha,\b)$ is the label for the saddle point,
and $n_{(\alpha,\b)}$ is an integer specifying the integration contour.
We again assume that the leading correction comes from the saddle point $(\alpha,\beta)=(\textrm{conj},0)$.
The saddle point here refers to the saddle point of the path integral with Wilson lines inserted,
and the Wilson lines in general affect the saddle point.
There are simplifications, however, 
when we consider the $K$-th antisymmetric representation $A_K$
the saddle point turns out to be the same regardless of the presence of Wilson lines.
This is because the original action is order $N^3$ (recall eq.~\eqref{F_H3}),
whereas the Wilson line is at at most of order $N^2$ and is subleading. The saddle point is still determined by 
$N^3$ piece, and hence we can safely assume that the saddle point is unmodified in the leading large $N$ 
limit.\footnote{The situation will be different when we consider $K$-th {\it symmetric} representation with $K$ large, say $K$ of order $N^2$ and higher.}

The prediction \eqref{co-dimension 4 in gravity} from gravity calculation imply the following perturbative expansion of the Wilson loop:%
\begin{align}
\begin{split}
&\lim_{N\rightarrow \infty}\frac{1}{N}   W^{(\textrm{conj},0)}_{0} (\CK_{\gamma},\Box )  = \frac{1}2 \ell(\gamma)\;,  
\\
&\lim_{N\rightarrow \infty}\frac{1}{N^2}   W^{(\textrm{conj},0)}_{0} (\CK_{\gamma},A_{\kappa N} )  = \frac{\kappa (1-\kappa)}2 \ell (\gamma) \;,
\\
&\lim_{N\rightarrow \infty}\frac{1}{N}  W^{(\textrm{conj},0)}_{1} (\CK_{\gamma},\Box )  =\frac{1}{4\pi }\ell(\gamma) \;, 
\\
&\lim_{N\rightarrow \infty}\frac{1}{N^2} W^{(\textrm{conj},0)}_{1} (\CK_{\gamma},A_{\kappa N} )   = \frac{\kappa (1-\kappa)}{4\pi }\ell(\gamma) \;, 
\\
&\lim_{N\rightarrow \infty}\frac{1}{N} W^{(\textrm{conj},0)}_{n>1} (\CK_{\gamma},\Box )   =\lim_{N\rightarrow \infty}\frac{1}{N^2} W^{(\textrm{conj},0)}_{n>1} (\CK_{\gamma},A_{\kappa N} ) =0 \;. \label{prediction on Wilson loop perturbation}
\end{split}
\end{align}
It is an interesting problem to check these results directly from the expressions of the partition functions
worked out in previous sections. We leave this question for future work, 
except to point out that the conjecture can easily be checked
as far as the classical part $W_0$ is concerned. For the classical part we only need to evaluate the Wilson loop at the classical saddle point $(\textrm{conj},0)$:
\begin{align}
\begin{split}
W^{(\textrm{conj},0)}_0 (\CK_\gamma, A_K)&=\textrm{Tr}_{A_K} P  e^{- \oint_{\CK_\g} \CA_N^{\textrm{conj}} }  
\\
&= \textrm{Tr}_{A_K} \bigg{[}[N]\cdot \left(\begin{array}{cc}\ell^*_{\mathbb{C}}/2 & 0 \\0 & -\ell^*_{\mathbb{C}}/2\end{array}\right) \bigg{]} \;\; \;(\because \textrm{eq}. \eqref{complex length})
\\
& =  \sum_{1 \leq i_1 <i_2\ldots <i_K \leq N} e^{\lambda_{i_1} +\lambda_{i_2}+\ldots + \lambda_{i_k}}\;\quad \left( \lambda_\ell := \left(\frac{N}{2}+\frac{1}{2}-\ell \right)\ell^*_{\mathbb{C}}(\gamma) \right) 
\\
&= \exp \left(\frac{K(N-K)}2 \ell^*_{\mathbb{C}}(\gamma) \right) + \cdots\;,
\label{Ak_comp}
\end{split}
\end{align}
where in the last equation, $(\cdots)$ represents terms exponentially suppressed than the first term 
in the large $N$ limit. This is consistent with the expectation \eqref{prediction on Wilson loop perturbation} since real part of complex hyperbolic length $\ell_\mathbb{C}(\gamma)$ is the hyperbolic length $\ell(\gamma)$. 
More generally, we show in App.~\ref{app.schur}
that the for the representation \eqref{lambda_general},
leading large $N$ answer gives eq.~\eqref{codim4_action_general},
thereby establishing the consistency with our previous proposal.

%%%%%%%%%%%%%%%%%%%%%%%%%%%%%%%%%%%%%%%%%%
\section{Discussion and Outlook}\label{sec.conclusion}
%%%%%%%%%%%%%%%%%%%%%%%%%%%%%%%%%%%%%%%%%%

In this paper, we systematically studied co-dimension 2 and 4 defects,
by consolidating results from a number of different approaches.
The methods are complementary in their scope, and 
whenever more than one results are available
we have checked the consistency between different approaches.

Our results on the one hand generalize the existing discussion of the 3d--3d correspondence by 
including supersymmetric defects.
On the other hand, our results shed light on several key aspects of the 3d--3d correspondence 
which has not been treated adequately in the literature.

The highlights of our work include the following:

\begin{itemize}[leftmargin=*]

\item In Sec.~\ref{sec : cluster partition function} and App.~\ref{app.cluster_derivation}, we obtained 
explicit integral expressions for the cluster partition function for 
a general quiver and a mutation sequence.
We also extended the result to include 
Wilson line insertions (in Sec.~\ref{app.cluster_with_Wilson}).
These results are rather general, and go well beyond 3d $\mathcal{N}=2$ theories
described by the 3-manifolds; they apply to the 3d $\mathcal{N}=2$ cluster theories of \cite{Terashima:2013fg}.
We expect that our results will be useful in 
such more general contexts.
 
\item
We have initiated the study of the 3d--3d correspondence for 
non-maximal punctures.
This includes the simple punctures for $N=3$ and $N=4$,
which we discussed in detail,
and we also commented on more general punctures (Sec.~\ref{subsec.examples}).
It seems, however, to be a challenging problem to generalize the discussion to completely general punctures.
Our results on non-maximal punctures should have a number of different applications, 
such as the discussion of loop operators in both 4d--2d and 3d--3d correspondence.

\item One missing ingredient in the existing 3d--3d setup is to 
better understand the consistency between 
Abelian and non-Abelian descriptions of $T_{N>2}[M]$ theories.
We have carried out quantitative consistency checks of the two
for the first time in the literature for the $N>2$ case.
This was made possible by our techniques to study the non-maximal punctures, as mentioned above.
We also pointed out that the non-Abelian description is crucial for the 
complete dictionary of co-dimension 4 defects in the 3d--3d correspondence;
such a non-Abelian description is currently not available, except for the cases discussed in Sec.~\ref{TSU3_index_glue}.

\item For co-dimension 4 defects, we proved the correspondence between Wilson loops in 5d $\mathcal{N}=2$ SYM
and those in CS theory, by explicit localization computation in 5d. Note that the proof applies to 
5d $\mathcal{N}=2$ SYM with any gauge group $G$, and is more general than the rest of the paper,
where $G$ is taken to be $SU(N)$.

\item For co-dimension 2 defects, we provided a Higgsing description relating different types of defects.
As a byproduct, this gives natural 1-parameter deformation of the partition function,
which is a certain `refinement' of the CS theory.

\item We obtained the supergravity duals of the supersymmetric defects in the large $N$ limit,
and worked out several large $N$ predictions.
This gives interesting set of predictions for the large $N$ behavior the partition functions,
which should be checked mathematically in the future works.

\end{itemize}

\begin{table}[htbp]
\caption{Summary and comparison between different approaches.
On the left of each entry, we indicate if our results are new: one smile means the result is 
known in the literature, while two means we obtained new results.
On the right, two smiles mean that story is at least partially complete, while one smile means 
there are still open problems. This is meant only as an illustration,
and readers are encourage to consult the main text for precise statements as to 
what is new/done/not done in this paper.
}
\begin{tabular}{c||c|c|c}
& $\rho=\textrm{maximal}$ & $\rho=\textrm{simple}$  & co-dimension 4 $R$  \\
\hline
\hline
cluster partition function & \smiley \smiley  /\smiley \smiley &\smiley\smiley/  \smiley  \smiley  &  \smiley  \smiley /\smiley  \\
\hline
state-integral model & \smiley /\smiley\smiley& \smiley\smiley/  \smiley & \smiley\smiley/\smiley  \\
\hline
domain wall $T[SU(N)]$  & --- & \smiley\smiley/ \smiley \smiley & \smiley\smiley/ \smiley \smiley \\
 \hline
 5d $\CN=2$ SYM   & \smiley\smiley / \smiley \smiley  & \smiley\smiley/ \smiley& \smiley\smiley/ \smiley \smiley   \\
 \hline
 Holographic dual  & \smiley\smiley/ \smiley & \smiley\smiley/ \smiley & \smiley\smiley/ \smiley \\
\end{tabular}
\end{table}

Let us close this paper by 
making two extra comments.

\begin{itemize}[leftmargin=*]

\item 

We can formulate our study of the co-dimension 2 and co-dimension 4 defect 
more mathematically, as follows.

First, the moduli space of flat connections on a 3-manifold $M$ with boundary should decompose into several
disconnected components, labeled by the conjugacy class of the meridian/longitude on the $\partial M$.
After quantization, it suffices to choose one of the say, say longitude, and we have
\begin{align}
\hat{\mathcal{M}}_{\rm flat}^N = \bigcup_{\rho} \hat{\mathcal{M}}^N_{\textrm{flat, }\rho}\;. \
\end{align}
Namely the choice of the co-dimension 2 defect is nothing but the choice of the connected component of the moduli space. In this respect, the Higgling proposal of Sec.~\ref{sec: Higgsing} is rather dramatic and maps one connected
component to another (however in these cases one component should be the closure of another).

We expect each of the moduli space allow for a descriptions in terms of quivers and mutations,
and has a set of nice coordinates associated with them: the moduli space is a cluster $\mathcal{X}$-variety.
Co-dimension 4 defects are represented as an element of the coordinate ring of 
the moduli space, and hence its quantization is part of the 
quantization of the coordinate ring $\mathbb{C}[\mathcal{M}_N (\hat{M}\backslash K)]$.
Such a quantization is related with the algebraic structure known as the
so-called Kauffman bracket skein module \cite{SkeinModule}.

\item

In this paper we treated the co-dimension 2 and co-dimension 4 defects separately.
However, we learned that both defects corresponds to the 
same line operators in $SL(N)$ CS theory.
Co-dimension 2 defects can be realized as boundary condition in the CS theory which fix the  holonomy around the knot $K$, whereas co-dimension 4 is realized as Wilson loop along $\gamma$ on $M$. 
However, in pure Chern-Simons theory, the Wilson line introduces a source term to the 
equation of motion, and hence can equivalently be formulated as a 
defect operator \cite{Deligne:1999qp,Beasley:2009mb}.\footnote{Recall that in this paper we assumed that 
the representation of the complex gauge group $SL(N)$ is obtained by natural complexification of a finite-dimensional unitary representation of $SU(N)$.}
This means that the two types of defects are equivalent, up to a proper identification of continuous mass-parameters of co-dimension 2 defects with discrete labels for unitary representations of co-dimension 4 defects, see \cite{Frenkel:2015rda} for related discussion
in the context of the 4d--2d correspondence.

\end{itemize}

%%%%%%%%%%%%%%%%%%%%%%%%%%%%%%%%%%%%%%%%%%%%%%%%%%%%%
\section*{Acknowledgements}
%%%%%%%%%%%%%%%%%%%%%%%%%%%%%%%%%%%%%%%%%%%%%%%%%%%%%

We would like to thank H.~Chung, T.~Dimofte, M.~Lackenby, D.~Xie and K.~Yonekura for discussion.
The contents of this paper was presented by th authors\footnote{DG: IPMU, May 2015; SNU, June, 2015; KIAS, July 2015, MR:``Workshop on Non-Abelian Gauged Linear Sigma Model and Geometric Representation Theory'', Peking University, June 2015, MY: Oxford, Oct.\ 2015.}, and we thank the audience for feedback.
The research of DG, MR and MY is supported in part by the WPI Research Center
Initiative (MEXT, Japan). 
DG is also supported by a Grant-in-Aid for Scientific Research on Innovative Areas 2303 (MEXT, Japan).
MY is also supported by  JSPS Program for Advancing Strategic
International Networks to Accelerate the Circulation of Talented Researchers,
by JSPS KAKENHI Grant Number 15K17634,
and by Institute for Advanced Study.
DG and MY would like to thank Simons Center for Geometry and Physics (2015 Summer Workshop) for hospitality.
MY would also like to thank Aspen Center for Physics (NSF Grant No.\ PHYS-1066293)
and the Mathematical Institute (University of Oxford) for providing refreshing environment.

%%%%%%%%%%%%%%%%%%%%%%%%%%%%%%%%%%%%%%%%%%%%%%%%%%%%%
\appendix
%%%%%%%%%%%%%%%%%%%%%%%%%%%%%%%%%%%%%%%%%%%%%%%%%%%%%

%%%%%%%%%%%%%%%%%%%%%%%%%%%%%%%%%%%%%
\section{Conditions on Boundary Holonomies} \label{app.boundary_holonomy}
%%%%%%%%%%%%%%%%%%%%%%%%%%%%%%%%%%%%%

In this appendix we comment in more detail the specification of the boundary holonomies for the co-dimension 2 defects
specified by $\rho$. This has been discussed in the literature in the other contexts (see {\it e.g.} \cite{Chacaltana:2012zy,Xie:2014pua}),
however little in the current subject of the 3d--3d correspondence. As in the main text let us discuss the holonomies for the meridian; holonomies for other boundary cycles,
say longitude, is completely parallel.

Let us first start with the case $N=2$. In this case,
the only non-trivial type of the co-dimension $2$ defect is $\rho=[1,1]$. If we turn on the mass parameters
$\frakM_{\alpha=1}=-\frakM_{\alpha=2}$, and if we assume that they are generic ({\it i.e.} $\frakM_{\alpha=1}\ne 0$),
then the two eigenvalues of the holonomy matrix are different, and hence we can always diagonalize the matrix.
This means that 
\begin{align}
\log \left(\textrm{Hol}(\mathfrak{m})\right) \in \textrm{orbit of } \left(
\begin{array}{cc}
\frakM_{1} & 0 \\
0 & -\frakM_1 
\end{array}
\right)\;,
\label{orbit_1}
\end{align}
where the orbit is defined by the adjoint action of the gauge group.

Let us next consider the limit $\frakM_{1}\to 0$. The most straightforward method is to 
take the limit $\frakM_{1}\to 0$. If we take the limit of the representative matrix 
in eq.~\eqref{orbit_1}, we obtain the matrix
\begin{align}
\bm{0}=\left(
\begin{array}{cc}
0 & 0 \\
0 & 0 
\end{array}
\right)\;.
\label{orbit_2}
\end{align}

However, this is not the only possibility. In fact, first note that for $\frakM_{1}\ne 0$
eq.~\eqref{orbit_1} can also be written as 
\begin{align}
\log \left(\textrm{Hol}(\mathfrak{m})\right) \in \textrm{orbit of } \left(
\begin{array}{cc}
\frakM_{1} & 1 \\
0 & -\frakM_1 
\end{array}
\right)\;,
\label{orbit_3}
\end{align}
and the limit $\mathfrak{M}_1\to 0$ for this representative matrix yields a nilpotent matrix
\begin{align}
\sigma_{+}=
 \left(
\begin{array}{cc}
0 & 1 \\
0 & 0 
\end{array}
\right)\;.
\label{orbit_4}
\end{align}
This subtlety arises since while the two representative matrices in eqs.~\eqref{orbit_1} and \eqref{orbit_3} are related 
by conjugation for $\mathfrak{M}_1\ne 0$, the matrix used in the conjugation becomes singular as $\mathfrak{M}_1\to 0$,
and hence $\bm{0}$ is only in the {\it closure} of the orbit of $\sigma_{+}$.
Since we should include all the limits of eq.~\eqref{orbit_1} (or equivalently eq.~\eqref{orbit_3}) in the limit $\mathfrak{M}_1\ne 0$,
we come to the conclusion that we should include both orbits. Equivalently, in this limit, the correct boundary condition 
for the co-dimension 2 defect should be 
\begin{align}
\log \left(\textrm{Hol}(\mathfrak{m})\right) \in \overline{O_{\sigma_{+} }}= O_{\bm{0}} \cup O_{\sigma_+} \;,
\label{orbit_5}
\end{align}
where $O_x$ denotes the orbit of $x$ by conjugation. This means that 
we should include Abelian flat connections for the descriptions of the co-dimension 2 defects.
As we will comment in the main text, neither the state-integral model (Sec.~\ref{sec.state_integral}) nor the cluster partition function (Sec.~\ref{sec :  cluster partition function}) contains
the contributions from the Abelian flat connections. This contrasts with the 
theories of Sec.~\ref{sec.TSUN}, which do contain the Abelian flat connections.

We can generalize the discussion for general $N$.
In the extreme case where all the eigenvalues are trivial, we obtain a closure of a nilpotent orbit,
as stated in  eq.~\eqref{co-dimension 2 defects as b.c. in CS theory-2}.
In fact, we can be more explicit and write
\begin{align}
\displaystyle
\log \left(\textrm{Hol}(\mathfrak{m} )\right) \in \overline{O_{ \rho^t(\sigma_{+})}} = \oplus_{\rho' \le \rho} ( O_{ \rho^t(e^{\sigma_{+}})})  \;,
\label{orbit_8}
\end{align}
where the partial ordering for two partitions $\rho, \rho'$ are defined by
\begin{align}
\sum_{i=1}^k \rho_i \ge \sum_{i=1}^k \rho'_i 
\end{align}
for all $k$ (we have taken $\rho_i=0$ if $i> \ell(\rho)$).
The right hand side of eq.~\eqref{orbit_8} is known to coincide with the Coulomb branch (or Higgs branch) of the mass-deformed $T^{\rho}[SU(N)]$ (or $T_{\rho}[SU(N)]$) theory.

%%%%%%%%%%%%%%%%%%%%%%%%%%%%%%%%%%%%%%%%%%%%%%%%%%
\section{Quantum Dilogarithm Function}\label{app.qdilog}
%%%%%%%%%%%%%%%%%%%%%%%%%%%%%%%%%%%%%%%%%%%%%%%%%%%

The non-compact quantum dilogarithm function \cite{FaddeevKashaevQuantum}
is defined by
\begin{align}
\psi_{\hbar}(x):=\begin{cases}
\displaystyle\prod_{n=0}^{\infty} \frac{1-q^{n+1} e^{-x} }{1-\tilde{q}^{-n} e^{-\tilde{x}}} & |q| <1 \ , \\
\displaystyle\prod_{n=0}^{\infty} \frac{1-\tilde{q}^{n+1} e^{-\tilde{x}} }{1-q^{-n} e^{-x}} & |q| >1 \ , 
\end{cases}
\end{align}
with 
$q:=e^{\hbar}=e^{2\pi i b^2}, \tilde{q}:=e^{\tilde{\hbar}}:=e^{2\pi i b^{-2}}, \tilde{x}:=b^{-2} x$.
For the value $|q|=1$, we would rather use the following integral expression
\begin{eqnarray}
\psi_{\hbar}(z)=\exp\left[\int_{\mathbb{R}+i0}\frac{dt}{4t}\frac{e^{(4\pi i  z +2\pi (b+b^{-1}))t}}{\sinh(t)\sinh(b^{-2}t)}\right]\;,
\end{eqnarray}
In the main text, we also used a version \eqref{octheron's CS wave ftn} where $\hbar$ and $\tilde{\hbar}$ are independent.
In the literature, sometimes a different notation $e_{b}(z)$ is used, which is related to 
$\psi_{\hbar}(z)$ by 
\begin{eqnarray}
\psi_{\hbar}(-2\pi b z+i\pi+ i\pi b^{2})=\psi_{\hbar}\left(-2\pi b z+\frac{\bhbar}{2}\right)=e_{b}(z)\;.
\end{eqnarray}
From the infinite product representation we can easily derive the 
following periodicity relations:
\begin{align}
\begin{split}
&\psi_{\hbar}(z+ 2\pi i b^2)=(1-e^{-z})\, \psi_{\hbar}(z)  \;, \\
&\psi_{\hbar}(z+2\pi i )=(1-e^{-z/b^2})\, \psi_{\hbar}(z) \;.
\end{split}
\label{psi_difference_eq}
\end{align}
An important property of the quantum dilogarithm is  
%\MY{Would rather write in $\psi$?}
%
\begin{align}
\int_{\mathbb{R}} dx \, e_b(x)e^{2\pi i w x} = e^{-i \pi w^2 + i \pi (1+\frac{(b+b^{-1})^2}{12})}e_b\left( w+i \frac{b+b^{-1}}2\right) \;.\label{integral identity of QDL}
\end{align}
%

%%%%%%%%%%%%%%%%%%%%%%%%%%%%%%%%%%%%%
\section{Derivation of Cluster Partition Function} \label{app.cluster_derivation}
%%%%%%%%%%%%%%%%%%%%%%%%%%%%%%%%%%%%%

One tool we heavily reply on in this paper is the cluster partition function, Sec.~\ref{sec : cluster partition function}. 
Here we derive an integral expression for the cluster partition function.
Our derivation, as well as the results, are the improvements over those of \cite{Terashima:2013fg};
for example we include dependence on central elements, and 
we are more explicit on the eliminating flat directions of the integral, and we also include Wilson lines in Sec.~\ref{app.cluster_with_Wilson}.

%%%%%%%%%%%%%%%%%%%%%%%%%%%%%%%%%
\subsection{Detailed Derivation}

%%%%%
\paragraph{Quiver Mutations and Cluster Algebras}

Let us first introduce \emph{quiver mutations} and \emph{cluster algebras} \cite{FominZelevinsky1}
(see \cite{KellerSurvey} for an introduction). Our notation here follows \cite{Terashima:2013fg}.

Let us begin with a \emph{quiver} $Q$, {\it i.e.}, a finite oriented graph.
We denote the set of the vertices of the quiver by $I$,
and its elements by $i,j, \ldots \in I$. We denote the number of vertices of $Q$ by $|Q|$.
For our purposes $Q$ is taken to be a quiver determined from a triangulation of a Riemann surface.

For vertices $i, j\in I$, we define 
\begin{align}
Q_{ij}:=\#\{\text{arrows from $i$ to $j$}\}-
\#\{\text{arrows from $j$ to $i$}\} \ ,
\end{align}
{\it i.e.} $|Q_{ij}|$ represents the number of arrows from the vertex $i$
to $j$, and the sign represents the chirality (orientation)
of the arrow.
The quivers discussed in this paper 
has no loops and oriented $2$-cycles,
and hence the quiver $Q$ is uniquely determined by the matrix
$Q_{ij}$.

Given a vertex $k$, we define a new quiver $\mu_k Q$ (\emph{mutation} of $Q$
at vertex $k$) by 
\begin{align}
(\mu_kQ)_{ij}:=
\begin{cases}
-Q_{ij} &  \text{($i=k$ or $j=k$)} \ , \\
Q_{ij} + [Q{}_{ik}]_+[Q_{kj}]_+ - [Q_{jk}]_+[Q_{ki}]_+ &
 \text{$(i,j\neq k$)} \ ,
\end{cases}
\label{Qmutate}
\end{align}
where we defined $[x]_+:=\textrm{max}(x,0)$.
For our purposes an appropriate mutation sequence will be determined from the change of the triangulation.
Given a quiver $Q=(Q_{ij})$, we can define a quantum-mechanical system by the commutation relation:
\begin{align}
[\mathsf{Y}_i, \mathsf{Y}_j] = \bhbar \, Q_{ji} \;,
\quad
[\bar{\mathsf{Y}}_i, \bar{\mathsf{Y}}_j] = \tilde{\bhbar} \, Q_{ji} \;.
\label{YYQ}
\end{align}
where we prepared a variable $\mathsf{Y}_i$ for each vertex $i$,
and a `Planck constants' $\bhbar, \tilde{\bhbar}$.
The value of $\bhbar, \tilde{\bhbar}$ is taken to satisfy $\bhbar^* = - \tilde{\bhbar}$, 
so that the $\mathsf{Y}^{\dagger}=\bar{\mathsf{Y}}$ 
without violating eq.~\eqref{YYQ};
the value of $\hbar$ is then analytically continued to 
other complex values after the computation. 
In terms of exponentiated variables $\sfy_i=\exp(\mathsf{Y}_i)$, 
this becomes 
\begin{align}
\scA_Q:=\{ \sfy_i, \bar{\sfy}_i \, _{(i \in I)} \, |\, \sfy_j \sfy_i=q^{Q_{ij}} \sfy_i
\sfy_j\;, \; \bar{\sfy}_j \bar{\sfy}_i=\tilde{q}^{Q_{ij}} \sfy_i
\sfy_j  \;, \; \bar{\sfy}_j \sfy_i = \sfy_i \bar{\sfy}_j\; \}\  \;,
\label{xCCR}
\end{align}
where $q:=e^{\hbar}$ and $\tilde{q}:=e^{\tilde{\hbar}}$. The variables $\sfy_i$ are the
so-called quantum $y$-variables \cite{FockGoncharovEnsembles,FockGoncharovQuantumCluster}. Mutation $\hat{\mu}_k$ acts on these variables as
\begin{align}
\begin{split}
&\hat{\mu}_{k}\, \sfy_{i}\, \hat{\mu}_k^{-1} =q^{\frac{1}{2} Q_{ik} [Q_{ik}]_{+}}\sfy_{i} \sfy_k^{[Q_{ik}]_{+}}
\prod_{m=1}^{|Q_{ki}|}\left(1+q^{\mathrm{sgn}(Q_{ki})(m-\frac{1}{2})}\sfy_{k}^{-1}\right)^{-\mathrm{sgn}(Q_{ki})} \;,
\\
&\hat{\mu}_{k}\, \bar{\sfy}_{i}\, \hat{\mu}_k^{-1} =\tilde{q}^{\frac{1}{2} Q_{ik} [Q_{ik}]_{+}}\bar{\sfy}_{i} \bar{\sfy}_k^{[Q_{ik}]_{+}}
\prod_{m=1}^{|Q_{ki}|}\left(1+\tilde{q}^{\mathrm{sgn}(Q_{ki})(m-\frac{1}{2})}\bar{\sfy}_{k}^{-1}\right)^{-\mathrm{sgn}(Q_{ki})} \;.
\label{quantum mutation}
\end{split}
\end{align}
The commutation relation \eqref{YYQ} has a central element of the form
\begin{align}
\frakL^A=\sum_{i=1}^{|Q|} \mathcal{C}^A_i\sfY_i  \ , \quad  \sum_{i=1}^{|Q|} \mathcal{C}^A_i Q_{ij}=0 \, ,\label{log_l}
\end{align}
where $A$ runs from 1 to dimension of Ker$(Q)$ and we included a logarithm on the left hand side for later convenience.
Since they commute with all other elements, the values of  
$\frakL^A$ can be taken to be fixed
constants (and hence we did not write 
$\frakL^A$ as an operator $\hat{\frakL}^A$.

Once eliminating  these central elements, $Q_{ij}$ is non-degenerate in other variables,
and hence we can choose linear combinations such that the commutation relation reduces to the 
canonical commutation relations. This has a standard representation in a Hilbert space,
which we denote by $\mathcal{H}_Q$. We will not describe this space in detail here since we 
would rather use a related but somewhat extended space $\hat{\mathcal{H}}_Q$, to be described momentarily.

In this quantization the mutation is promoted to an operator, sending an element of $\scA_Q$ to $\scA_{\mu_k Q}$:
\footnote{This can also be written as 
$\hat{\mu}_k = e_b \left(\frac{-\mathsf{Y}_k}{2\pi b}\right) \hat{P}_k$,
where $e_b(z)$ is defined in App.~\ref{app.qdilog}.
}
\begin{align}
\hat{\mu}_k =\psi_{\hbar}\left(\mathsf{Y}_k+i \pi b^2+i \pi \right) \hat{P}_k\;,
\label{mu_k}
\end{align}
where $\psi_{\hbar}(x)$ is a quantum dilogarithm function defined in App.~\ref{app.qdilog},
and in particular satisfy the difference equations of eq.~\eqref{psi_difference_eq}. The mutation operator is chosen to satisfy the operator equations  \eqref{quantum mutation}. The mutation operator is unitary operator.
The hermitian operator $\hat{P}_k$ give a transformation properties of (the logarithm of) the so-called tropical version of $y$-variables:
\begin{align}
\hat{P}_k (\mathsf{Y}_i):=\hat{P}_k \mathsf{Y}_i \hat{P}_k^{-1} = 
\left\{\begin{array}{cc} -\mathsf{Y}_k & \quad\quad i=k \\ \mathsf{Y}_i+[Q_{ik}]_+ \mathsf{Y}_k & \quad  \quad  i\neq k \end{array}\right.  \;,
\end{align}
or equivalently
\begin{align}
\hat{P}_k (\mathsf{y}_i):=\hat{P}_k \mathsf{y}_i \hat{P}_k^{-1} = 
\left\{\begin{array}{cc} \mathsf{y}_k^{-1} & \quad\quad i=k \\ 
q^{\frac{1}{2} Q_{ik} [Q_{ik}]_{+}} \mathsf{y}_i \mathsf{y}_k^{[Q_{ik}]_+}   & \quad  \quad  i\neq k \end{array}\right.  \;.
\end{align}
\paragraph{Cluster Partition Function}

In the following we consider a sequence of quiver mutations
$(\mu_{m_0}, \ldots, \mu_{m_{L-1}})$ and permutations $(\sigma_0, \ldots, \sigma_{L-1})$, specified by 
a set $\bm{m}=(m_0, \ldots, m_{L-1})$ of vertices. 

We define the quiver at ``time'' $t$ by 
\begin{align}
Q(t): = \hat{\sigma}_{t-1} \hat{\mu}_{m_{t-1}}\ldots \hat{\sigma}_{0} \hat{\mu}_{m_0} Q\ ,
\quad
Q(0):=Q \ .
\end{align}
Permutation $\sigma$ acts on quiver $Q$  in the following way 
\begin{align}
\begin{split}
&\hat{\sigma} \cdot Q:= \sigma^T Q \sigma\;,
\\
&(\sigma)_{ij}:= 
\left\{\begin{array}{cc} 1 & \quad\quad i=\sigma(j) \\ 
0  & \quad  \quad  i\neq \sigma(j) \end{array}\right.  \;.
\end{split}
\end{align}

We can then define the \emph{cluster partition function} $Z^{\rm cluster}_{Q, \bm{m},\bm{\sigma}}$ by\footnote{Here 
we inserted a permutation $\sigma_t$ for each mutation. We can easily commute the permutations with the other operators, and hence can choose to do a permutation only in the last step. It is technical useful, however,
to allow for this flexibility.}
\begin{align}
Z^{\rm cluster}_{Q, \bm{m},\bm{\sigma}}:=\big{\langle} \textrm{in} \big{ |} \hat{\mu}_{m_0}\hat{\sigma}_0 \ldots
\hat{\mu}_{m_{L-1}} \hat{\sigma}_{L-1} \big{|} \textrm{out} \big{\rangle} \ ,
\label{Zeb}
\end{align}
for the initial and final states $|\textrm{in}\rangle \in \scH_{Q(0)}$
and $|\textrm{out} \rangle \in \scH_{Q(L)}$.\footnote{In our convention we read the product from left to right.}
This partition function depends on the choice of initial and final
states.  We will compute  the matrix element   \eqref{Zeb} using quantization with $k=1$ in eq.~\eqref{Quantization for several (k,s)}.

\bigskip

For the explicit computation of the expectation value of eq.~\eqref{Zeb},
it is useful to double the degrees of freedom, namely to replace $\sfY_i$ by 
two variables $\sfp_i, \sfu_i$, and write \cite{2003math.....11149F}
\begin{align}
&\sfY_i = \sfp_i - Q_{ij}\sfu_j\;.
\label{Ypu}
\end{align}
Here repeated index $j$ is assumed to be  summed from 1 to $|Q|$.
The advantage of this trick is that the (exponentiated version of) commutation relation \eqref{YYQ} is then is reproduced from the 
canonical commutation relations: 
\begin{align}
[\sfu_i, \sfu_j]=[\sfp_i, \sfp_j]=0 \;, \quad [\sfu_i, \sfp_j]=i \pi b^2 \, \delta_{ij} \;.
\end{align}
This commutation relation, of course, has a simple representation in a Hilbert space $\hat{\mathcal{H}}_Q$, 
spanned by position basis $|u \rangle$ (or momentum basis $|p \rangle$)
\begin{align}
\begin{split}
& \langle u | \sfu_i = \langle u|u_i  \;, \quad  \langle u|  \sfp_i =-i \pi b^2 \frac{\partial}{\partial u_i} \langle u | \;,\\
& \langle p | \sfu_i = i \pi b^2 \frac{\partial}{\partial p_i} \langle p | \;, \quad \langle p| \sfp_i  =\langle p|  p_i \;,\\
&\langle u | p \rangle=\exp \left(\frac{i}{\pi b^2} u\cdot p\right) \;.
\end{split}
\end{align}
We have following Hermiticity and the completeness relation\footnote{Here we define $\int du := \int \prod_{i=1}^{|Q|}du_i$ and $\int d p := \int \prod_{i=1}^{|Q|} \frac{dp_i}{(2i\pi b^2)}$. }
\begin{align}
\sfu_i^\dagger= \sfu_i \;, \quad \sfp_i^\dagger = \sfp_i \;, \quad \mathbb{I} = \int  du |u\rangle \langle u|   = \int dp |p\rangle \langle p|\;. 
\end{align}
The action of the operator $\hat{P}_k$ naturally extends to the variable $\sfp_i, \sfu_i$: 
\begin{align}
\begin{split}
&\hat{P}_k (\sfp_i):=\hat{P}_k \sfp_i \hat{P}_k^{-1} = 
\begin{cases} -\sfp_k & (i=k) \\ \sfp_i+[Q_{ik}]_+ \sfp_k & (i\neq k) \end{cases} \;,
\\
&\hat{P}_k (\sfu_i):=\hat{P}_k \sfu_i \hat{P}_k^{-1} = 
\begin{cases} -\sfu_k+\sum_j [Q_{jk}]_+ \sfu_j & (i=k) \\ 
\sfu_k & (i\neq k) \end{cases} \;.
\end{split}
\label{hatP_def}
\end{align}
It follows from the second relation that 
$\mathsf{w}_i:=\sum_j Q_{i,j} \mathsf{u}_j$ transforms in the same manner with $\sfp_i$,
and hence eq.~\eqref{hatP_def} is compatible with eq.~\eqref{Ypu}.
The operator $\hat{P}_k$ is Hermite:
\begin{align}
\hat{P}_k^{\dagger} = \hat{P}_k\;, \quad  \big{\langle} p \big{|} \hat{P}_k \big{|}u \big{\rangle} = \big{\langle} \hat{P}_k (p) \big{|}u\big{\rangle} = \big{\langle} p \big{|} \hat{P}_k (u) \big{\rangle}\;.
\end{align}

%%%%%%%%%%%%%%%%%%%%%%%%%%%%%%%%%%

In this basis, the cluster partition function \eqref{Zeb} becomes
\begin{align}
Z^{\textrm{cluster}}_{Q,\bm{m},\bm{\sigma}}\big{(}p(0),p(L) \big{)}:=\big{\langle} p(0)\big{|}\hat{\mu}_{m_0} \hat{\sigma}_{0} \ldots \hat{\mu}_{m_{L-1}} \hat{\sigma}_{L-1} \big{|}p(L)\big{\rangle}\;,
\end{align}
where the operator $\hat{\sigma}$ is a permutation operator associated with a permutation $\sigma$,
acting on the $|p\rangle$ basis by
\begin{align}
\hat{\sigma}|p_i\rangle = |p_{\sigma(i)}\rangle\;.
\end{align} 
The the cluster partition function is then computed to be
 (\cite{Terashima:2013fg}, see also \cite{KashaevNakanishi})
\begin{align}
&Z^{\textrm{cluster}}_{Q,\bm{m},\bm{\sigma}}\big{(}p(0),p(L) \big{)}  =\int  \left[ \prod_{t=0}^{L-1}du(t) \prod_{t=1}^{L-1}dp(t) \right]  \big{\langle} p(0) \big{|}\psi_{\hbar} \left(\sfY_{m_0} +i \pi b^2+i \pi\right)\big{|} u(0) \big{\rangle}
\nn\\
&  \qquad
\times  \big{\langle} u(0) \big{|} \hat{P}_{m_0} \sigma_0 \big{|}p(1)\big{\rangle}  
\ldots  \big{\langle} p(L-1)\big{|}\psi_{\hbar} \left(\sfY_{m_{L-1}}+i \pi b^2+ i \pi \right) \big{|}u(L-1)\big{\rangle} \nn\\
& \qquad\times \big{\langle} u(L-1)| \hat{P}_{m_{L-1}} \sigma_{L-1} |p(L) \big{\rangle} \nn
\\
&=\int  \left[ \prod_{t=0}^{L-1}du(t) \prod_{t=1}^{L-1}dp(t) \right] \prod_{t=0}^{L-1}\psi_{\hbar} \left( p_{m_t}(t)-Q_{m_t,j} u_j(t)+ i \pi b^2+i \pi  \right) \\
& \qquad \times e^{\frac{1}{i \pi b^2} \big{[} u(t)\cdot p(t) - \hat{P}_{m_t}\big{(}u(t)\big{)} \cdot \big{(}\sigma_t  \cdot p(t+1) \big{)} \big{]}}  \;. \nn
\end{align}
Integration over $p_{i\ne m_t}(t)$ simply gives $\delta$-functions and $p_{m_t}(t)$ also can be integrated using the identify \eqref{integral identity of QDL}:
\begin{align}
&Z^{\textrm{cluster}}_{Q,\bm{m},\bm{\sigma}} \big{(}p(0),p(L) \big{)} =\int  \left[ \prod_{t=0}^{L-1} du(t) \right] \psi_{\hbar} \left( p_{m_0}(0)-Q_{m_0,j} u_j(0)+i \pi b^2+i \pi  \right) \nn
\\
&   \times e^{\frac{1}{i \pi b^2} \big{[} u(0)\cdot p(0) -\hat{P}_{m_{L-1}} \big{(}  u(L-1)\big{)}\cdot \big{(}\sigma_{L-1}\cdot p(L) \big{)} \big{]}  }  \prod_{t=1}^{L-1}\psi_{\hbar} \big{(}Z(t) \big{)} e^{- \frac{1}{4\pi i b^2} Z(t)Z''(t)}  \delta (\textrm{eq.~}\eqref{remember})\;.
\label{Zcluster}
\end{align}
Here we ignore the overall factor independent on $p(0)$ and $p(L)$.
For $1 \leq t \leq L-1$, we defined\footnote{The names $Z(t), Z''(t)$ originates from the fact that when we discuss
cluster partition functions associated with a 3-manifold, these
parameters coincide with the moduli of ideal tetrahedra,
which are often denoted by the same names.}
\begin{align}
\begin{split}
&Z(t):=2 \bigg{[} \bigg{(}\sigma_{t-1}^{-1}\cdot \hat{P}_{m_{t-1}}\big{(}u(t-1)\big{)} \bigg{)}_{m_t} - u_{m_t}(t)  \bigg{]}\;, 
\\
&Z''(t):=2 \bigg{[} -\bigg{(}\sigma_{t-1}^{-1}\cdot \hat{P}_{m_{t-1}}\big{(}u(t-1) \big{)}\bigg{)}_{m_t}  + u_{m_t}(t) + \sum_j Q_{m_t,j} (t)  u_j(t)    \bigg{]}\;. 
\end{split}
\label{Z,Z'}
\end{align}
The arguments of the delta function constraints in \eqref{Zcluster} are
\begin{align}
u_i(t) = \bigg{(}\sigma_{t-1}^{-1}\cdot \hat{P}_{m_{t-1}}\big{(}u(t-1)\big{)} \bigg{)}_{i}\;, \quad \textrm{for $i\neq m_t$}\;,\quad t=1,\ldots, L-1\;. \label{remember}
\end{align}
We would also like to impose extra constraints coming from \eqref{log_l}:
This means that inside the delta functions we should also have additional constraints
\begin{align}
&\sum_i \mathcal{C}^A_i Y_i (0)   = \frakL^A
 \quad \overset{\eqref{Ypu}}{\Longrightarrow } \quad \sum_i \mathcal{C}^A_i p_i (0) = \frakL^{A}
 \;.
\label{remember2p}
\end{align}

%-----------------------------------------------------------------------------------------------
\paragraph{Trace}

To this point we have followed the results of \cite{Terashima:2013fg}.
For our application in this paper, there are still some points to clarified.
First, what we wish to compute is the trace
\begin{align}
\textrm{Tr}_{Q,\bm{m}, \bm{\sigma}} (\frakL):= \textrm{Tr}_{\hat{\CH}_{Q}}(\hat{\mu}_{m_0} \hat{\sigma}_0 \ldots \hat{\mu}_{m_{L-1}} \hat{\sigma}_{L-1})\;, \label{TrQsm}
\end{align}
Of course, in order to this trace to be well-defined, the Hilbert space at $t=0$ and that at $t=L$
should be the same. We therefore impose the following two constraints
on $(Q,\bm{m},\bm{\sigma})$.
First, we obviously need 
\begin{align}
Q(L)=Q(0) \label{conditions on (Q,m,sigma) 1} \;.
\end{align}
Second, we choose central elements $\{ \frakL^\a=\sum_{i=1}^{|Q|}  c^\a_i  \sfY_i \}_{\a=1}^{n_c}$ commuting with $\hat{\varphi}$
\begin{align}
\frakL^\a (t=L):=  \hat{\varphi}^{-1} (\frakL^\a) \hat{\varphi} =\frakL^\a \;, \quad
(\hat{\varphi}:=\hat{\mu}_{m_0} \hat{\sigma}_0 \ldots \hat{\mu}_{m_{L-1}} \hat{\sigma}_{L-1}) \;,
\label{conditions on (Q,m,sigma) 2}
\end{align}
and we  impose the constraints in \eqref{remember2p} only for these central elements:
\begin{align}
\sum_{i=1}^{|Q|}c^\a_i p_i (0) = \frakL^{\a}\;. \label{remember2}
\end{align}
The trace in \eqref{TrQsm} depends on the $n_c$ central elements.
Since we are potentially identifying the puncture parameters when taking the trace, 
$n_c$ is not greater than dimension of kernel of $Q$ in general. For example, 
in the example of Fig.~\ref{time_evolution} we have $|\textrm{Ker}(Q)=4|$, since we have a fourth-punctured sphere;
however, when after closing the braids we have only  two independent link components, and hence we have
$n_c=2$.
The second constraint means that we in general have to identify some of the central elements $\frakL^A$,
and not all the $\frakL^A$ will be independent after taking a trace.
This is needed for the identification of the two Hilbert spaces at $t=0$ and $t=L$. For certain choices of $(Q, \bm{m}, \bm{\sigma})$, the trace 
$\textrm{Tr}_{Q,\bm{m}, \bm{\sigma}} (\frakL)$  can be considered as the $SL(N)$ CS partition function on a 3-manifold where 
$\frakL^\a$ are position variables in a certain polarization choice of boundary phase space of the 3-manifold.

This identification has a natural geometrical interpretation
in the 3-manifold setup of Sec.~\ref{sec: cluster partition function  for 3-manifold}. There
a central element corresponds to the holonomy around a puncture in the 2d surface,
which looks like a braid in 3d.
When we close the mapping cylinder $\Sigma\times [0,1]$ into a mapping torus $(\Sigma\times S^1)_{\varphi}$, 
the ends of the braids are identified, giving rise to a link component,
and the number of link components after identification is smaller than the number of braids (see Fig.~\ref{time_evolution}).

Going to the trace means to start with the final formula \eqref{Zcluster}, identify 
initial state and the final state by setting $p(L)=p(0)$, and integral over the state $|p(0)\rangle$.
Then, we immediately have
\begin{align}
&\textrm{Tr}_{Q,\bm{m}, \bm{\sigma}} (\frakL)  \nn
\\
&= \int dp(0)\int  \prod_{t=0}^{L-1} du(t) \bigg{[} \psi_{\hbar} \left( p_{m_0}(0)-Q_{m_0,j} u_j(0)+i \pi b^2+i \pi \right) \prod_{t=1}^{L-1}\psi_{\hbar} \big{(}Z(t) \big{)} e^{- \frac{1}{4\pi i b^2} Z(t)Z''(t)}
\nn
\\
&\qquad \qquad \qquad \qquad   \times   e^{\frac{1}{\pi i b^2 } \big{[} u(0)\cdot p(0) - \hat{P}_{m_{L-1}}\big{(}u(L-1)\big{)} \cdot \big{(} \sigma_{L-1}\cdot p(0) \big{)} \big{]}}  \delta \big{(}\textrm{eq.}\eqref{remember} \textrm{ and } \eqref{remember2}\big{)} \bigg{]}\;.
\end{align}
where the $(Z(t),Z''(t))$ are defined in \eqref{Z,Z'}. %

For our purposes, it is useful to take a Fourier transform from $\frakL_{\alpha}$ to 
another set of variables $\bfrakM_{\alpha}$. 
The constraint \eqref{remember2}
could then easily be dealt with, since we can trivially integral over the $\frakL_{\a}$s.
We then obtain
\begin{align}
&\mathrm{F. T. } \left[\textrm{Tr}_{Q,\bm{m}, \bm{\sigma}}\right] (\bfrakM):=\int \prod_{\a=1}^{n_c}
d\frakL^\a \;e^{\frac{1}{2\pi i b^2}\sum_{\alpha}\frakL^{\alpha} \bfrakM^{\alpha}}\textrm{Tr}_{Q,\bm{m}, \sigma} (\frakL) \nn
\\
&= \int  \prod_{t=0}^{L-1} du(t) \bigg{[} \prod_{t=0}^{L-1}\psi_{\hbar} \big{(}Z(t) \big{)} e^{- \frac{1}{4 \pi i b^2} Z(t)Z''(t)}      \delta \big{(}\textrm{eq.}\eqref{remember} \textrm{ and eq.}\eqref{remember3}\big{)} \bigg{]}\;. \label{F.T of cluster ptn}
\end{align}
Here the delta functions gives constraints in \eqref{remember},
and the following additional constraints:
\begin{align}
u_{i} (0)+\frac{1}2 \sum_{\alpha} c_{i}^\alpha \bfrakM_\alpha = \bigg{(}\sigma_{L-1}^{-1}\cdot \hat{P}_{m_{L-1}}\big{(}u(L-1)\big{)} \bigg{)}_{i}\;, \;\textrm{for $i\neq m_0$}  \label{remember3}
\end{align}
and  with  \eqref{Z,Z'}
\begin{align}
&Z(0):=2 \bigg{[}\bigg{(}\sigma_{L-1}^{-1}\cdot \hat{P}_{m_{L-1}}\big{(}u(L-1)\big{)} \bigg{)}_{m_0} - u_{m_0}(0) -\sum_{\alpha} \frac{1}2c^{\alpha}_{m_0} \bfrakM_\alpha  \bigg{]}\;, \nn
\\
&Z''(0):=2 \bigg{[}- \bigg{(}\sigma_{L-1}^{-1}\cdot \hat{P}_{m_{L-1}}\big{(}u(L-1)\big{)} \bigg{)}_{m_0} + u_{m_0}(0)+\sum_{\alpha} \frac{1}2 c^{\alpha}_{m_0} \bfrakM_\alpha  + \sum_j Q_{m_0,j} (0)  u_j(0)  \bigg{]}\;.
\end{align}
Collecting these results, we obtain \eqref{trace_result_1}.

%%%%%%%%%%%%%%%%%%%%%%%%%%%%%%%%%%%%%%%%%%%%%%%%%%%%%%%%%%%
\subsection{Inclusion of Wilson Lines}\label{app.cluster_with_Wilson}
%%%%%%%%%%%%%%%%%%%%%%%%%%%%%%%%%%%%%%%%%%%%%%%%%%%%%%%%%%%
Here we derive the cluster partition function from the previous section but, adding an extra ingredient: a Wilson loop insertion. We will focus on loops on the Riemann surface $\Sigma_{g,h}$ of the mapping torus/cylinder. Then, the Wilson loop can be written as a linear combination of the operators:
\begin{eqnarray}
\exp \left(\sum_i a_{i}\sfY_{i}\right)\qquad a_{k}\in \mathbb{Q} \;.
\end{eqnarray}
(More precisely, from periodicity conditions the constants $a_{k}$ should be quantized, but, for the derivation of the formula, this is not important). Our starting point is\footnote{Here we inserted a Wilson line at time $t=0$. Since we are taking a trace, the time is cyclic we do not lose generality in assuming this.}
\begin{eqnarray}
Z_{Q,\bm{m},\bm{\sigma},a}^{\mathrm{cluster}}=\langle p(0)|e^{\sum_i  a_i\sfY_i(0)}\hat{\mu}_{m_{0}}
\hat{\sigma}_0 \cdots \hat{\mu}_{m_{L-1}} \hat{\sigma}_{L-1} |p(L)\rangle \;.
\end{eqnarray}
When then insert complete sets, we need an extra complete set of the form $|p'(0)\rangle\langle p'(0)|$ in the first factor:
\begin{align}\label{identityWL}
\begin{split}
&\langle p(0)|e^{\sum_i a_{i}\sfY_{i}(0)}|p'(0)\rangle\langle p'(0)| \, \psi_{\hbar}\left(\sfY_{m_{0}}(0) +\pi i b^2+i\pi \right)|u(0)\rangle \\
& \qquad \times \langle u(0)|\widehat{P}_{m_{0}} \hat{\sigma}_0|p(1)\rangle\langle p(1)| \;.
\end{split}
\end{align}
It is not difficult to evaluate this expression. Since 
$\sum_i a_i \sfY_i=\sum_i a_i \sfp_i+\sum_{l,j} a_l Q_{l,j} u_j$ and 
$\left[\sum_i a_{i}\sfp_{i}, \sum_{l,j}a_{l}Q_{lj}\sfu_{j}\right]=0$,
\begin{eqnarray}
\langle p(0)|e^{\sum_i a_i\sfY_i (0)}|p'(0)\rangle &=& e^{\sum_i a_i p_i(0)}\langle p(0)|e^{-\sum_{i,j}a_{i}Q_{ij}\sfu_{j}}|p'(0)\rangle \nn
\\
&=&e^{\sum_i a_i p_i(0)}\Big\langle p(0)-\pi i b^2 \sum_{i,j}a_{i}Q_{ij}\hat{e}_{j}\Big|p'(0)\Big\rangle\nonumber
\\
&=& e^{\sum_i a_i p_i (0)}\delta\left( p(0)- \pi i b^2 \sum_{i,j} a_i Q_{ij}\hat{e}_{j}-p'(0)\right) \;.
\end{eqnarray}
Then, doing the integration over $p'(0)$ in (\ref{identityWL}) we obtain:
\begin{align}
\begin{split}
&e^{a\cdot p(0)+\frac{1}{\pi i b^2}\left(u(0)\cdot p(0)-\pi i b^2 a \cdot Q\cdot u(0)\right)} \\
&\quad \times \psi_{\hbar}\left(Y_{m_{0}}(0)+\pi i b^2 (a \cdot Q)_{m_0} +\pi i b^2 + i \pi \right)\langle p(1)| \;.\;\;\;\;
\end{split}
\end{align}
This is the only change, so we are left with
\begin{align}
&Z_{Q,\bm{m},\bm{\sigma},a}^{\mathrm{cluster}}(p(0),p(L))=\int\left(\prod_{t=0}^{L-1}du(t)\right)e^{a\cdot p(0)}\psi_{\hbar}\left(Y_{m_{0}}(0)+\pi  i b^2 (a\cdot Q)_{m_0}+\pi i b^2+ i\pi  \right) \nn
\\
&\times e^{\frac{1}{\pi i b^2}[u(0)\cdot(p(0)+\pi i b^2 Q \cdot a)-\widehat{P}_{m_{L-1}}(u(L-1))\cdot p(L)]} \prod_{t=1}^{L-1}\psi_{\hbar}\left(Z(t)\right)e^{-\frac{1}{4\pi i b^2}Z(t)Z''(t)}
\delta (\textrm{eq.~}\eqref{remember})\;.
\end{align}
Taking trace with insertion of delta functions related to central elements $\vec{c}$ and the doing Fourier transformation as before,  we finally have:
\begin{align}\label{partitionWL}
\begin{split}
&\left\langle \exp \left(\sum_i a_{i}\sfY_{i}\right) \right\rangle_{Q,\bm{m},\bm{\sigma}} :=\mathrm{F.T.}\left[  \textrm{Tr}_{Q,\bm{m},\bm{\sigma},a}\right] (\bfrakM_{\alpha}) \nn\\
&=\int  \left(  \prod_{t=0}^{L-1} d\vec{u}(t) \right)  e^{-\sum_{k,j}a_{k}Q_{kj}u_{j}(0)- \frac{1}{2} \sum_k a_k Q_{km_0} Z(0)} \prod_{t=0}^{L-1}\psi_{\hbar}(Z(t))e^{-\frac{1}{4\pi ib^2}Z(t)Z''(t)} \delta (\textrm{eq.~}\eqref{remember}) \\
&  \times \prod_{i\neq m_{0}}\delta\left(\frac{1}{2}c_{i}\cdot \bfrakM+u_{i}(0)-\bigg{(} \sigma_{L-1}^{-1}\cdot\widehat{P}_{m_{L-1}}(u(L-1))\bigg{)}_{i}+\pi i b^2 a_{i}\right) \prod_{\alpha=1}^{n_{C}}\delta\left(\sum_i c^{\alpha}_{i}u_{i}(0)\right) \;.
\end{split}
\end{align}
Note we have already entered the factor $\prod_{\alpha=1}^{n_{C}}\delta(\sum_i c^{\alpha}_{i}u_{i}(0))$ to quotient the flat directions. Here, the $Z$ variables are slightly modified due to the Wilson loop insertion:
\begin{align}
\begin{split}
&Z(0)=2\left[-\frac{1}{2}c_{m_{0}}\cdot \bfrakM-u_{m_{0}}(0)+\bigg{(} \sigma_{L-1}^{-1}\cdot\widehat{P}_{m_{L-1}}(u(L-1))\bigg{)}_{m_{0}}-\pi i b^2 a_{m_{0}}\right]\;,
\\
&Z''(0)=2\left[\frac{1}{2}c_{m_{0}}\cdot \bfrakM+u_{m_{0}}(0)-\bigg{(} \sigma_{L-1}^{-1}\cdot\widehat{P}_{m_{L-1}}(u(L-1))\bigg{)}_{m_{0}}+ \sum_j Q_{m_0 j}u_j(0)\right]\;.
\end{split}
\end{align}

%%%%%%%%%%%%%%%%%%%%%%%%%%%%%%%%%%%%%%%%%%%%%%%%%%
\section{Proof of \texorpdfstring{\eqref{kernel_transf}}{An Identity}}\label{app.kernel}
%%%%%%%%%%%%%%%%%%%%%%%%%%%%%%%%%%%%%%%%%%%%%%%%%%%

In this appendix we prove \eqref{kernel_transf}. 

As we change the `time' from $t$ to $t+1$,
the coefficients $c_i^{\alpha}(t)$ transform as in \eqref{c_alpha_def}:
\begin{align}
\begin{split}
c_i^{\alpha}(t+1)  
&=  \sigma_t^{-1} \hat{P}_{m_t}(c_i^{\alpha}(t))  \\
&=\begin{cases}
-c^{\alpha}_{M_t} (t)+ \sum_{i \ne m_t} [Q_{I, M_t}(t)]_{+} c_{I}^{\alpha}(t)  & (i=  m_t) \;,\\
c_{I}^{\alpha}(t) & (i\ne m_t) \;,
\end{cases}
\end{split}
\end{align}
where to save spaces we denoted $I:=\sigma_t^{-1}(i), J:=\sigma_t^{-1}(j), M_t:=\sigma_t^{-1}(m_t)$.
The quiver $Q$ transforms as
(recall \eqref{Qmutate})
\begin{align}
Q_{i,j}(t+1)=
\begin{cases}
-Q_{I,J}(t) &  \text{($i=m_t$ or $j=m_t$)} \ , \\
Q_{I,J}(t) + [Q{}_{I, M_t}(t)]_+[Q_{M_t, J}(t)]_+ - [Q_{J, M_t}(t)]_+[Q_{M_t, I}(t)]_+ &
 \text{$(i,j\neq m_t$)} \ .
\end{cases}
\end{align}

Suppose that we have $\sum_{i }  c_i^{\alpha}(t) Q_{i,j}(t) =0$
for all $j$. Note that this can also be written as
\begin{align}
\sum_{I}  c_i^{\alpha}(t) Q_{I,J}(t) =0 
\label{kernel_assumption}
\end{align}
for all $J=\sigma_t^{-1}(j)$.
For $j\ne m_t$, we can compute
\begin{align}
\begin{split}
&\sum_i c_i^{\alpha}(t+1) Q_{i,j}(t+1)  \\
&\qquad= c_{m_t}^{\alpha}(t+1) Q_{m_t,j}(t+1)  + \sum_{i\ne m_t}  c_i^{\alpha}(t+1) Q_{i,j}(t+1)    \\
&\qquad=\left(-c^{\alpha}_{M_t} (t)+ \sum_{i \ne m_t} [Q_{I, M_t}(t)]_{+} c_{I}^{\alpha}(t) \right) (-Q_{M_t,J}(t))  \\
&\qquad\qquad  + \sum_{i\ne m_t}  c_I^{\alpha}(t) \left( Q_{I,J}(t) + [Q{}_{I, M_t}(t)]_+[Q_{M_t, J}(t)]_+ - [Q_{J, M_t}(t)]_+[Q_{M_t, I}(t)]_+  \right) \\
&\qquad= \left( c^{\alpha}_{M_t} (t) Q_{M_t,J}(t)   + \sum_{i\ne m_t}  c_I^{\alpha}(t) Q_{I,J}  \right)  \\
& \qquad\qquad + \sum_{i\ne m_t}  c_I^{\alpha}(t) \left( [Q{}_{I, M_t}(t)]_+[Q_{M_t, J}(t)]_+ - [Q_{J, M_t}(t)]_+[Q_{M_t, I}(t)]_+  -[Q_{I, M_t}(t)]_{+} Q_{M_t,J}(t)\right) \;.
\end{split}
\end{align}
The expression inside the bracket simplifies, with the help of 
$x=[x]_{+}-[-x]_{+}$,
\begin{align}
\begin{split}
& [Q{}_{I, M_t}(t)]_+[Q_{M_t, J}(t)]_+ - [Q_{J, M_t}(t)]_+[Q_{M_t, I}(t)]_+  -[Q_{I, M_t}(t)]_{+} Q_{M_t,J}(t)\\
&\qquad =[Q{}_{I, M_t}(t)]_+([Q_{M_t, J}(t)]_{+}-Q_{M_t,J}(t)
) - [Q_{J, M_t}(t)]_+[Q_{M_t, I}(t)]_+ \\
&\qquad =[Q{}_{I, M_t}(t)]_+ [Q_{J, M_t}(t)]_{+}
) - [Q_{J, M_t}(t)]_+[Q_{M_t, I}(t)]_+  \\
&\qquad =Q{}_{I, M_t}(t) [Q_{J, M_t}(t)]_{+} \;,
\end{split}
 \end{align}
 and hence
 \begin{align}
\begin{split}
&\sum_i c_i^{\alpha}(t+1) Q_{i,j}(t+1)   \\
&\qquad= \sum_{i }  c_I^{\alpha}(t) Q_{I,J}   + \sum_{i\ne m_t}  c_I^{\alpha}(t) Q{}_{I, M_t} [Q_{J, M_t}]_{+}\\
&\qquad= \sum_{I}  c_I^{\alpha}(t) Q_{I,J}   +  \left( \sum_{i }  c_I^{\alpha}(t) Q{}_{I, M_t} \right) [Q_{J, M_t}]_{+} =0 \;,
\end{split}
\end{align}
where we used \eqref{kernel_assumption}.
Similarly, for $j=m_t$, 
\begin{align}
\begin{split}
\sum_i c_i^{\alpha}(t+1) Q_{i,j}(t+1)  
&=\sum_{i \ne m_t} c_i^{\alpha}(t+1) Q_{i,m_t}(t+1)  \\
%Fig.~
&= \sum_{I\ne m_t}  c_I^{\alpha}(t) (-Q_{I,M_t}(t))  =0 \;.   \\
\end{split}
\end{align}
This proves \eqref{kernel_transf}. 

%%%%%%%%%%%%%%%%%%%%%%%%%%%%%%%%%%%%%%%%%%%%%%%%%%%%%
\section{Direct computation  of  \texorpdfstring{Tr($\hat{\varphi}$) on $\CH^{k=0}_{N=3}(\Sigma_{1,1}, \textrm{simple})$}{Tr(hat(varphi)) on H(k=0,N=3)(Sigma(1,1), simple)}}\label{app.rho_simple}
%%%%%%%%%%%%%%%%%%%%%%%%%%%%%%%%%%%%%%%%%%%%%%%%%%%%%

In this appendix we present an alternative method to compute the partition function,
for the example discussed in Sec.~\ref{sec.Xe_rule}.

The purpose of this appendix is threefold. First, we present a more direct derivation of the cluster partition function which does not rely on the `doubling trick' of the coordinates described in App.~\ref{app.cluster_derivation}.
Second, we work out the consistency of the quiver for the simple puncture case,
by explicitly working out the representation of the mapping class group $SL(2, \mathbb{Z})$. Third, we explicitly  confirm that using  gluing equations  derived from the cluster partition function for $k=1$  we can  reproduce the cluster partition function for other quantizations (for $k=0$ here).

As discussed in  sec.\ref{sec.Xe_rule}, the two generators $\hat{\bm{S}}, \hat{\bm{T}}$ of $SL(2,\mathbb{Z})$ can be generated by following sequence of mutations
and permutations :
\begin{align} 
\begin{split}
&\hat{\bm{S}}\cdot \big{\{}\sfy_{1\pm},\sfy_{2\pm},\sfy_{3\pm},\sfy_{4\pm},\sfy_{5\pm} \big{\}}\cdot \hat{\bm{S}}^{-1}   
\\
&\qquad= \big{\{}\hat{\mu}_5 (\sfy_{3\pm}),\hat{\mu}_5 (\sfy_{4\pm}),\hat{\mu}_{5} (\sfy_{2\pm}),\hat{\mu}_5 (\sfy_{1\pm}),\hat{\mu}_5 (\sfy_{5\pm}) \big{\}}  
\\
&\qquad= \big{\{} q^{\pm \frac{1}{2}} \sfy_{3\pm}   \sfy_{5\pm} \left(1+q^{\mp \frac{1}{2}}\sfy_{5\pm}^{-1}\right)\;,  \; q^{\pm \frac{1}{2}}  \sfy_{4\pm}   \sfy_{5\pm} \left(1+q^{\mp \frac{1}{2}}\sfy^{-1}_{5\pm}\right) \;,
\\
&\qquad\quad \quad  \sfy_{2\pm} \left(1+q^{\pm \frac{1}{2}}\sfy_{5\pm}^{-1}\right)^{-1}\;,\;   \sfy_{1\pm} \left(1+q^{\pm \frac{1}{2}}\sfy_{5\pm}^{-1}\right)^{-1}\;,\; \sfy^{-1}_{5\pm}\}  \ , 
\\  
&\hat{\bm{T}} \cdot \big{\{}\sfy_{1\pm},\sfy_{2\pm},\sfy_{3\pm},\sfy_{4\pm},\sfy_{5\pm} \big{\}}\cdot \hat{\bm{T}}^{-1}   
\\
&\qquad= \big{\{}\hat{\mu}_3 \hat{\mu}_4 (\sfy_{3\pm}),\hat{\mu}_3\hat{\mu}_4 (\sfy_{4\pm}),\hat{\mu}_3\hat{\mu}_4 (\sfy_{1\pm}),\hat{\mu}_3\hat{\mu}_4 (\sfy_{2\pm}),\hat{\mu}_1\hat{\mu}_2 (\sfy_{5\pm}) \big{\}}  
\\
&\qquad= \big{\{} \sfy^{-1}_{3\pm} \;,\;   \sfy^{-1}_{4\pm}\;,\;
\\
&\qquad  \qquad q^{\pm 2} \sfy_{1\pm} \sfy^2_{3\pm}  (1+q^{\pm \frac{1}{2}}\sfy_{4\pm}^{-1})^{-1} (1+q^{\mp \frac{3}{2}}\sfy_{3\pm}^{-1}) (1+q^{\mp \frac{1}{2}}\sfy_{3\pm}^{-1})\;,  
\\
&\qquad\quad \quad q^{\pm 2} \sfy_{2\pm} \sfy^2_{4\pm}  \left(1+q^{\mp \frac{3}{2} }\sfy_{4\pm}^{-1}\right)   \left(1+q^{\mp \frac{1}{2}}\sfy_{4\pm}^{-1}\right) \left(1+q^{\pm \frac{1}{2}}\sfy_{3\pm}^{-1}\right)^{-1} \;,  
\\
&\qquad\quad \quad    \sfy_{5\pm} \left(1+q^{\pm \frac{1}{2}}\sfy_{4\pm}^{-1}\right)^{-1} \left(1+q^{\pm \frac{1}{2}}\sfy_{3\pm}^{-1}\right)^{-1}\} \;.
\end{split}
\label{ST_expr}
\end{align}
Here the suffix $\pm$ denotes the complex pairs  ($\sfy_+ :=\sfy, \sfy_-:=\bar{\sfy}$) and we use the fact that $\tilde{q}=q^{-1}$ for $k=0$, see eq.~\eqref{real_S1S2}. We can check 
the unitarity of the representation \eqref{ST_expr}.
For $\bm{S}$, we compute
\begin{align}
(\bm{S}^{-1})^\dagger  \sfy_{3\pm} \bm{S}^{\dagger} =(\bm{S}  \sfy_{3\pm} \bm{S}^{-1})^{\dagger} = \left(\sfy_{2\pm} \left(1+q^{\pm \frac{1}{2}}\sfy_{5\pm}^{-1}\right)^{-1}\right)^{\dagger} = \sfy_{2\pm} \left(1+q^{\pm \frac{1}{2}}\sfy_{5\pm}^{-1}\right)^{-1} \;.
\end{align}
By repeating the similar computations for $\sfy_{i=2,3,4}$ we find
\begin{align}
(\bm{S}^{-1})^\dagger  \sfy_{i\pm} \bm{S}^{\dagger} = \bm{S} \sfy_{i\pm} \bm{S}^{-1} 
\end{align}
for all $\sfy_{i}$. Schur's lemma then tells us that $(\bm{S}^{-1})^\dagger=\bm{S}$.
We can similarly show the unitarity of $\bm{T}$, and 
since arbitrary $\hat{\varphi} \in SL(2,\mathbb{Z})$ can be generated by $\bm{S}$ and $\bm{T}$, this implies that every element in $SL(2,\mathbb{Z})$ is unitary. 

We can decompose the operators $\bm{S}, \bm{T}$ as\footnote{This is similar to the decomposition \eqref{mu_k}. One difference is that the $\mathbf{P}_{S,T}$ here contain not only $\hat{P}_k$ but also permutations $\hat{\sigma}_{S,T}$. More explicitly,  ${\bf P}_T = \hat{P}_{3}\hat{P}_4 \hat{\sigma}_T$ and ${\bf P}_S=\hat{P}_5 \hat{\sigma}_S$.}
\begin{align}
\begin{split}
&\hat{\bm{S}}=\hat{\psi}_5 \cdot  \mathbf{P}_{S} \;,
\\
&\hat{\bm{T}} =\hat{\psi}_3 \hat{\psi}_4\cdot  \mathbf{P}_{T}  \;, 
\end{split} 
\end{align}
where
\begin{align}
\hat{\psi}_i := \prod_{r=1}^{\infty} \frac{1+q^{r-\frac{1}{2}} \sfy^{-1}_{i+}}{1+q^{r- \frac{1}{2}} \sfy^{-1}_{i-}} \;,
\end{align}
and $\mathbf{P}_{S,T}$ are operators satisfying
\begin{align}
\begin{split}
&\mathbf{P}_S\cdot (\sfY_1, \sfY_2, \sfY_3, \sfY_4, \sfY_5)\cdot \mathbf{P}^{-1}_S= ( \sfY_3+\sfY_5, \sfY_4+\sfY_5, \sfY_2, \sfY_1,-\sfY_5) \;. 
\\
&\mathbf{P}_T\cdot (\sfY_1, \sfY_2, \sfY_3, \sfY_4, \sfY_5)\cdot \mathbf{P}^{-1}_T= ( -\sfY_3, -\sfY_4, \sfY_1+2\sfY_3, \sfY_2+2\sfY_4,\sfY_5) \;, 
\end{split}
\end{align}
To quantize the system, we choose canonical variables as
\begin{align}
\Pi=(\frakX_1, \frakX_2, \frakP_1, \frakP_2)
= \left( Y_1+Y_2, Y_1-Y_2, \half(Y_3+Y_4), \frac{1}6 (Y_3- Y_4)\right) \;,
\label{XP_def}
\end{align}
such that the commutations relations take the canonical form
\begin{align}
\begin{split}
&\{ \frakX_i, \frakP_j\}_{P.B}= -\bhbar \,\delta_{ij}\;, \quad \{ \bar{\frakX}_i, \bar{\frakP}_j\}_{P.B}= \bhbar\,  \delta_{ij}\ \;, \quad \textrm{or equivalently} 
\\
&\{\textrm{Re}(\frakX_i), \textrm{Im}(\frakP_j)\}_{P.B} =\frac{ i \bhbar}{2}\, \delta_{ij}\;,  \quad \{\textrm{Im}(\CX_i), \textrm{Re}(\frakP_j)\}_{P.B} =\frac{ i \bhbar}{2}\, \delta_{ij}\;. 
\end{split}
\end{align}
After the coordinate transformation \eqref{XP_def}, periodicity of the variables becomes
\begin{align}
\textrm{Im}[\frakX_1\pm \frakX_2] \sim \textrm{Im}[\frakX_{1}\pm \frakX_2 ]+ 4 i  \pi \mathbb{Z}\;, \quad \textrm{Im}[2\frakP_{1}\pm 6 \frakP_{2}] \sim \textrm{Im}[2\frakP_{1}\pm 6 \frakP_{2}]+ 4i  \pi \mathbb{Z} \;.
\end{align}
Quantizing the phase-space with $k=0$, we obtain a Hilbert space
\begin{align}
&\CH_3 (\Sigma_{1,1}, \textrm{simple})= \textrm{Span}\big{\{}  \big{\langle} m_1,  m_2 ,e_1 , e_2 \big{|} : (e_1\pm e_2 ) \in \mathbb{Z}\;,\; ( 3m_1 \pm  m_2) \in 6\mathbb{Z}  \; \big{\}}\;,
\end{align}
where the position basis associated to the $\Pi$ are defined as
\begin{align}
&\big{\langle}m_1, m_2, e_1, e_2 \big{ |}:=\big{\langle}m_1, m_2, e_1, e_2 ;\Pi\big{ |}  \nn
\\
&:=  \left\langle \textrm{Re}[\frakX_1] = \frac{\bhbar}2 m_1, \textrm{Re}[\frakX_2]= \frac{\bhbar}2 m_2, \textrm{Re}[\frakP_1] = \frac{\bhbar}2 e_1, \textrm{Re}[\frakP_2] = \frac{\bhbar} 2 e_2 \right|\;.
\end{align}
The action of the quantized operators $(\hat{\frakX}_\pm , \hat{\frakP}_{\pm})$ on the basis is given in eq.~\eqref{(X,P) on charge basis} and the completeness relation is 
\begin{align}
&\langle m_1, m_1,e_1, e_2|m_1',m_2',e_1',e_2'\rangle = \delta(m_1- m_1')\delta(e_1,-e_1')\delta(m_2- m_2')\delta(e_2-e_2')\;, \nn
\\
&
\sum_{(m_1,m_2,e_1,e_2)} |m_1, m_2, e_1 ,e_2 \rangle \langle m_1, m_2, e_1,e_2|=\mathbb{I}\;. \label{completeness relation-2}
\end{align}
In this basis, the matrix element for $\bm{S}$ and $\bm{T}$ 
are computed to be
\begin{align}
\begin{split}
&\langle  m_1,m_2, e_1, e_2|\hat{\bm{S}}(\frakL)|m_1',m_2',e_1',e_2\rangle\; 
\\
&=\delta\left(e_2'+\frac{1}6 m_2\right)\delta\left(e_2-\frac{1}{6} m_2'\right)\delta\left(m_1+2e_1+m_1'+2e_1'-2m_\eta\right)   
\\
&\quad \times \eta^{2e_1'}\left( -q^{\half} \eta\right)^{-e_1-2e_1'-\frac{1}2 m_1'+m_\eta} \CI^C_{\Diamond}\left(-m_1-2e_1+m_\eta, -e_1+m_\eta-\frac{1}2 m_1'-2e_1'\right)\;,  \label{matrix element for S}
\end{split}
\end{align}
and
\begin{align}
\begin{split}
&\langle m_1, m_2, e_1, e_2|\hat{\bm{T}}(\frakL) | m_1',m_2',e_1', e_2'\rangle 
\\
&=\left(-q^{-\frac{1}{2}}\right)^{m_1-2e_1'-2m_1'}  \CI_\Diamond^{C} \left(e_1-3e_2, e_1'-e_2'-\frac{m_1}2 +m_1'+\frac{m_2}6 - \frac{m_2'}3\right) 
\\
 &\times \delta\left(e_1+\frac{m_1'}2\right)\delta \left(e_2+\frac{m_2'}6\right) \CI_\Diamond^{C} \left(e_1+3e_2, e_1'+e_2'-\frac{m_1}2 +m_1'-\frac{m_2}6 + \frac{m_2'}3\right)  \;. \label{matrix element for T}
\end{split}
\end{align}
Here $\frakL:=\frac{\hbar}2 m_\eta + \log \eta$ denote the  central element, see eq.~\eqref{central element for Q_{2,1}}.  
\paragraph{Projectivity of the $SL(2,\mathbb{Z})$ Representation} First, note that
\begin{align}
\begin{split}
&\langle m_1, m_2, e_1, e_2|\bm{S}(\frakL+\hbar ) | m_1',m_2',e_1', e_2'\rangle 
\\
&\qquad=q^{\frac{1}4 (m_1+2e_1')} \left\langle m_1, m_2, e_1-\frac{1}2, e_2|\bm{S}(\frakL ) | m_1'-1,m_2',e_1', e_2'\right\rangle  \;,
\\
&\langle m_1, m_2, e_1, e_2|\bm{T}(\frakL+\hbar ) | m_1',m_2',e_1', e_2'\rangle = \langle m_1, m_2, e_1, e_2|\bm{T}(\frakL ) | m_1',m_2',e_1', e_2'\rangle  \;.
\end{split}
\end{align}
This implies that
\begin{align}
\begin{split}
&e^{-\hat{\bfrakM}} \hat{\bm{S}}(\frakL)\, e^{\hat{ \bfrakM}}  = e^{\frac{1}2 \hat{\frakX}_1}  \hat{\bm{S}}(\frakL) \,  e^{\hat{\frakP}_1} = e^{\frac{1}2 (\sfY_1+\sfY_2)}  \hat{\bm{S}}(\frakL)\, e^{\frac{1}2 (\sfY_3+\sfY_4)}\;,
\\
&e^{-\hat{ \bfrakM}} \hat{\bm{T}}(\frakL)\, e^{\hat{ \bfrakM}}  =   \hat{\bm{T}}(\frakL) \;, \label{S,T under shift of L}
\end{split}
\end{align}
where $ \hat{\bfrakM} $ is defined as 
\begin{align}
\begin{split}
&e^{ -\hat{\bfrakM}}\frakL \, e^{ \hat{\bfrakM}}=\frakL+\hbar\;,\; e^{-\hat{\bfrakM}}\bar{\frakL}\, e^{\hat{\bfrakM}}=\bar{\frakL}\;,
\\
&\Longleftrightarrow e^{- \hat{\bfrakM}} : (m_\eta,  \eta)\rightarrow (m_\eta+1,  \eta  q^{\half} )\;.
\end{split}
\end{align}
Using eqs.~\eqref{ST_expr} and \eqref{S,T under shift of L}, one can compute how $(\sfy_i, e^{\frakL} , e^{ \hat{\bfrakM}})$ operators transformation under the $SL(2,\mathbb{Z})$. For a central element $e^{\frakL}$, we  know that it is invariant under all $SL(2,\mathbb{Z})$ :
\begin{align}
\hat{\varphi} \, e^{\frakL}  = e^{\frakL} \hat{\varphi} \;, \quad \textrm{for all $\varphi \in SL(2,\mathbb{Z})$} \;.
\end{align}
For $\hat{\varphi}=\hat{\bm{T}}\hat{\bm{S}}\hat{\bm{T}}\hat{\bm{S}}\hat{\bm{T}}\hat{\bm{S}}$, one can easily check that%
\begin{align}
\sfy_i\, \hat{\varphi} = \hat{\varphi}\, \sfy_i \;, \quad e^{- \hat{\bfrakM}} \hat{\varphi}= \hat{\varphi} \, e^{ -\hat{\bfrakM}+3 \frakL} \;,  \quad \textrm{at $q=1$} \;.
\end{align}
Actually the above is correct even at $q\neq 1$. It means that $\hat{\bm{T}}\hat{\bm{S}}\hat{\bm{T}}\hat{\bm{S}}\hat{\bm{T}}\hat{\bm{S}}$ is not actually identity operator but it acts as
\begin{align}
\begin{split}
\hat{\bm{T}}\hat{\bm{S}}\hat{\bm{T}}\hat{\bm{S}}\hat{\bm{T}}\hat{\bm{S}}  &:  \hat{\bfrakM}  \;\rightarrow \;   \hat{\bfrakM}-3 \frakL\;, \;\mathfrak{L}\rightarrow \mathfrak{L}\;, \; \sfy_i \rightarrow \sfy_i\;.
\end{split}
\end{align}
Doing similar computation for $\hat{\varphi}=\hat{\bm{S}}\hat{\bm{S}} \hat{\bm{S}} \hat{\bm{S}}$\;,
\begin{align}
\begin{split}
\hat{\bm{S}} \hat{\bm{S}} \hat{\bm{S}} \hat{\bm{S}} &: \hat{\bfrakM}  \;\rightarrow \;   \hat{\bfrakM}- 2 \frakL\;, \;\mathfrak{L}\rightarrow \mathfrak{L}\;, \; \sfy_i \rightarrow \sfy_i\;.
\end{split}
\end{align}
It means that 
\begin{align}
\begin{split}
&\hat{\bm{T}}\hat{\bm{S}}\hat{\bm{T}}\hat{\bm{S}}\hat{\bm{T}}\hat{\bm{S}}  = \textrm{exp}\left(\frac{3}{2\hbar} \mathfrak{L}^2 +\frac{3}{2\tilde{\hbar}} \mathfrak{\bar{L}}^2 \right) = \eta^{3m_\eta}\;,
\\
&\hat{\bm{S}}\hat{\bm{S}}\hat{\bm{S}}\hat{\bm{S}} = \textrm{exp} \left(\frac{2}{2\hbar} \mathfrak{L}^2 +\frac{2}{2\tilde{\hbar}} \mathfrak{\bar{L}}^2 \right) = \eta^{2 m_\eta}\;. \label{projectivity for N=3 simple}
\end{split}
\end{align}
This is compatible with eq.~\eqref{projectivity of MCG reps}.

For $\varphi =\bm{L}\bm{R}=\bm{S}\bm{T}^{-1}\bm{S}^{-1} \bm{T}$  (the corresponding mapping torus is  figure eight knot complement) 
\begin{align}
&\textrm{Tr} (\hat{\bm{S}}\hat{\bm{T}}^{-1} \hat{ \bm{S}}^{-1}\hat{\bm{T}})(m_\eta, \eta) \nn
\\
& =\sum (-q^{\frac{1}{2}})^{-e_1'+m_1+\frac{m_1'}{2}-m_\eta}\eta^{-e_1'- \frac{m_1'}{2}+m_\eta} \nn
\\
&\times \CI_{\Diamond}^{C}\left(e_1-3e_2,-2e_1-e_1'+2e_2-m_1-\frac{m_1'}{2}+2m_\eta\right)   \nn
\\
&\times \CI_{\Diamond}^{C}\left(e_1+3e_2,-2e_1-e_1'-2e_2-m_1-\frac{m_1'}{2}+2m_\eta\right) \nn
\\
&\times \CI_{\Diamond}^{C}\left(2e_1+2e_1' +m_1+m_1'-3m_\eta,-e_1-e_1'\right) \nn
\\
& \times \CI_{\Diamond}^{C} \left(-e_1-\frac{m_1}{2}-\frac{m_1'}{2}-\frac{m_2}{2}+m_\eta, 2e_1+e_1'+m_1+\frac{m_1'}{2}+\frac{m_2}{3}-2m_\eta\right) \nn
\\
&\times \CI_{\Diamond}^{C} \left(-e_1-\frac{m_1}{2}-\frac{m_1'}{2}+\frac{m_2}{2}+m_\eta, 2e_1+e_1'+m_1+\frac{m_1'}{2}-\frac{m_2}{3}-2m_\eta\right)  \nn
\\
&\times \CI_{\Diamond}^{C} \left(-2e_1-m_1+m_\eta, e_1+m_1+\frac{m_1'}{2}-m_\eta\right)\;,
 \label{index for figure eight knot with simple}
\end{align}
where the summation ranges are
\begin{align}
2 e_1, m_1 \in \mathbb{Z}\;,\quad e_2, e_1', \frac{m_1'}{2} \in e_1 +\mathbb{Z}\;, \quad m_2 \in 6\mathbb{Z}+3m_1\;.
\end{align}
From $q$-expansion, we have
\begin{align}
\begin{split}
&\textrm{Tr} (\hat{\bm{S}}\hat{\bm{T}}^{-1} \hat{ \bm{S}}^{-1}\hat{\bm{T}})(m_\eta=0, \eta)  =1+ \left(2\eta+\frac{2}{\eta} \right) q^{\frac{3}{2}}+\left(8+2 \eta^2 + \frac{2}{\eta^2}\right) q^2 
\\
& \qquad\qquad +\left(6 \eta+\frac{6}\eta \right)q^{\frac{5}{2}}+\left(2- 3 \eta^2- \frac{3}{\eta^2}\right)q^3+\ldots  \;,
\\
&\textrm{Tr} (\hat{\bm{S}}\hat{\bm{T}}^{-1} \hat{ \bm{S}}^{-1}\hat{\bm{T}})(m_\eta=1, \eta) =   \left(\frac{1}{\eta^2}+\frac{1}{\eta} + \eta +\eta^2 \right) q+ \left(6 +3\eta +  \frac{3}{\eta}\right) q^2 
\\
&\qquad \qquad+ \left(-6 - \frac{1}{\eta^3}- \frac{3}{\eta^2}-\frac{5}{\eta}-5 \eta-3\eta^2 -\eta^3\right)q^3+\ldots \;. 
\label{STSinvTinv}
\end{split}
\end{align}
These indices exactly match the  indices in eq.~\eqref{N=3 4_1 simple from cluster} obtained using the gluing equations derived from cluster partition function for $k=1$ and  the indices  in eq.~\eqref{N=3 4_1 simple from cluster} obtained by gluing  $T[SU(3)]$ theories up to a framing factor \eqref{projectivity for N=3 simple}.

 %%%%%%%%%%%%%%%%%%%%%%%%%%%%%%%%%%%%%%%%%%%%%%%%%%%%%%%%%
\section{Holonomies and Snakes} \label{sec : snakes}
%%%%%%%%%%%%%%%%%%%%%%%%%%%%%%%%%%%%%%%%%%%%%%%%%%%%%%%%

In this appendix we explain the `snake' rules ,
which will be necessary for the expressing of the holonomies along cycles  of flat $SL(N)$ connections on 2d/3d manifolds in terms of FG coordinates/octahedron's vertex variables hence for the discussion of co-dimension 4 defects in sections \ref{sec.codim_4_state} and \ref{sec.codim_4_cluster}.
For 2d Riemann surface case, the holonomy computation was developed in the original paper by Fock and Goncharov \cite{2003math.....11149F}.
By generalizing the 2d case,  snake rule for 3d is developed in  \cite{Dimofte:2013iv}. Here we review general 3d snake first and then  explain how to apply the 3d snake to holonomy computations in 2d.  

\begin{figure}[htbp]
 \centering
  \includegraphics[width=2.3in]{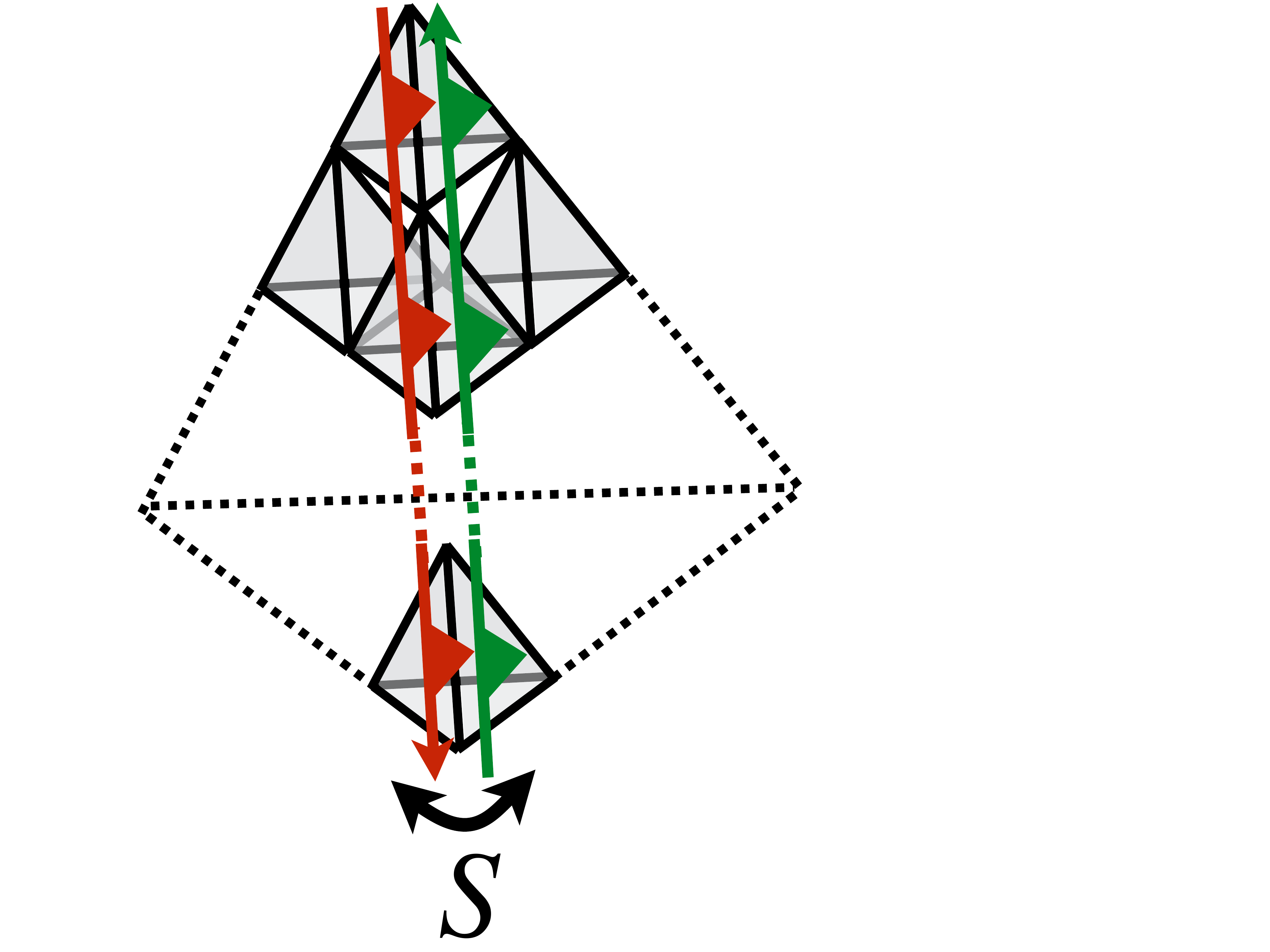}
   \includegraphics[width=2.3in]{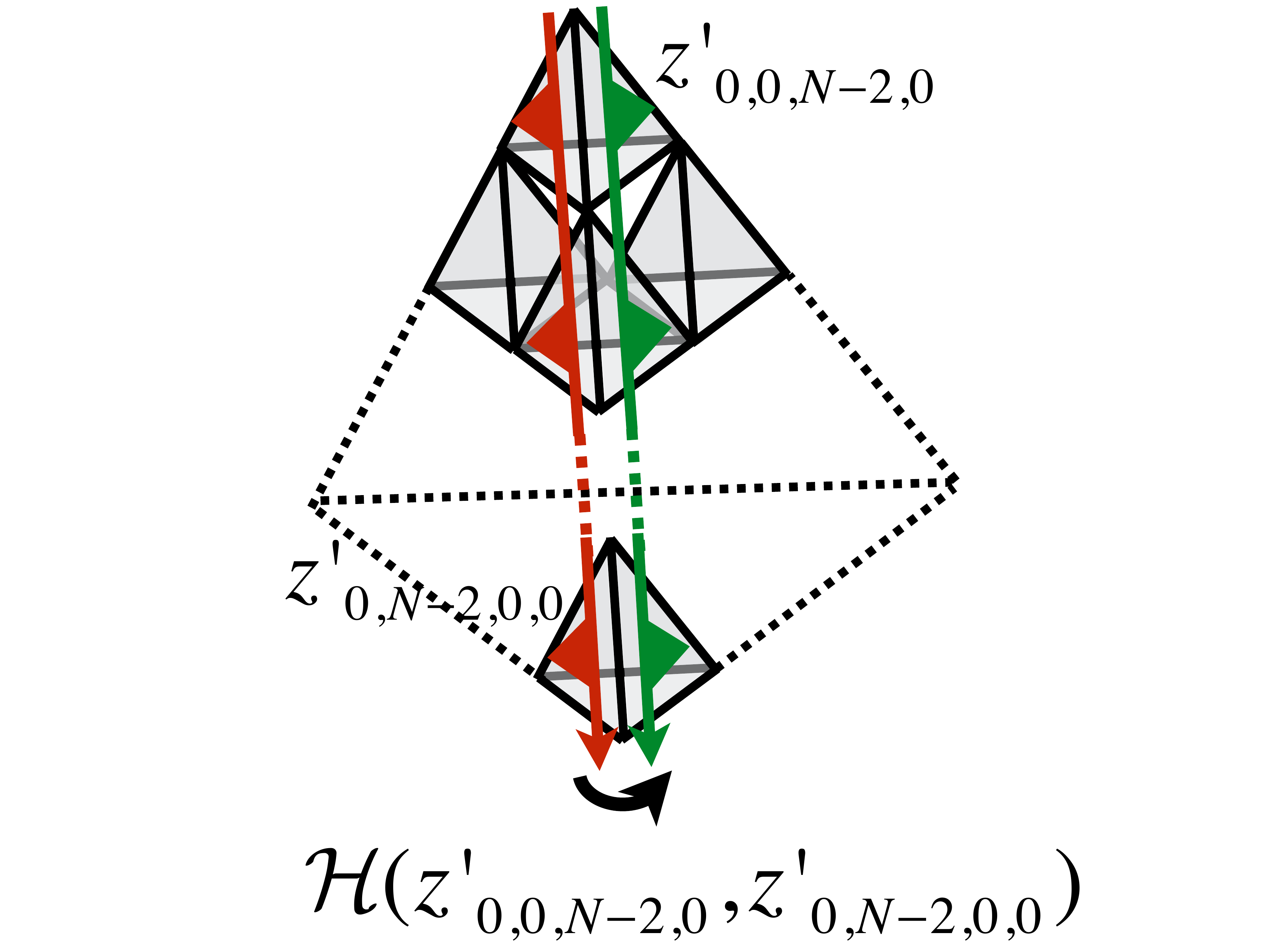}
    \includegraphics[width=2.3in]{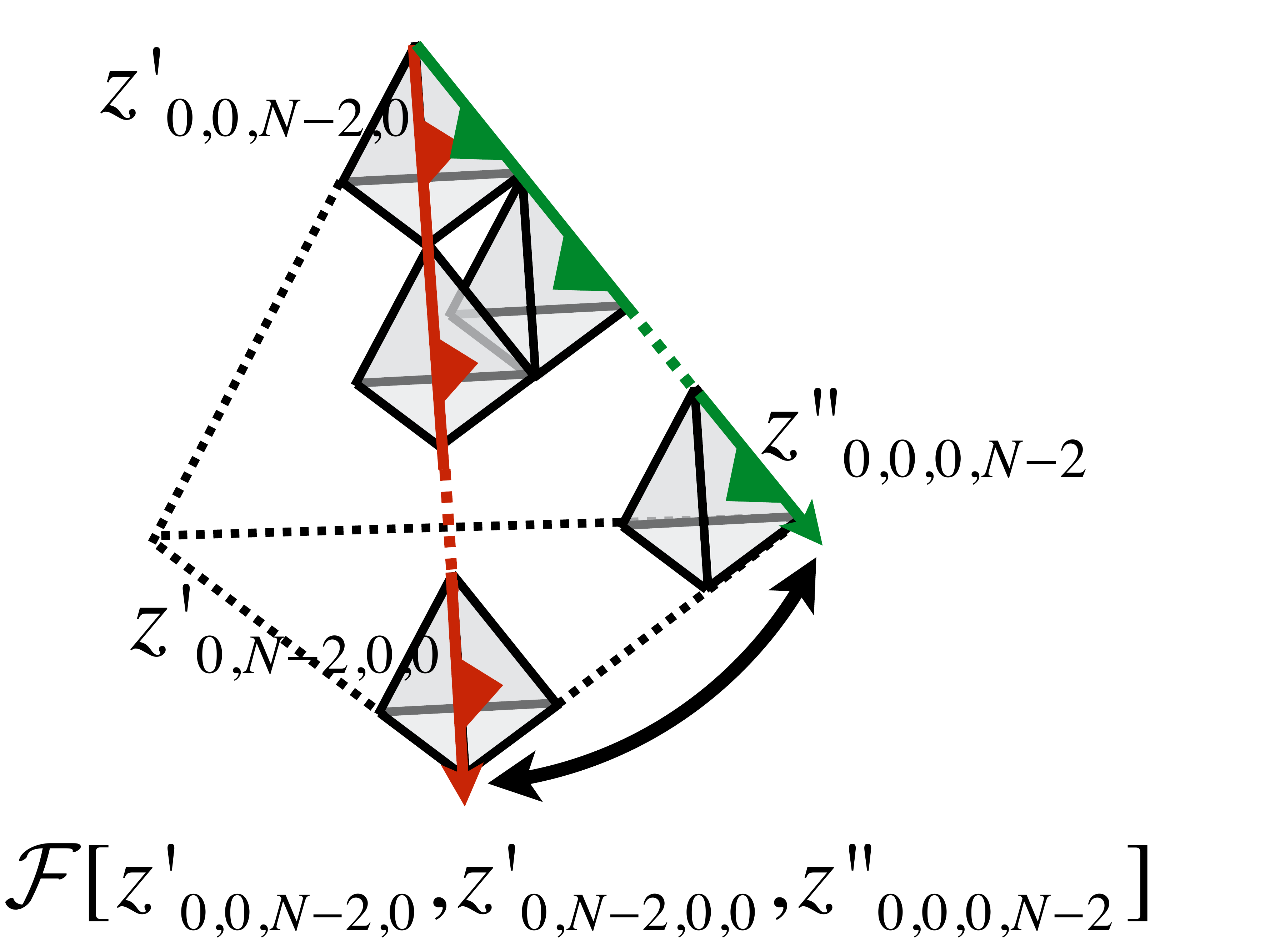}
  \includegraphics[width=2.3in]{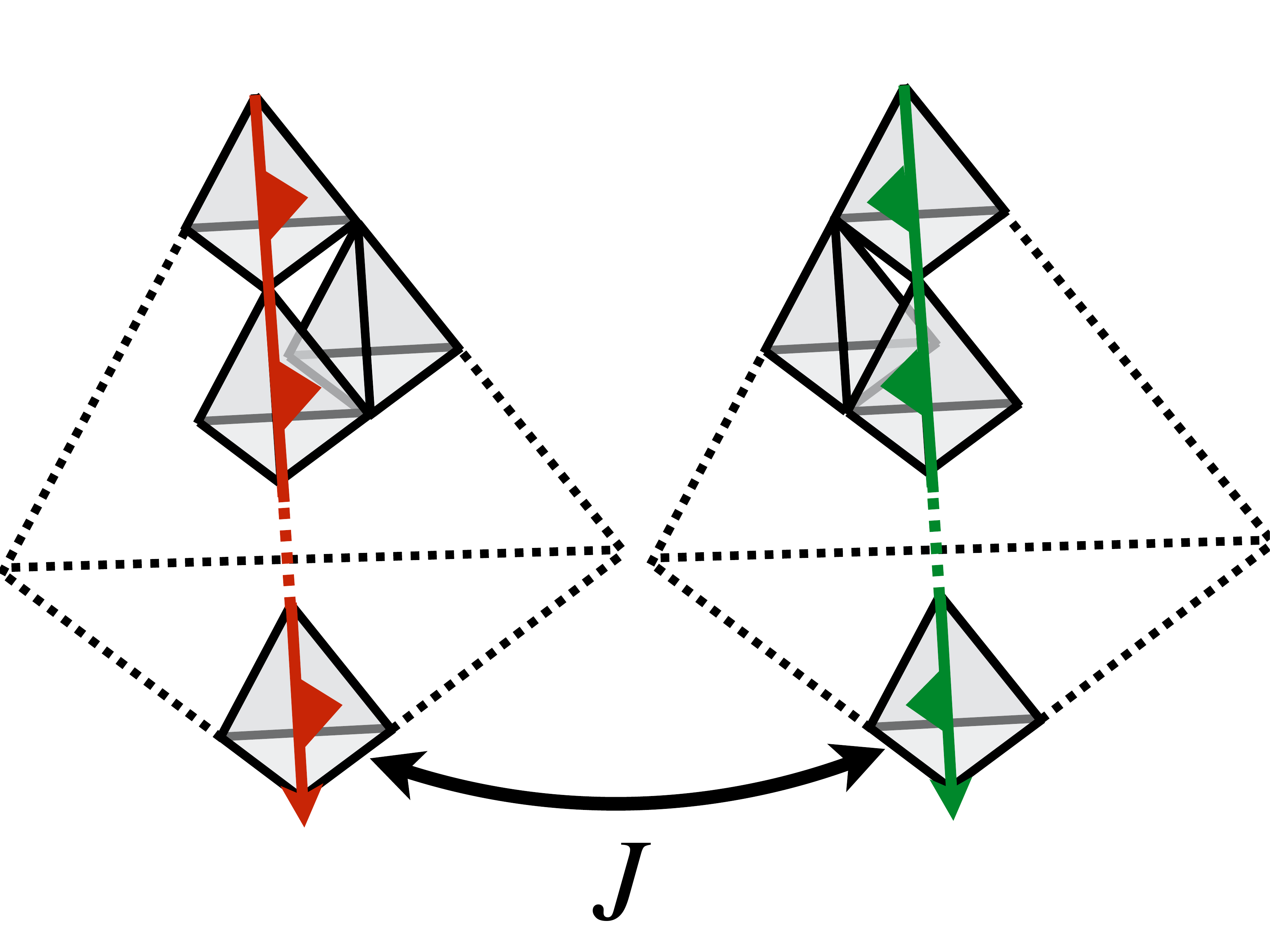}
 \caption{Four basic big 3d snake moves.  $S$: reversing the orientation of snake,  $\CH$: Flipping  fins of snake, $\CF$: moving along the face of ideal tetrahedron, $J$: moving from one ideal tetrahedron to another one.  
 }
 \label{fig_snake-basic}
 \end{figure}

\begin{figure}[htbp]
\begin{center}
   \includegraphics[width=.45\textwidth]{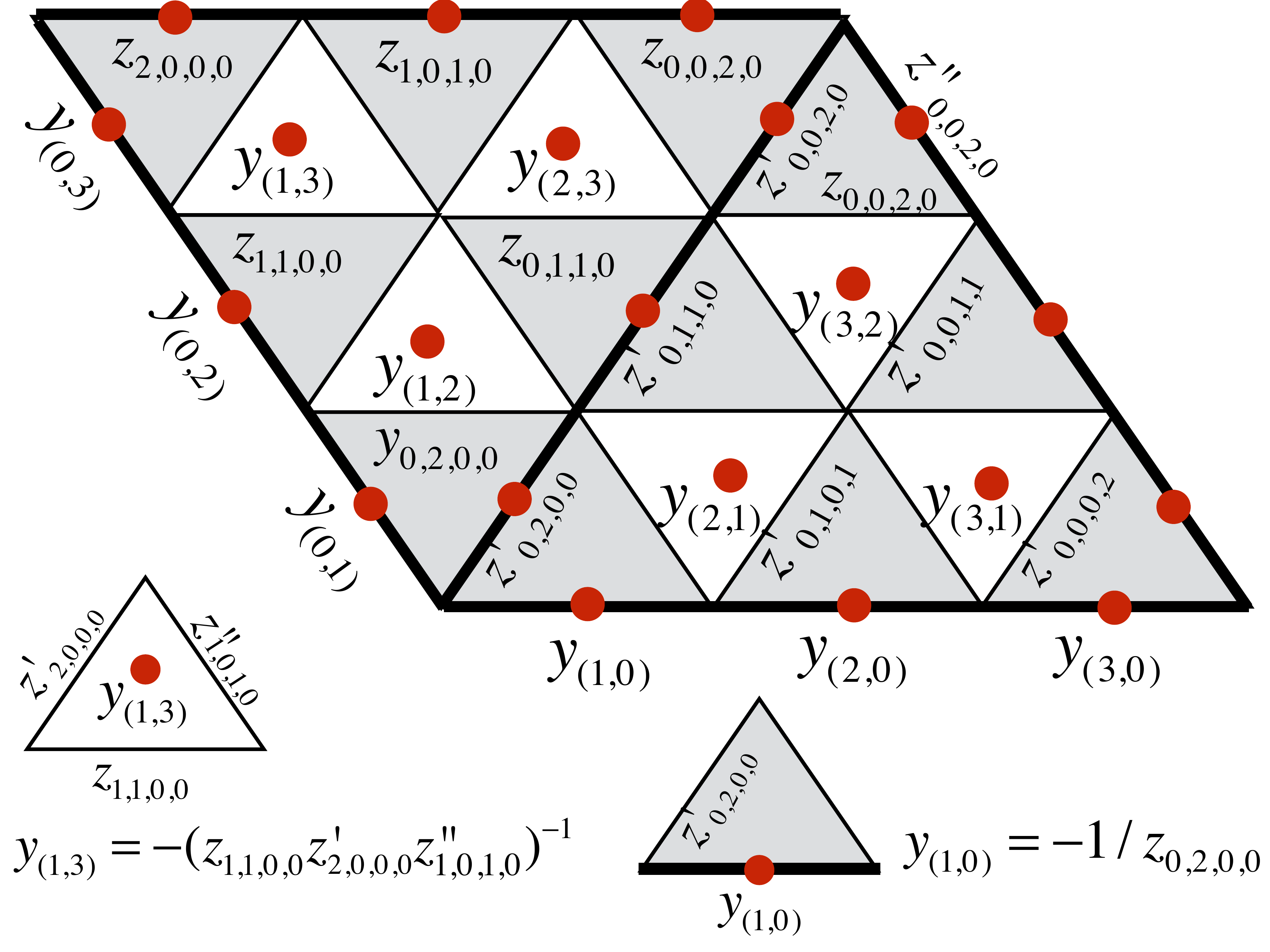}
   \end{center}
   \caption{Relation  between boundary vertex variables $\{z_{a,b,c,d},z'_{a,b,c,d},z''_{a,b,c,d} \}$ (associated to edges of  small gray triangles) of $N$-decomposition with FG coordinates  $y_{(a,b)}:=e^{Y_{(a,b)}}$ (associated to red points) of  triangulated Riemann surface. For a FG coordinate whose node is located on a edge of a triangle, it is identified with inverse of  vertex variable associated to the small edge where the node is located up to sign. For a FG coordinate located on a white small triangle, it is identified (minus of) products of inverse of vertex variables associated to three edges of the triangle. }
    \label{fig:vertex-to-FG}
\end{figure}

\paragraph{Basic Snake Moves}
Let us now summarize 3d snake rule \cite{Dimofte:2013iv}. 
We consider an ideal $N$-triangulation of the 3-manifold $M$, where we replace each tetrahedron in an ideal triangulation by a pyramid of  $\frac{1}6 N(N^2-1)$ small tetrahedrons, see Fig.~\ref{fig_snake-basic}. In the $N$-triangulation, octahedron's vertex variables are associated to edges of small tetrahedrons. To compute the $SL(N)$ holonomy matrices along  cycles $\gamma\in \pi_1(M)$,
we  move the snakes along $\gamma$. 
The segment of the snake is located on the edge, 
and we need to remember which triangle around the edge does the snake segment belongs to.
This is represented by the `fin'.
Any snake move can be decomposed into a sequence of four fundamental snake moves introduced in \cite{Dimofte:2013iv}.
For each fundamental move,  $GL(N)$ matrix is assigned and  the $SL(N)$ holonomy is obtained by multiplying all these matrices along the snake move and dividing it by a proper overall factor. 
Instead of reviewing details of four fundamental moves,  we only introduce four big basic moves (fig.~\ref{fig_snake-basic}) which are  more useful for actual computation. These move can be obtained by a sequence of fundamental moves. 
Three of them are the natural lifts of the 2d snake moves, and consequently they become equivalent to 2d snakes in \cite{2003math.....11149F} after giving a proper  relation between 3d vertex variables and 2d FG coordinates. The new move represent the move of the snake from 
one tetrahedron to another. Following the rules in \cite{Dimofte:2013iv}, $GL(N)$ matrices assigned to each basic move are given by 
 \begin{align}
 \begin{split}
 & S=
  \begin{cases}
    (-1)^{i-1} & \text{if } j=N+1-i\\
    0 & \text{otherwise }
  \end{cases} \;,
 \\
&
 \CH [z'_{0,0,N-2,0}, z'_{0,N-2,0,0}] :=\prod_{i=i}^{N-1} H_i (-z'_{0,i-1,N-1-i,0})\;, \quad
H_i (x):= \textrm{diag} (\overbrace{1,\ldots, 1}^{i}, \overbrace{x,\ldots,x}^{N-i}) \;,
 \\
 &\CF [z'_{0,0,N-2,0},z'_{0,N-2,0,0},z''_{0,0,0,N-2}]\\
& \quad := \prod_{k=1}^{N-1} \Big[ \prod_{i=1}^{k} H_{N-k+i}\left(- z'_{0,N-1-k,k-1-i,i} \right)
 F_{N-k-1+i}  H_{N-k+i}\left(-z_{0,N-1-k,k-i,i-1} \right) 
 \\
 & \qquad \qquad \times \prod_{i=1}^{k-1}  H_{N-i }(-z''_{0,N-k,i-1,k-1-i})  \Big]\;,
 \\
 &J:=\textrm{diag}(1,-1,1,-1,\ldots)\;. 
\end{split}
 \end{align}
In the expressions above, ordering in the products of non-commuting matrices are fixed as
 \begin{align}
 \prod_{i=1}^k A_i := A_k A_{k-1}\ldots A_1\;.
 \end{align}
 Matrix $F_{i}$ is given by:
\begin{eqnarray}
(F_{i})_{k,l}=  \begin{cases}
    1 & \text{if } k=l \textrm{ or }(k,l)=(i+1,i)\\
    0 & \text{otherwise }
  \end{cases} \;.
\label{Fi_def}
\end{eqnarray}
In the expressions above, the ordering in the products of non-commuting matrices are fixed as
 \begin{align}
 \prod_{i=1}^k A_i := A_k A_{k-1}\ldots A_1\;.
 \end{align}
Other $\CH$ and $\CF$ matrices are give in a similar expression with proper change of their arguments. For example, 
\begin{align}
&\CF[z''_{0,0,0,N-2},z''_{0,0,N-2,0},z_{0,N-2,0,0}]  \nn
\\
&= \CF [z'_{0,0,N-2,0},z'_{0,N-2,0,0},z''_{0,0,0,N-2}]\big{|}_{z_{0,a,b,c} \rightarrow z'_{0,c,a,b},z'_{0,a,b,c}\rightarrow z''_{0,c,a,b},z''_{0,a,b,c}\rightarrow z_{0,c,a,b}}\;.
\end{align}
The snake can also be used for computation of $SL(N)$ holonomy on a Riemann surface in terms FG coordinates. 
We consider ideal triangles in a triangulation of the 2d Riemann surface as  faces in the boundary of ideal tetrahedron.
In the computation, Fock-coordinates of the Riemann surface is related to octahedrons' vertex variables of vertices  located  on the boundary faces, see 
Fig.~\ref{fig:vertex-to-FG}.
Explicit relation is
\begin{align}
\begin{split}
&y_{n,0}=-z^{-1}_{0,N-1-n,0,n-1}\;, \quad y_{0,n}=-z''^{-1}_{n-1,N-1-n,0,0}\;, \quad y_{n,n}=-z'^{-1}_{0,N-1-n,n-1,0} \quad (1\leq n\leq N-1) \;,
\\
& y_{N,n} = -z''^{-1}_{0,0,n-1,N-1-n}\;, \quad y_{n,N} = -z^{-1}_{N-1-n,0,n-1,0}\quad (1\leq n\leq N-1) \;,
\\
&y_{a,b}=-z^{-1}_{0,N-1-a,b,a-b-1}z'^{-1}_{0,N-1-a,b-1,a-b}z''^{-1}_{0,N-a,b-1,a-b-1} \quad (0< b <a\leq N-1) \;,
\\
&y_{a,b}= -z^{-1}_{b-a-1,N-b,a-1,0}z'^{-1}_{b-a,N-b-1,a-1,0}z''^{-1}_{b-a-1,N-b-1,a,0} \quad (0<a<b\leq N-1) \;. \label{Explicit vertex to FG}
\end{split}
\end{align}
The relation between 2d and 3d variables is  not one-to-one. 
Nevertheless, we can express the $SL(N)$ holonomies along the boundary faces in terms of 2d FG coordinates because the holonomies depend on only some  combinations of vertex variables, which always can be written as FG coordinates. 
%

\begin{figure}[htbp]
\begin{center}
   \includegraphics[width=.8\textwidth]{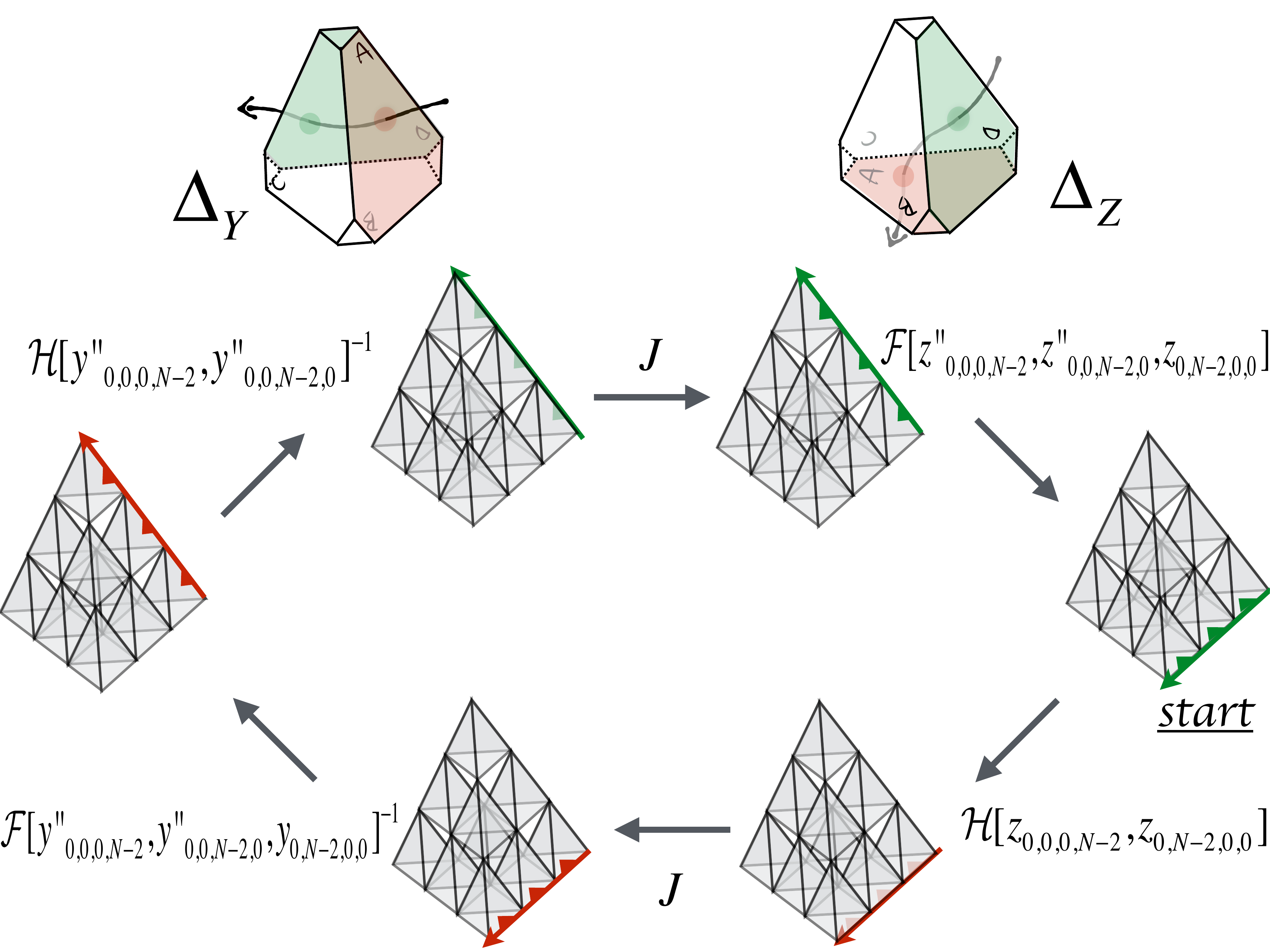}
   \end{center}
   \caption{Snakes' move for one cycle $a$ in figure-eight knot complement with $N=4$. The 3-manifold can be triangulated by two ideal tetrahedra $\Delta_Y$ and  $\Delta_Z$. The one cycle $a$ start from a point inside $\Delta_Z$  then it moves to $\Delta_Y$ through a face $A$ and then it comes back to $\Delta_Z$ through a face $D$.}
    \label{fig:a-snake}
\end{figure}

\begin{figure}[htbp]
\begin{center}
   \includegraphics[width=.5\textwidth]{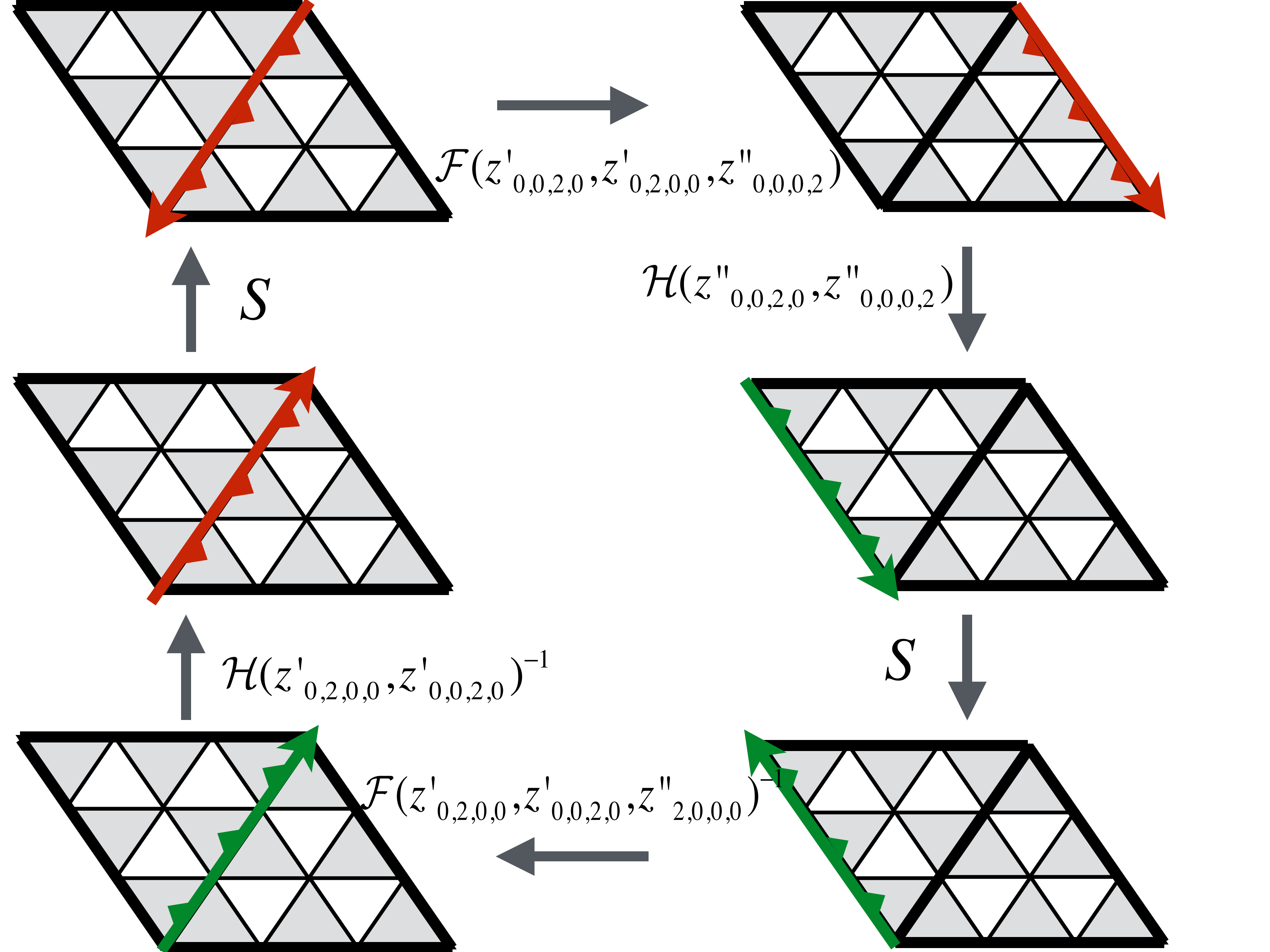}
   \end{center}
   \caption{Snake's move for  $\gamma_x$ on once-punctured torus with $N=4$.}
    \label{2d snake}
\end{figure}

 %%%%%%%%%%%%%%%%%%%%%%%%%%%%%%%%%%%%%%%%%%%%%%
\paragraph{Holonomies for \texorpdfstring{$S^3\backslash \bm{4}_1$}{S3\ 4_1}}
The snake move for $a\in \pi_1 (S^3\backslash \bm{4}_1)$ is given the Fig.~\ref{fig:a-snake}. From the snake move, we can compute the holonomy along the cycle :
\begin{align}
\begin{split}
&\textrm{Hol}(a)=\CF[z''_{0,0,0,N-2},z''_{0,0,N-2,0},z_{0,N-2,0,0}]J \, \CH[y''_{0,0,0,N-2},y''_{0,0,N-2,0}]^{-1}
\\
&\qquad \qquad \times \CF[y''_{0,0,0,N-2},y''_{0,0,N-2,0},y_{0,N-2,0,0}]^{-1}J\, \CH[z_{0,0,N-2,0},z_{0,N-2,0,0}] \;.\label{hol(a)}
\end{split}
\end{align}
Similarly for $b$ and $c$ cycle of $\pi_1 (S^3\backslash \bm{4_1})$, the corresponding $SL(N)$ holonomies are
\begin{align}
\begin{split}
&\textrm{Hol}(b)=J\, \CH[y'_{0,0,0,N-2},y'_{N-2,0,0,0} ] \CF[y_{0,0,0,N-2},y_{0,N-2,0,0},y'_{N-2,0,0,0}]
\\
&\qquad \qquad \times S\, J\,  \CH[z'_{0,N-2,0,0},z'_{0,0,N-2,0}] \CF[z_{0,N-2,0,0},z_{0,0,0,N-2},z'_{0,0,N-2,0}]S\;,
\\
&\textrm{Hol}(c)= S\, \CF[z_{0,N-2,0,0},z_{0,0,0,N-2},z'_{0,0,N-2,0}] S\,  J\,  \CH[ y_{0,0,N-2,0},y_{N-2,0,0,0}]^{-1}  
\\
&\qquad \qquad   \times  \CF[y_{0,0,N-2,0},y_{N-2,0,0,0},y'_{0,N-2,0,0}]^{-1} S\, J\,   \CH[z''_{0,0,0,N-2},z''_{0,0,N-2,0}]^{-1}
\\
&\qquad \qquad \times \CF[z''_{0,0,0,N-2},z''_{0,0,N-2,0},z_{0,N-2,0,0}]^{-1} \;. \label{hol(b,c)}
\end{split}
\end{align}
%
%\begin{figure}[htbp]
%\begin{center}
   %\includegraphics[width=.8\textwidth]{b-snake.pdf}
      %\includegraphics[width=.8\textwidth]{c-snake.pdf}
   %\end{center}
   %\caption{Snakes' move for $b$(above) and $c$(below) in figure-eight knot complement. (\DG{[DG: This figure will be erased later if it's too much.]})  }
    %\label{fig:a-snake}
%\end{figure}
%
\paragraph{Holonomies for $\Sigma_{1,1}$} 
Let $(\gamma_x, \gamma_y)$ be the two cycles in once-punctured torus, see Fig.~\ref{fig:two-cycles in torus}. Snake move for $\gamma_x$ is given in Fig.~\ref{2d snake} .
From the snake move (and similar snake move for $\gamma_y$),  holonomy matrices for $(\gamma_x,\gamma_y)$ are %
\begin{align}
\begin{split}
&\textrm{Hol}(\gamma_x) = S\,  \CH[z'_{0,N-2,0,0}, z'_{0,0,N-2,0}]^{-1} \CF[z'_{0,N-2,0,0},z'_{0,0,N-2,0},z''_{N-2,0,0,0}]^{-1}
\\
&\qquad \qquad \times S\,  \CH[z''_{0,0,N-2,0},z''_{0,0,0,N-2}]\CF[z'_{0,0,N-2,0},z'_{0,N-2,0,0},z''_{0,0,0,N-2}] \;,
\\
&\textrm{Hol}(\gamma_y)= S \,  \CF[z_{0,N-2,0,0},z_{0,0,0,N-2},z'_{0,0,N-2,0}  \CH[z_{N-2,0,0,0},z_{0,0,N-2,0}] 
\\
&\qquad \qquad \times S\,   \CF[z_{0,0,N-2,0},z_{N-2,0,0,0},z'_{0,N-2,0,0}]^{-1}  \CH[z'_{0,0,N-2,0},z'_{0,N-2,0,0}]^{-1} 
\label{hol(gammax,gammy)}
\end{split}
\end{align}
These holonomies can be written as 2d FG coordinate using the relation \eqref{Explicit vertex to FG}.

%%%%%%%%%%%%%%%%%%%%%%%%%%%%%%%%%%%%%%%%%%%%%%%%%%%%%
\section{Proof of eq.~\texorpdfstring{\eqref{XC_lemma}}{}}\label{app.delta_proof}
%%%%%%%%%%%%%%%%%%%%%%%%%%%%%%%%%%%%%%%%%%%%%%%%%%%%%
In this appendix we prove eq.~\eqref{XC_lemma}, by working out the 
action of the operators $\hat{\bfrakM}_{\a}, \hat{C}_I$ and $\hat{\frakL}_{\a}$.
\\
For $\hat{\bfrakM}_\a$, we compute\footnote{In the computation, the dummy integration variable $\{Z''(t)\}$ should be distinguished from the operators $\{\hat{Z}''(t) \}$ acting on $\CH(\partial \Diamond)^{\otimes L}$, $\langle Z| e^{\hat{Z}''} = \langle Z+\hbar|$. We  check eq.~\eqref{position basis} for only holomorphic part and checking anti-holomorphic part can be done in a similar way.}
\begin{align*}
&\langle \bfrakM , C=0| e^{\hat{\bfrakM}_\a} |Z(t)\rangle = \langle\bfrakM , C=0| e^{A^x_\a \cdot \hat{Z}} |Z(t)\rangle \nn
\\
&=
\int \left( \prod dZ'' \right)  e^{ A^x_\a \cdot Z}  e^{- \frac{1}{2\hbar} Z \cdot Z''}  \delta (\textrm{eq.}\eqref{delta-fuctions-to-gluing})= e^{\bfrakM_\a}   \langle \bfrakM , C=0|Z(t)\rangle\;. \nn
\end{align*}
For $\hat{C}_I$,
\begin{align*}
&\langle \bfrakM , C=0| e^{\hat{C}_I} |Z(t)\rangle = \langle \bfrakM , C=0| e^{A^c_I \cdot \hat{Z}+B^{c}_I \cdot \hat{Z}''} |Z(t)\rangle \nn
\\
&=  \langle \bfrakM, C=0| q^{-\frac{1}2 A^c_I \cdot B^c_I}  e^{A^c_I \cdot Z} |Z(t) -2\pi i b^2 B^c_I(t) \rangle 
\nn
\\
&=  
\int \left( \prod_{t=0}^{L-1}dZ'' \right) q^{-\frac{1}2 A^c_I \cdot B^c_I}    e^{A^c_\a \cdot Z}  e^{- \frac{1}{4\pi i b^2} (Z- 2\pi i b^2 B_I^c )\cdot Z''}  \delta (\textrm{eq.}\eqref{delta-fuctions-to-gluing})|_{ Z \rightarrow Z-2\pi i b^2 B^c_I} \nn
\\
&=
\int \left( \prod dZ'' \right)q^{-\frac{1}2 A^c_I \cdot B^c_I}   e^{A^c_I \cdot Z}  e^{- \frac{1}{4\pi i b^2}(Z- 2\pi i b^2 B_I^c )\cdot (Z''+2\pi i b^2 A^c_I)}  \delta (\textrm{eq.}\eqref{delta-fuctions-to-gluing})\big{|}_{\genfrac{}{}{0pt}{}{Z \rightarrow Z- 2\pi i b^2 B^c_I, }{Z''\rightarrow Z''+2\pi i b^2 A^c_I}}  \nn
\\
&=
 \int \left( \prod dZ'' \right) e^{\frac{1}2 A^c_I \cdot Z + \frac{1}{2} B^c_I \cdot Z''}  e^{- \frac{1}{4\pi i b^2} Z \cdot Z''}  \delta (\textrm{eq.}\eqref{delta-fuctions-to-gluing}) \quad  ( \because \eqref{A,B,C,D matrices} )\nn
\\&=
\int \left( \prod dZ'' \right)   e^{- \frac{1}{4\pi i b^2} Z \cdot Z''}  \delta (\textrm{eq.}\eqref{delta-fuctions-to-gluing}) =    \langle \bfrakM, C=0 |Z(t)\rangle\;. \nn
\end{align*}
Finally, for $\hat{\frakL}_{\a}$,
\begin{align*}
&\langle \bfrakM , C=0| e^{\hat{\frakL}_{\a}} |Z(t)\rangle = \langle \bfrakM , C=0| e^{C^p_\a \cdot \hat{Z}+D^{p}_\a \cdot \hat{Z}''} |Z(t)\rangle \nn
\\ 
&=  \langle \bfrakM , C=0| q^{-\frac{1}2 C^p_\a \cdot D^p_\a}  e^{C^p_\a \cdot Z} |Z(t) -2\pi i b^2 D^p_\a(t) \rangle
\nn
\\
&=
  \int \left( \prod_{t=0}^{L-1}dZ'' \right) q^{-\frac{1}2 C^p_\a \cdot D^p_\a}    e^{C^p_\a \cdot Z}  e^{- \frac{1}{4\pi i b^2} (Z- 2\pi i b^2 D_\a^p )\cdot Z''}  \delta (\textrm{eq.}\eqref{delta-fuctions-to-gluing})\big{|}_{ Z \rightarrow Z-2\pi i b^2 D^p_\a} \nn
\\
&= 
 \int \left( \prod dZ'' \right)q^{-\frac{1}2 C^p_\a \cdot D^p_\a}   e^{C^p_\a \cdot Z}  e^{- \frac{1}{4\pi i b^2}(Z- 2\pi i b^2 D_\a^p )\cdot (Z''+2\pi i b^2 C^p_\a)}  \delta (\textrm{eq.}\eqref{delta-fuctions-to-gluing}) \big{|}_{\genfrac{}{}{0pt}{}{Z \rightarrow Z- 2\pi i b^2 D^p_\a,}{Z''\rightarrow Z''+2\pi i b^2 C^p_\a}  }\nn
\\
&= 
\int \left( \prod dZ'' \right) e^{\frac{1}2 C^p_\a \cdot Z + \frac{1}{2} D^p_\a \cdot Z''}  e^{- \frac{1}{4\pi i b^2} Z \cdot Z''}  \delta (\textrm{eq.}\eqref{delta-fuctions-to-gluing})|_{\bfrakM_\a \rightarrow \bfrakM_\a+ \hbar}\qquad ( \because \eqref{A,B,C,D matrices}) \nn
\\&=
\int \left( \prod dZ'' \right)   e^{- \frac{1}{4\pi i b^2} Z \cdot Z''}  \delta (\textrm{eq.} \eqref{delta-fuctions-to-gluing})|_{\bfrakM_\a \rightarrow \bfrakM_\a+ \hbar} =    \langle \bfrakM , C=0 |Z(t)\rangle \big{|}_{\bfrakM_\a \rightarrow \bfrakM_\a+ \hbar}\;.
\end{align*}
This proves eq.~\eqref{XC_lemma}.

%%%%%%%%%%%%%%%%%%%%%%%%%%%%%%%%%%%%%%%%%%%%%%%%%%%%%
\section{Index for \texorpdfstring{$T[SU(3)]$}{T[SU(3)]}}\label{app.TSU3}
%%%%%%%%%%%%%%%%%%%%%%%%%%%%%%%%%%%%%%%%%%%%%%%%%%%%%

In this appendix we present explicit details on the 
index for the $T[SU(3)]$ theory.

Following the prescription in the literature \cite{Kim:2009wb,Imamura:2011su}, it is straightforward to
write down the index for $T[SU(3)]$ theory (see the main text for notations):\footnote{Notice however there are subtleties in the choice of the sign dependent on the monopole charges for flavor/gauge symmetries \cite[Appendix A]{Aharony:2013dha}. 
For consistency
it is extremely important to take into account appropriate signs.}
\begin{align}
\begin{split}
&\CI_{T[SU(3)]} (m_1, m_2,w_1, w_2 |n_1, n_2, v_1, v_2; m_\eta, \bareta) 
\\
&\quad\quad =\sum_{\sigma, (s_1, s_2)}\oint \frac{d\zeta}{2\pi i \zeta} \frac{d z_1}{2\pi i  z_1}\frac{d z_2}{2\pi  i z_2} \Delta_2 \times   \CI^{\textrm{CS}} \times  \CI_0  \times \textrm{PE} (f^{\textrm{single}})\;.
\label{TSU3_index}
\end{split}
\end{align}
Let us discuss each of these factors in turn.
First, $\Delta_2$ denote the measure coming from an $\CN=2$ $U(2)$ vector multiplet
\begin{align}
\Delta_2  (z_1, z_2, s_1, s_2 ;q):= \frac{1}{\textrm{sym}(s_1,s_2)}q^{-\frac{|s_1-s_2|}{2}} \left(1- \frac{z_1}{z_2} q^{\frac{|s_1- s_2|}{2}}\right)\left(1- \frac{z_2}{z_1} q^{\frac{|s_1- s_2|}{2}}\right)\;.
\end{align}
where the symmetric factor is
\begin{align}
\textrm{sym}(s_1, s_2): = \begin{cases}
2 & \; (s_1= s_2) \\
1 & \;(s_1\neq s_2)
\end{cases} \;.
\end{align}
The factor $I^{\rm CS}$ denotes the classical contribution from the mixed CS terms
\begin{align}
\begin{split}
\CI^{\rm CS} &=\left( \frac{w_1}{w_2}(-1)^{m_1-m_2}\right)^\sigma \left( \zeta(-1)^{\sigma}\right)^{m_1- m_2} \\
& \qquad \times  \left(\frac{1}{w_1^2 w_2} (-1)^{-2m_1-m_2}\right)^{s_1+s_2} \big{(}z_1 z_2 (-1)^{s_1+s_2}\big{)}^{-2m_1 - m_2}\;.
\end{split}
\end{align}
Here we choose a particular linear combination of two topological $U(1)_J$ symmetries, such  that the fugacities $(w_1, w_2)$ are conjugate to  $\textrm{diag}(H_1, H_2, -H_1- H_2) \in \mathfrak{su}(3)_{\rm top}$.
The expression $f^{\textrm{single}}$ denote the single particle index from $\mathcal{N}=2$ chiral multiplets (we have $v_3=1/(v_1 v_2), n_3=-n_1-n_2$ since we have $SU(3)$ symmetry)
\begin{align}
\begin{split}
f^{\textrm{single}}&= \frac{q^{\frac{1}{4}}}{1-q} \sum_{i=1,2} \left( q^{\frac{|m_\eta/2 + \sigma - s_i|}{2}} \bareta^{\frac{1}{2}} \frac{\zeta}{z_i}  +  q^{\frac{|m_\eta/2 - \sigma + s_i|}{2}} \bareta^{\frac{1}{2}} \frac{z_i}{\zeta}  \right) 
\\
&+  \frac{q^{\frac{1}{4}}}{1-q}  \sum_{i=1,2 : j=1,2,3} \left(  q^{\frac{|m_\eta/2 +s_i - n_j|}{2}} \bareta^{\frac{1}{2}} \frac{z_i}{v_j}+ q^{\frac{|m_\eta/2 -s_i + n_j|}{2}} \bareta^{\frac{1}{2}} \frac{v_j}{z_i} \right)  
\\
&-  \frac{q^{\frac{3}{4}}}{1-q} \sum_{i=1,2} \left( q^{\frac{|m_\eta/2 + \sigma - s_i|}{2}} \bareta^{-\frac{1}{2}} \frac{z_i}{\zeta}   +  q^{\frac{|m_\eta/2 - \sigma + s_i|}{2}} \bareta^{-\frac{1}{2}} \frac{\zeta}{z_i}   \right) 
\\
&-  \frac{q^{\frac{3}{4}}}{1-q}  \sum_{i=1,2 : j=1,2,3} \left(  q^{\frac{|m_\eta/2 +s_i - n_j|}{2}} \bareta^{-\frac{1}{2}} \frac{v_j}{z_i}+ q^{\frac{|m_\eta/2 -s_i + n_j|}{2}} \bareta^{-\frac{1}{2}} \frac{z_i}{v_j} \right)  
\\
&+ \frac{q^{\frac{1}{2}}}{1-q} \left(3 q^{\frac{| m_\eta|}{2}} \bareta^{-1} + q^{\frac{|-m_\eta + s_1-s_2|}{2}} \bareta^{-1} \frac{z_1}{z_2} +  q^{\frac{|-m_\eta -s_1+s_2|}{2}}\bareta^{-1} \frac{z_2}{z_1} \right) 
\\
&-  \frac{q^{\frac{1}{2}}}{1-q} \left(3 q^{\frac{|m_\eta|}{2}} \bareta +  q^{\frac{|m_\eta -s_1+s_2|}{2}} \bareta \frac{z_2}{z_1} +q^{\frac{|m_\eta +s_1-s_2|}{2}} \bareta\frac{z_1}{z_2} \right)\;.
\end{split}
\end{align}
The first two terms are from scalars  in the $\CN=4$ bi-fundamental hyper multiplets, the next two terms from fermions in the $\CN=4$ hyper multiplets, and final two terms from the $\CN=2$ adjoint chiral multiplet in the $\CN=4$ vector multiplet. PE denote the plethystic exponential,
\begin{align}
\textrm{PE}[f(\cdot)] := \exp \left( \sum_{n=1}^{\infty}\frac{1}{n} f(\cdot^n) \right)\;.
\end{align}
For example,
\begin{align}
\textrm{PE} \left(\pm \frac{q^{\frac{|m_\eta/2 + \sigma- s |}{2}} \bareta^{\frac{1}{2}} \frac{z}{\zeta}}{1-q}\right) = \prod_{\a=0}^{\infty}\left(1-q^{\frac{|m_\eta/2 +  \sigma-s|}{2}+\a} \bareta^{\frac{1}{2}} \frac{z}{\zeta}\right)^{\mp 1}\;.
\end{align}
$I^{(0)}$ is contributions for zero-point shift and sign factors
\begin{align}
\begin{split}
I^{(0)} & = q^{\epsilon_0} \big{(}\zeta(-1)^{\sigma}\big{)}^{\zeta_0 } \big{(}z_1 (-1)^{s_1}\big{)}^{(z_1)_0}  \big{(}z_2 (-1)^{s_2}\big{)}^{(z_2)_0}   
\\
&\quad \times \big{(}v_1 (-1)^{n_1}\big{)}^{(v_1)_0}  \big{(}v_2 (-1)^{n_2}\big{)}^{(v_2)_0}  \big{(}v_3 (-1)^{n_3}\big{)}^{(v_3)_0} (\bareta (-1)^{m_\eta})^{\eta_0}\;.
\end{split}
\end{align}
These zero point shifts can be obtained from single particle index: 
\begin{align}
\begin{split}
&\frac{1}2 \partial_q f^{\textrm{single}}\
\overset{\zeta=1, z_i=1, v_i=1, \bareta=1, q\rightarrow 1}{\xrightarrow{\hspace*{2.5cm}}} \epsilon_0 + \mathcal{O}(q-1)  \;,
\\
&\frac{1}2 \partial_\zeta f^{\textrm{single}}
\overset{\zeta=1, z_i=1, v_i=1, \bareta=1, q\rightarrow 1}{\xrightarrow{\hspace*{2.5cm}}}
\frac{(\textrm{constant})}{q-1} + \textrm{(constant)}+\zeta_0 + \mathcal{O}(q-1) \;,
\\
&\frac{1}2\partial_{z_1} f^{\textrm{single}}
\overset{\zeta=1, z_i=1, v_i=1, \bareta=1, q\rightarrow 1}{\xrightarrow{\hspace*{2.5cm}}}
\frac{\textrm{(constant)}}{q-1} + \textrm{(constant)}+(z_1)_0+ \mathcal{O}(q-1)  \;,
\\
& \quad \quad \quad \vdots 
\end{split}
\end{align}
where $\textrm{constant}$ refers to numerical constants independent of fugacities and 
magnetic fluxes (and hence is only an overall constant factor for the index).
For example,
\begin{align}
\begin{split}
\epsilon_0 = &\frac{1}{8}\left(
     \sum_{\genfrac{}{}{0pt}{}{i=1,2,3,}{j=1,2}} |\frac{m_\eta} 2 +n_i -s_j| + |\frac{m_\eta}2 -n_i+s_j|  \right.  \\
& \qquad \qquad\qquad    \left.  + \sum_{i=1,2} |\frac{m_\eta}2 +s_i-\sigma| +   |\frac{m_\eta}2-s_i+ \sigma | 
    \right) \;. 
\end{split}
\end{align}
Note that thanks to 3d $\mathcal{N}=4$ mirror symmetry (exchanging for example
$SU(3)_{\rm top}$ and $SU(3)_{\rm bot}$), we have
\begin{align}
\begin{split}
&\CI_{T[SU(3)]}(m_1,m_2, w_1, w_2|n_1, n_2, v_1, v_2; m_\eta, \bareta) \\
&\quad = \CI_{T[SU(3)]}(n_1, n_2, v_1, v_2|m_1,m_2, w_1, w_2; -m_\eta, \bareta^{-1})\; .
\end{split}
\end{align}
Finally, in eq.~\eqref{TSU3_index}
the range of summation is  over $\sigma, s_2\geq s_1$ satisfying
\begin{align}
s_2,  s_1  \in \mathbb{Z} +\frac{m_\eta}2 +n_1\;, \quad \sigma \in \mathbb{Z}+n_1 \;,
\end{align}
and the quantization conditions of $(n_1, n_2,m_1,m_2, m_\eta)$ are given by
\begin{align}
m_i, n_i \in \mathbb{Z}/3\;, \quad n_i -n_j,  \;m_i-m_j,\; n_i-m_j \in \mathbb{Z}\;, \quad m_\eta \in \mathbb{Z}\;.
\end{align}
These summation range/quantization for monopole fluxes are fixed by Dirac quantization condition.

%%%%%%%%%%%%%%%%%%%%%%%%%%%%%%%%%%
\section{Derivation of eq.~\texorpdfstring{\eqref{Z_TSUN_largeN}}{Large N Limit of T[SU(N), varphi] Theory}}\label{app.deriv_TSUN}

In this appendix we present a derivation of eq.~\eqref{Z_TSUN_largeN}, concerning
the large $N$ behavior of the $S^3_{b=1}$ partition function of the domain wall theory for 
$\varphi=\bm{ST}^k$.

Let us start with eq.~\eqref{I_wish_evaluate}.
Since the integral is Gaussian, we can easily evaluate it
in terms of determinants:
\begin{align}
\begin{split}
&Z^{\textrm{simple}}_N (\varphi):=Z_{S^3_{b=1}}\left[T_N[(\Sigma_{1,1}\times S^1)_\varphi, \textrm{simple}]\right]
\\
&= \frac{1}{N!} \sum_{\sigma\in \mathfrak{S}_N} (-1)^{\sigma} \int d\vec{\mu} \,\, e^{2\pi i \left( \frac{k \vec{\mu}^2}{2} + \vec{\mu}\cdot \sigma(\vec{\mu}) \right)}
=(\pi i)^{-\frac{N}{2}} \frac{1}{N!}\sum_{\sigma\in \mathfrak{S}_N} (-1)^{\sigma} \frac{1}{\sqrt{\det M_{k,\sigma}}} \ ,
\end{split}
\end{align}
where we defined a $N\times N$ matrix $(M_{k, \sigma})_{i,j}:= k \delta_{i,j}+ \delta_{i, \sigma(j)}+\delta_{\sigma(i),j}$.

The summation involves $N!$ terms, but can be 
simplified by noting that the summand depends only on the 
conjugacy class of the permutation $\sigma$.
We can then represent $\sigma$ into a sum over the conjugacy classes,
which in practice can be given by a product of cyclic permutations:
module the relabeling of the indices (which keep our expression invariant), 
we have a product of cyclic permutations, labeled by a partition $\lambda=[n_1, \ldots, n_s]$:
\begin{align}
\sigma=\sigma_{\lambda}:=(1, \ldots, n_1)\,\, (n_1+1, \ldots, n_2) \,\, \left(\sum_{i=1}^{s-1} n_i+1 , \ldots,  \sum_{i=1}^{s} n_i\right) \;.
\label{sigma_lambda}
\end{align}
Note that the number of $i$-cycles in $\sigma_i$ is given by the number of time the $i$ appears in $n_1, \ldots, n_s$;
we denote this number by $N_i$, and hence $\lambda$ is given by
\begin{align}
\lambda= [\underbrace{N, \ldots, N}_{N_N}, \ldots, \underbrace{1\ldots, 1}_{N_1}] \;,
\end{align}
with $\sum_{i=1}^N i N_i=N$.
Then, the matrix $M_{k,\sigma}$ can be written into a 
product of the following matrix: $(m_{k, \sigma})_{1\le i,j\le n}= k \delta_{i,j}+ \delta_{i, j-1}+ \delta_{i,j+1}$ (with identification $i\sim i+n$).
The determinant of $m_{k,\sigma}$ can be evaluated by the formula for the circulant
determinant:
\begin{align}
\det (m_{k,\sigma})=\left(c_k^{n}+(-1)^{n-1} c_k^{-n} \right)^2\;, 
\label{det_m}
\end{align}
where the $c_k$ ($k\geq 3$) is determined by 
\begin{align}
c_k^2+c_k^{-2}:=k\;\quad \textrm{and } |c_k|>1\;,
\end{align}
and more concretely,
\begin{align}
c_k^2=\frac{k+\sqrt{k^2-4}}{2}
\label{c_k}
\end{align}
for $k\ge 3$. We can easily check that this is the largest eigenvalues of the $2\times 2$ matrix $\bm{ST}^k$.
From eq.~\eqref{det_m} we obtain
\begin{align}
\det (M_{k,\sigma_{\lambda}})
=\prod_{i=1}^N \left(c_k^{n_i}+(-1)^{n_i-1}c_{k}^{-n_i} \right)^2 \;.
\end{align}
We therefore have
\begin{align}
Z^{\textrm{simple}}_N (\varphi = \bm{ST^k}) = (\pi i)^{-\frac{N}{2}}\frac{1}{N!}\sum_{\lambda}  \frac{N!}{\prod_j N_j! j^{N_j}}  \prod_{j=1}^N \frac{(-1)^{n_j-1}}{c_k^{n_j}+(-1)^{n_j-1}c_k^{-n_j}} \;,
\end{align}
which can be rewritten as
\begin{align}
\begin{split}
&(\pi i)^{-\frac{N}{2}}\frac{1}{N!}\sum_{\lambda}  \frac{N!}{\prod_j N_j! j^{N_j}}  \prod_{j=1}^N   \frac{(-1)^{(j-1)N_j}}{(c_k^{j}+(-1)^{j-1}c_k^{-j})^{N_j}}  \\
&\qquad= (\pi i)^{-\frac{N}{2}}(-1)^{N}\sum_{\lambda}  \frac{1}{\prod_j N_j! j^{N_j}} \prod_{j=1}^N   \frac{(-1)^{N_j}}{(c_k^{j}+(-1)^{j-1}c_k^{-j})^{N_j}}  \;.
\end{split}
\end{align}
The useful trick to sum this up, as is familiar from statistical mechanics, is to consider the grand canonical partition function. Introducing chemical potential $\mu$ for $N$, we obtain
\begin{align}
G(\mu) &:=\sum_{N=1}^{\infty} \mu^N  Z^{\textrm{simple}}_N   \nonumber
\\
&= \sum_{N=1}^{\infty} \left(  \sum_{N_1=0}^{\infty} \sum_{N_2=0}^{\infty} \ldots  \right) \delta \left(\sum_{j=1}^\infty j N_j-N \right)  \nn\\
& \qquad \qquad \qquad \qquad \times
 \left(- \frac{\mu}{\sqrt{\pi i}}\right)^N  \prod_{j=1}^{\infty}  \frac{(-1)^{N_j}}{ N_j! j^{N_j}(c_k^{j}+(-1)^{j-1}c_k^{-j})^{N_j}} \nonumber
\\
&=  \sum_{\{N_j\}=0}^{\infty}  \prod_{j=1}^{\infty}  \frac{(-1)^{(j-1)N_j} \left(\frac{\mu}{\sqrt{\pi i}}\right)^{jN_j}}{ N_j! j^{N_j}(c_k^{j}+(-1)^{j-1}c_k^{-j})^{N_j}}  
= \prod_{j=1}^\infty   \sum_{N_j=1}^{\infty}  \frac{(-1)^{(j-1)N_j}\left(\frac{\mu}{\sqrt{\pi i}}\right)^{jN_j}}{ N_j! j^{N_j}(c_k^{j}+(-1)^{j-1}c_k^{-j})^{N_j}} \nonumber
\\
&= \prod_{j=1}^\infty \exp\left(\frac{ \left( \frac{\mu}{\sqrt{\pi i}}\right)^j  }{j(c_k^{-j}+(-1)^{j-1} c_k^{j})} \right) =\exp \left(\sum_{j=0}^\infty  \frac{ \left( \frac{\mu}{\sqrt{\pi i}}\right)^j }{j(c_k^{-j}-(- c_k)^{j})} \right) \nonumber
\\
&= \prod_{n=0}^\infty \left( 1-(-1)^{n+1} c_k^{-2n-1} \frac{\mu}{\sqrt{\pi i}} \right) = \left(-c_k^{-1}\frac{\mu}{\sqrt{\pi i}} ; -c_k^{-2}\right)_\infty
 \nonumber
\\
&= \sum_{N=1}^\infty \frac{\left( \frac{\mu}{\sqrt{\pi i}} 
\right)^N}{c_k^{N^2} (-c_k^{-2};-c_k^{-2})_{N}}\;.
\end{align}
where the $q$-Pochhammer symbol $(a;q)_n$ was defined in eq.~\eqref{q_Pochhammer}.
When expanded in powers of $\mu$, we obtain eq.~\eqref{Z_TSUN_largeN}.

%%%%%%%%%%%%%%%%%%%%%%%%%%%%%%%%%%%%%%%%%%%%%%%%%
\section{Verification of eq.~\texorpdfstring{\eqref{codim4_action_general}}{} in Chern-Simons Theory}\label{app.schur}

In this section, we generalize the computation of the VEV of the Wilson line \eqref{Ak_comp}
to a more general representation.

The VEV of the Wilson line in the representation $R$ of $SU(N)$ is given 
as a specialization of the character of the representation, namely the Schur function
$s_{\lambda}(x_1, \ldots, x_{N})$:
\begin{align}
W^{(\textrm{conj})}_0 (\gamma, R)&:=\textrm{Tr}_{R} \, P  e^{- \oint_\gamma \mathcal{A^{\textrm{conj}}}}
=s_{\lambda} \left(x_i=q^{\frac{N+1}{2}-i} \right)  \;,
\end{align}
where $\lambda$ is the partition associated with the representation $R$,
and we defined $q:=e^{\ell^*_{\mathbb{C}}(\gamma)}$. 

We can now appeal to two formulas. One is \cite[Section I.3, Example 1]{MacdonaldBook}
\begin{align}
s_{\lambda}\left(x_i=q^{-(i-1)}\right)= q^{-n(\lambda)}
\prod_{(i,j)\in \lambda}
\frac{1-q^{-(N+j-i)} } 
 {1-q^{-h_{i,j}}} \;,
\end{align}
where $n(\lambda):=\sum_i (i-1) \lambda_i=\sum_i \binom{(\lambda^t)_i}{2}$, and the hook length
is given by $h_{i,j}=\lambda_i+\lambda_j^t-i-j+1$.
Another is the homogeneity of the Schur function:
\begin{align}
s_{\lambda}(c x_i)=c^{|\lambda|} s_{\lambda}(x_i) \;.
\end{align}
From these equations, we learn that
\begin{align}
s_{\lambda} \left(x_i=q^{\frac{N+1}{2}-i} \right) 
=q^{(\frac{N-1}{2}) |\lambda| -n(\lambda)}
\prod_{(i,j)\in \lambda}
\frac{1-q^{-(N+j-i)} }
 {1-q^{-h_{i,j}}}
 \;.
 \label{hook_result}
\end{align}%
In the large $N$ limit the product in eq.~\eqref{hook_result} can be neglected,
and the  leading order is given by
\begin{align}
\langle W_{\lambda} \rangle \sim \exp\left[ 
      \left( \frac{N}{2} |\lambda| - n(\lambda)  \right) 
      \ell^*_{\mathbb{C}}(\gamma)
\right] \;.
\end{align}
As an example, consider the representation \eqref{lambda_general}.
In this case, $|\lambda|=\sum_{i=1}^N K_i$ and $n(\lambda)=\sum_i \binom{K_i}{2}$, and hence
\begin{align}
\langle W_{\lambda=\eqref{lambda_general}} \rangle \sim
\exp\left[\sum_{i=1}^N \left( \frac{N}{2} K_i - \binom{K_i}{2} \right) \ell^*_{\mathbb{C}}(\gamma)\right]\;,
\end{align}
which is consistent with eq.~\eqref{codim4_action_general} in the large $N$ limit ($K_i \sim O(N)$).

%%%%%%%%%%%%%%%%%%%%%  bibtex  %%%%%%%%%%%%%%%%%%%%%%%%%%%%%
\bibliographystyle{nb}
\bibliography{w3c}

%%%%%%%%%%%%%%%%%%%%%%%%%%%%%%%%%%%%%%%%%%%%%%%%%%%%%%%
\end{document}